\documentclass[runningheads,natbib]{svjour2}
\bibpunct{(}{)}{,}{a}{}{,} 
\smartqed  
\usepackage{graphicx}
\usepackage{mathptmx}      
\newcommand{\msun}{\mbox{$M_{\odot}$}}
\newcommand{\lsun}{\mbox{$L_{\odot}$}}
\newcommand{\rsun}{\mbox{$R_{\odot}$}}
\newcommand{\Zsun}{\mbox{$Z_{\odot}$}}
\newcommand{\Teff}{\mbox{$T_{\rm eff}$}}
\newcommand{\vinf}{\mbox{$v_{\infty}$}}
\newcommand{\vinfeight}{\mbox{$v_{\infty,8}$}}
\newcommand{\Cinf}{\mbox{$C_{\infty}$}}
\newcommand{\mdot}{\mbox{$\dot{M}$}}

\newcommand{\ratio}{\mbox{$v_{\infty}$/$v_{\rm esc}$}}

\newcommand{\msunyr}{\mbox{$M_{\odot} {\rm yr}^{-1}$}}
\newcommand{\Mdu}{\mbox{$\cdot 10^{-6}\,M_{\odot} {\rm yr}^{-1}$}}
\newcommand{\mdu}{\mbox{$10^{-6}\,M_{\odot} {\rm yr}^{-1}$}}

\newcommand{\beq}{\begin{equation}}
\newcommand{\eeq}{\end{equation}}
\newcommand{\beqa}{\begin{eqnarray}}
\newcommand{\eeqa}{\end{eqnarray}}

\newcommand{\ii}{{\rm i}}

\newcommand{\eu}{\mbox{${\rm e}$}}

\newcommand{\kms}{\mbox{${\rm km}\,{\rm s}^{-1}$}}

\newcommand{\half}{\mbox{$\frac{1}{2}$}}

\newcommand{\rarrow}{\rightarrow}
\newcommand{\dd}{{\rm d}}

\newcommand{\HI}  {H\,{\sc i}}
\newcommand{\HII} {H\,{\sc ii}}
\newcommand{\HeI} {He\,{\sc i}}
\newcommand{\HeII}{He\,{\sc ii}}
\newcommand{\HeIII}{He\,{\sc iii}}
\newcommand{\NIII}{N\,{\sc iii}}
\newcommand{\NIV}{N\,{\sc iv}}
\newcommand{\NV}{N\,{\sc v}}
\newcommand{\CIII}{C\,{\sc iii}}
\newcommand{\CIV}{C\,{\sc iv}}
\newcommand{\OVI}{O\,{\sc vi}}
\newcommand{\PV}{P\,{\sc v}}
\newcommand{\FeII}{Fe\,{\sc ii}}
\newcommand{\FeIII}{Fe\,{\sc iii}}
\newcommand{\FeIV}{Fe\,{\sc iv}}

\newcommand{\Ha} {H$_{\rm \alpha}$}
\newcommand{\Lya}{L$_{\rm \alpha}$}
\newcommand{\Bra}{Br$_{\rm \alpha}$}

\newcommand{\logLL}{\mbox{$\log (L/L_{\odot})$}}

\newcommand{\Rstar}{\mbox{$R_{\ast}$}}

\newcommand{\Mstar}{\mbox{$M_{\ast}$}}
\newcommand{\Lstar}{\mbox{$L_{\ast}$}}

\newcommand{\Dmom}{\mbox{$D_{\rm mom}$}}

\newcommand{\logg}{\mbox{$\log g$}}
\newcommand{\YHe}{\mbox{$Y_{\rm He}$}}

\newcommand{\Rp}{\mbox{$R_{\rm p}$}}
\newcommand{\Req}{\mbox{$R_{\rm eq}$}}
\newcommand{\Reqmax}{\mbox{$R_{\rm eq}^{\rm max}$}}

\newcommand{\vcrit}{\mbox{$v_{\rm crit}$}}
\newcommand{\vcrito}{\mbox{$v_{\rm crit,1}$}}
\newcommand{\vcritt}{\mbox{$v_{\rm crit,2}$}}
\newcommand{\rcrit}{\mbox{$r_{\rm crit}$}}
\newcommand{\vesc}{\mbox{$v_{\rm esc}$}}
\newcommand{\vth}{\mbox{$v_{\rm th}$}}

\newcommand{\vrot}{\mbox{$v_{\rm rot}$}}
\newcommand{\vsini}{\mbox{$v{\thinspace}\sin{\thinspace}i$}}

\newcommand{\vphi}{\mbox{$v_\phi$}}
\newcommand{\vtheta}{\mbox{$v_\theta$}}

\newcommand{\dvdr}{\mbox{$\dd v/\dd r$}}
\newcommand{\fcl}{\mbox{$f_{\rm cl}$}}
\newcommand{\fv}{\mbox{$f_{\rm vol}$}}
\newcommand{\xne}{\mbox{$n_{\rm e}$}}
\newcommand{\taur}{\mbox{$\tau_{\rm Ross}$}}
\newcommand{\kl}{\mbox{$k_{\rm L}$}}
\newcommand{\se}{\mbox{$s_{\rm e}$}}

\newcommand{\chibar}{\mbox{$\bar{\chi}$}}
\newcommand{\nl}{\mbox{$n_{\rm l}$}}
\newcommand{\taus}{\mbox{$\tau_{\rm Sob}$}}

\newcommand{\grad}{\ensuremath{g_{\rm rad}}}

\newcommand{\Neff}{\ensuremath{N_{\rm eff}}}

\newcommand{\qb}{\mbox{$\langle q \rangle$}}
\newcommand{\rhob}{\mbox{$\langle \rho \rangle$}}
\newcommand{\rhobtwo}{\mbox{$\langle \rho^2 \rangle$}}

\newcommand{\la}{\mathrel{\mathchoice {\vcenter{\offinterlineskip\halign{\hfil
$\displaystyle##$\hfil\cr<\cr\sim\cr}}}
{\vcenter{\offinterlineskip\halign{\hfil$\textstyle##$\hfil\cr
<\cr\sim\cr}}}
{\vcenter{\offinterlineskip\halign{\hfil$\scriptstyle##$\hfil\cr
<\cr\sim\cr}}}
{\vcenter{\offinterlineskip\halign{\hfil$\scriptscriptstyle##$\hfil\cr
<\cr\sim\cr}}}}}
\newcommand{\ga}{\mathrel{\mathchoice {\vcenter{\offinterlineskip\halign{\hfil
$\displaystyle##$\hfil\cr>\cr\sim\cr}}}
{\vcenter{\offinterlineskip\halign{\hfil$\textstyle##$\hfil\cr
>\cr\sim\cr}}}
{\vcenter{\offinterlineskip\halign{\hfil$\scriptstyle##$\hfil\cr
>\cr\sim\cr}}}
{\vcenter{\offinterlineskip\halign{\hfil$\scriptscriptstyle##$\hfil\cr
>\cr\sim\cr}}}}}

\newcommand\bgam{\hbox{\rm Br$_{\gamma}$}}
\newcommand\hap{\hbox{\rm H$_{\alpha}$}}

\newcommand\pap{\hbox{\rm Pa$_{\alpha}$}}

\newcommand\OIV{O\,{\sc iv}}
\newcommand\OV{O\,{\sc v}}
\newcommand\MgII{Mg\,{\sc ii}}

\newcommand\SiII{Si\,{\sc ii}}

\newcommand\SVI{S\,{\sc vi}}

\newcommand\ie{\hbox{i.e.,}}

\newcommand\puv{P\,{\sc v}\,$\lambda\lambda$1118-1128}
\newcommand\ouv{O\,{\sc v}\,$\lambda$1371}

\newcommand\nivuv{N\,{\sc iv}\,$\lambda$1718}

\newcommand\civuv{C\,{\sc iv}\,$\lambda$1550}
\newcommand\ciiiuv{C\,{\sc iii}\,$\lambda$1175}
\newcommand\siivuv{Si\,{\sc iv}\,$\lambda$1397}

\journalname{Astronomy and Astrophysics Review}
\begin{document}

\title{Mass loss from hot massive stars
}


\author{Joachim Puls  \and
        Jorick S. Vink \and \\
	Francisco Najarro
}

\authorrunning{J. Puls, J.S. Vink \& F. Najarro} 

\institute{Joachim Puls \at
              Universit\"atssternwarte M\"unchen, Scheinerstr. 1, D-81679
M\"unchen, Germany \\
              \email{uh101aw@usm.uni-muenchen.de}           
           \and
           Jorick S. Vink \at
              Armagh Observatory, College Hill, Armagh BT61 9DG, 
	      Northern Ireland \\
              \email{jsv@arm.ac.uk}
	   \and
	   Francisco Najarro \at
	   Instituto de Estructura de la Materia, CSIC, Serrano 121, 
	   28006 Madrid, Spain\\
	   \email{najarro@damir.iem.csic.es}
}

\date{Received: date}

\maketitle

\begin{abstract} Mass loss is a key process in the evolution of massive stars,
and must be understood {\it quantitatively} if it is to be successfully included in
broader astrophysical applications such as galactic and cosmic evolution and
ionization. In this review, we discuss various aspects of radiation driven
mass loss, both from the theoretical and the observational side. We focus on
developments in the past decade, concentrating on the winds from OB-stars,
with some excursions to the winds from Luminous Blue Variables (including
super-Eddington, continuum-driven winds), winds from Wolf-Rayet stars,
A-supergiants and Central Stars of Planetary Nebulae. After recapitulating
the 1-D, stationary {\it standard model} of line-driven winds, extensions
accounting for rotation and magnetic fields are discussed. Stationary wind
models are presented that provide theoretical predictions for the mass-loss
rates as a function of spectral type, metallicity, and the proximity to the
Eddington limit. The relevance of the so-called bi-stability jump is
outlined. We summarize diagnostical methods to infer wind properties from
observations, and compare the results from corresponding campaigns
(including the VLT-{\sc flames} survey of massive stars) with theoretical
predictions, featuring the mass loss-metallicity dependence. Subsequently,
we concentrate on two urgent problems, {\it weak winds} and {\it
wind-clumping}, that have been identified from various diagnostics and that
challenge our present understanding of radiation driven winds. We discuss
the problems of ``measuring'' mass-loss rates from weak winds and the
potential of the NIR \Bra-line as a tool to enable a more precise
quantification, and comment on physical explanations for mass-loss rates
that are much lower than predicted by the standard model.  Wind-clumping,
conventionally interpreted as the consequence of a strong instability
inherent to radiative line-driving, has severe implications for the
interpretation of observational diagnostics, since derived mass-loss rates
are usually overestimated when clumping is present but ignored in the
analyses. Depending on the specific diagnostics, such overestimates can amount
to factors of 2 to 10, and we describe ongoing attempts to allow
for more uniform results. We point out that independent arguments from
stellar evolution favor a moderate reduction of present-day mass-loss
rates. We also consider larger scale wind structure, interpreted in terms of
co-rotating interacting regions, and complete this review with a discussion of
recent progress on the X-ray {\it line} emission from massive stars. Such
emission is thought to originate both from magnetically confined winds and
from non-magnetic winds, in the latter case related to the line-driven
instability and/or clump-clump collisions. We highlight as to how far the
analysis of such X-ray line emission can give further clues regarding an
adequate description of wind clumping.

\keywords{hydrodynamics \and stars: atmospheres \and stars: early-type \and stars: mass loss
 \and stars: winds, outflows}
\end{abstract}

\setcounter{tocdepth}{3}

\section{Introduction}
\label{intro}

Within the last decade, our understanding of the physics of massive stars
has significantly improved. It has been realized that massive stars are
critical agents in galactic evolution, during both the present epoch as well
as in the early Universe, and they are regarded as ``cosmic engines''
\citep{bresolin08}. For instance, a population of very massive, First Stars (for a
recent review, see \citealt{Johnson08}) is thought to play a dominant role
in the reionization of the Universe and its first enrichment with metals.
Already with the next generation of telescopes, the integrated light from
these First Generations of Stars might become observable in the
near-infrared (NIR), by means of Ly$_\alpha$ \citep{Barton04} and
\HeII\,$\lambda1640$ \citep{Kud03a} emission from the surrounding interstellar
medium, {\it if} the corresponding initial mass function (IMF) is indeed
top-heavy \citep{Abel00, Bromm01}. A proper knowledge of the mass-loss
mechanisms during such early stages is crucial to understand the interplay
of these First Stars with their environments. Another highlight concerns the
long gamma-ray bursters (GRBs), which are likely the result of the terminal
collapse of massive stars at low metallicity. GRBs may also become important
tracers of the star formation history of the Universe at high redshift.
Again, mass loss plays a dominant role in the evolution of angular-momentum
loss, thus controlling whether the star will become such a burster.

There are three factors which have contributed to the aforementioned
progress: (i) new observational facilities such as ground-based 10-m class
telescopes equip\-ped with multi-object spectrographs, and space-born
observatories with spectroscopic capabilities in the ultraviolet ({\sc hst,
fuse, galex}), infrared ({\sc spitzer}), X-ray ({\sc xmm-Newton, chandra}),
and in the $\gamma$-ray ({\sc hete-2, swift}) domains; (ii) the development
of simulations of complex physical processes in the interior and outer
envelopes of massive stars, such as rotation and magnetic fields; and (iii)
the advancement of model atmospheres and radiative transfer techniques. 

Modeling the atmospheres of hot stars is a tremendous challenge due to 
severe departures from Local Thermodynamic Equilibrium (LTE) because of the
intense radiation, low densities, and the presence of supersonic velocity
fields initiated by the transfer of momentum from the stellar radiation
field to the atmospheric plasma. Radiation-driven winds are fundamentally
important, in providing energy (kinetic and ionizing radiation) and momentum
input into the ISM, in creating wind-blown bubbles and circumstellar shells,
and in triggering star formation. They can affect stellar evolution by
modifying evolutionary timescales, surface abundances, and stellar
luminosities.  The mass loss does not only influence the stellar evolution
but also the atmospheric structure, and winds need to be properly modeled 
to derive the correct stellar parameters by means of {\it quantitative
spectroscopy}. 

The last review on radiation-driven winds from hot stars was given by
\citet{KP00} who concentrated on the derivation and calibration of the
so-called {\it wind-momentum luminosity relation} (WLR) and its potential
application as an extra-galactic distance indicator. Thereafter, efficient
methods have been developed that allow for {\it theoretical predictions} of
mass-loss rates as a function of stellar parameters in the complete upper
HRD \citep{Vink00, Vink01}, and these can be used to check our standard
picture of radiation-driven winds through comparison with empirical results.
Owing to the enormous improvements in observational capabilities and
diagnostic methods, a number of puzzling phenomena have meanwhile emerged,
and this has resulted in exciting speculations and theoretical developments
which are by no means settled. To name two of the most prominent, there is
the so-called {\it weak-wind problem} which indicates that the mass-loss
rates from late O-/early B-type dwarfs might be a factor of 10 to 100 lower
than theoretically expected, challenging our understanding of
radiation-driven winds. Second, there is the issue of {\it wind-clumping} which
refers to small-scale density inhomogeneities distributed across the wind.
Because clumping ``contaminates'' almost all our mass-loss sensitive
diagnostics, it might result in severe down-sizing of previously derived
mass-loss rates, and consequently impact significantly our understanding of
stellar and galactic evolution. 

It is the goal of this review to provide an overview of these developments,
to outline their implications, and to indicate possible directions for
future work. Because of the breadth of the topic and its complexity, we
cannot cover all aspects. Instead, we primarily focus our review on the
winds from {\it single}, ``normal'' hot massive stars in well-established
evolutionary stages such as OB-dwarfs, giants, and supergiants, and on some
of the most relevant aspects of the winds from Luminous Blue Variables
(LBVs) and Wolf-Rayet (WR) stars (detailed in a recent review by
\citealt{Crowther07}). We also consider the winds from Central Stars of
Planetary Nebulae (CSPN) which are thought to obey similar physical rules as
their massive O-star counterparts. For a review on physical processes
related to wind-wind collisions in binary systems, we refer to
\citet{DeBecker07}.

This review is organized in such a way that most chapters can be read 
independently from each other. Sect.~\ref{sec:standardmodel} provides the
basic line-driven wind theory in terms of a 1-D, stationary model with
statistically distributed lines. In Sect.~\ref{sec:rotmag} this ``standard 
model'' is extended by accounting for the interaction with rotation and
magnetic fields, which are not only important when the field strengths are
large, as it is the case for few ``normal''\footnote{as opposed to the
chemically peculiar Ap/Bp-stars.} OB stars, but also when the wind density is
quite low and relatively modest fields are present.

In Sect.~\ref{sec:statmodels}, we compare and contrast the various
stationary wind models providing theoretical predictions for the mass-loss
rates as a function of spectral type, metallicity, and the proximity to the
Eddington limit. We focus on predictions obtained by Monte Carlo methods for
stars in various evolutionary phases (OBA, LBV, WR) and for various chemical
mixtures.  These prescriptions are used in most modern evolutionary
calculations and are generally referred to when observational findings are
compared to theory. An issue of particular relevance is the so-called {\it
bi-stability} jump.  Objects below a specific effective temperature (around
25~kK, kK = kilo Kelvin) should have systematically larger mass-loss rates
and lower terminal wind velocities than objects above this jump temperature,
due to an abrupt change in the ionization equilibrium of iron, the most
important element with respect to line-driving in the inner wind.

After a brief discussion of various wind-diagnostics and their pros and
cons, recent observational results and their comparisons with theoretical
predictions are summarized in Sect.~\ref{sec:obswindpara}. We first
demonstrate how the continuing development of NLTE model atmospheres has led
to severe parameter changes due to the inclusion of metal-line and
wind-blanketing. Most affected is the effective temperature scale of
Galactic OB-stars, with the new temperatures significantly below the older
ones. We subsequently present observationally derived wind-parameters for
OBA-stars, LBVs, and WRs, and we discuss their implications. Special
emphasis is given to the VLT-{\sc flames} survey of massive stars in the Galaxy
and the Magellanic Clouds \citep{Evans05, Evans06} which allowed for a
derivation of stellar and wind parameters of a large sample of objects,
enabling a detailed {\it empirical} study of the mass loss-metallicity
dependence.  We also cover topics such as the winds from BA-supergiants
beyond the Magellanic Clouds, the winds from CSPN, and we end with an
excursion to super-Eddington, {\it continuum-driven} winds which might be
relevant for the giant eruptions of LBVs. 

At first glance, Sect.~\ref{sec:obswindpara} might give the impression that
theory and observations largely agree and that only a few items remain to be
clarified before hot-star winds can be regarded as ``understood''. Indeed
this was a widely held belief at the end of the 1990's. In the meantime, 
however, the issues of wind clumping and weak winds (see above) have now
emerged as urgent.  

In Sect.~\ref{sec:weakwinds} we specify the ``weak wind
problem'' and discuss various physical processes which might be responsible
for the apparent dilemma. We demonstrate the challenges for conventional
mass-loss diagnostics that become rather insensitive for very low mass-loss
rates.  As an alternative diagnostic, we discuss the potential of the NIR
\Bra-line as a tool to enable more precise measurements and that might allow
for a quantification of the mass-loss ``deficit'' with respect to
theoretical predictions. 

Sect.~\ref{sec:inhomowinds} is devoted to another central topic of current
activity in stellar wind research, comprising various aspects of
wind-structure and time-dependence -- neglected in the standard model. After
a summary of observational findings, we provide theoretical arguments which 
show that line-driven winds are subject to a strong instability inherent to 
the line acceleration itself. This line-driven (or de-shadowing) instability
has been proposed to be the origin of {\it wind-clumping}, triggering the
formation of {\it small-scale} structure. In Sect. \ref{sec:clumping}, we
discuss this in detail. Clumping has severe implications for the
interpretation of observed line-profiles, particularly with respect to the
derived mass-loss rates. Conventional recombination-based $\rho^2$-processes
such as \Ha, IR, and radio-excess become very sensitive to clumping,
resulting in overestimated mass-loss rates when clumping is present but has
been neglected in the analyses.  We discuss various diagnostic methods to
investigate the clumping properties of stellar winds and to derive ``true''
mass-loss rates. Corresponding results disagree with respect to the
estimated degree of mass-loss reduction, ranging from factors 
in between 2 and 10, and even more. Potential problems with these diagnostics have been
identified by different research groups and are likely related to the {\it
porosity} of a clumped wind medium and the corresponding velocity space, or
{\it vorosity}. We finish the section with some independent insights from
stellar evolution, which suggest that the reduction of the mass-loss rates
is more likely to be moderate.

Turning to larger-scale structure, Sect.~\ref{sec:cirs} covers the {\it
discrete absorption components} (DACs) and {\it modulation features}, as
well as corresponding time-dependent models of co-rotating interaction zones
(CIRs) and co-rotating, azimuthally extended structures which might be
responsible for these other structural phenomena. 

Since the first X-ray measurements of massive stars became available, it was
clear that they are strong X-ray emitters. In Sect.~\ref{sec:xraylines}, we
review the status of the X-ray wind {\it line emission}, as observed by {\sc
xmm-Newton} and {\sc chandra}. This emission has been interpreted to either
result from magnetically confined winds or to result
from shocks embedded in the wind, likely related to the line-driven
instability or clump-clump collisions. Recent diagnostics has enabled to
``measure'' the onset of this emission, and current analyses utilize the
line emission as a tool to deduce the degree of wind-clumping, its radial
distribution, and the shape of the clumps. We report on corresponding
results and implications.

In Sect.~\ref{sec:summary}, we summarize the most important achievements
obtained during the last decade of wind research and provide some future
prospects.

\section{Line driven winds of hot stars -- theoretical background}
\label{sec:ldw}

As each photon carries momentum, $h \nu/c$, it was speculated as early as in
the 1920s (e.g., \citealt{Milne26}) that radiative acceleration by spectral
lines might be capable of ejecting metal ions from stellar atmospheres.
However, it was not until UV balloon flights in the late 1960s that the
theory of radiation line driving became the leading theory for
the steady outflows from hot luminous OB stars. \citet{LS70} and \citet{CAK}
realized that if the acceleration exerted on the metal ions could be shared
with the more abundant hydrogen and helium species in the plasma, it could
result in a significant mass-loss rate \mdot, of the order of \mdu, which
might affect the evolution of massive stars in a substantial manner.

\subsection{The standard model (1-D, stationary, smooth)}
\label{sec:standardmodel}

\paragraph{Momentum transfer by line scattering.~~} In the following, we
will concentrate on those lines that contribute to most of the total
acceleration, i.e., resonance = scattering lines from ionic ground states
(C,N,O,{\ldots}) or low-lying meta-stable levels (iron group elements). The
principal concept of momentum transfer by line-scattering is that the
initial photon originates from a particular direction in the stellar
photosphere, whereas the subsequent re-emission is more or less isotropic,
or, to be more precise, at least fore-aft symmetric with respect to the
radial direction. This change in direction angle $\theta$ leads to a radial 
transfer of momentum, $\Delta P = h/c (\nu_{\rm in} \cos \theta_{\rm in} -
\nu_{\rm out} \cos \theta_{\rm out})$ (for details, see, e.g.,
\citealt{Puls87}), and this angle change (together with the Doppler-effect,
see below) is the key of the momentum transfer
and associated radiative line acceleration. 

When integrating this momentum transfer over all scatterings, the losses or
gains due to re-emission processes cancel because of fore-aft symmetry,
resulting in a radially directed, total line acceleration, $g_{\rm rad}^{\rm
line}$. This acceleration acts on the total wind plasma, as long as the
momentum gained by the metal ions can be shared with the more abundant ions
of H and/or He which are (usually) less accelerated due to far fewer
lines close to the stellar flux maximum.

\paragraph{Coulomb coupling.~~} \label{coulomb_coupling} For denser winds, the momentum transfer from
the active metal ions to the passive hydrogen and helium ions is easily
fulfilled via Coulomb collisions, but this may no longer hold for tenuous
winds. The one-fluid approximation is applicable as long as the active ions
are slowed down efficiently through electric charge interactions with the
passive particles, i.e., if the relevant time scale for momentum transfer
due to collisions is small in comparison to the time the active ions
(accelerated by line acceleration) need to drift apart from the passive ones.
For a one-component medium, this drift velocity, $v_{\rm d}$, scales with
$\rho^{-1}$ (with density $\rho$; for details, see \citealt{Springmann91}).
Thus, decoupling ($v_{\rm d} > v_{\rm th}$, with $v_{\rm th}$ the thermal
velocity, here for the dominant passive ion) and subsequent ionic runaway
can be expected to become most relevant for stars with small mass-loss rates
and large terminal wind velocities \vinf\, \citep{SP92, babel95, babel96,
krticka00, OP02, Krticka03}.

\paragraph{Basic assumptions, equation of motion.~~} \label{sec:assump} In
order to predict the physical structure of a line-driven wind, let us first
employ the so-called standard model. This model assumes the wind to be
stationary, homogeneous (e.g., no shocks or clumps) and spherically
symmetric, which is justified as long as rotation and magnetic fields 
(covered in Sect.~\ref{sec:rotmag}) can be neglected, since all external
forces are purely radial.

In this case, the total mass flux (or mass-loss rate) through a spherical
shell with radius $r$ surrounding the star is conserved, as may be seen from
the equation of continuity
\begin{equation}  
\label{eq_const}
\dot{M} =  4\,\pi\, r^{2}\, \rho\,(r)\,  v\,(r) = \mbox{constant}.
\end{equation}
The corresponding (stationary) momentum equation reads
\begin{equation}
v \frac{\dd v}{\dd r}~=~- \frac{G M}{r^2}~-~\frac{1}{\rho}\frac{\dd p}{\dd
r}~+~g_{\rm rad},
\label{eq_mom}
\end{equation}
with inwards directed gravitational acceleration and outwards directed
pressure ($p$) term  and radiative acceleration. 
The necessary condition for initiating a stellar wind is that the total
radiative acceleration, $g_{\rm rad}$ = $g^{\rm line}_{\rm rad}$ $+$ $g^{\rm
cont}_{\rm rad}$, exceeds gravity beyond a certain point in the outer
photospheric layers. With the equation of state, $p = a^{2}\,\rho$,
where $a$ is the isothermal speed of sound, and using Eq.~\ref{eq_const}, the
equation of motion is given by
\begin{equation}
\bigl(1-\frac{a^2}{v^2}\bigr)~v \frac{\dd v}{\dd r}~=~
\frac{2a^2}{r}~-~\frac{\dd a^2}{\dd r}~-~\frac{G M}{r^2}~~+~g_{\rm rad}.
\label{eq_motion}
\end{equation}

\paragraph{The radiative acceleration.~~} The prime challenge lies in
accurately predicting the $g_{\rm rad}$ term in the equation of motion.  For
free electrons (the Thomson acceleration), this is simply proportional to the
Thomson opacity, $\sigma_{\rm e} = s_{\rm e} \rho$ ($s_{\rm e}$
proportional to cross section) times the flux:
\begin{equation}
g^{\rm Th}_{\rm rad}~=~\frac{1}{c \rho} \frac{\sigma_{\rm e} L}{4 \pi r^2}~=~
g_{\rm grav}\,\Gamma,
\label{eq_elec}
\end{equation}
with Eddington's $\Gamma$. However in reality line scattering is the more
dominant contributor to the overall radiative acceleration.  The reason is
that (i) line scattering is intrinsically much stronger than electron
scattering, due to the resonant nature of bound-bound transitions
\citep{Gayley95}, and (ii) that although photons and matter are only allowed
to interact at specific frequencies, they can be made to resonate over a
wide range of stellar wind radii via the Doppler effect (see reviews by
\citealt{Abbott84, Owocki94}). 

For a single line at frequency $\nu$ and line optical depth $\tau$
illuminated by radially streaming photons, the line acceleration can be
approximated 
in terms of purely local quantities within the so-called Sobolev
\label{sobo} theory \citep{Sobo60}. Under typical conditions, this
approximation applies for the supersonic part of the wind, and, to a lesser
extent, also for the transonic region as long as opacity, source function
and velocity gradient do not change significantly over a velocity interval
$\Delta v = \vth$, corresponding to a geometrically region of width $\Delta
r \approx \vth/(\dvdr) =: L^{\rm Sob}$, the so-called Sobolev length.
\label{sobolength} Insofar, the line optical depth in expanding atmospheres
has an entirely different character than in hydrostatic photospheres: in
winds, the interaction between photon and absorbing atom is restricted to a 
region of width $L^{\rm Sob}$, which is usually very narrow, and 
the line optical depth becomes an (almost) {\it local quantity}. 

Using this approximation, the line acceleration can be expressed as
\begin{equation}
g^{\rm line}_{\rm rad, i}~=~\frac{L_\nu \nu}{4 \pi r^2 c^2}~(\frac{\dd v}{\dd
r})~ \frac{1}{\rho}~(1~-~e^{-\tau}),
\label{eq_gline}
\end{equation}
where $L_\nu$ is the luminosity at line frequency, and the (radial) line optical
depth, in the Sobolev approximation, is 
\beq
\tau \approx \tau^{\rm Sob} = \bar \chi \lambda /(\dvdr), 
\label{tausob}
\eeq
with $\bar \chi$ the frequency
integrated line-opacity and $\lambda$ the transition wavelength. Note that
for optically thin lines, $\tau < 1$, the acceleration becomes indendent of
the velocity field and recovers the same functional form as Eq.~\ref{eq_elec},
with $\sigma_{\rm e} L$ replaced by $\bar \chi L_\nu$, whereas for optically
thick lines, $\tau > 1$, the line acceleration depends on the {\it spatial
velocity gradient} \dvdr. This dependence (unique in physics and being a
consequence of the Doppler effect) is the origin of a number of peculiar
features related to radiative line driving, as we will see in
Sect.~\ref{sec:inhomowinds}.

\begin{figure*}
\hspace{-0.5cm}
\begin{minipage}{6.3cm}
\resizebox{\hsize}{!}
   {\includegraphics{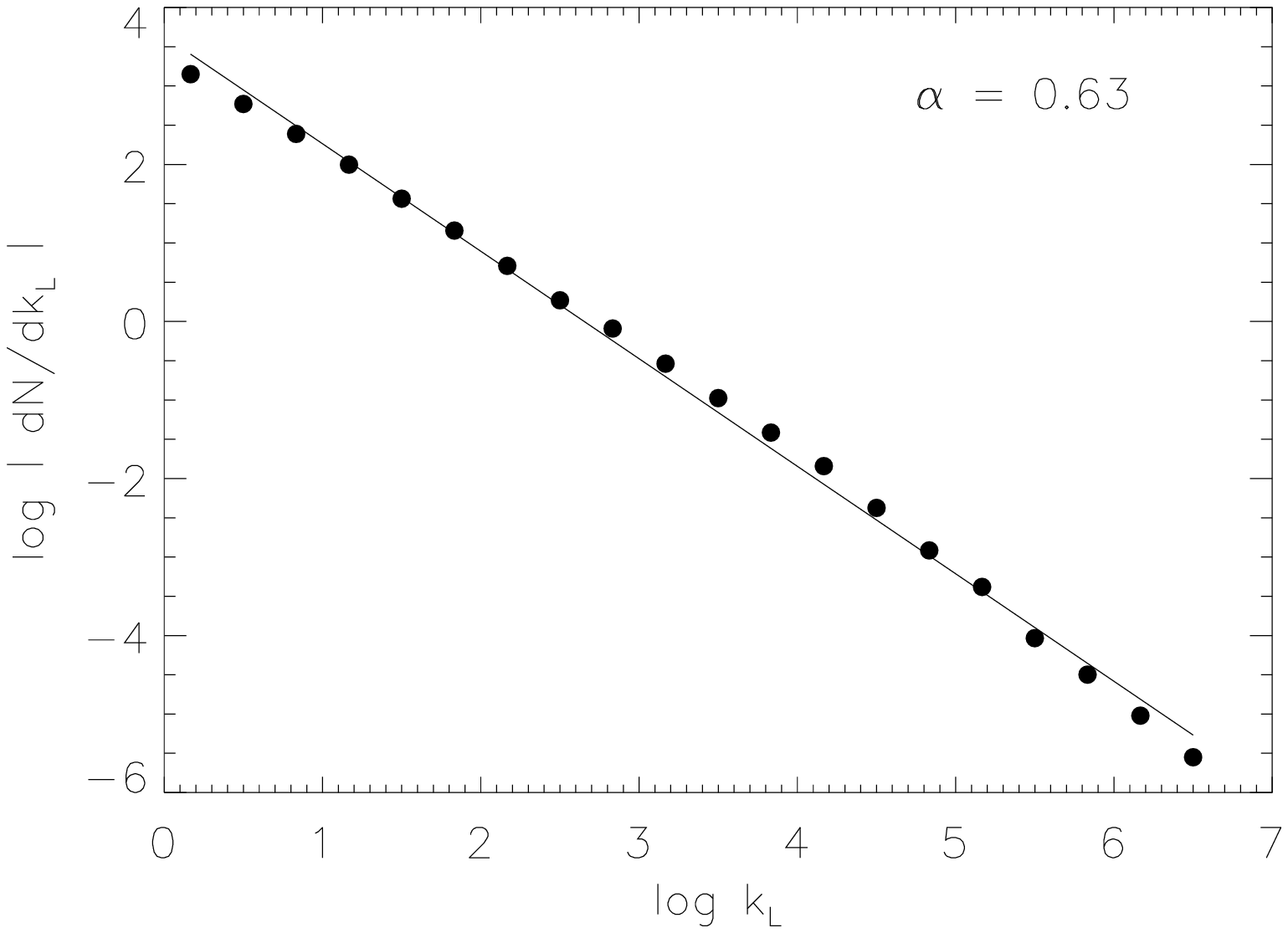}}
\end{minipage}
\hspace{-0.3cm}
\begin{minipage}{6.3cm}
   \resizebox{\hsize}{!}
   {\includegraphics{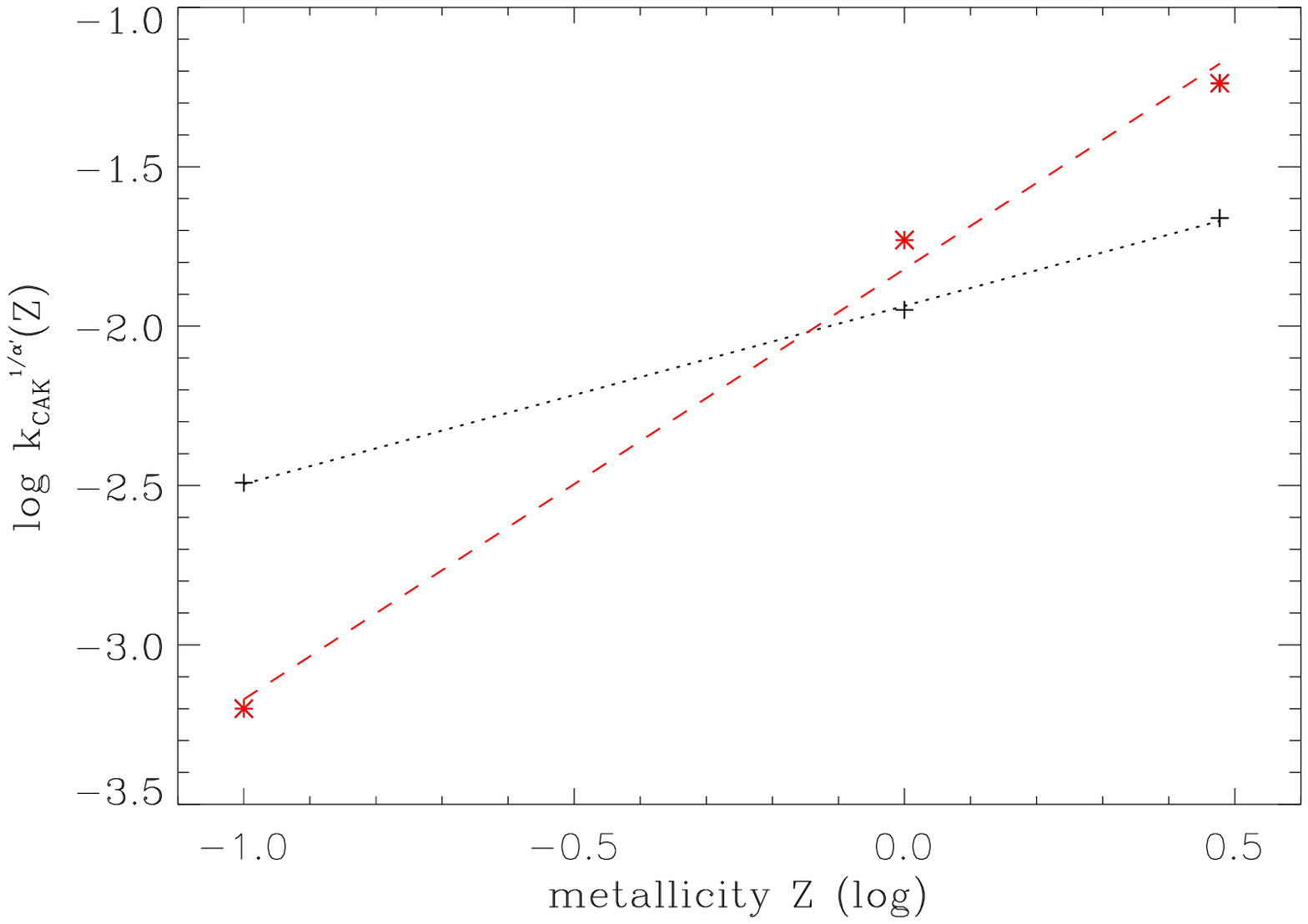}}
\end{minipage}
\caption{Left: Logarithmic plot of the line-strength distribution function
for an O-type wind of 40~kK (solar abundance), and corresponding
power-law fit.\newline
Right: Predictions from line-statistics. Dependence of mass loss (via
$k^{1/\alpha'}$, cf. Eq.~\ref{eq_mdotzstat}) on metallicity, for \Teff =
40~kK (dotted) and \Teff = 10~kK (dashed). The slopes are 0.56 and 1.35,
respectively. Data from \citet[ Table 3]{Puls00}.} 
\label{dndk}
\end{figure*}

The final exercise is now to sum over all contributing lines, which,
following the ideas by \citet[ ``CAK'']{CAK}, is conventionally done by using
so-called line-strength distribution functions, which describe the
statistical dependence of the number of lines on frequency position and
line-strength, \kl,
\beq
\kl=\frac{\bar \chi \lambda}{\rho} \frac{1}{\se \vth}.
\label{linestr}
\eeq
\kl\ measures the line opacity in units of the Thomson opacity and, for
resonance lines and frozen in ionization, remains constant throughout the
wind. From investigating an ensemble of \CIII\, lines (in LTE), CAK modeled 
the corresponding distribution function by a power law, 
\beq
dN(\nu, \kl)= -N_o f_\nu(\nu) \kl^{\alpha-2} \dd \nu \dd \kl
\label{dn}
\eeq
with $0 < \alpha < 1$, $N_o$ a normalization constant and $f_\nu(\nu)$ the
frequency distribution, which is rather independent from the distribution
in line-strength. More elaborate calculations including ions from all
participating elements have confirmed this approach (see Fig.~\ref{dndk},
left panel).

Combining the radiative line acceleration (Eq.~\ref{eq_gline}) and the
distribution of lines, (Eq.~\ref{dn}), the {\it total} line acceleration can be
calculated by integration, and, following CAK, expressed in terms of the
Thomson acceleration (Eq.~\ref{eq_elec}) multiplied by a factor $M$, the
so-called force-multiplier,
\begin{equation}
\frac{g^{\rm line}_{\rm rad}}{g^{\rm TH}_{\rm rad}}~=~M(t)~=~k~t^{-\alpha}
\propto \bigl(\frac{\dvdr}{\rho}\bigr)^\alpha 
\label{eq_cak}
\end{equation}
(for details see, e.g., \citealt{Puls00}, and for an alternative, sometimes
advantageous, formulation \citealt{Gayley95}). The optical depth parameter, 
$t$, 
\beq 
\tau = t\,\kl; \qquad t=\frac{\se \vth \rho}{\dvdr} = \frac{1}{\kl(\tau=1)} 
\eeq 
is the optical depth of a line with an opacity equal to Thomson-scattering,
or, alternatively, corresponds to the inverse of the line-strength of a line
with unit optical depth. Because of the power-law distribution, the
line-acceleration is dominated by lines distributed around a line-strength
$\kl(\tau=1) = t^{-1}(r)$ ($\pm$ 1 to 2 dex above and below this value,
depending on $\alpha$).

The force-multiplier parameter $k$ is proportional
to the effective, i.e., flux-weighted number of lines stronger than
Thomson-scattering, $N_{\rm eff}$,\footnote{
For the above power-law distribution,
$N_{\rm eff} = N_o/(1-\alpha)\times \int_0^\infty (L_\nu \nu/L) f_\nu(\nu)
\dd \nu$, $k=(\vth/c) \Gamma(\alpha)\times N_{\rm eff}$, with $\vth$ the
thermal speed of a representative ion (chosen as hydrogen by CAK) and
Gamma-function $\Gamma(\alpha) \approx 1/\alpha$ for the values considered
here.}
and can be interpreted as the fraction of flux that would be blocked
already in the photosphere if all lines were optically thick. $\alpha$ does
not only correspond to the slope of the line-strength distribution, 
but, more generally, quantifies also {\it the ratio of the line acceleration
from optically thick lines to the total one.} 
In case of deviations from a power-law distribution, the
radiative acceleration at location $r$ is controlled by its {\it local}
slope $\alpha$ around $\kl(\tau=1)$ \citep{Puls00}.  

This line force description can easily be extended with an
additional dependence via $(\xne_{11}/W)^\delta$ 
\label{deltaterm}
to account for ionization
effects \citep{Abbott82}, in the realistic case that the ionization is {\it
not} frozen in ($\xne_{11}$ is the electron density in units of $10^{11}$
cm$^{-3}$, and $W$ the dilution factor of the radiation field). 
Under typical O-star conditions, $\delta$ is small (of the order of 0.1, see
below), but can reach larger (positive or even negative) values if either
the ionization equilibrium changes significantly throughout the wind (which
might happen in dense winds) or the wind is mostly driven by hydrogen and
helium lines, in case of extremely metal-poor stars (see page
\pageref{largedelta}). Note that in any case $\delta$ is not a free
parameter, but has to be determined -- in parallel with the other two
force-multiplier parameters $k$ and $\alpha$ -- in a self-consistent way
from the (NLTE-) line opacities and the radiation field (see, e.g.,
\citealt{Pauldrach94}).

Finally, the so-called finite cone angle correction factor has to be
included to account for non-radially streaming photons \citep{PPK, FA86}.

\paragraph{Solving the equation of motion.~~} Also for the complete line
sample, the radiative acceleration depends on $\dvdr$ (now through a power
of $\alpha$, Eq.~\ref{eq_cak}), and it is conventionally assumed that this term has a similar
meaning as the velocity gradient entering the inertial term on the left hand
side (hereafter lhs) of
Eq.~\ref{eq_motion} (but see \citealt{Lucy98}).  In this case, the equation
of motion becomes non-linear in the variable $y=r^2 v \dvdr$, and can be
approximated by
\beq
y=-GM(1-\Gamma)+ f\, k\, L \mdot^{-\alpha} y^\alpha
\label{simple_eom}
\eeq
where $f$ is either constant (in the CAK case) or depends on $(r,v, \dvdr,
\mdot)$ if one includes all subtleties. As the enthalpy term in the energy
equation is much smaller than the potential and kinetic energy, the above
equation has been simplified by neglecting the gas pressure. For constant
$f$, this equation can be easily solved, and also the complete equation
provides no real difficulty (see CAK and \citealt{PPK}). Note that these
solutions involve the presence of a critical \label{critpoint} point which
controls the mass-loss rate. For models including non-radial photons, this
critical point is located close to the stellar photosphere and connects a
somewhat shallow (with respect to $y$), sub-critical flow smoothly with a
steeper, super-critical one. For further discussion regarding solution 
topologies and non-Sobolev line-forces, we refer the reader to \citet{POC90, 
OP99, FeldmeierShlosman00, FeldmeierShlosman02} and \citet{Madura07}.

As a final result, one obtains the following velocity law and scaling
relations for \vinf\, and \mdot:
\begin{equation}
v(r) = \vinf (1~-~R/r)^{\beta}, \qquad \left\{ \begin{array}{l} 
\beta = 0.5 \mbox{ (CAK case)}\\
\beta = 0.8 \mbox{ (O-stars) {\ldots}  2 (BA-supergiants)}\\ 
\end{array} \right.
\label{eq_vr}
\end{equation}
\begin{equation}
\vinf = \Cinf \bigl(\frac {2 G M (1-\Gamma)}{\Rstar} \bigr)^{\half} = \Cinf
\vesc, \qquad \left\{ \begin{array}{l} 
\Cinf  = (\frac{\alpha}{1-\alpha})^{0.5}  \mbox{\quad CAK case}\\
\Cinf \approx 2.25\, \frac{\alpha}{1-\alpha} \mbox{\quad incl. all details}\\ 
\end{array} \right.
\label{eq_vinf}
\end{equation}
\beq
\dot{M} \propto (kL)^{1/\alpha'}~(M(1-\Gamma))^{1-1/\alpha'}
\label{eq_mdot}
\eeq
with photospheric escape velocity \vesc\ 
(already corrected for the Thomson acceleration!) 
and $\alpha'=\alpha-\delta$. For
O-stars, $\delta$ is usually small (order of 0.1), $\alpha' \approx 0.6$ and
$k$ of the order of 0.1. From Eq.~\ref{eq_vr}, we can also derive the
behaviour of the optical depth parameter, $t$. It is proportional to
$(v(r)/\vinf)^{1/\beta-2}$, i.e., remains constant throughout the wind for
the CAK case, and increases from outside to inside for $\beta>0.5$.
\label{t(r)}

These relations (e.g., \citealt{Kud89}; for first order corrections to
Eq.~\ref{eq_vinf} see \citealt{KP00}, their Eq. 15) compare very well with
numerical results from integrating the complete equation, but are only valid
for spatially constant force multiplier parameters, $k, \alpha$ and
$\delta$. \citet{Kud02} provides elaborate additions to consider depth
dependent parameters as well.

Further assumptions involve the adoption of a core-halo structure that
neglects continuum formation in the wind (important for dense winds), and
the neglect of multi-line effects: in the approach as described above, {\it
each} line is allowed to interact with {\it unattenuated} photospheric
continuum radiation, irrespective of its frequency separation from other
lines. This neglect leads to overestimated radiation forces in most cases 
\citep{FC83, AL85, Puls87, LA93, Springmann94, Gayley95a, Vink00}. In
Sect.~\ref{sec:statmodels}, most of these assumptions will be relaxed and
more quantitative predictions will be provided. Note, however, that the
basic scaling relations remain rather unaffected by these improvements.

\paragraph{The wind-momentum luminosity relation.~~} Using the above scaling
relations, one can construct a {\it modified} wind momentum rate,
$\Dmom\,=\,\mdot\,\vinf\,(\Rstar/\rsun)^{1/2}$. Given that $\vinf$ scales
with the escape velocity (Eq.~\ref{eq_vinf}), \Dmom\ scales with luminosity
and effective line number alone,
\begin{equation}
\Dmom~=~\mdot~\vinf~\Bigl(\frac{\Rstar}{\rsun}\Bigr)^{\half}~\propto 
\Bigl(k \frac{L}{L_\odot}\Bigr)^{\frac{1}{\alpha'}}
\bigl(\frac{M}{\msun}(1-\Gamma)\Bigr)^{\frac{3}{2}-\frac{1}{\alpha'}} \propto
\Bigl(k \frac{L}{L_\odot}\Bigr)^{\frac{1}{\alpha'}},
\label{eq_wlr}
\end{equation}
as long as $\alpha'$ $\simeq$ $2/3$ (remember that $\alpha' \approx 0.6$ for
O-star). In this case, the effective stellar mass $M(1-\Gamma)$ cancels in
the product $\mdot \, \vinf$ when applying the scaling relations, which is
rather convenient since masses are difficult to determine from observations
and, moreover, seem to disagree when compared with evolutionary predictions
(e.g., \citealt{Puls08} and references therein). Taking the logarithm of Eq.~(\ref{eq_wlr})
and assuming $\alpha'$ $\simeq$ $2/3$, we find
\begin{equation}
\log \Dmom~\approx~x~\logLL + D
\label{wlr}
\end{equation}
(with slope $x$ and offset $D$, depending on the flux-weighted number of
driving lines), which is known as the ``Wind momentum
luminosity relationship (WLR)'' \citep{Kud95, Puls96} and can be considered
to be one of the big triumphs of the radiation-driven wind theory. This
relationship can, e.g., be utilized to determine extragalactic distances by
purely spectroscopic means \citep[ but see Sect.~\ref{sec:obastars}]{Kud99,
KP00}, and it also played an instrumental role in determining the
empirical mass-loss metallicity dependence for O stars in the Local Group,
which will be discussed in Sect.~\ref{sec:obswindpara}. Equally important,
observed and theoretically predicted WLRs can be compared to test the
predictions and validity of the radiation driven wind theory for various
spectral types, and to identify present shortcomings (e.g.,
Sect.~\ref{sec:weakwinds}).

\paragraph{Predictions from line-statistics. What determines the slope?~~}
\label{sec:linestat}
One may now ask what controls the slope of the line-strength distribution
function (Eq.~\ref{dn}), and in particular, why is $\alpha \approx 2/3$?
Since the line-strength \kl\, is proportional to the product of absorber
density per unit mass (depending on abundance, ionization fraction and
excitation) and oscillator-strength, the answer to this question is rather
difficult, in particular since a large fraction of the line acceleration, at
least in the lower wind, is due to lines from iron group elements (see
below) with a complex electronic configuration. In the following, we will
highlight certain aspects (for details, see \citealt{Puls00}).

Most interestingly, it is rather easy to show that $\alpha$ would be almost
exactly 2/3 if {\it all} lines would originate from hydrogenic ions and
would be resonance lines, i.e., excitation effects could be neglected. Then 
the line-strength distribution function depends solely on the distribution
of the oscillator strengths $f$, and from the well known Kramers-formula for
hydrogenic ions with transitions $1 \rarrow n$ (with principal quantum
number $n$) and corresponding selection rules, one finds that $f(1,n)
\approx \mbox{const}/n^3$ and $\dd N/\dd f \propto f^{-4/3}$. Since, under
the discussed conditions, this slope corresponds to $\alpha-2$, we
immediately obtain $\alpha = 2/3$. 

\label{largedelta}
Similarly important is the result that for {\it trace} ions one stage below
the major ionization stage the equality $\alpha + \delta \approx 1$ is valid
throughout the wind. Thus, for {\it hot, metal-poor} stars driven
predominantly by \HI\, and \HeII, the $\delta$-term can become much more
influential on the wind solution than in ``conventional'' winds driven by
ions of the dominant stage, where $\delta \approx 0.1$. This analytical
result has been proven by \citet{Kud02} by means of consistent wind
calculations, and is of relevance in the context of radiation driven winds
from the First Stars. Note that in these extreme cases the scaling relations
for \mdot\, and \vinf\, as provided above become invalid, since they are
based on small values for $\delta$.

For non-hydrogenic ions (with many and low lying levels), the situation is
more complicated. Per ion, the typical line-strength distribution consists
of two parts, a steeper one dominated by excitation effects and a flatter
one following the oscillator strength distribution, which, after summing up
over all participating ions, becomes almost irrelevant.  Most important for
the shape of the {\it total} distribution is the difference in
line-statistics between iron group and light ions as well as their different
(mean) abundances. Since the former group comprises a large number of
meta-stable levels, the line number from iron group elements is much higher,
especially at intermediate and weak line-strengths. Additionally, this
number increases significantly with decreasing temperature (more lines from
lower ionization stages), which becomes important with respect to the
so-called bi-stability mechanism (see below).

For solar abundances, iron group elements dominate the distribution at low
and intermediate values of line-strengths, i.e., they dominate the
acceleration of the inner wind (large $t$) and thus determine the mass-loss
rate. Lines from light ions (including hydrogen under A-star conditions), on
the other hand, populate the high $\kl$ end and accelerate the outer wind
(low $t$), i.e., determine \vinf. Typically, this part of the distribution
is steeper than the rest (e.g., Fig.~\ref{dndk}), due to excitation effects,
and only a small number of lines (\citealt{Krticka06}: ``a few dozens'') is 
responsible for this acceleration!  Insofar, the rather large observed
variance in $\vinf/v_{\rm esc}$ even for stars of similar spectral type
(Sect.~\ref{sec:obastars}) is not surprising, since subtle effects
(small variations in temperature, density, composition - cf.
\citealt{Pauldrach90}, their Fig.~8 - and also X-rays, see
Sect.~\ref{sec:inhomoobs}) can have a significant effect on the population and
line-strengths of light ions.

Note finally that the above results, which are based on the investigation of
line-strength distribution functions alone, have been confirmed by direct
wind calculations, see \citet{Pauldrach87, Vink99, Vink01} and
\citet{Krticka06}. 

\paragraph{Predictions from line-statistics. Dependence on temperature,
wind-density and metallicity.~~} Summarizing the above paragraphs,
deviations from a perfect power law are to be expected, mainly due to the
different distributions of ``light'' and ``heavy'' ions, their specific
dependence on temperature and the fact that the distribution at the high
$\kl$ end is controlled by excitation effects for the predominant ion
species, inducing a steeper slope. Translated to the effective value of
$\alpha$ (i.e., evaluated at $\kl=t^{-1}$) which controls the line-force
(and the slope of the WLR), this quantity should decrease with decreasing
\Teff, decreasing wind density and decreasing metallicity $Z$. In all these
cases, the WLR is expected to become steeper than in the ``standard''
situation encountered for O-stars where $\alpha$ is of the order of
0.6{\ldots}0.7. For A-supergiants, e.g., line-statistics predict $\alpha
\approx 0.45$ such that in addition to a steeper WLR also the scatter
introduced by a non-vanishing mass term should become significant.

Besides the variation of $\alpha$ with $Z$, the major consequence of a
different metallicity is a variation of $k$, i.e., a different effective
line-number. Roughly speaking, the higher the metallicity, the more effective
lines are to be expected, and vice versa. Since $\kl$ varies proportionally to
$Z$, as long as the ionization equilibrium remains rather unaffected from
the different metallicity, the
assumption of a power-law line-strength distribution function (Eq.~\ref{dn})
immediately implies $N_o(Z) \propto Z^{1-\alpha}$, with similar scalings for
$k$ and $N_{\rm eff}$. By means of Eq.~\ref{eq_mdot}, the mass-loss 
rate should then scale according to
\beq
\mdot(Z) \propto Z^\frac{1-\alpha}{\alpha'}.
\label{eq_mdotzstat}
\eeq
From Fig.~\ref{dndk}, right panel, 
\label{alpha_asg}
we see that the predicted dependence of
$k(Z) \propto Z^{1-\alpha}$ is actually present, and the derived slopes 
($s=(1-\alpha)/(\alpha-\delta)$) are consistent with prototypical values,
namely ($\alpha,\alpha'$)=(0.68, 0.58) for the O-type and
($\alpha,\alpha'$)=(0.48, 0.38) for the A-type models, respectively (lower
$\alpha$ because of lower \Teff: more Fe lines from lower stages). 
Moreover, these numbers are also consistent with results derived from the
analysis of a large sample of O-stars in the Galaxy and the Magellanic
Clouds (Sect.~\ref{sec:obastars}). Detailed {\it measurements} of the
metallicity dependence of A-supergiant winds are still missing though 
(but see page~\pageref{aextra}).

\paragraph{The bi-stability mechanism.~~} 
\label{sec:bistab} 
Most of the previous paragraphs may have given the reader the impression
that -- save for their intrinsic luminosity dependence -- the mass-loss
properties from the winds of massive stars are more or less identical, but
nothing is further from the truth: the wind properties of massive stars are
intriguingly diverse! This diversity is, of course, related to the variation
of effective line number and $\alpha'$ as a function of spectral type and
metallicity and to the different efficiencies of multi-line effects as a
function of wind density.

Of particular interest is the so-called bi-stability mechanism, which was
first encountered in the model calculations of the famous Luminous Blue
Variable (LBV) P Cygni by \citet{PP90}. At a critical effective temperature
around 21\,000 K \citep{Lamers95}, the overall properties of $\vinf$ and
$\dot{M}$ are expected to change drastically. \citet{Vink99} showed that
when a massive star evolves from hot temperatures to lower ones, the
dominant wind driving element of iron (Fe) recombines from Fe{\sc iv} to
\FeIII, which is predicted to result in an increase of $\dot{M}$ by a
factor 5 and a drop of $\vinf$ by a factor 2. 

As was pointed out already by \citet{PP90}, such a recombination can be
triggered by the optical depth of the hydrogen Lyman continuum.\footnote
{At least for dense
winds such as from P~Cygni considered in the original explanation.  For
most ``normal'' massive stars as studied by \citet{Vink99, Vink00}, the Lyman 
continuum remains optically thin between photosphere and critical point, and
the abrupt change of the Fe ionization equilibrium is rather due to conventional
temperature effects and the runaway in density as described below. For these
objects we consider $\tau_{\rm Ly}$ as an alias for the \FeIII/\FeIV\
ionization equilibrium.}
As long as this is
optically thin (e.g., for hotter temperatures), the EUV continuum is rather
strong, such that most iron resides in \FeIV, which has fewer lines than
\FeIII, resulting in a lower density, faster wind. When the Lyman continuum
becomes optically thick, there is almost no EUV flux left, and iron
recombines to \FeIII\ with more lines, inducing a dense, low velocity wind.
Since the optical depth of the Lyman continuum is closely coupled to wind
density via recombination effects, 
\beq \tau_{\rm Ly} \propto \rho^2 \propto \Bigl(\frac{\mdot}{\vinf}\Bigr)^2
\propto \bigl(M(1-\Gamma)\bigr)^{1-2/\alpha'} \approx
\bigl(M(1-\Gamma)\bigr)^{-2{\ldots}-3},  
\eeq 
the abrupt (see below) change in wind conditions does not only depend on
\Teff, but also, and severely, on effective mass (e.g., Figs. 4 and 5 in
\citealt{PP90}). As we will see in Sect.~\ref{sec:rot}, this can play an
important role in understanding the B[e]-star phenomenon.
 
The reason that the wind should react to the different accelerating
conditions by an actual ``jump'' in \mdot\ and \vinf\ is that the physics of
the line driving implies that when the density increases due to an
increasing $\dot{M}$, the inner wind is anticipated to recombine even more, 
leading to a run-away mass-loss increase until all iron has accumulated in
\FeIII. The line-driving in the outer wind is expected to react to an
increased inner density by lowering its terminal wind velocity (remember
that $g^{\rm line}_{\rm rad} \propto \rho^{-\alpha}$, Eq.~\ref{eq_cak}).  In
Sect.~\ref{sec:obastars} we will discuss the empirical constraints on the
bi-stability mechanism.

Also here, the spectral-type dependence of the mass-loss rate and terminal
velocity occurs in conjunction with changes in $N_{\rm eff}$ and $\alpha'$,
which could subsequently be reflected in the slope and offset of the WLR.
Let us emphasize that the spectral-type dependence of the WLR represents a
most active aspect of hot-star wind research \citep{Benaglia07, MP08}.

\subsection{The impact of rotation and magnetic fields}
\label{sec:rotmag}

All previous results have been derived under the assumption of spherical
symmetry, with gravity and radiative acceleration as the only external forces.
In the following sections, we will summarize alterations due to rotation
and magnetic fields, if present.

\subsubsection{Rotation and stellar winds}
\label{sec:rot}

(Almost) all massive stars start their evolution as rapid rotators, and
remain rapidly rotating during the largest part of their life-time, at least
with respect to the most decisive quantity, namely the so-called ``critical
ratio'' between surface rotation \vrot\ and critical velocity
\citep{MeynetV}. \vrot\ itself decreases (almost) continuously during the
main sequence and evolved phases, where the rate of this decrease is larger
for the more massive stars due to the larger mass loss and accompanying larger
removal of angular momentum. For supergiants, the reduction is due to an
increasing stellar radius as well.

The role of rotation as a fundamental stellar parameter of massive stars has
been highlighted by \citet{Langer97}, and effects on stellar structure and 
evolution were detailed by the Geneva group (A. Maeder, G. Meynet and
co-workers) in a series of publications.  For corresponding reviews, we
refer to \citet{MM00, MM03} (see also material presented in 
\citealt{IAUColl169} and \citealt{IAUS215}).

Rotation affects the physics and diagnostics of line-driven winds from
massive stars at least in three ways: (i) the dynamics is modified, (ii) the
diagnostics, particularly for \mdot, needs to consider rotation due to
deviations from spherical symmetry, and (iii) line-profile variability is
induced due to an interplay between disturbances in the photosphere (stellar
spots, non-radial pulsations), rotation and wind-dynamics (co-rotating
interaction regions, CIRs, see Sect.~\ref{sec:cirs}). 

\paragraph{1-D models, scaling relations.~~} First attempts to include
rotation into the stellar wind dynamics were based on a 1-D solution in the
equatorial plane, and have been performed by \citet{FA86} and \citet{PPK},
with applications regarding outflowing disks of Be-stars by \citet{Araujo95},
and investigations of the solution topology by \citet[ see also
\citealt{Cure04} and \citealt{Madura07}]{CureRial04}.  

Under the assumption of purely radial line-forces, the equation of motion is
governed by central forces, i.e., the angular momentum is conserved, $L
\propto r \vphi = \mbox{const}$, and the azimuthal velocity in the
equatorial plane is given by $\vphi(r) = \vrot(\Rstar) \Rstar /r$, with
polar velocity $\vtheta(r, \theta=90) = 0$ and co-latitude $\theta$. Thus,
the equation of motion remains almost identical to Eq.~\ref{eq_motion} or
the simplified version Eq.~\ref{simple_eom}, except that the effective
gravity (here
defined as the difference between gravitational and Thomson acceleration) 
has to be corrected for the centrifugal acceleration $\vphi^2/r$, 
\beq
-\frac{GM(1-\Gamma)}{r^2}~\rarrow~-\frac{GM(1-\Gamma)}{r^2} \bigl(1-\Omega^2\frac{\Rstar}{r}\bigr)
\qquad \Omega=\frac{\vrot(\Rstar)}{\vcrit}, 
\eeq
with \vcrit\ the 1-D critical velocity\footnote{for the radiatively
reduced gravity as defined above and neglecting gravity darkening.}, $\vcrit = \vesc/\sqrt 2$. In the following, we will mostly consider
models not too close to critical rotation ($\Omega \la 0.75$). For a
discussion of more extreme situations in one dimension, see \citet{Cure04,
CureRial04} and \citet{Madura07}. Since in the former case the critical
point (page \pageref{critpoint}) of the non-rotating solution remains almost
unaffected by the centrifugal term due to $\rcrit \approx \Rstar$ (e.g.,
\citealt{PPK, OuD04, Madura07}), the effective mass can be replaced by
$M(1-\Gamma)(1-\Omega^2)$, and \mdot\ scales, to a very good approximation,
with 
\beq 
\mdot(\Omega) \approx \mdot(0)(1-\Omega^2)^{1-1/\alpha'}.
\label{eq_mdot_eq}
\eeq
Similarly, since \vinf\ scales with \vesc, a
rather good approximation for the dependence of \vinf\ on $\Omega$ is
\beq 
\vinf(\Omega) \approx \vinf(0)(1-\Omega^2)^{1/2}.
\label{eq_vinf_eq}
\eeq
For independent arguments leading to this equation, see \citet{Gayley00}. 
Comparisons of exact vs. approximate solutions for \mdot\ and \vinf\ as a
function of $\Omega$ have been given, e.g., by \citet{PP99}. 

\paragraph{Wind compressed disks and zones.~~} In a seminal paper,
\citet{BC93} included the latitude dependence of the centrifugal forces into
the previous 1-D description in order to generalize the approach. Since (at that
time) it was still undebated to assume line-forces which are purely radially
directed, under this assumption the angular momentum is conserved for {\it
each} particle starting at a given co-latitude $0 < \theta_0 < \pi$, and its
trajectory is restricted to an orbital plane tilted by $\theta_0$ with
respect to the equatorial one.  Thus, a 1-D solution is still possible per
orbital plane, if one accounts for the centrifugal acceleration as a
function of co-latitude, 
\beq 
g_{\rm cent} = (\vrot(\Rstar) \sin \theta_0)^2 \frac{\Rstar^2}{r^3}, 
\eeq
and the scaling relations for the equatorial flow (Eqs.~\ref{eq_mdot_eq},
\ref{eq_vinf_eq}) can be generalized with respect to $\theta_0$,
\beq 
\mdot(\Omega,\theta_0) \propto (1-\Omega^2 \sin^2\theta_0)^{1-1/\alpha'}, \qquad
\vinf(\Omega,\theta_0) \propto (1-\Omega^2 \sin^2\theta_0)^{1/2}.
\label{mdotvinfwcd}
\eeq
In this description, \mdot\ increases and \vinf\ decreases towards the
equator, respectively. Since the particle is restricted to its orbital plane, the 
corresponding azimuthal angle, $\Phi'(r)$, increases for increasing $r$ and
can be calculated from integrating $\dd \Phi'(r)/\dd r =
\vphi(r,\theta_0)/(r v_r(r,\theta_0))$, using a $\beta$-velocity law
(Eq.~\ref{eq_vr}) in combination with  $\vinf(\theta_0)$ for the radial
velocity component $v_r$.

\begin{figure*}
\begin{minipage}{3.8cm}
\resizebox{\hsize}{!}
   {\includegraphics{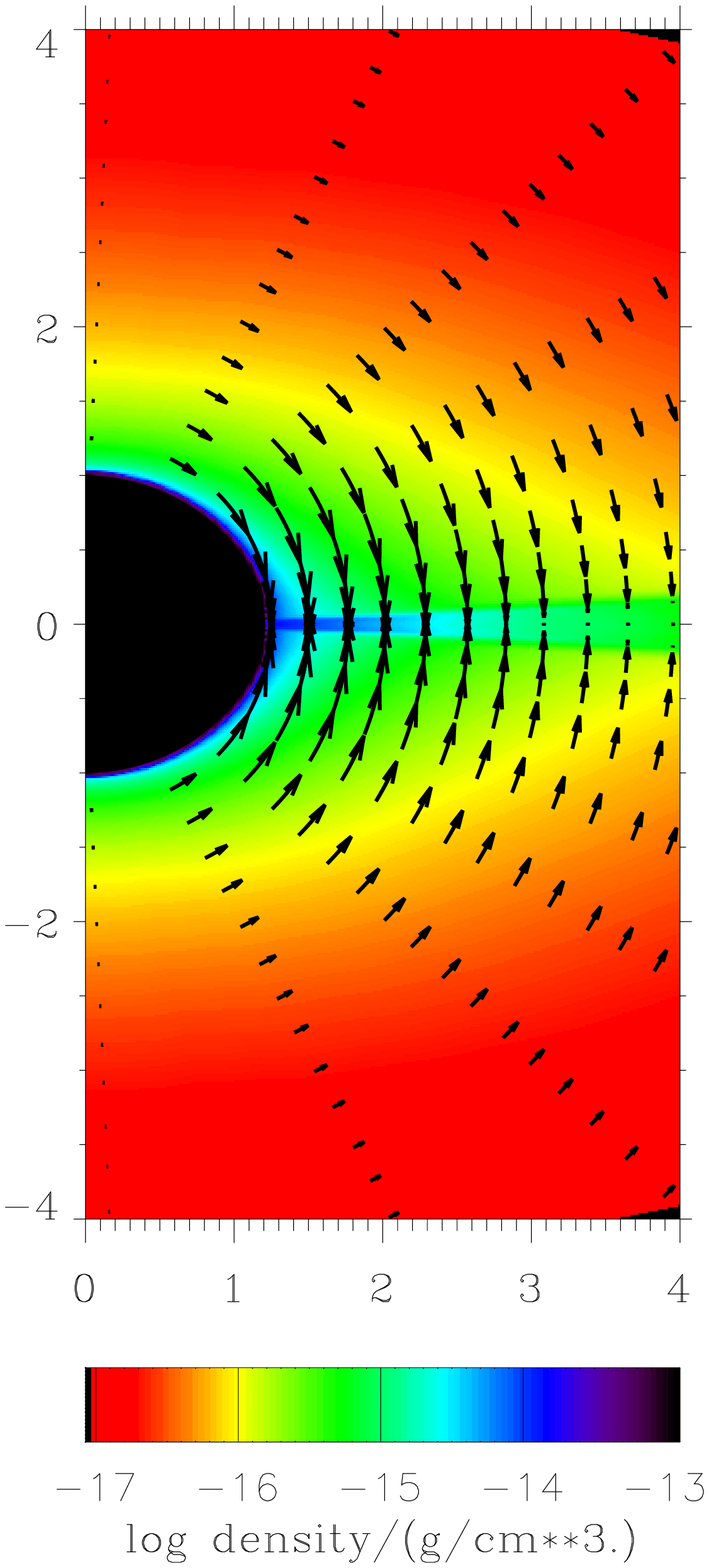}}
\end{minipage}
\begin{minipage}{3.8cm}
\resizebox{\hsize}{!}
   {\includegraphics{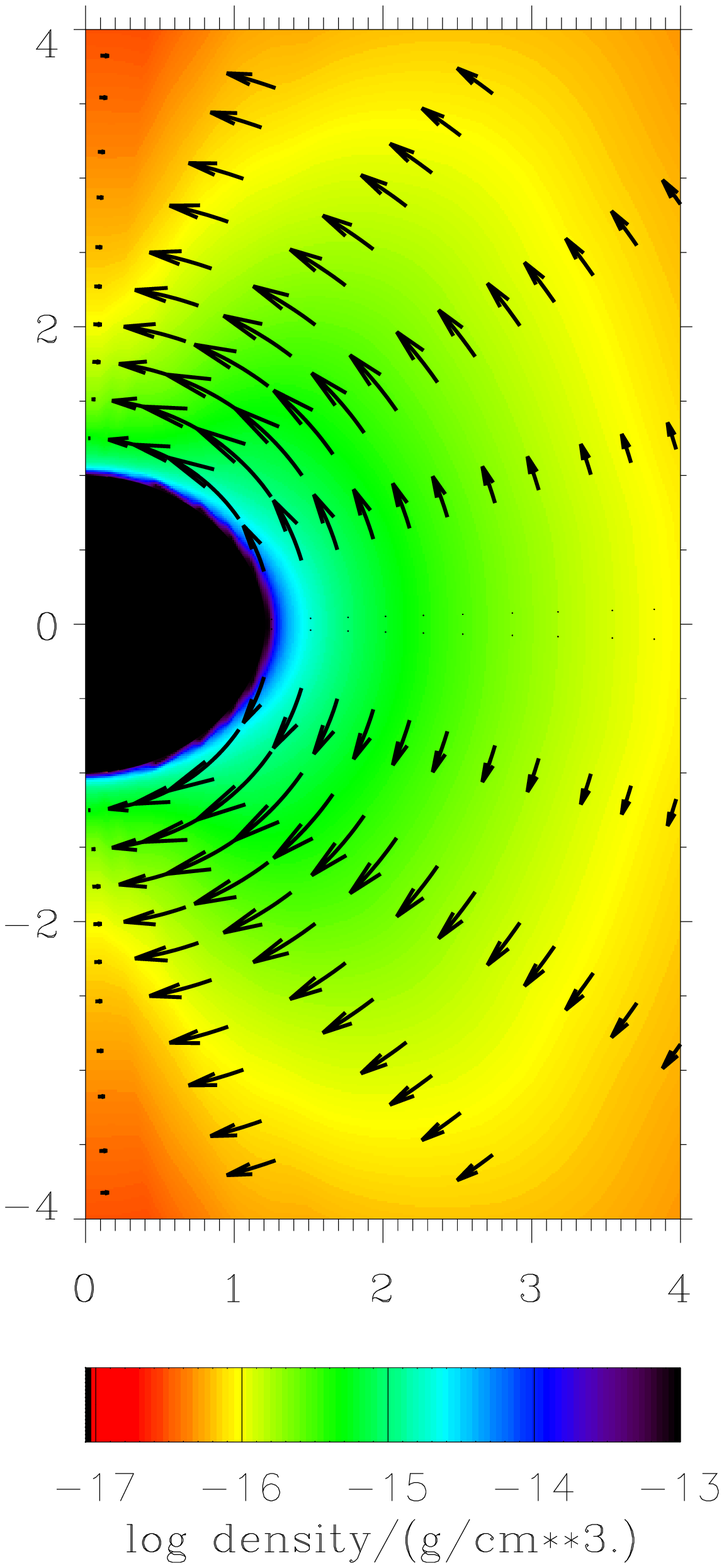}}
\end{minipage}
\begin{minipage}{3.8cm}
\resizebox{\hsize}{!}
   {\includegraphics{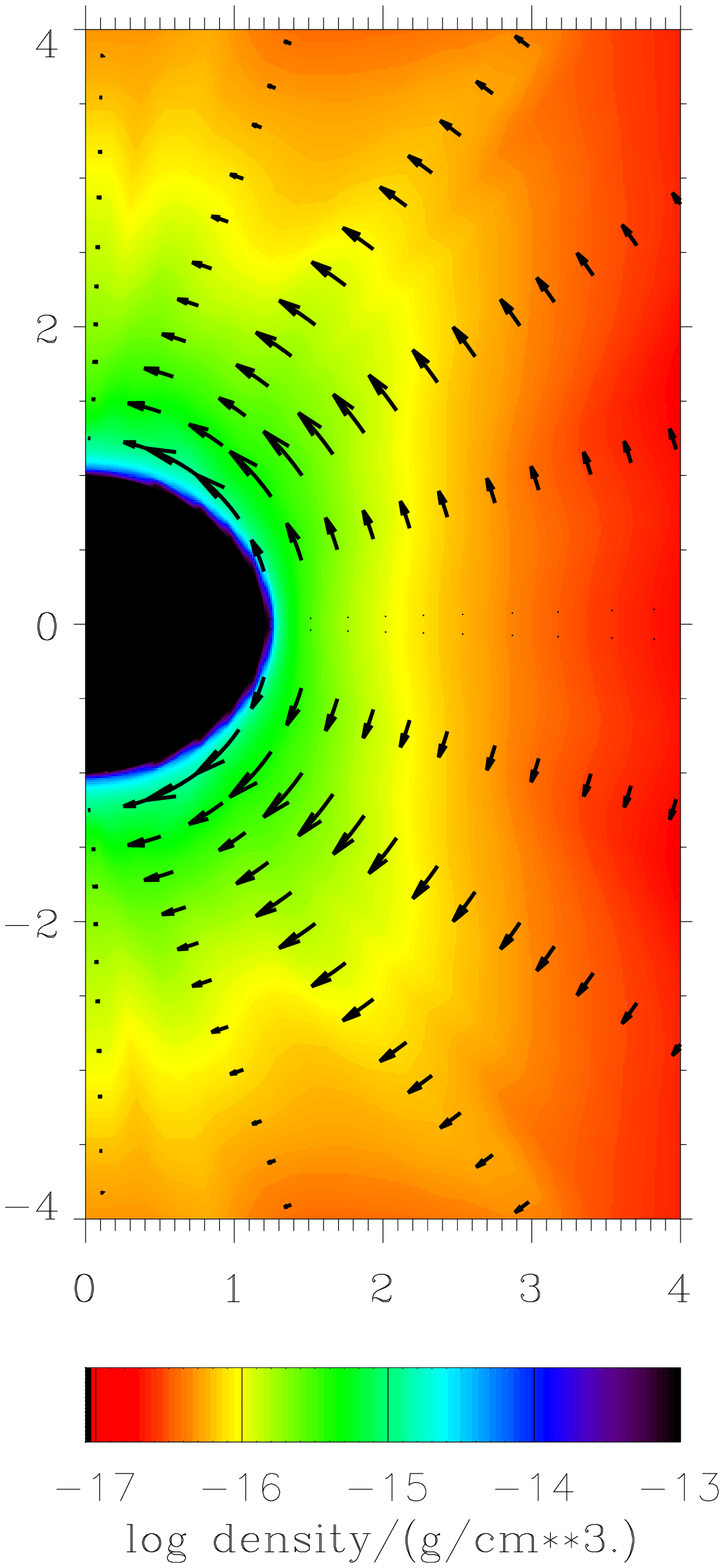}}
\end{minipage}
\vspace{0.5cm}
\caption{Density and polar velocity of radiation driven wind models 
for a rapidly rotating B-dwarf, using
different assumptions. The rotation axis is directed along the ordinate.\newline 
Left: {\it radial} line force $\rightarrow$ wind compressed disk. 
Middle: {\it non-radial} line force, uniform brightness temperature. 
Right: {\it non-radial} line force $+$ gravity darkening.\newline
Model parameters (polar values): \Teff = 20~kK, \logg = 4.11, \Rstar = 4
\rsun, M = 7.5 \msun. \vrot = 350~\kms, \vesc(pole) = 843~\kms, 
\vesc(equator) = 585~\kms, \vcrit = 487~\kms (see Eq.~\ref{vcrit1}).
From \citet{Petrenz99}.} 
\label{2dmodels}
\end{figure*}

$\Phi'(r)$ may become large if either \vrot\ is large, \vinf($\theta_0$) is
small, the velocity field is flat ($\beta$ large) or the particles
start from equatorial regions. In any case, however, the particles move
towards the equator and, if $\Phi'(r)$ became $\pi/2$, would collide in
the equatorial plane with particles from the opposite hemisphere which have
started at similar (absolute) latitudes. Since the equatorwards directed
motion corresponds to a non-vanishing polar velocity component in the
stellar frame, and this component is small, but supersonic for large
$\Phi'$, a shock develops, and a ``wind-compressed disk (WCD)'' will be
created, confined by the ram pressure of the wind. The WCD model has been
confirmed in all aspects by numerical models using identical assumptions
\citep[ and Fig.~\ref{2dmodels}, left panel]{OCB94}. 

Even if the deflection in $\Phi'$ is too small to allow for the creation of a
disk, i.e., the equatorial plane cannot be reached by certain trajectories
(because of too low rotation or too steep a velocity field) the above scaling
relations (Eq.~\ref{mdotvinfwcd}) remain valid, and a non-spherical, {\it
oblate} wind is predicted, in this case denoted as a ``wind-compressed zone
(WCZ)'' model. For such models, \citet{PetrenzPuls96} have calculated \Ha\
profiles as a function of \vrot\ and \vsini, and concluded that standard 1-D
diagnostics might lead to severe overestimates of \mdot\, (50 to 70\%) in
those cases were $\vsini$ is small and the star is observed pole-on.

\paragraph{Non-radial line forces: the inhibition effect.~~} After
publication, the WCD model was regarded as a major step towards an
explanation of disk phenomena in hot massive stars (e.g., Be-stars) until 
it became clear that it was based on an insufficient assumption
\citep{Owocki96, Owocki98}. Since, in a more-D situation, the optical depth
depends on the {\it directional} derivative of the local velocity field (the
radial gradient \dvdr\ in Eq.~\ref{tausob} has to be replaced by $\vec n
\cdot \vec \nabla (\vec n \cdot \vec v)$), and the radiative acceleration of
strong lines is proportional to the integral over solid angle of this
derivative times direction $\vec n$, the line force does not only depend on
the radial velocity gradient, but also on other velocity gradients present
in the flow. Most importantly, the dependence of $v_r$ on polar angle
(increasing towards the pole, see Eq.~\ref{mdotvinfwcd}) leads to a small
but substantial polar acceleration, which is directed towards the pole as
long as $\partial v_r/\partial \theta <0$,\footnote{more precisely, if $(r
\frac{\partial}{\partial r} \frac{v_\theta}{r} + \frac{1}{r} \frac{\partial
v_r}{\partial \theta})~<~0$} i.e., as long as the equatorial outflow is
slower than the polar one, which is (almost) inevitable for a rotating
line-driven wind
(due to the decrease of the effective, centrifugally corrected gravity
towards the equator and $\vinf \propto \vesc$).
This polar component of the line force, 
\beq \frac{g_\theta^{\rm rad}}{g_r^{\rm rad}} \approx \frac{\alpha}{4}
\frac{\partial v_r/\partial \theta}{\partial v_r/\partial r}
\Bigl(\frac{\Rstar}{r}\Bigr)^2 \eeq 
is sufficient to stop the small, equatorwards directed $v_\theta$ component of the WCD
model and to induce a polewards directed velocity field,
as shown by \citet{Owocki98}. This is the so-called ``inhibition'' effect
\citep{Owocki96}, which can be seen in all numerical models including
non-radial line forces (e.g., \citealt{Owocki96} and Fig.~\ref{2dmodels},
middle panel).

\paragraph{Gravity darkening: The $\Omega\Gamma$ limit and prolate winds.~~} 

Because of the centrifugal forces, not only the wind is disturbed, but also
the stellar surface distorted. 
The corresponding shape can be calculated from a Roche
potential\footnote{First observations of such a surface distortion have been
obtained by \citet{deSouza03} for the rapidly rotating Be-star $\alpha$ Eri
with the VLTI, resulting in a ratio between equatorial and polar radius of
1.56$\pm$0.05. The shape of the observed distortion seems to indicate that
the conventional assumption of a uniform rotation might not apply for this
star.} (see \citealt{Collins63, Collins66}, and for applications
\citealt{CO95, PetrenzPuls96}). In this case, and with angular velocity
$\omega$, 
\beq
\Omega = \frac{\omega}{\omega_{\rm crit}}=
\frac{\vrot}{\vcrit}\frac{\Reqmax}{\Req(\Omega)},
\eeq
$\Omega$ becomes a function of equatorial radius $R_{\rm eq}$ (which itself
is a function of $\Omega$). $\Omega=1$ (i.e., ``total gravity'' = 0 at the
equator, see below) occurs when the equatorial radius has reached its
maximum extent, \Reqmax\ = 1.5 \Rp, with polar radius \Rp. 

The critical 
velocity, \vcrit, requires special attention, due to the problem of {\it
gravity darkening}. 
After some controversial discussions arising from the suggestion of an
$\Omega$ limit by \citet{Langer97a, Langer98} which has been criticized by
\citet{Glatzel98} (because of disregarding gravity darkening),
\citet{MaederVI} performed a detailed study on the issue. According to this
study, for not too large $\Gamma$ the critical speed has to be calculated
from the actual mass (independent from $\Gamma$) and from the maximum
equatorial radius, consistent with the arguments provided by Glatzel, 
\beq
\vcrito=\Bigl(\frac{GM}{\Reqmax}\Bigr)^{\half} =
\Bigl(\frac{2 GM}{3 \Rp}\Bigr)^{\half} \qquad \Gamma < 0.639.
\label{vcrit1}
\eeq
Note that this value is lower by a factor of $\sqrt(2/3)$ than in the
corresponding 1-D case, due to the difference of polar and equatorial radius, 

For such a distorted surface with latitude dependent {\it effective} gravity
(in the following defined as gravity corrected for centrifugal terms), one
has also to consider gravity darkening as a consequence of the von
Zeipel-theorem \citep{Zeipel24} which states that the surface flux of a
rotating star is proportional to the local effective
gravity.\footnote{Though the original version of this theorem was restricted
to the case of rotational laws that can be derived from a potential (e.g.,
cylindrical rotation), more recent investigations by \citet{MaederIV} have
proven that this theorem is also applicable (within a 10\% error) to the
case of ``shellular'' rotation ($\omega = \omega(r)$ for not too fast
rotation) which has been proposed as the relevant rotational law at the
surface levels of non-magnetic stars \citep{Zahn92}.} Thus, the effective
temperature depends on the latitude (in the example of Fig.~\ref{2dmodels},
\Teff\ (pole) = 21.3~kK and \Teff\ (equator) = 16.9~kK), which has to be
accounted for in the calculation of the occupation numbers (through the
radiation field). 

The principle presence of gravity darkening has been recently 
confirmed by \citet{Monier07}, who combined the NIR
radiation of the rapidly rotating A7V star Altair collected from four
telescopes of the CHARA interferometric array to obtain an intensity image
of the surface of this star. Interestingly, also their measurements
(remember the problem for $\alpha$ Eri) showed a somewhat different
temperature distribution compared to the standard approach assuming a
uniformly rotating star and the von Zeipel theorem. 

In any case, since the flux is modified by gravity darkening, 
also the $\Gamma$-term needs to be modified \citep{MaederIV, MaederVI}.
Within the standard approach and to a good approximation, it turns out that 
\beq
\Gamma_\Omega \approx
\frac{\Gamma}{\bigl(1-\frac{4}{9}\frac{\vrot^2}{\vcrito^2} f(\Omega)\bigr)},
\label{GammaOmega}
\eeq
if one expresses the ``total'' gravitational acceleration (centrifugally
corrected gravity, $g_{\rm eff}$, 
minus radiative acceleration) as $g_{\rm tot} = g_{\rm eff} (1-\Gamma_\Omega)$, to
preserve the formulations obtained so far. The function $f(\Omega)$ varies
from 1 at low \vrot\ to $0.813 \approx \sqrt(2/3)$ at critical rotation, and
can be neglected in most cases. For large $\Gamma$ then, the critical speed 
needs to be redefined, and can be approximated (for exact
expressions and details, see \citealt{MaederVI}) by
\beq
\vcritt \approx \vcrito \, \bigl(\frac{3}{2}\bigr)^{1.25}(1-\Gamma)^{\half} \qquad \Gamma > 0.639.
\label{vcrit2}
\eeq
which results from the condition $g_{\rm tot} = 0$ when $\Gamma_\Omega =1$,
the so-called $\Omega \Gamma$ limit (\citealt{MaederVI}): For large
$\Gamma$, the critical velocity is significantly reduced by the proximity to
the Eddington-limit. Thus, there is indeed a lower critical speed (as
suggested by \citealt{Langer97a, Langer98}), at least when $\Gamma >0.64$,
but it is important to emphasize that the $\Omega \Gamma$ limit occurs over
all latitudes, not just near the equator, as assumed in the original idea of
the $\Omega$ limit. 

Note that the redefined $\Gamma_\Omega$-term needs to be considered in fast
rotating models close to the Eddington-limit, a problem which has been
neglected in all hydrodynamical {\it wind} calculations performed so far.
In the present context, however, the most important effect of gravity
darkening concerns the line force, due to the modified illumination of the
lines. In a rotating wind illuminated by a uniformly bright disk, the
mass flux increases from pole to equator according to
Eq.~\ref{mdotvinfwcd}.\footnote{Note that the inhibition effect does not
change this basic result, since it affects ``only'' the small polar velocity
component and thus the disk formation.} The inclusion of gravity darkening
into the luminosity dependence of the mass-loss rate (Eq.~\ref{eq_mdot}),
\[
L^{1/\alpha'}
\rarrow (F(\theta) \Rstar^2(\theta))^{1/\alpha'} \stackrel{\mbox{\small von
Zeipel}}{\propto} (g_{\rm ef\/f}(\theta)
\Rstar^2(\theta))^{1/\alpha'},
\] 
thus leads to an {\it additional} dependence on $(1-\Omega^2 \sin^2
\theta)]^{1/\alpha'}$, such that in total
\beq
\mdot(\theta) \propto L(\theta)^{1/\alpha'} M_{\rm eff}(\theta)^{1-1/\alpha'}
\propto (1-\Omega^2 \sin^2 \theta)^{+1} \quad \mbox{\small (gravity darkening
included)}
\label{eq_mdot_dark}
\eeq
the situation becomes reversed and the mass flux is predicted to increase
from equator to pole, in parallel with the corresponding terminal velocity, 
inducing a {\it prolate} wind structure (cf.
\citealt{Owocki98, Owocki98bmw, Pelupessy00} and Fig.~\ref{2dmodels}, right 
panel). The present-day stellar wind of $\eta$ Car has indeed been found to 
be elongated \citep{Smith03shape, Boekel03} with a position angle equal to
that of its ejecta. 

Though the predicted angular dependence of the mass-loss rate is not
affected by the modified $\Gamma_\Omega$-term (except for its influence on
$\Omega$ via $\vcrit$), this modification plays an important role when
comparing the total, polar-angle integrated mass-loss rate with mass-loss
rates from non-rotating models. To a similar degree of precision as valid
for Eq.~\ref{GammaOmega}, one obtains \citep{MaederVI}
\beq
\frac{\mdot(\vrot)}{\mdot(0)} = \frac{(1-\Gamma)^{1/\alpha'-1}}
{(1 - \frac{4}{9}\frac{\vrot^2}{\vcrito^2} -\Gamma)^{1/\alpha'-1}}.
\label{mdotrot}
\eeq
Thus, no dramatic influence of rotation on the {\it total} mass loss is to
be expected in most cases.
\footnote{The formal divergence at the $\Omega \Gamma$ limit is artificial,
since in this case the material can no longer be lifted from the
gravitational potential.}  
 
\begin{figure}
\centerline{\includegraphics[width=8cm]{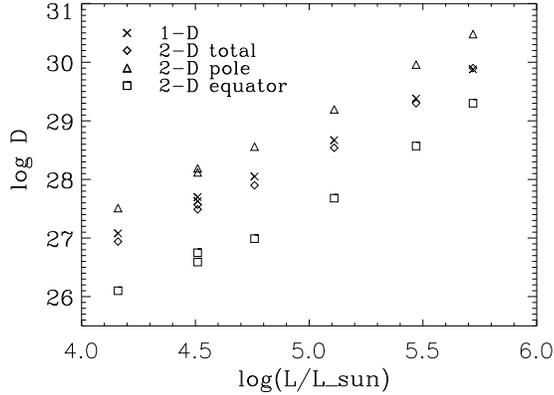}}
\caption{Predicted WLR for rapidly rotating B-type stars (\vrot = 0.85 \vcrit) with
\Teff = 20~kK and different radii. Compared are the modified wind-momenta
\Dmom\ calculated from total (diamonds), equatorial (squares) and polar
(triangles) mass-loss rates and average terminal velocities resulting from 2-D, NLTE simulations, and from
corresponding 1-D models with $\vrot=0$ (crosses). From
\citet{PetrenzPuls00}.} 
\label{wlr_rot}
\end{figure}

\paragraph{2-D NLTE models and the WLR of rapidly rotating stars.~~}
Strictly speaking, the scaling relations derived in the previous sections
are only valid if the line-statistics remains unaffected from the modified
wind structure. In order to investigate the influence of density, velocity and
radiation field on the occupation numbers and line-forces as a function of
$(r, \theta)$, \citet{PetrenzPuls00} calculated self-consistent, 2-D, NLTE
models for rapidly rotating B-star winds\footnote 
{computational feasible at that time because they applied a Sobolev line
transfer and assumed an optically thin continuum.}, 
to investigate a spectral regime where
the ionization structure is most sensitive to local conditions and
variations of the radiation field. For all models, a {\it prolate} wind
structure has been found, in accordance with the principal predictions from above. 

As obvious from Fig.~\ref{wlr_rot}, the {\it total} wind momenta are hardly
affected by rotation, since the increase in \mdot, Eq.~\ref{mdotrot}, is
compensated by a decrease in the average \vinf. Of course, the wind-momenta 
differ significantly when observed either pole or equator on. This problem
can be mitigated when \mdot\, diagnostics is used which scans mostly the
lower wind region, e.g., \Ha, since in these regions the density contrast
between polar and equatorial regions is lower. Consequently, WLRs derived
from 1-D analyses should remain rather unaffected from deviations from
spherical symmetry, when samples of significant size are used, and objects
with low \vsini\ (likely to be observed pole-on) are avoided. The scatter in
the WLR increases due to rotational effects, of course.

\paragraph{B[e]-supergiants and the bi-stability.~~} B[e] supergiants (for
an improved classification scheme, see \citealt{Lamers98}) show a hybrid
spectrum (narrow Balmer lines and low ionization metal lines in emission,
but also UV and Balmer P~Cygni lines and a strong IR excess), which has been
explained by an outflowing disk with high density, low velocity and low
ionization and a fast polar wind with low density and high ionization
\citep{Zickgraf86, Zickgraf89}.

\begin{figure*}
\begin{minipage}{5.8cm}
\resizebox{\hsize}{!}
   {\includegraphics{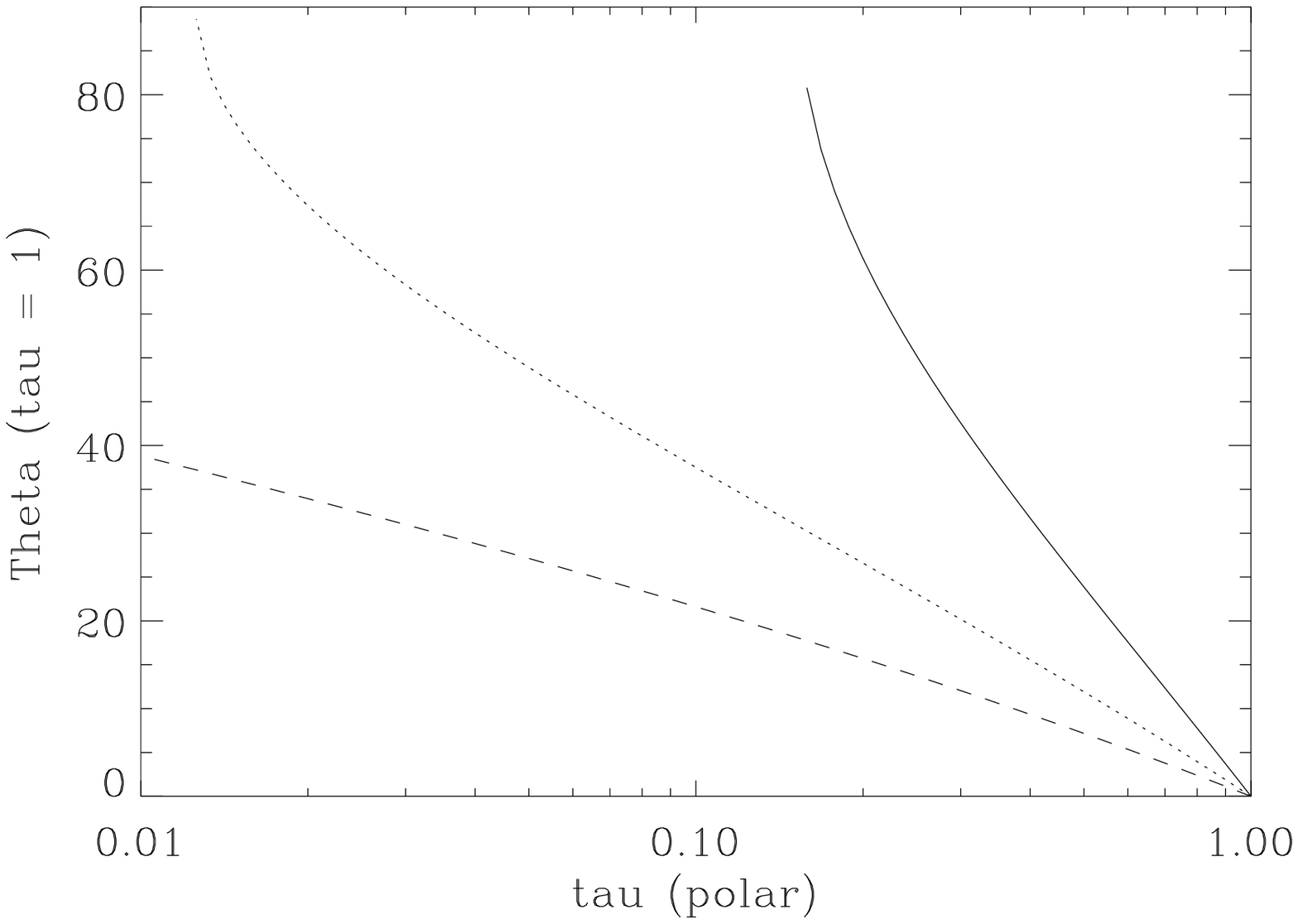}}
\end{minipage}
\hfill
\begin{minipage}{5.8cm}
   \resizebox{\hsize}{!}
   {\includegraphics{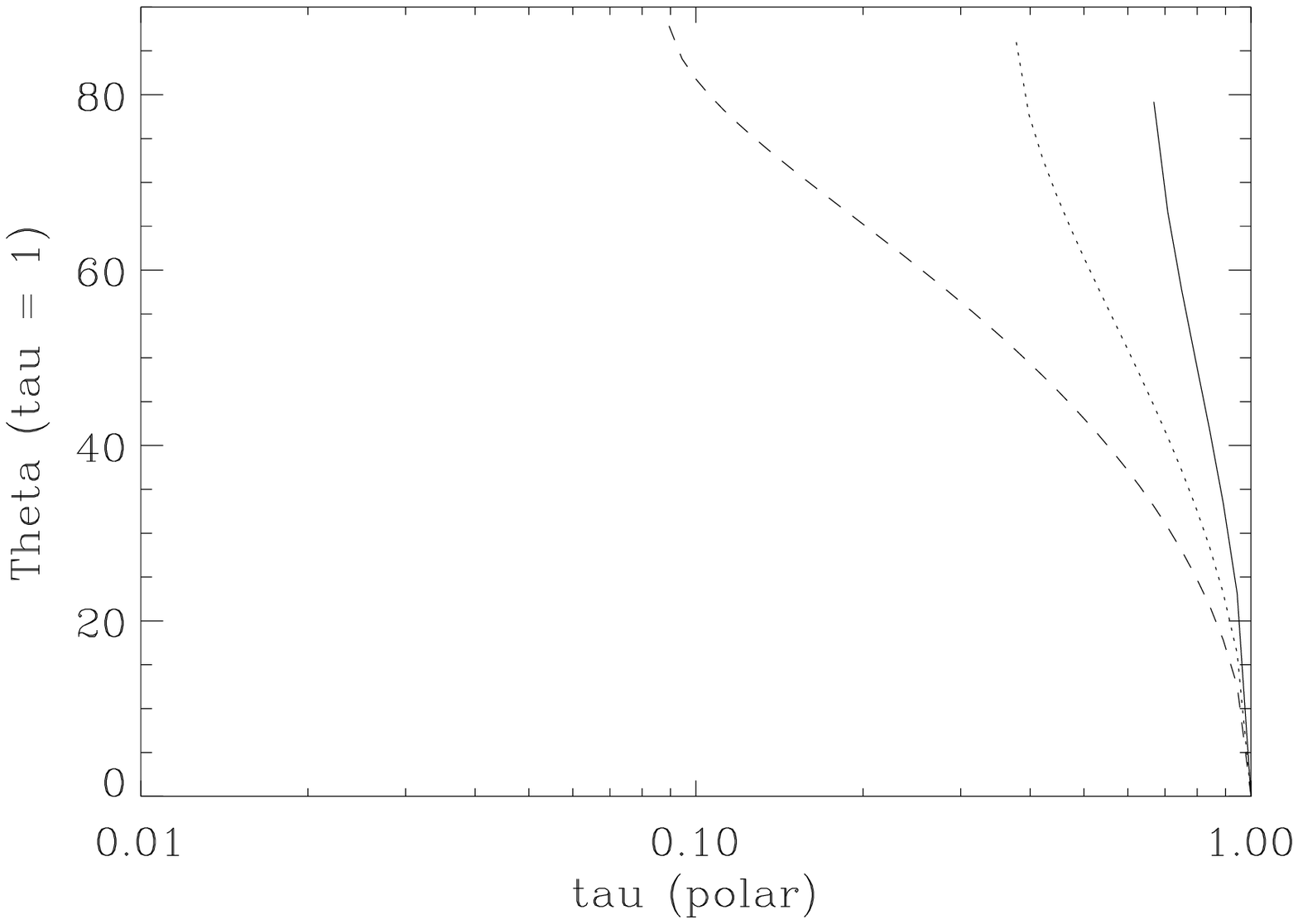}}
\end{minipage}
\caption{Onset of bi-stability (co-latitude $\theta$ where $\tau_{\rm Ly} =
1$ is reached) as a function of $\tau_{\rm Ly}$(polar) and three rotational
speeds, $\Omega = 0.5, 0.7, 0.9$ (solid, dotted, dashed), for a B-supergiant
with \Teff = 20~kK. Optical depth of Lyman continuum calculated according to
\citet{LP91}. Left: original version. Right: mass-loss rate calculated 
including gravity darkening. The formation of a disk requires much larger
polar mass loss and rotation than in the original version, see text.} 
\label{bedisk}
\end{figure*}

The origin of this phenomenon has been suggested by \citet[ see also
\citealt{Lamers99}]{LP91} to rely on a combination of rotation and
bi-stability. Since, for a rotating wind, \mdot\ was thought to increase and
\vinf\ to decrease towards the equator (the impact of gravity darkening 
on \mdot\ was
yet unknown), the optical depth of the Lyman continuum
\beq
\tau_{\rm Ly} = f\bigl[(\frac{\mdot}{\vinf})^2,T_{\rm rad}\bigr]
\eeq
becomes a strongly increasing function of $\theta$, particularly if
including the von Zeipel theorem when calculating $T_{\rm rad}(\theta)$
which establishes the ionization equilibrium.\footnote{In a {\it
consistent} calculation, the von Zeipel theorem has be employed both for the
NLTE-aspect (occupation numbers) and for the flux-dependent
line-acceleration.}
\citet{LP91} were able to parameterize the optical depth as a function of
latitude-dependent gravity, $\tau_{\rm Ly} \propto$ $(g_{\rm pole}/g(\theta))^{6.5}$.
Because of the dependence of the bi-stability jump on this optical depth,
the latitude where $\tau_{\rm Ly}$ reaches unity separates two regions, a
polar one with low density and high velocity (predominantly driven by
\FeIV) and an equatorial one with high density and low velocity, driven by
\FeIII\, (see Sect.~\ref{sec:standardmodel}). The contrast between both
regions results in an equatorial outflowing disk. From Fig.~\ref{bedisk},
left panel, one can see that even a thin polar wind with
$\tau_{\rm Ly} = 0.01$ (corresponding to \mdot = 0.35~\Mdu) can switch almost
abruptly to $\tau_{\rm Ly} >1$ at $\theta \approx 40^{\circ}$ for
$\Omega = 0.9$ in the outlined scenario.

After the impact of gravity darkening on the mass-loss rate had been
realized, \citet{Owocki98} pointed out that the creation of B[e] star disks
is much more difficult than previously thought since \mdot\ decreases 
equatorwards (Eq.~\ref{eq_mdot_dark}), in contrast to the original approach.
Thus, the increase of $\tau_{\rm Ly}$ towards equatorial regions is much
weaker, $\tau_{\rm Ly} \propto (g_{\rm pole}/g(\theta))^{1.5}$, and even for
highest rotational speeds a disk formation becomes impossible as long as the
polar wind is not considerably dense (Fig.~\ref{bedisk}, right panel). 

Based on their previous investigations, \citet{Cure05} suggest an
alternative possibility to explain the B[e] phenomenon. They propose that
for {\it near-critical} rotation the wind can switch to a slower,
shallow-acceleration solution which can lead to a density enhancement and,
again in combination with the bi-stability mechanism, to the formation of a
dense equatorial disk. From analytic considerations and (1-D) time-dependent
hydrodynamics, \citet{Madura07} confirmed most of these results, in
particular the slow acceleration for $\Omega > 0.85$ and a corresponding
increase of density. However, they argue that an inclusion of gravity
darkening and 2-D flow effects still renders the formation of dense disks as
unlikely.

Let us also note that \citet{Zsargo08a} have identified an additional
problem. From 2-D, NLTE calculations by means of their axial-symmetric, NLTE
code {\sc astroroth} they concluded that hydrogen recombination in the 
disks of B[e]-supergiants is almost
impossible, due to strong ionization from excited levels. Thus, the chance
that $\tau_{\rm Ly}$ becomes much larger than unity and enables line driving
by \FeIII\ is rather low, even if the problem of gravity darkening was not
present. 

On the other hand, \citet{Pelupessy00} solved the momentum equation for
sectorial line-driven winds from rotating stars including both the stellar
oblateness and gravity darkening, as well as force-multiplier parameters
specifically computed for the bi-stability regime. In their computations
for stellar parameters far from the bi-stability limit, they found the
mass-loss rate and the terminal wind velocity to increase from the equator
to the pole 
(thus confirming the reliability of their approach),
however when the counteracting effect of wind bistability was
included, they were able to produce a density contrast between the equator
and the pole of a factor 10 for a relatively modest rotational velocity of
$\Omega$ = 0.6. One of the attractions of the scenario remains that it may
naturally account for the expectation to find outflowing disks only around
B-type stars. 

From the above discussion it should be evident that more work (in 2-D) is
needed to prove whether or not the bistability mechanism is the root cause
for the existence of disks around B[e]-supergiants.

\subsubsection{Winds and magnetic fields}
\label{sec:mag}

In contrast to the Sun and other cool stars where magnetic fields are
thought to be generated through a dynamo mechanism related to convective
motions in the H/He recombination zones, hot stars have been considered to
be free of strong, {\it dynamo-created} 
surface magnetic fields until
recently, because of the absence of {\it strong} outer convection zones (H
remains completely ionized even throughout the photosphere). There are,
however, several other possibilities to generate surface magnetic fields,
(i) either related to a thin convection zone due to recombination of \HeIII;
(ii) due to the strongly convective cores of massive stars which might give
rise to dynamo-generated flux tubes diffusing to the surface
(\citealt{CG00}, but see \citealt{MacMullan04}; (iii) by a dynamo that is
operating in shear-unstable gas in the radiative stellar envelope
\citep{MacMullan04}; (iv) formed already in an early convective phase during
stellar formation; (v) or even through compression of interstellar magnetic
flux during the initial collapse. It has been shown by \citet{Moss01,
BraithwaiteSpruit04} and \citet{BraithwaiteNordlund06} that a dynamically
stable configuration of such {\it fossil} fields on long time-scales is
actually possible.

Another promising mechanism is the Tayler-Spruit dynamo (\citealt{Spruit99,
Spruit02} and generalizations by \citealt{MM04}; see also
\citealt{MullanMac05}) which does not need a convection zone but is based on
a strong instability for generating magnetic fields in radiative layers of
differentially rotating stars. This dynamo has been incorporated into recent
stellar structure/evolution calculations \citep{MM03B, MM04, MM05, Heger05,
Brott08} to understand the internal angular momentum transport (though the
induced mixing processes might be too strong, e.g., \citealt{Brott08}).
Note, however, that this dynamo is able to create magnetic fields (in the
azimuthal direction, of the order of a few $10^4$~Gauss (G)) only outside
the stellar core and {\it not} in the outermost layers of the star (e.g.,
\citealt{MM05}).

In recent years, considerable effort has been spent to measure {\it surface}
magnetic fields, by means of spectropolarimetry (exploiting the circular
polarization of the Zeeman components) and a so-called least square
deconvolution \citep{Semel89, Donati97} which allows to measure the
longitudinal component of the $B$-field averaged over the stellar disk.
Though large samples of OB-stars ($>$ 45) have been analyzed (see the
material presented by \citealt{Schnerr07} and references therein and
\citealt{Petit08}), significant field strengths above the present detection
threshold (1-$\sigma$: 40-100~G) were detected for only ten stars 
(excluding chemically peculiar Ap/Bp stars), where in most cases the
observations could be fitted by an oblique dipole (but see
\citealt{Donati06} for the complex magnetic configuration in $\tau$~Sco).
Among these ten stars, there are three OB-stars with well known
peculiarities in their spectra ($\theta^1$~Ori~C (O4-6V), HD\,192612
(O6.5f?pe-O8fp) and $\tau$~Sco (B0.2V)), with polar field-strengths from
500-1500~G (assuming a dipole), and three $\beta$~Cep stars which
incidentally were found to be nitrogen rich \citep{Morel06}.
\citet{Verdugo03} report the non-detection of surface magnetic fields in a
sample of 12 A-supergiants. \citet{Donati06} conclude that in {\it
non-peculiar} hot stars, $B$ is weak and/or acts on small scales (spots?).

As should be clear from these introductory remarks, magnetic fields might be
large in the stellar interior, but only few ``normal'' OB-stars show
indications of strong {\it surface} fields, at least on larger scales.  Even
weak fields, however, can affect the wind, as we will see in the following.
We will report on recent progress on magnetic winds, at first for a
simplified case (no rotation) and then for the problem of field aligned
rotation and the limiting case of an {\it oblique} rotator with very high
field-strength.

\paragraph{Magnetically channeled line-driven winds - The confinement
parameter.~~}

In order to explain the observed X-ray emission from the A0p star IQ Aur,
\citet{Babel97a} suggested that the confinement of the wind by
the strong magnetic field leads to a collision from the wind components
of the two hemispheres in the closed magnetosphere and thus to the formation of
a strong shock. Subsequently, \citet{Babel97b} used this ``magnetically
confined wind shock (MCWS) model to explain the periodic variations of the
X-ray emission from the O-star $\theta^1$~Ori~C (see above),
requiring a polar field strength at the stellar surface of $B_{\ast} \approx$
300 G (present knowledge: $\approx$ 1100 G, \citealt{Donati02}), where the
variability should be caused by a circumstellar disk predicted by
the MCWS model. In these investigations, a prescribed, {\it fixed} 
magnetic field configuration had been used.

This approximation has been relaxed by \citet[ see also \citealt{OuD04} for a
comprehensive analytic description]{udDoula02}, who investigated the {\it
interaction} between $B$-fields and winds in detail, based on 2-D
time-dependent magneto-hydrodynamics for the simplified case of no rotation
and a purely radial line-force. Thus, the external forces in the momentum
equation are (effective) gravity, line-acceleration and
Lorentz-acceleration, $(4 \pi \rho)^{-1} (\vec \nabla \times \vec B) \times
\vec B$, and one has to solve for the $B$-field\footnote{$\vec \nabla \cdot
\vec B = 0$ and $\partial \vec B/\partial t = \vec \nabla \times (\vec v
\times \vec B)$ in the MHD approximation of infinite conductivity.} in
parallel with the wind-dynamics. In the following, we will summarize the
major findings by \citet{udDoula02} and subsequent work.

As it turned out, the most important parameter controlling the structure of
the modified wind, $\eta_*$ (see below), is related to the ratio of
magnetic to wind energy, 
\beq
\eta (r,\theta ) = :\frac{E_{\rm B}}{E_{\rm wind}} \approx 
\frac{B^2 /8\pi }{\rho v^2 /2} = \frac{B^2 (\Rstar, \theta )\Rstar^2 }
{\mdot \vinf}\;\frac{(r/\Rstar)^{2 - 2q}}{(1 - \Rstar/r)^\beta},
\eeq
assuming a $\beta$-velocity field for the wind and $B(r) \propto
(\Rstar/r)^q$ (e.g., $q = 3$ for a dipole field). The scaling factor in
front of the last expression, evaluated at the magnetic equator under the
assumption of a dipole configuration
($B_{\rm{dipol}}(\Rstar,\theta) = B_{\rm p} (\cos^2 \theta +
1/4\sin^2 \theta)^{1/2}$) has been denoted as the ``confinement parameter'',
\beq
\eta_* = \frac{B^2(\theta  = 90^\circ)\Rstar^2}{\mdot \vinf} = 
\frac{(B_{\rm p}/2)^2 \Rstar^2}{\mdot \vinf} \approx 0.19 
\frac{B_{100}^2 R_{10}^2}{\dot M_{-6} \vinfeight }. 
\eeq
In the latter expression, the polar field strength is measured in units of
100~G, the stellar radius in units of 10~\rsun, the mass-loss rate in units
of \mdu\ and the terminal velocity in units of 1000~\kms. 

As we will see below, a value of $\eta_* = 1$ relates to the onset of larger
structural modifications due to $B$. To reach this, e.g., for the
prototypical O4(f) supergiant $\zeta$~Pup ($R_{10} \approx 2, \dot M_{-6}
\approx 4, v_8 = 2$), a field of $B_{\rm p} \approx 320\,{\rm G}$ would be
required, whereas for the sun (where the magnetic field has a dominating
influence on the wind) one finds (with $\dot M_{-6} \approx 10^{-8}, v_8 = 0.5,
B_{\rm p} \approx 1\,{\rm G})$ a considerable value for the confinement
parameter, $\eta_*(\odot) \approx 40$.  

For so-called weak winds (Sect.~\ref{sec:weakwinds}), on the other hand, a
magnetic field of only $B_{\rm p} \le 30\,{\rm G}$ (well below present
detection limits) is needed to reach $\eta_* = 1$ (assuming $\mdot \le
10^{-8}$~\msun yr$^{-1}$ and \vinf = 2000~\kms). We note that $\tau$ Sco
(a prototypical weak-winded star) reaches $\eta_* \approx 80$
for the measured field strength of 500~G \citep{Donati06}, i.e., extreme
effects are to be expected.

The reason to denote $\eta_*$ as a ``confinement'' parameter relates to the
Alfv\'en radius of the wind. Since MHD waves propagate with the Alfv\'en speed,
\beq
v_{\rm A} = \frac{B}{(4\pi\rho)^{1/2}} \quad \Rightarrow \quad M_{\rm A} =
\frac{v}{v_{\rm A}} = \frac{1}{\sqrt \eta}
\eeq
the Alfv\'en radius can be calculated from the corresponding Mach number
$M_{\rm A}(R_{\rm A}) = 1$, i.e., from the condition $\eta(r, \theta) = 1$.
For a dipole field, one obtains 
\beq
R_{\rm A} \rarrow \Rstar \quad (\eta_*  \ll 1), \quad
R_{\rm A} \approx \eta_*^{0.25} \Rstar \quad (\eta_* \gg 1).
\eeq
Since the Alfv\'en radius corresponds roughly to the maximum radius of closed
loops (see the discussion in \citealt{udDoula02}), the confinement parameter
controls this size, i.e., controls whether the wind is {\it confined} in
such loops.

\paragraph{Wind-structure as a function of $\eta_*$.~~} In their
calculations, \citet{udDoula02} derived the wind-structure from evolving
a model with initial conditions of a dipole field superposed upon a
spherically symmetric wind. In dependence of the confinement parameter, 
three exemplary topologies have been found:

\noindent 
(i) For moderately small confinement, $\eta_*= 0.1$, the surface magnetic
fields are extended by the wind into an open, nearly radial configuration.
There is still a noticeable global influence from $B$ on the wind, enhancing
the density and decreasing the flow speed near the magnetic equator.

\noindent 
(ii) At intermediate confinement, $\eta_*= 1$, the field lines are still
opened by the wind, but retain near the surface a significant tilt
which channels the flow towards the magnetic equator. The polar velocity
components can become significant, of the order of a few 100~\kms, and a thin
disk begins to develop.

\noindent (iii) For strong confinement, $\eta_*= 10$, the field remains
closed in loops near the equatorial surface, and the wind is accelerated and
{\it channeled} upwards from footpoints of opposite polarity. These flows
collide near the loop tops, with shock velocity jumps of up to 1000~\kms,
leading to hard X-ray emission in the keV band. These findings are in
qualitative agreement with the MCWS model by \citet{Babel97a}, and also in
more quantitative agreement with X-ray observations for $\theta^1$~Ori~C
(\citealt{Gagne05} and Sect.~\ref{sec:xraylinesmagnetic}).
Since the shocked material at the loop tops is very
dense, its support by magnetic tension becomes unstable, leading to complex
infall patterns along the loop lines down to the star.
Even for large $\eta_*$, the faster
radial decline of the magnetic energy (compared to the kinetic energy of the
wind) leads to a final dominance of flow effects, and the field lines are extended
into an open configuration at larger radii. 

Globally, the mass flux of the {\it outer} wind increases towards the
equator due to the tendency of the field to divert the flow into this
direction, whilst the {\it base} mass flux in equatorial regions (and beyond
for large $\eta_*$) is lower than the field-free case (see \citealt{OuD04}
for further details). The terminal velocity, on the other hand, is larger
than for a non-magnetic wind at all latitudes except for the equatorial
plane, where it becomes significantly reduced due to the higher mass flux.
Similar to the case of rotating winds, the total mass flux is barely
affected as long a $\eta_* \la 10$. For further scaling relations, see
\citet{udDoula08}.

In summary, non-rotating magnetic winds develop a geometrically thin, slow
and dense disk, superimposed by a fast, thin polar wind, quite different
from rotating non-magnetic winds. Non-radial line forces should be
irrelevant since the polar velocities are much larger than in the WCD case.
X-rays are to be expected from the channeled flows colliding at loop tops 
and from shocks neighboring the compressed disk. Finally, it might be
speculated whether {\it oblique} magnetic rotators can explain part of the
observed UV variability (Sect.~\ref{sec:inhomoobs}) and induce CIRs
(Sect.~\ref{sec:cirs}), because of the large density/velocity contrast which
would occur between the magnetic and rotational equator in such stars.

\paragraph{Field aligned rotation and the rigidly rotating magnetosphere
model.~~} In a follow-up study, \citet[ see also
\citealt{Owocki05}]{udDoula08} included rotation (aligned with the magnetic
field to avoid 3-D calculations) and investigated the question whether
magnetic fields could spin up the stellar wind outflow into a ``magnetically
torqued disk'', as suggested by \citet{Cassinelli02}. The closed loops
present for larger $\eta_*$ tend to keep the outflow in {\it rigid body}
rotation (i.e., $v_\phi(r) \propto r$), and it might be possible to propel
material into a Keplerian disk if the Alfv\'en radius is somewhat larger
than the Keplerian co-rotation radius, 
$R_{\rm K} \approx (\vrot/v_{\rm orb})^{-2/3} \Rstar$, where rigid-body
rotation would yield a balance between centrifugal and gravitational
acceleration. $v_{\rm orb} = \sqrt{GM/\Rstar}$ is the orbital speed near the
equatorial surface.\footnote{This approximation for $R_{\rm K}$ assumes the
star to be far away from the $\Omega \Gamma$ limit.} 
Note that strong
confinement {\it and} rapid rotation are required to obtain $R_{\rm A} >
R_{\rm K}$. Although various combinations of rotational speeds and magnetic
field strengths have been considered, and the analysis has been extended to
$\eta_* =100$, no stable {\it Keplerian} disk has been found to be formed in
any of these cases, simply because most of the material does not have the
appropriate velocity for a stationary orbit. However, there were clear
indications of a "quasi-steady" rigid-body disk, as discussed further below.
Note also that in the opposite case of slow rotation (more precisely, for
$R_{\rm A} < R_{\rm K}$), the basic results as obtained for the non-rotating
case remain valid.

For the limiting case of very high field strength, $\eta_* \rarrow \infty$,
\citet{Townsend05a} developed the alternative ``rigidly rotating
magnetosphere'' model, originally designed to investigate the prototypical
Bp star $\sigma$~Ori~E with $B \sim 10^4$~G. In this model, the field lines
behave like rigid tubes, and the outflowing material is constrained along
corresponding trajectories fixed by the prescribed field geometry (see also
the rigid field hydrodynamics approach by \citealt{Townsend07}).

For sufficient rotation then, minima in the effective potential along each
field line should gradually fill with plasma, forming an almost steady, 
co-rotating magnetosphere. For the oblique rotator $\sigma$~Ori~E, this
should occur in the form of two co-rotating clouds. Time-series of synthetic
\Ha\ profiles for such a geometry are consistent with the observed behaviour
\citep{Townsend05b}. As mentioned above, the time-dependent calculations for
large $\eta_*$ by \citet{udDoula08} have confirmed the basic picture of a
rigid-body disk (for aligned rotation), but with a complex dynamic behaviour
with infall and outflow limiting the growth of the disk. E.g., the eventual
centrifugal breakout of such material might suggest a new heating mechanism
(via magnetic reconnection) to explain the hard X-ray flares, as observed in
Bp-stars \citep{GrooteSchmitt04} and simulated by \citet{udDoula06}.

\section{Stationary models of radiation driven winds}
\label{sec:statmodels}

\subsection{Predictions from the improved CAK approach}
\label{sec:mcakmodels}

As was noted earlier, the {\it original} CAK approach to compute the wind
structure has been massively improved \citep{FA86, PPK, Pauldrach87,
Pauldrach94, pauldrach01}. Whilst the CAK models were computed in local
thermodynamical equilibrium (LTE), departures from LTE have been included
because of the strong influence of the radiation field and the low densities
in the atmospheres of hot massive stars. Another significant improvement
involved the inclusion of millions of lines (predominantly from iron-group
elements) from some 150 different ions -- in contrast to the single \CIII\
list used in the original work of CAK. Within the current state-of-the-art
modified CAK modeling, the rate equations, radiative transfer equations,
energy equation, and approximate hydrodynamic equations, using
force-multiplier parameters, are all accounted for. More simplified models
relying on an approximate solution for the rate-equations and an analytic
approach to solve the hydrodynamic equations (with consistent, depth
dependent force-multiplier parameters) have been presented by \citet{Kud02}.

A complementary approach currently in use concerns the hydrodynamical method
of \citet{Krticka06}.  As in the models by Pauldrach and co-workers
(``WM-basic'', see Sect.~\ref{sec:obastars}), the various equations are
consistently solved, but the concept of force multipliers is dropped, 
though the Sobolev approximation is still applied. The major advantage of
this method is that it allows for a multi-component description of the
hydrodynamical equations, which enables an investigation of potential
metal-ion decoupling from the H/He plasma. A drawback of the
\citet{Krticka06} models is that line-blocking and multi-line effects are
not accounted for -- limiting their applicability to weaker winds.

\subsection{Predictions using a Monte Carlo radiative transfer approach}

A different approach involves the ``Monte Carlo'' method developed by
\citet{AL85}, where photons are emitted at an inner boundary, and their
scattering history is tracked on their journey outwards. At every
interaction some amount of momentum and energy is transferred from the
photon to the ion under consideration. From a macroscopic point of view, the
overall integrated mass-loss rate is obtained from global energy
conservation:
\begin{equation}
\frac{1}{2} \mdot (\vinf^2 + \vesc^2) = \Delta L
\label{energy}
\end{equation}
where $\Delta L$ is the total amount of energy that has been transferred per
second from the radiation field to the scattering ions in the wind. The main
drawback of this approach is that one first needs to establish an
appropriate velocity law, $v(r)$, generally using an empirically
pre-determined $\vinf$, although in principle it is possible to iterate and
obtain dynamically consistent mass-loss rates (see
\citealt{SpringmannPuls98, Vink99}).

One of the major advantages of the Monte Carlo method is that it easily
allows for multi-line scattering, which becomes important in denser winds,
as already outlined by \citet[but see also \citealt{Puls87}]{FC83}. Before
the year 2000, theoretical mass-loss rates fell short of the observed rates
for dense O star and WR winds, whilst for weak winds the oft-used single
line approach overestimated the mass-loss rates considerably. Although early
Monte Carlo simulations (e.g. \citealt{AL85}) only treated line {\it
scattering}, this approximation has been relaxed in the mass-loss rate
determinations by \citet{Vink00,Vink01}, although further improvements to
the method, e.g., in terms of line branching \citep{Sim04} and wind clumping
need to be considered in future model generations.

Here we wish to mention that the high mass-loss rates derived for dense
winds when accounting for multi-line scattering (as in the Monte Carlo
method) do not violate the principle of momentum conservation. An appealing
picture is that of ``hemispheric scattering'' where a single photon can be
launched in one direction, transfer its momentum to an ion, and travels
towards the opposite hemisphere -- repeating the same process over and over
again, thereby pumping up the momentum transfer in the wind. One may note
that all what really happens is that at each iteration the photon is
slightly red-shifted, until it eventually runs out of energy. In reality, we
know that hemispheric scattering will not occur. It has been demonstrated
(e.g., \citealt{Springmann94, Gayley95a}) that when one employs realistic
line lists, photons follow paths that are less extreme, but the essential
point is that multiply scattered photons are always adding radially outward
momentum to the wind, and the wind momentum may easily exceed the so-called
single-scattering limit, i.e., $\mdot \vinf/(L/c)$ can become larger than
unity.
 
\citet{Vink00,Vink01} used the Monte Carlo method over a wide range of
stellar parameters and derived a ``mass-loss recipe'' using multiple linear
regressions to their model results. On the hot side of the {\it predicted}
bi-stability jump at $\sim 25~000$ K (but see Sect.~\ref{sec:obastars}), the
rates of OB supergiants should roughly scale as
\begin{equation}
\mdot~\propto~L^{2.2}~M^{-1.3}~\Teff~(\ratio)^{-1.3}.
\label{eq_formula}
\end{equation}
when $L$ is measured in $10^5~\lsun$, $M$ in $30~\msun$ and \Teff\ in 40~kK.
This relation predicts that the mass-loss rates should strongly depend on
the luminosity ($L^{2.2}$), steeper than \mdot\ $\propto$ $L^{1.6}$ which is
often quoted and relies on the simplified scaling relation Eq.~\ref{eq_mdot}
with $\alpha' \approx 0.6$.  The reason is that the efficiency of multi-line
effects accounted for in Monte Carlo simulations increases with increasing
wind density, i.e., for more luminous stars, whereas in (modified) CAK
models these effects are normally neglected, particularly in the scaling
relations presented in Sect.~\ref{sec:standardmodel}. Because the mass-loss
rate {\it also} scales with stellar mass as $M^{-1.3}$, and applying a
typical $M-L$ ratio of $L$ $\propto$ $M^x$ with $x$ = 2 - 3 for massive
stars, one obtains an overall $\mdot$ scaling as $L^{1.55-1.77}$,
in agreement with observational findings.

\begin{figure}
\begin{center}
   {\includegraphics[width=8cm]{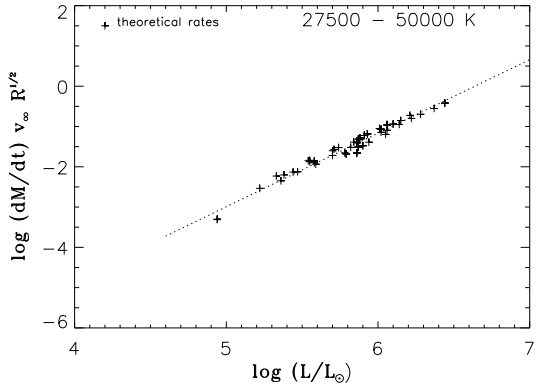}}
\end{center}
\caption{Predicted WLR for O stars hotter than 27~kK for 
a range of $(L,M)$-combinations in the upper HRD. From \citet{Vink00}.}
\label{wlr_vink}
\end{figure}

The basic success of the Monte Carlo method is exemplified when comparing
Figs.~1 and ~3 from \citet{Vink06}, which display the degree of agreement 
between modified CAK models and observations on the one side and Monte Carlo
models and observations on the other: by properly including multiple
scatterings in the predictions, the results are equally successful for 
relatively weak (with \mdot\ $\sim$ $10^{-7}$ \msunyr) as well as denser
winds (with \mdot\ $\sim$ $10^{-5}$ \msunyr).  The predictions can be
conveniently expressed via the WLR.  For O-stars hotter than 27\,500 K, the
relation is shown in Fig.~\ref{wlr_vink} and given by Eq.~(\ref{wlr}) with
$x$ $=$ 1.83, corresponding to $\alpha'$ $=$ $0.55$.  For objects on the
cool side of the bi-stability jump, the mass-loss rate increases and the
slope of the WLR becomes $x$ $=$ 1.91, corresponding to $\alpha'$ $=$
$0.52$.

Monte Carlo predictions for hot stars of lower mass were computed by
\citet{VinkCas02}, whilst modified CAK predictions for Central Stars of
Planetary Nebulae were computed by Pauldrach et al. (1988, 2004), see
page~\pageref{cspn}.

\subsection{Predictions for models close to the Eddington limit}
\label{sec:modelsclosetoedd}

The predictions of Eq.~(\ref{eq_formula}) are valid for models that are
located at a sufficient distance from the Eddington limit, with $\Gamma$
$\le$ 0.5. There are two regimes where this is no longer the case: (i)
stars that have formed with large initial masses and luminosities, i.e. very
massive stars (VMS) with $M$ $\ga$ 100 $\msun$, and (ii) less extremely
luminous ``normal'' stars that approach the Eddington limit when they have
lost a substantial fraction of their initial mass. Examples of such objects
are the LBVs and the more common evolved WR stars. 

\citet{Vink02} and \citet{Smith04} showed with their Monte Carlo
computations that when lowering the mass for constant luminosity, the
mass-loss rate increases more rapidly than Eq.~(\ref{eq_formula}) indicates.
For these LBV models, \mdot\ $\propto$ $M^{-1.8}$ instead of $M^{-1.3}$ as
for normal OB supergiants. In other words, when stars start to lose mass,
not only does the mass-loss rate increase due to a lower stellar mass, but
as the mass-loss rate increases more rapidly this leads to a strong positive
feedback.

An illustration of this feedback-effect for VMS in the mass range $100-300$
\msun\ was provided by \citet{Vink06} where it was noted that VMS mass loss 
drastically increases when the objects approach the Eddington limit.
\citet{Vink05} also simulated mass loss close to the Eddington limit when
performing a pilot study for evolved late-type WR (WN and WC) winds, as a
function of metallicity.  They showed that these mass-loss rates were a
factor 6-7 higher than using the \citet{Vink00} recipe for OB supergiants --
for objects with similar stellar parameters, as a result of the closer
proximity to the Eddington limit.

\begin{figure}
\begin{center}
   {\includegraphics[width=8cm]{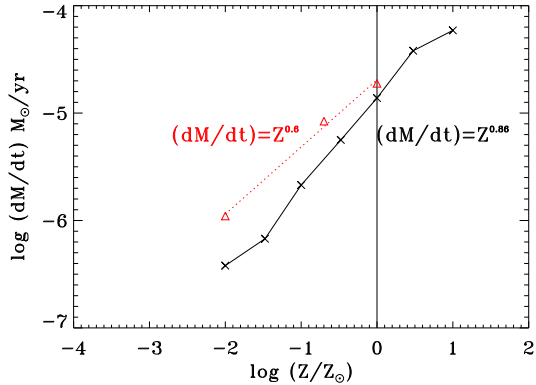}}
\end{center}
\caption{Mass-loss predictions as a function of $Z$, for objects with
$Z/\Zsun$ $>$ 1/100. The solid line indicates the $\mdot(Z)$ dependence from
Monte Carlo simulations by \citet{Vink01} and the dotted line corresponds to
the predictions by \citet{Kud02} . See text.}
\label{f_kudrcomp}
\end{figure}

A different approach to the problem of winds close to the Eddington limit 
was employed by \citet{Graf08}. One should note that their WR wind-models
were not applied to evolved WR stars, but to WR stars in which the broad
emission lines form due to a large intrinsic luminosity. For WR wind-models,
the CAK formalism is no longer applicable, as a myriad of weak iron lines
provide massive line overlap, and the wind has become optically thick.  

The framework of optically thick winds has been employed in studies of
\citet{Kato85} and \citet{NL02} where the basic assumption is that the
critical point of the problem is the sonic point, similar to recent
suggestions by \citet{Lucy07a, Lucy07b} regarding ``normal'' O-star
winds\footnote{A discussion of this assumption is beyond the scope of this
review.}, and that the corresponding conditions can be described in the
diffusion limit. In the transonic region, the Rosseland mean opacity shows
an outwards increasing behaviour and, through the use of the equation of
continuity, \mdot\ follows from the specific conditions at the sonic
point. In the more advanced studies of \citet{Graf05, Graf08} the momentum
equation of the entire wind is accounted for and a strong dependence of
$\dot{M}$ on the proximity to the Eddington limit is also predicted for
these models (see also Sect.~\ref{sec:wrstars}).

\subsection{Predictions for very low Z and Pop III stars}

\subsubsection{The observable massive stars at low Z}
\label{sec:lowzmodels}

\citet{Vink01} applied their Monte Carlo models to OB-stars within a $Z$-range
representative for the observable Universe, $Z/\Zsun$ $>$ 1/100. The
corresponding results are compared to those from \citet{Kud02}
in Fig.~\ref{f_kudrcomp}. Over the last decades the modified CAK models
predicted $\dot{M}$ $\propto$ $Z^{m}$ with $m$ in the range 0.47-0.94 
\citep{Abbott82,Kud87, Kud02}. The main reason for this scaling is that the CAK
$k$ parameter is strongly $Z$-dependent, whilst second order effects that
relate to how $\alpha$ varies with $Z$ have a larger effect on the terminal
wind velocity instead (see pages~\pageref{sec:linestat}ff.).

As can be noted from the figure, the \citet{Vink01} predictions exhibit a
steeper slope ($m$ = 0.85) than the \citet{Kud02} predictions, with $m$ =
0.5 - 0.6. Again, the reasons are thought to be related to the neglect of
multi-scattering in CAK-type models, i.e., the corresponding single-line
approach could result in both an $\mdot-L$ dependence and an $\mdot$-$Z$
dependence that are too shallow. On the other hand, most results computed
using the Monte-Carlo method have not accounted for a $\vinf$-$Z$
dependence, which is anticipated to result in an overestimate of $m$.
Therefore, a value of $m$ $=$ $0.7$ has been recommended as the appropriate
value for the mass-loss scaling with $Z$ for OB-stars
\citep{Vink01,Krticka06}. The {\it observed} metallicity dependence of
OB-stars will be discussed in Sect.~\ref{sec:obastars}.

\subsubsection{Objects at very low Z and Pop III stars}

Extending the predictions to very low $Z/\Zsun$ $\la$
$10^{-2}$, where stars can no longer be observed individually, it has been
shown that $\dot{M}$ keeps dropping until the winds reach a regime where
they become susceptible to ion-decoupling and multi-com\-po\-nent effects
\citep{Krticka03}.  The only way to maintain a one-fluid wind model is 
through an increase of the Eddington factor by pumping up the stellar mass
and luminosity.  Such an increase in stellar mass is by no means
artificial, as the earliest generations of massive stars are anticipated to
be more massive than today's population since stars formed with fewer metals 
to provide the cooling -- at the very least resulting in a larger Jeans
mass. 

In case of Pop III stars with truly ``zero'' metallicity, i.e. only H and He
present in the atmosphere, it seems unlikely that these stars would develop
line-driven winds of significant strength \citep{Kud02, KrtickaKubat06}.
Nonetheless, other effects may contribute to the driving.  Interesting
possibilities include stellar rotation and pulsations, although pure
vibration models for Pop III stars also indicate little mass loss via
pulsations alone \citep{Baraffe01}. Perhaps a combination of several effects
would result in large mass loss close to the Eddington limit. Moreover, we
know that even in the present-day Universe a significant amount of mass is
lost in LBV type eruptions (potentially driven by {\it continuum} radiation
pressure, see page~\pageref{tiring}) which might be also relevant for the
First Stars \citep{Vink05,Smith06}.

\subsubsection{Winds enhanced by self-enrichment}
\label{sec:selfenrichment}

Given that the First Stars are potentially massive, luminous, and rapidly
rotating, it is not inconceivable that, despite the fact that the first
generation(s) of massive stars start their evolutionary clock with few
metals, the objects may enrich their outer atmospheres with nitrogen and
carbon due to rotational mixing \citep{Meynet06}, inducing a strong
line-driven wind \citep{Vink06,Graf08}. 

In a first attempt to investigate the effects of self-enrichment on the
total wind strength, \citet{Vink05} performed a pilot study of WR mass loss
versus $Z$. The prime interest in WR stars here is that these objects,
especially those of WC subtype, show the products of core burning in their
outer atmospheres. 

Although the last few decades provided increasing evidence that WR winds are
radiatively-driven \citep{LA93,Gayley95a,SpringmannPuls98,NL02,Graf05}, the
question of a WR $Z$-dependence remained controversial, with some stellar
modelers extrapolating $Z$ dependencies from O stars, whilst others assumed
WR winds to be $Z$-independent.

\begin{figure}
\begin{center}
   {\includegraphics[width=8cm]{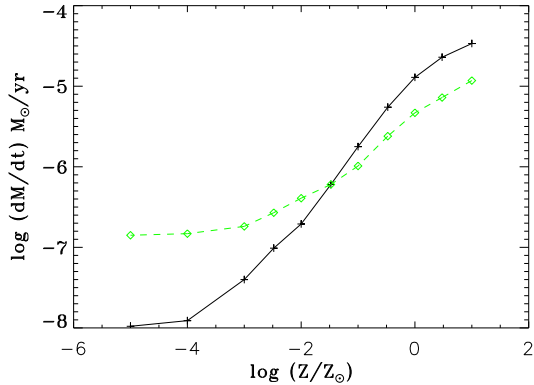}}
\end{center}
\caption{Monte Carlo WR mass-loss predictions as a function of $Z$. The dark
line represents the late-type WN stars, whilst the lighter dashed line
shows the results for late-type WC stars. The slope for the WN models is
similar to the predictions for OB-supergiants, whilst the slope is shallower
for WC stars. At low $Z$, the slope becomes smaller, whilst it flattens off
entirely at $Z/\Zsun$ $=$ $10^{-3}$. The computations are from
\citet{Vink05}.} \label{wcwn}
\end{figure}

The reasoning behind the assertion that WR winds may not be $Z$-dependent
was that WR stars enrich themselves by burning He into C, and it could
be the large C-abundance that is the most relevant ion for the WC wind
driving, rather than the sheer number of Fe lines. Figure~\ref{wcwn} shows
that despite the fact that the C ions overwhelm the amount of Fe, both
late-type WN (dark line) and WC (light line) show a strong $\dot{M}$-$Z$
dependence, basically because Fe has such a complex electronic structure.

The implications of Fig.~\ref{wcwn} are two-fold. The prediction that mass
loss no longer decreases when $Z/\Zsun$ drops below $\sim$$10^{-3}$ (due to
the final dominance of driving by carbon lines) suggests that once
massive stars enrich their outer atmospheres, radiation-driven winds might
still exist, even if the object started its life with negligible amounts of
metals.  Whether the rate of mass loss is high enough to seriously alter the
evolutionary tracks of the First Stars also depends on other physical
factors, such as the proximity to the Eddington- or the $\Omega \Gamma$-limit.

The second point to address regarding Fig.~\ref{wcwn} is the new result that
the mass-loss rates of WR stars drop steeply with $Z$ at subsolar $Z$.  This
is anticipated to have important consequences for black hole formation and
the progenitor evolution of long duration gamma-ray bursts. The collapsar
model of \citet{Woosley93} requires a rapidly rotating core before collapse,
but at solar $Z$ stellar winds are anticipated to remove significant 
angular momentum \citep{Zahn92}. The WR $\dot{M}$-$Z$ dependence from
Fig.~\ref{wcwn} provides the intriguing possibility to maintain rapid
rotation because WR winds are predicted to be weaker at lower $Z$.

Interestingly, the WR $\dot{M}-Z$ dependence also appears to be able to
reproduce the drop in the WC/WN ratio at low $Z$ \citep{Eldridge06}. 
Although there are various reasons to assume that additional effects 
such as rotational mixing and binarity play an important role for the
formation of WR stars, the results suggest that the mass-loss-$Z$ dependence 
is a first order effect as far as the WC/WN ratio is concerned.

\section{Observed wind parameters}
\label{sec:obswindpara}

As we have seen in the preceeding sections, radiation driven wind theory
provides precise predictions for the {\it two global parameters} of hot
stellar winds, namely the terminal velocity, \vinf, and the mass-loss rate,
\mdot, as a function of stellar parameters (including abundances, rotation
rates, magnetic field-strengths). The most prominent prediction concerns the
presence of a wind-momentum luminosity relation (if ``contaminating''
effects do not dominate), which should allow for an easy check of the theory
when the global wind parameters are known from observations. 

Later on, we will summarize recent results on such comparisons. In
order to do so, we need to know how and to which precision these wind
parameters can be ``measured''.

\paragraph{A few caveats.~~}

Most of the following investigations (still) rely on the assumptions of the 
standard model (Sect.~\ref{sec:standardmodel}), partly extended to allow for
the presence of inhomogeneities (clumping) and X-rays (see below).  Within
this model, the mass-loss rate follows from the continuity equation
(Eq.~\ref{eq_const}), and practically all diagnostics base on the approach
that certain observational features (spectra or SEDs) are fitted by
corresponding radiative transfer calculations {\it on top of a wind model}
with a density expressed in terms of the mass-loss rate, $\rho(r) = \mdot/(4
\pi r^2 v(r))$, and a $\beta$-velocity field, 
\beq
v(r) = \vinf \bigl(1 - b \frac{\Rstar}{r}\bigr)^\beta.
\label{vlaw}
\eeq
Remember that this functional form relies on the theoretical predictions
(Eq.~\ref{eq_vr}). The constant $b$ fixes the onset of the wind, usually
obtained from requiring a smooth transition between photospheric
density-/velocity-stratification and wind conditions, or by imposing a
certain minimum velocity $v(\Rstar) = v_{\rm min}$, of the order of the
isothermal speed of sound. 

By varying $\mdot, \vinf$ and $\beta$, an optimum fit is aimed at and the 
so-called ``observed wind parameters'' are just the outcome of such
fitting procedures. Insofar, already their determination relies on a
consistent description of the wind {\it and} the photosphere, since not only
the wind properties but also additional atmospheric conditions such as \Teff,
\logg\ and abundances affect the radiative processes which are modeled to
fit the observations. Whenever ``observed wind parameters'' are presented,
one should be aware of the fact that already these are the result of
diagnostic techniques based on a substantial amount of theoretical modeling.
Depending on the degree of sophistication of the underlying model but also 
on the part of the spectrum and the specific process which is used
for the determination, the reliability of the results can vary drastically.

Even worse, by investigating the scaling properties of particular solutions,
it turns out that such fits are not unique, but depend, subject to the
considered process, on certain {\it combinations} of wind parameters and
stellar radius. Important quantities are the ``optical depth invariants''
$Q_{\rm res}$ for resonance lines with a line opacity $\propto \rho$ and $Q$
for recombination based line processes (such as H$_{\alpha}$) with opacities
$\propto \rho^2$,
\beq Q_{\rm res} = \frac{\mdot}{\Rstar \vinf^2}, \qquad \qquad Q =
\frac{\mdot}{(\Rstar \vinf)^{1.5}} 
\label{qq} 
\eeq 
(details are given in the following). Individual mass-loss rates can be
obtained only when the terminal velocity and the stellar radius (depending
on distance which is problematic for Galactic objects) are known, or, at
least in principle, if resonance and recombination processes are fitted
simultaneously by means of NLTE atmosphere codes. Otherwise, the result of
such fits is not the mass-loss rate but one of the above scaling invariants.
Thus, even if the fit looks perfectly, the same needs not to be true for the
derived mass-loss rate.

The standard model is certainly an (over?)-simplification. Significant
evidence for deviations from spherical symmetry, non-stationarity,
clumpiness and shocks has been found from the beginning of wind
observations on. Aspects related to rotation and magnetic fields have 
already been discussed in Sect.~\ref{sec:rotmag}, and we will turn to
problems related to time-dependence and inhomogeneity in
Sect.~\ref{sec:inhomowinds}. Though these phenomena had been widely ignored 
with respect to diagnostical methods until the late 1990's, increasing
observational and computational feasibilities have lead to the conclusion
that certain aspects must be included to avoid obvious inconsistencies. In
due course, simplified approaches have been developed to account for these
aspects, where the inclusion of wind-clumping and the X-ray/EUV emission
from shocked material into the present generation of NLTE codes are the most
prominent examples of such updates. Note, e.g., that the optical depth
invariant $Q$ as defined in Eq.~\ref{qq} needs to be modified with respect
to the so-called clumping factor if the winds are considered as
inhomogeneous (Sect.~\ref{sec:clumping}, Eq.~\ref{qclump}). 

Obviously, a {\it detailed} inclusion of deviations from the standard model
requires significant effort regarding both diagnostics and the development
of new models and methods, a task which is just at the beginning to date.
Examples are the developments of multi-dimensional NLTE codes such as {\sc
astroroth} \citep{Georgiev06} mentioned already in the context of B[e] stars
(Sect.~\ref{sec:rot}) and a 2-D description of rotational effects within the
spectrum synthesis based on NLTE model atmospheres, as reported by
\citet{Bouret08}. Further progress can be expected in the near future.

\subsection{Diagnostic methods}
\label{sec:diagnostics}

Diagnostic methods can be roughly divided into two different classes.
Approximate methods concentrate on the analysis of one specific process and
try to deduce specific (wind-)parameters by a simplified description,
usually by considering the wind physics alone and requiring additional
information from external {\it photospheric} diagnostics.

Alternatively, modern diagnostics rely on the use of ``unified'' NLTE
model atmospheres. Stellar and wind parameters are derived simultaneously by
optimizing synthetic spectra and SEDs for a large wavelength range,
comprising diagnostics of a multitude of processes with different 
dependencies and scaling relations. In order to draw correct conclusions,
these processes have still to be understood though.

\subsubsection{Approximate methods}
\label{sec:approxdiag}

A detailed discussion and comparison of the various approximate methods to
derive global wind parameters has been given by \citet{KP00}, and we will 
summarize only important aspects. 

The two most prominent line types formed in a stellar wind are P~Cygni
profiles with a blue absorption trough and a red emission peak, and 
pure emission profiles (or absorption lines refilled by wind-emission). Their 
different line shapes are caused by different population mechanisms of
the upper (emitting) energy level of the transition. 

In a P~Cygni line, the upper level is populated by the interplay between
absorption from and spontaneous decay to the lower level, usually the ground
state of an ion. This process is called line-scattering. If, on the other
hand, the upper level is populated via independent processes, e.g., by
recombination into this level or decay from levels above, a pure emission line
(or an absorption line weaker than formed by purely photospheric processes)
will result. The former line type as seen in the (F)UV is majorly used to
determine the velocity field in the wind, particularly \vinf, and the latter
(\Ha, but also \HeII\ 4686, \Bra) to derive \mdot, or, more precisely, $Q$.

\paragraph{Analysis of UV P~Cygni profiles.~~}

(F)UV P-Cygni lines from hot stars (e.g., \CIV, \NV, Si\,{\sc iv}, O\,{\sc vi},
\PV) are usually analyzed by means of the so-called ``SEI'' method (Sobolev
plus exact integration, see \citealt{Lamersetal87}), based on a suggestion
by \citet{Hamann81a}. The central quantity which controls the line formation
process is the (radial) Sobolev optical depth of the specified line (cf. page
\pageref{sobo})
\beq
\taus(r)  =  \frac{\chibar(r) \lambda}{\dvdr} \frac{\Rstar}{\vinf},  \qquad \qquad
\chibar(r) =  \frac{\pi e}{m_{\rm e} c} f \nl(r).
\label{taus}
\eeq

\begin{figure*}
\begin{minipage}{2.8cm}
\resizebox{\hsize}{!}
   {\includegraphics{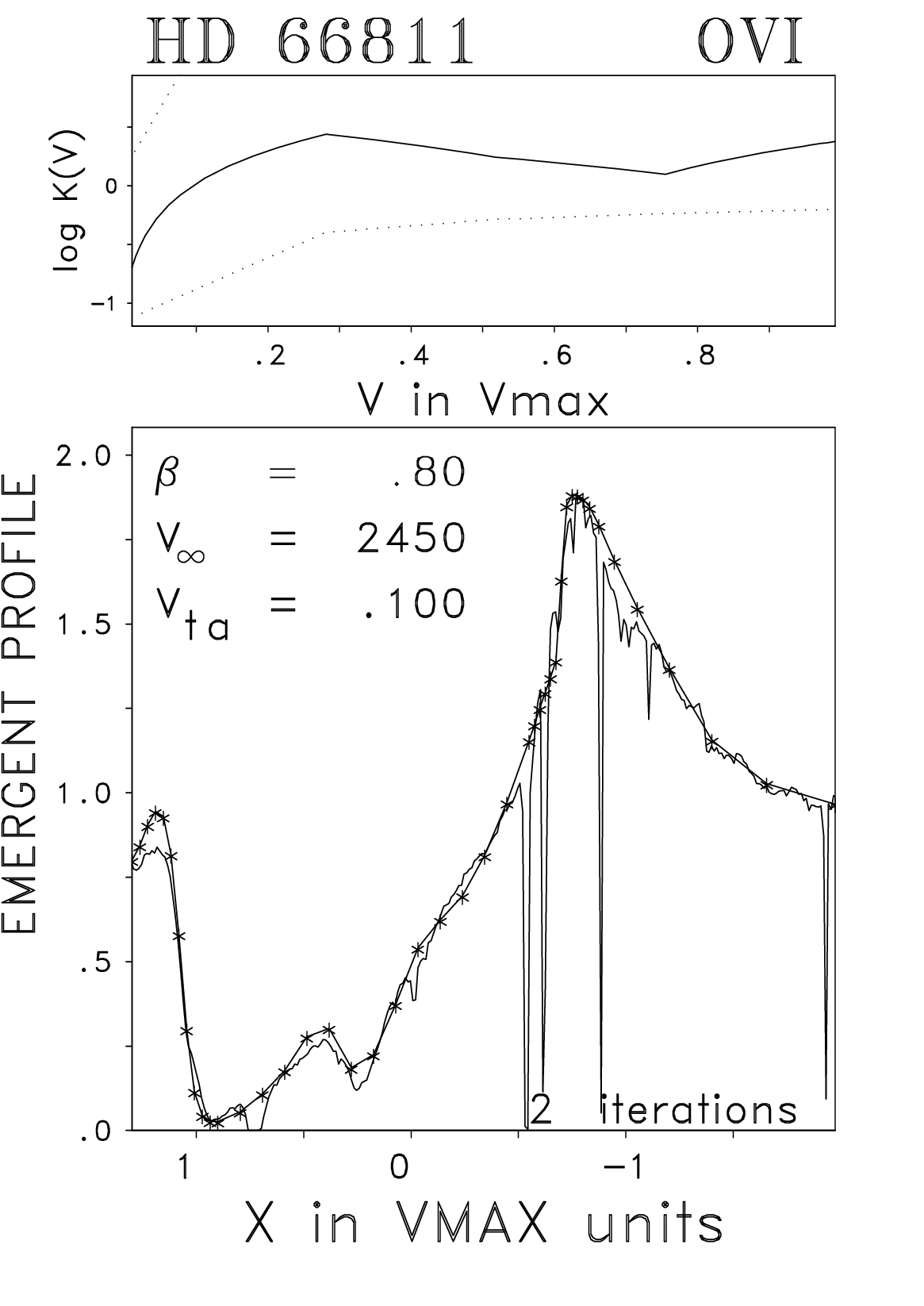}}
\end{minipage}
\begin{minipage}{2.8cm}
\resizebox{\hsize}{!}
   {\includegraphics{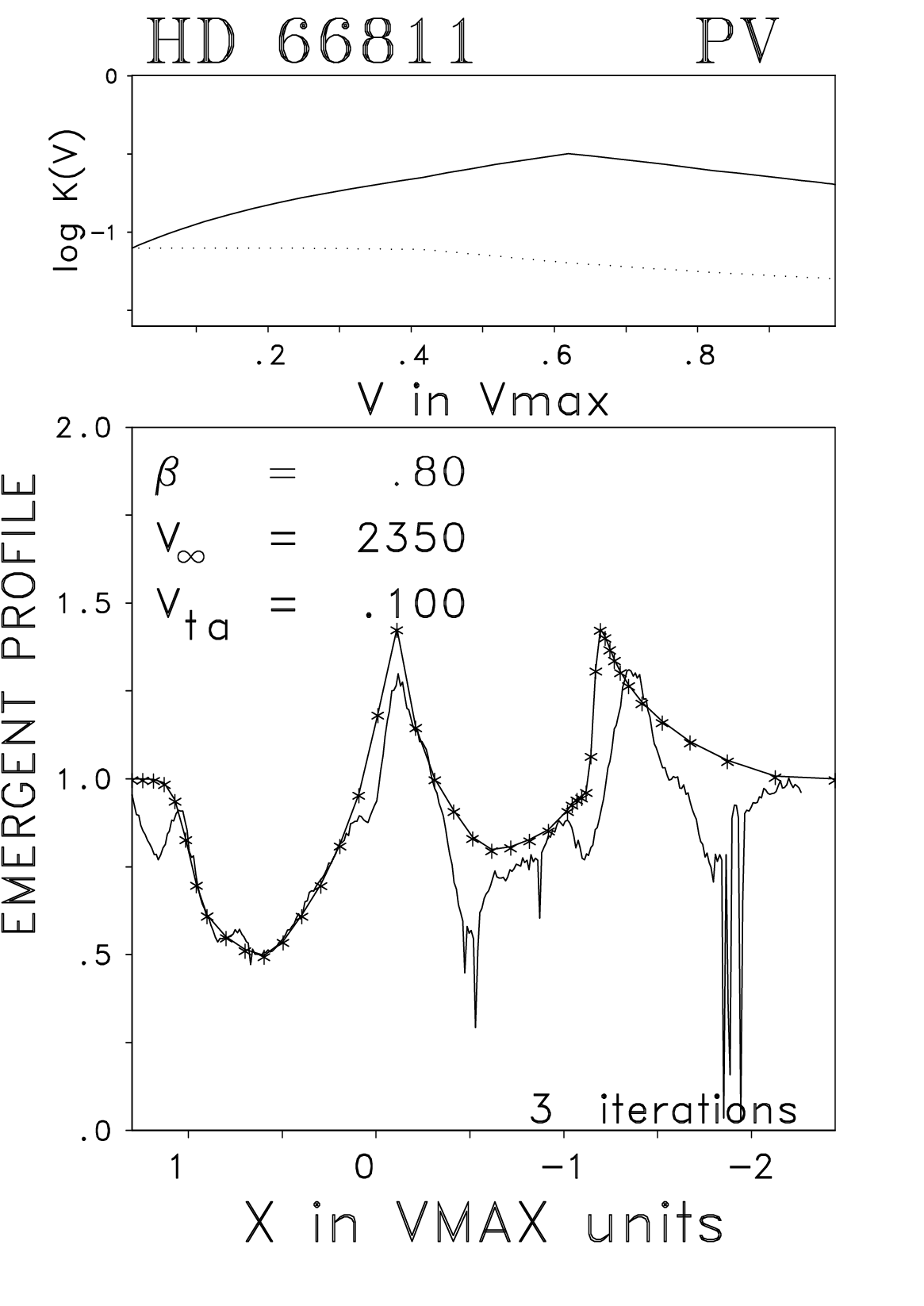}}
\end{minipage}
\begin{minipage}{2.8cm}
   \resizebox{\hsize}{!}
   {\includegraphics{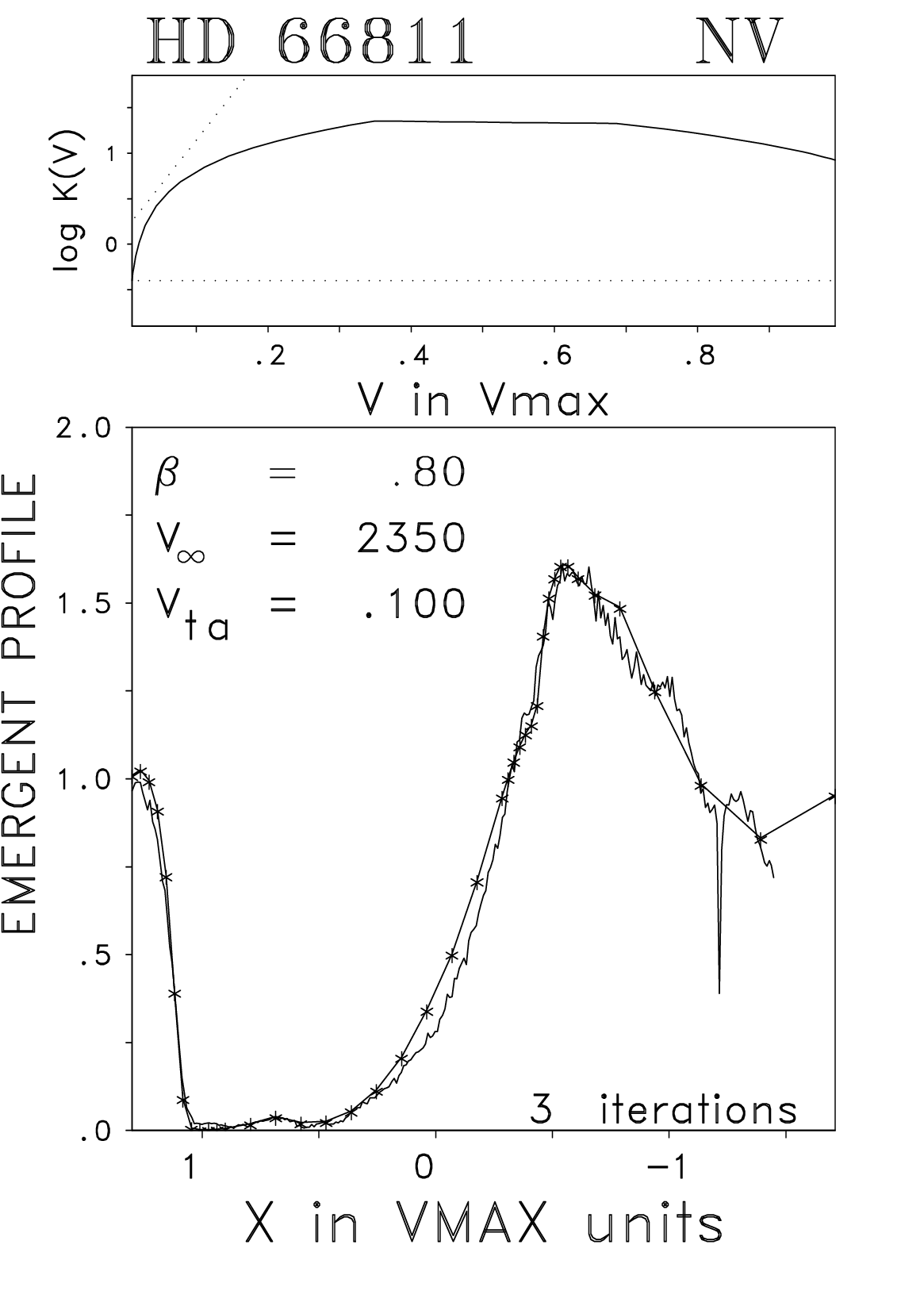}}
\end{minipage}
\begin{minipage}{2.8cm}
   \resizebox{\hsize}{!}
   {\includegraphics{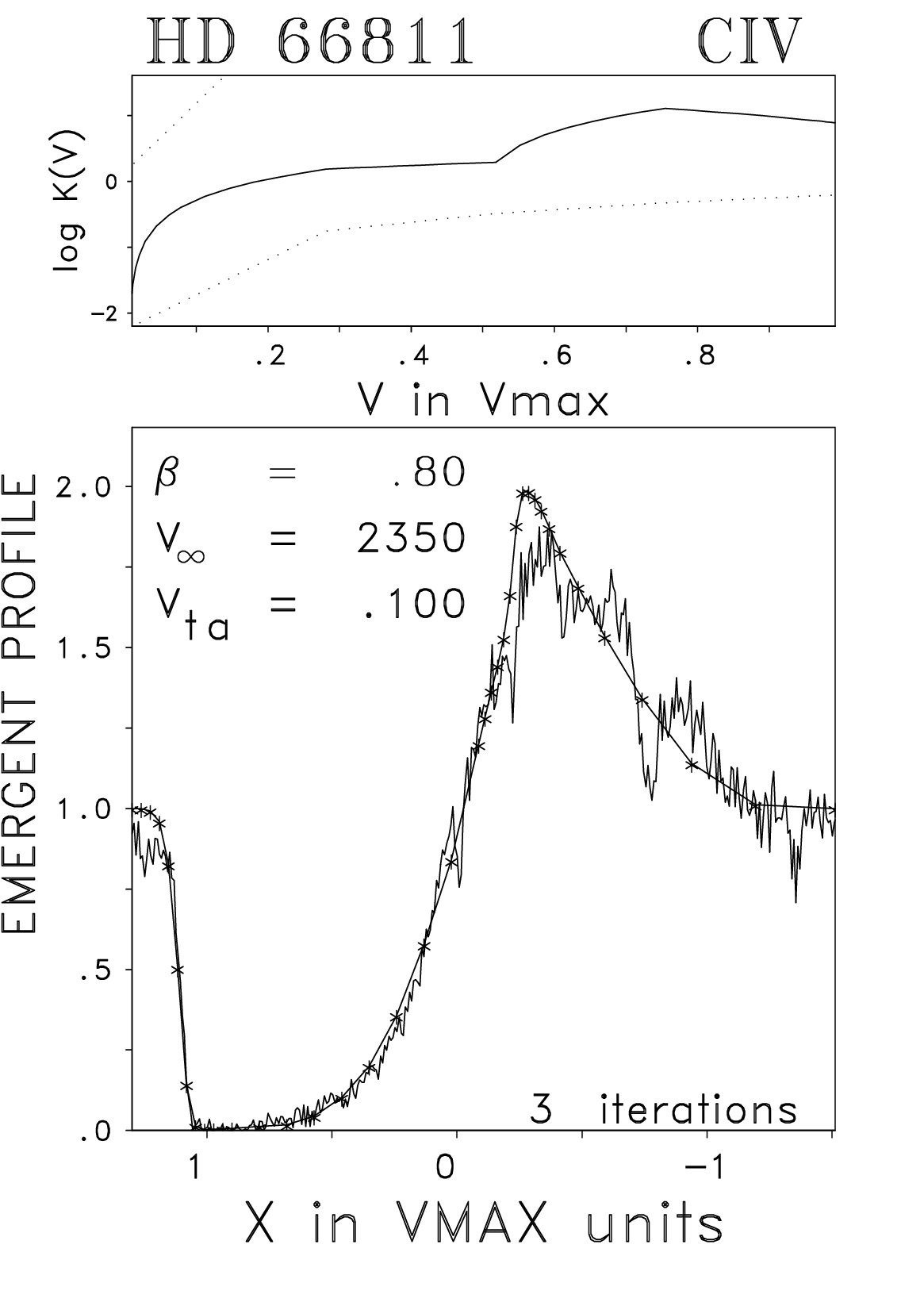}}
\end{minipage}
\caption{SEI fits to the P~Cygni profiles of \OVI, \PV, \NV\ and \CIV\ of $\zeta$
Pup (O4I(f)) as observed by {\sc copernicus} \citep{mortonunderhill77}.  An
almost unique solution with \vinf\ = 2350 \kms\ and $\beta=0.8$ is found for all
four profiles, with a ``micro-turbulence'' of 10\% of \vinf. The
upper panels show the derived run of the quantity $ k(v)\propto \mdot q$
(Eq.~\ref{kdef}). Note the black troughs in \NV\ and \CIV. From
\citet{Haser95}.} 
\label{seifits}
\end{figure*}

$f$ is the oscillator-strength and $\nl$ the lower occupation number of the
transition, neglecting stimulated emission. Here and in the following, radii
$r$ and velocities $v$ are measured in units of the stellar radius and the
terminal velocity, respectively (remember that the line optical depth in a
stellar wind is an almost purely {\it local} quantity, cf.
page~\pageref{sobolength}).  Relating the occupation number, \nl, to the
local density, this quantity can be expressed by
\beq
\label{kdef}
\taus(r)  = \frac{k(r)}{r^2 v \dvdr},\,\,
k(r) = E(r)q(r) \, \frac{\mdot}{\Rstar\vinf^2} \,
\frac{(\pi e^2)/(m_{\rm e}c)}{4\pi m_{\rm H}} \,
\frac{A_{\rm k}}{1+4\YHe} \, f \lambda
\eeq
where $E$ is the excitation factor of the lower level ($\equiv 1$ for
resonance lines coupled to the ground-state), $q$ the ionization
fraction, $A_{\rm k}$ the abundance of the element with respect to hydrogen
and $\YHe$ the helium abundance. Obviously, this expression is scaling
invariant with respect to the quantity $Q_{\rm res} = \mdot /(\Rstar
\vinf^2)$ (cf. Eq.~\ref{qq}) as long as the ground-state population of the
considered ion is proportional to $\rho$ throughout the wind, e.g., if the
ion is a major one. From the latter equation and the assumed
$\beta$-velocity law, it should be clear that the maximum set of deducible
parameters from a fit to the observations consists of $(\vinf$, $\beta$,
$k(r))$, see Fig.~\ref{seifits}.

Thus, information on the mass-loss rate is hidden in the quantity $k(r)$, 
and mass-loss rates can be derived from the P~Cygni profiles of resonance
lines (if at all) only when {\it external} knowledge about ionization,
abundance and stellar radius is available. Exemplarily, \citet{HP89} and
\citet{Haser95} inferred the product $\mdot \qb$ for large O-star samples
from {\it unsaturated} profiles , where \qb\ is a suitable (spatial) average of
the corresponding ionization fraction. 

\label{mdotq} Most of the ``strategic'' P~Cygni lines, however, are
saturated under typical conditions (C\,{\sc iv}, N\,{\sc v} in early O-stars
and Si\,{\sc iv} in late O-supergiants). In these cases, the derivation of
mass-loss rates (or more precisely, of the product $\mdot \qb$) becomes
impossible, and only lower limits are accessible. The reason that these
lines are often saturated has to do with the particular dependence of the
line optical depth on the product $(r^2 v \dvdr)^{-1}$, which, as already
stated on page~\pageref{t(r)}, varies according to $v(r)^{1/\beta-2}$. For
lines with a rather constant ionization fraction throughout the wind and
typical O-star parameters, this variation is only mild, and a line most
likely remains optically thick throughout the wind (i.e., the P~Cygni
profile is saturated) when it has a large optical depth already in the sonic
region. In this case, ``only'' terminal velocities (from the
frequency position of the blue absorption edge) and velocity field
parameters $\beta$ (from the shape of the emission peak) can be measured
(e.g., \citealt{HP89, Groenewegenetal89} and \citealt{Haser95}). Typical
values $\beta=0.7\ldots 1.5$ have been found for OB-stars, where supergiants
show a clear tendency towards higher $\beta$. For A-supergiants, values as
high as $\beta=3\ldots 4$ have been derived (e.g., \citealt{Stahl91,
Kud99}).

\label{veldisp}
The most severe problem of P~Cygni line diagnostics is related to the
presence of so-called black troughs, which are
extended regions in the absorption part of saturated profiles with almost zero
residual intensity (see Fig.~\ref{seifits}). These troughs cannot be
explained without further assumptions. \citet{Hamann81a, Hamann81b}
introduced a highly supersonic ``microturbulence'' (of the order of 10\% of
\vinf) throughout the entire wind to overcome this and related problems,
Since in this case the {\it intrinsic} absorption profile is extremely broad,
the observed troughs can be simulated, in parallel with the red-shifted
emission peaks (see also \citealt{ GL89} and \citealt{Haser95} for the
inclusion of a {\it depth-dependent} velocity dispersion). 

\citet{Lucy82, Lucy83} suggested an alternative, more physical explanation
for the generation of black troughs due to enhanced back-scattering in {\it
multiply non-monotonic velocity fields}, which most likely are present
according to observations and time-dependent hydrodynamical simulations (see
Sect.~\ref{sec:inhomowinds}). Detailed UV line formation calculations have
shown that black troughs can be actually created in the profiles of such
models \citep{POF93, Owocki94b}.

Note finally that until very recently UV resonance lines have been
considered as being (almost) uncontaminated by clumping effects. As we will
see in Sect.~\ref{sec:clumping}, this might not be true, due to the effects
of {\it porosity} or {\it ``vorosity''} (= velocity porosity).

\paragraph{\Ha\ diagnostics.~~}

To avoid the problems inherent to the \mdot-determination from UV~P Cygni
lines (which, moreover, require space-bound observations), the commonly
accepted standard diagnostics to derive \mdot\ relies on the wind emission in
H$_{\alpha}$, where the influence of ionization (which is almost complete)
and abundance becomes almost negligible. The idea to use \Ha\ as a mass-loss indicator
goes back to \citet{KC78}, and \citet{Leitherer88, Drew90, Gabler90,
Scuderi92} and \citet{LL93} applied this method to derive first results for
a variety of hot stars. \citet{Puls96} extended the method to obtain a {\it
fast} analysis tool and to eliminate some systematic errors inherent to
previous approaches. 

The basic difference between \Ha\ (and similarly behaving lines such as
\HeII\ 4686 and \Bra) and P-Cygni lines is related to the source function,
which in the former case is rather constant throughout the wind, as long as
the involved levels are predominantly fed by recombinations or decays from
upper levels. This condition is met in O- and early B-stars. For cooler
stars, the population by pumping due to Lyman lines becomes equally
important, until, in the A-supergiant domain, these lines and the Lyman
continuum become optically thick and the corresponding transitions reach
detailed balance. Consequently, the second level of hydrogen ($n_2$, the
lower level of \Ha) becomes the effective ground state of H{\sc i} and the
line behaves as a scattering line, thus displaying a typical P-Cygni
signature when the winds are sufficiently strong. In such cases, the profile
is usually visible until a blue edge frequency corresponding to $\vinf$,
which allows to derive also this quantity and thus enables a complete wind
analysis in the optical alone (see \citealt{Kud99}). 

Since neutral hydrogen is almost always a trace ion in the winds of OBA
stars, the opacity of \Ha\ scales with $\rho^2$, 
\beq
\chibar(r) \propto b_2(r) \bigl(\frac{\mdot}{r^2 v(r)}\bigr)^2, \quad
\taus(r) = \frac{\chibar(r) \lambda}{\dvdr} \frac{\Rstar}{\vinf} \propto
\frac{\mdot^2}{(\Rstar \vinf)^3} \frac{b_2(r)}{r^4 v^2 \dvdr},
\eeq
i.e., the scaling invariant is $Q^2$ (Eq.~\ref{qq}), and $b_2$ is the NLTE
departure coefficient of $n_2$ which can be rather easily
calculated.\footnote{When $n_2$ becomes the effective ground state, the
opacity scales with $\rho^2/W$ with dilution factor $W$, see
\citealt{KP00}.} Although similar arguments hold for the formation of \HeII\ 
4686, the corresponding departure coefficients have a more complicated (and
more uncertain) stratification which favors the use of \Ha.

Scaling relations for the equivalent width of \Ha\ in O-stars have been
provided by \citet{Puls96}, and typical errors for $Q$ are of the order of
30\%, where these errors can reach values below 10\% for strong wind
emission. Also here, the velocity field parameter $\beta$ can (and has to)
be determined from the shape of the emission peak or wing, resulting in
typical values $\beta \approx 1$ for O-type supergiants \citep{Puls96}. For
weaker winds, on the other hand, the wind emission is low and ``hidden''
within the (rotationally broadened) {\it photospheric} profile, and accurate
determinations of \mdot\ become more difficult. Detailed NLTE calculations
based on unified model atmospheres (Sect.~\ref{sec:nltediagnostics}) are
required to derive meaningful results in such cases. But even if doing so,
uncertainties up to a factor of two in \mdot\ can arise if the photospheric
absorption profile is only marginally refilled. Such uncertainties are 
rooted in the strong dependence of the mass-loss rate on $\beta$ and the
fact that this parameter can no longer be derived (because of the hidden
wind emission) but has to be assumed instead (see, e.g., the corresponding
error analysis in \citealt{repolust04}).

At even lower wind-densities (roughly below a few times $10^{-8}$ \msunyr\ in
the case of O-/early B-stars), \Ha\ completely looses its sensitivity to
mass loss, due to a vanishing wind emission. For such low \mdot\ (which is
present, e.g, in many SMC O-stars and so-called weak-winded stars, see
Sect.~\ref{sec:weakwinds}), other diagnostics must be used, either UV
resonance lines or \Bra\ (see below).

The major problem of \Ha, however, is its $\rho^2$ dependence. Any
significant degree of clumping will lead to an {\it overestimate of the
mass-loss rates if this is neglected in the analysis}. Somewhat advantageous,
however, is the fact that \Ha\ remains optically thin in the largest part of
the emitting wind-volume, such that porosity/vorosity effects can be
neglected, contrasted to the analysis of (strong) UV resonance lines. Further
details and consequences are discussed in Sect.~\ref{sec:clumping}.

\paragraph{Thermal radio and FIR continuum emission.~~} A conceptually
different approach to ``measure'' mass-loss rates is to exploit the
information contained in the (F)IR, (sub)millimeter and radio {\it
continua}. Actually, this approach provides a remarkably reliable and
easy-to-use tool, since it is based on rather simple processes.

The basic idea is to measure the {\it excess} relative to the flux predicted
by photospheric models, which is emitted by free-free
(``Bremsstrahlung'') and bound-free processes in the wind. This excess flux
becomes significant at longer wavelengths, due to the $\lambda^2$ dependence
of the corresponding opacities and the corresponding increase of the {\it
effective} IR/radio-photosphere, and can be translated into a mass-loss
rate. In the following, we will concentrate on purely thermal emission,
whereas non-thermal effects are discussed further below.

The approach to exploit the {\it radio} emission (here, the ``excess''
corresponds to the totally emitted flux, due to negligible photospheric
emission) was independently developed by
\citet{WrightBarlow75}\footnote{these authors considered IR emission as
well.} and \citet{PanagiaFelli75}, and applied to larger O-star samples by
\citet{ABC80, ABC81} and \citet{LL93}. A generalization towards shorter
IR-wavelengths has been provided in a series of papers by \citet{LW84a,
LW84b} and \citet{WL84}, who expressed the flux excess in
terms of a curve of growth, from which the velocity law and the mass-loss
rate can be derived simultaneously. A first application of this method was
the analysis of {\sc iras} observations at 12, 25 and 60 $\mu$m of
$\zeta$~Pup \citep{Lamersetal84}. Further references (regarding radio, submm
and IR observations and analyses) can be found in \citet{KP00}.  

Following \citet{LW84a}, the combined free-free and bound free opacity  (in
units of cm$^{-1}$) at frequency $\nu$ can be written as
\beq 
\kappa_\nu = 3.692 \cdot 10^8\,\bigl(1-\exp(-\frac{h\nu}{kT})\bigr) \,
\bar{z^2} \, \bigl(g(\nu,T)+b(\nu,T)\bigr) \frac{\gamma n_i^2}{T^{\half}
\nu^3}, \eeq
with electron temperature T, $\bar{z^2}$ the mean value of the squared
atomic charge, $g(\nu,T)$ and $b(\nu,T)$ the Gaunt factors for free-free and
bound-free emission, $\gamma$ the ratio between electron and ion density,
and $n_i$ the ion density, which relates to the mass density via $\rho = n_i
\mu m_{\rm H}$ with atomic weight $\mu m_{\rm H}$. 
For long wavelengths
and (almost) completely ionized H and He, the major dependencies are
\beq
\kappa_\nu \propto \frac{g(\lambda,T) \lambda^2 \rho^2}{T^{3/2}},
\eeq
which increases strongly with $\lambda$ and $\rho$. Inserting typical
values, it is easy to show that from the FIR on the continuum becomes 
optically thick already in the wind, and the emitting wind volume increases
as a function of $\lambda$. E.g., the radio photosphere of typical
O-supergiants is located at 100 stellar radii or even further out.
Exploiting this knowledge, one can approximate $v(r) \approx \vinf$ at radio
wavelengths, and an analytical solution of the radiative transfer problem
becomes possible for an isothermal wind:
\beq
\label{fnuradio}
F_\nu \approx 23.2 \Bigl(\frac{\mdot}{\vinf}\Bigr)^{4/3}\, 
\frac{\bigl(\nu g(\nu,T)\bigr)^{2/3}}{d^2}\, \Bigl(\frac{\gamma
\bar{z^2}}{\mu}\Bigr)^{2/3},
\eeq
when $F_\nu$ is the observed radio flux measured in Jansky (10$^{-26}$
Wm$^{-2}$ Hz$^{-1}$), \mdot\ in units of \msunyr, \vinf\ in \kms, distance
$d$ to the star in kpc and frequency $\nu$ in Hz. Thus, the spectral index
of {\it thermal} wind emission is 0.6, since
\beq
F_\nu \propto \bigl((\nu g(\nu,T)\bigr)^{2/3} \propto \nu^{0.6}.
\label{specindradio}
\eeq
The scaling invariant of both radio and IR emission is $\mdot/\Rstar^{1.5}$,
rather similar to the quantity $Q$, if the object's angular diameter
$\propto d/\Rstar$ remains fixed during the analysis. For typical O-stars
located at 1 kpc, $F_\nu$ is of the order of 0.1 mJy in the radio range (2 -
20 cm). Thus, OB-stars are generally weak radio sources (which prohibits the
use of this method for extragalactic work!) and useful radio observations
can be obtained only for nearby stars with rather dense winds. Moreover,
since the spectral index of a non-thermal component (see below) is negative
(e.g., $= -0.7$ for synchrotron emission), the radio domain can be more
easily contaminated than shorter wavelengths such as the (sub)millimeter
regime.

Again, due to the scaling with $\rho^2$, the major problem is the impact of
clumping, already discussed by \citet{ABC81} and \citet{LW84b}.
\citet{Puls06} have used this additional dependence to derive the radial
stratification of the clumping factor, see Sect.~\ref{sec:clumping}.
 
\paragraph{Non-thermal (NT) emission.~~} \label{nonthermal} So far, we have
concentrated on the thermal emission. However, there is ample evidence of
non-thermal emission in hot stellar winds. First observational findings have
been reported by \citet{WhiteBecker83} and \citet{ABC84}, and
\citet{Bieging89} concluded that roughly 30\% of all massive stars were 
non-thermal emitters. To date, there are 16 O-type stars confirmed as
non-thermal radio sources (\citealt{vanLoo06} and references therein).  
A recent review on non-thermal emission from massive stars (with emphasis on 
massive binaries) has been given by \citet{DeBecker07}, which includes the
most up-to-date census of non-thermal radio emitters (the
aforementioned 16 O-stars plus all known WR sources) including their
multiplicity status. 

Although it seems to be clear that the non-thermal emission is due to 
synchrotron radiation from relativistic electrons, different agents
responsible for the acceleration are discussed in the literature: wind
accretion onto compact objects \citep{ABC84}, magnetic reconnection in
single or colliding winds (\citealt{Pollock89}, see also
\citealt{Litvinenko03}) 
and first-order Fermi acceleration \citep{Fermi49}, which, in the presence
of {\it strong} hydrodynamic shocks, is also referred to as the Diffusive
Shock Acceleration (DSA) mechanism \citep{Blandford78, Bell78a, Bell78b}. With
respect to massive stars, the latter process has been invoked by
\citet[ for important modifications to the original idea see
\citealt{ChenWhite94}]{White85} to explain the NT emission in suspected single stars,
where the required shocks should be those embedded in the wind and created
by the line-driven instability (Sect.~\ref{sec:lineinstab}).
\citet{Eichler93}, on the other side, showed that also the shocks created by 
stellar wind collisions in early type binaries are potential sources of the
observed NT emission, in line with even earlier suggestions by
\citet{Williams87, Williams90} regarding the non-thermal emission from the
WC7+05 binary WR140. More recently, \citet{Pittard06} and \citet{Reimer06} have
developed models applying the DSA mechanism to the corresponding physical
conditions of hydrodynmical shocks in colliding wind binaries.

The presently favorized hypothesis is that non-thermal emission from massive
stars is generally procuded by the latter mechanism. Two elements have
simultaneously led to this idea. On the observational side, a number of
systematic observational studies of NT emitters to investigate their
multiplicity have revealed that the vast majority of them are confirmed (or
at least suspected) binaries (see \citealt{DeBecker07}). Note, e.g., that
already \citet{DoughertyWilliams00} found strong
indications that at least for WR-stars a massive companion is a prerequisite
for the appearance of NT emission.
On the theoretical side, \citet[ based on \citealt{vanLoo05}]{vanLoo06}
showed that the observed spectral shape of non-thermal emission {\it cannot}
be reproduced by current (1-D) hydro-simulations applying the wind-instability
scenario in presumed single O-stars. They suspected that also all O-stars
with non-thermal radio emission should be members of binary or multiple
systems, where the non-thermal component is created in the shocks associated
with colliding stellar winds.

In the present context, NT emission is particularly important because of its
impact on the radio spectrum and flux. If present, it can strongly perturb
mass-loss rate determinations based on the thermal radio flux, as, e.g., pointed out
by \citet{Stevens95}. In this context and assuming the above hypothesis of
NT emission created in colliding winds, high angular resolution observations
allowing for a direct disentangling of purely thermal (single star winds) and
non-thermal (colliding-winds) emission in wide binaries will prove as highly
valuable. However, let us also note that an additional thermal radio
contribution may be produced by the shocked gas in the colliding-wind
region itself \citep{Stevens95, Pittard06}, therefore contributing to
some additional confusion.

\subsubsection{Methods based on NLTE atmospheres}
\label{sec:nltediagnostics}

As outlined above, more recent wind diagnostics is mostly based on NLTE
model atmospheres. At least three different methods can be envisaged:

\smallskip \noindent (i) Synthetic spectra from self-consistent atmospheric
models (comprising a NLTE + hydro-description, where the wind-structure
relies on a line force resulting from self-consistent NLTE occupation
numbers and the corresponding radiation field, see, e.g., Fig.~1 in
\citealt{Pauldrach94}) are fitted to observations which cover a significant
spectral range and are sensitive to both photospheric and wind conditions. 
The only input (= fit) parameters are \Teff, \logg, \Rstar (or different
combinations, e.g., \Teff, \Lstar, \Mstar), the individual elemental
abundances, and, if necessary, (stratified) clumping factors and a
description of the X-ray/EUV emission. Besides the fact that such an
approach is very time-consuming, it relies on the assumption that the
complete underlying wind-physics is correctly treated. This method does not
allow for an independent test of the theory itself. All
inaccuracies/problems of the theory will, of course, show up in deviations
between synthetic spectra and observations. Unfortunately, however, these
will be interpreted in terms of peculiar abundances, clumping factors, radii
etc., since for a given set of stellar parameters only one specific
wind-structure (i.e., a specific combination of \mdot, \vinf\ and velocity
field) is possible, which may not be correct. The better the theory, the
better the results from such an approach, of course. An interesting
application of method (i) will be presented in the context of determining
stellar parameters of Central Stars of Planetary Nebulae via wind
diagnostics.

\smallskip \noindent (ii) Synthetic spectra from consistent atmospheric
models (NLTE + hydro-de\-scrip\-tion) are fitted to observations. In
contrast to method (i), \mdot\ and \vinf\ are adapted to fit the wind-lines,
but not directly (as in method (iii) discussed below), but by varying the
force-multiplier parameters k, $\alpha$, $\delta$ (or equivalent quantities)
in a reasonable\footnote{within the range predicted by line-statistics, see
pages~\pageref{sec:linestat}ff.} way. The fit parameters are as above,
augmented by the force-multiplier parameters. The advantage of this method
relies in the possibility to obtain a {\it physically} justified
stratification of the velocity field (and the indirect possibility to check
the wind-theory, by checking whether the observations {\it allow} for
reasonable force-multiplier parameters), but this advantage can also become
a disadvantage when the actual velocity field {\it cannot} be matched by the
hydrodynamic approach, due to shortcomings in the description. Method (ii)
is usually applied in investigations using the atmosphere/wind-code WM-Basic
(see Sect.~\ref{sec:obastars}), and impressive fits of the complete UV
region have been obtained for a (small) number of stars, e.g., $\zeta$~Pup,
$\alpha$~Cam (O9.5Ia) \citep{pauldrach01} and HD\,93129A (O2If$^{*}$,
\citealt{Taresch97}), in the latter case including the optical region.
Similar analyses of a larger sample of Galactic O supergiants and dwarfs 
have been performed by \citet{BG02} and \citet{GB04} (see below). Note that
in almost all these cases the inclusion of the observed X-ray/EUV emission
(interpreted as due to the emission from cooling zones of shocks embedded in
the wind, see Sect.~\ref{sec:inhomoobs}) has been crucial to explain the
ionization of highly ionized species such as \OVI\ and \SVI, see
\citet{Pauldrach94, pauldrach01} and \citet{Macfarlane93}. These ions which
show strong P~Cygni lines in the FUV (e.g., Fig.~\ref{seifits}) cannot be
created in ``standard'' NLTE calculations based on a wind with an electron
temperature of the order of \Teff, a problem realized already at the
beginning of hot star UV spectroscopy and denoted as {\it superionization}
(e.g., \citealt{LamersMorton76, Castor79, Hamann80}).

\smallskip \noindent (iii) The last method enumerated here is the
most frequently used. Synthetic spectra from a NLTE model are fitted to the
observations, where the wind-structure is analytically described via a
$\beta$ velocity law (Eq.~\ref{vlaw}), and a smooth transition between the
analytical wind structure and a quasi-hydrostatic photosphere is adopted.
Input-parameters (in addition to those from method (i)) are \mdot, \vinf\
and $\beta$.  Such models (and those calculated by method (ii)) are called
{\it Unified Model Atmospheres} \citep{Gabler89}, and are standard nowadays. 
Corresponding codes will be discussed in the next section.

Let us finally note that not only different methods can be (and are!)
combined with (different) atmospheric codes, but that also the fit-optimization
method can be different. Even in recent investigations, a ``fit-by-eye''
method has been frequently used, but more objective and automatic methods
such as the optimization by genetic algorithms \citep{Mokiem05} or using
predefined grids in combination with a $\chi^2$/maximum
likelihood optimization \citep{lefever07a} or via a {\it principal component analysis} 
\citep{Urbaneja08} have also entered the spectral analysis of hot stars with
winds.

\paragraph{NIR spectroscopy.~~} 
\label{nirspec} The dramatic progress of IR astronomy in the
last two decades has opened a completely new window for the systematic
investigation of hot star winds, particularly for the analysis of
dust-enshrouded, very young objects and samples in highly reddened clusters.
Note also that the current generation of 10m class ground-based 
telescopes has been optimized to observe in this wavelength range. 

Also the IR spectra from hot stars show pure emission lines,
wind-con\-tami\-nated absorption lines and sometimes even P~Cygni profiles,
and these lines have to be analyzed by means of NLTE models due to strong
NLTE effects in certain transitions. Since the IR continuum of objects with
strong winds is formed already in the wind, however, IR lines may sample
different depths inside the wind and provide additional information about the
shape of the velocity field and particularly on the clumping properties.

The winds of hot stars of extreme luminosity and with strong IR emission
characteristics in the Galactic Center have been investigated by different
groups (e.g., \citealt{najarro97a,najarro06,Martins07}), and
\citet{Lenorzer04}, \citet{Repo05} and \citet{geballe06} have demonstrated
in how far stellar and wind parameters of ``normal'' OB-stars may be
constrained by IR-spectroscopy alone. One of the most important applications
of IR line diagnostics, however, will be the measurement of mass-loss rates
from stars with very weak winds by means of \Bra, as detailed in
Sect.~\ref{sec:weakwinds}.

\subsection{Recent results}
\label{sec:results}

\subsubsection{OBA-stars}
\label{sec:obastars} Most of the relevant results in the field of
quantitative spectroscopy of massive hot stars that have emerged with the
new millennium are the outcome of more than ten years of considerable
efforts made by different research groups to incorporate, in a realistic
way, the effects of line-blanketing in NLTE-modeling of the atmospheres of
massive stars. Indeed, the possibility to include the role of several
thousands (up to millions) of metal lines was intimately coupled to the
revolution of computer facilities.  First approaches were made by
\citet{abbotthummer85} who modeled the effect of ``wind blanketing'' by
introducing a wavelength dependent albedo representing the radiation
reflected back onto the photosphere by electron and line scattering from the
wind. Abbott and Hummer showed that the main result of ``wind
blanketing'' was to heat the photosphere from the surface to the depth of
continuum formation (``back-warming''), so that the resulting spectrum would
resemble that of a hotter star. Thus, depending on the strength of the
stellar wind (\mdot), the albedo could lead to a net reduction of the
derived stellar \Teff\ by up to 20\%.  The main caveat of this approach was
the neglect of photospheric ``line blanketing'', so that a unified picture
could not be provided. As well, effort has been undertaken to incorporate
the effects of photospheric ``flux blocking'' alone
\citep{herrero00,villamariz01}. However, since only blocking was considered
but no back-warming, revised \Teff\ values were significantly {\it higher}
than models without blocking.  

One of the major steps in the blanketing affair was presented by
\citet{anderson85,anderson89} who introduced the so-called ``superlevels''.
Instead of dealing with all individual energy levels of metal species,
several states with close enough energies were packed into one superlevel.
Thus, the number of statistical equilibrium equations could be drastically
reduced saving an enormous amount of computing time (but see, e.g., 
\citealt{Lucy01} on potential problems inherent to this approach). After
some pioneering work \citep{haus92}, by the end of the nineties, four major
codes, PHOENIX \citep{haus92}, CMFGEN \citep{hilliermiller98}, WM-Basic
\citep{pauldrach98,pauldrach01} and TLUSTY (\citealt{hubeny98},
plane-parallel, no wind) could address line blanketing in a realistic way.
Just a couple of years later two more codes, PoWR \citep{graefener02} and
FASTWIND \citep{santo97,puls05} became available. For a brief
comparison of the features of these codes, see \citet{Puls08}.
 
\paragraph{Effects of line blanketing.~~}              
\label{sec:linelb}

\begin{figure*}
\begin{minipage}{6.0cm}
\resizebox{\hsize}{!}
   {\includegraphics[angle=90]{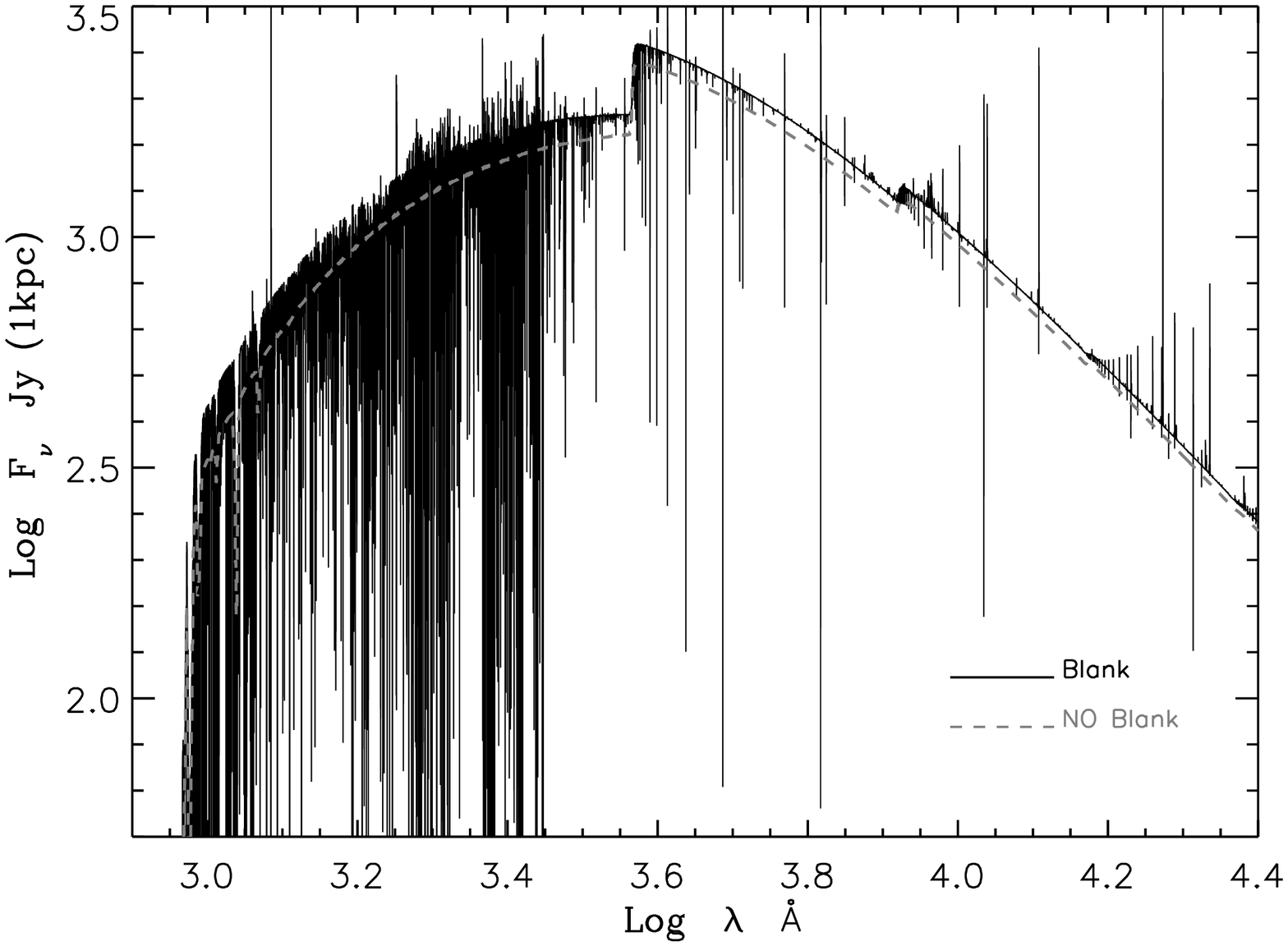}}
\end{minipage}
\hspace{-.5cm}
\begin{minipage}{6.0cm}
   \resizebox{\hsize}{!}
   {\includegraphics[angle=90]{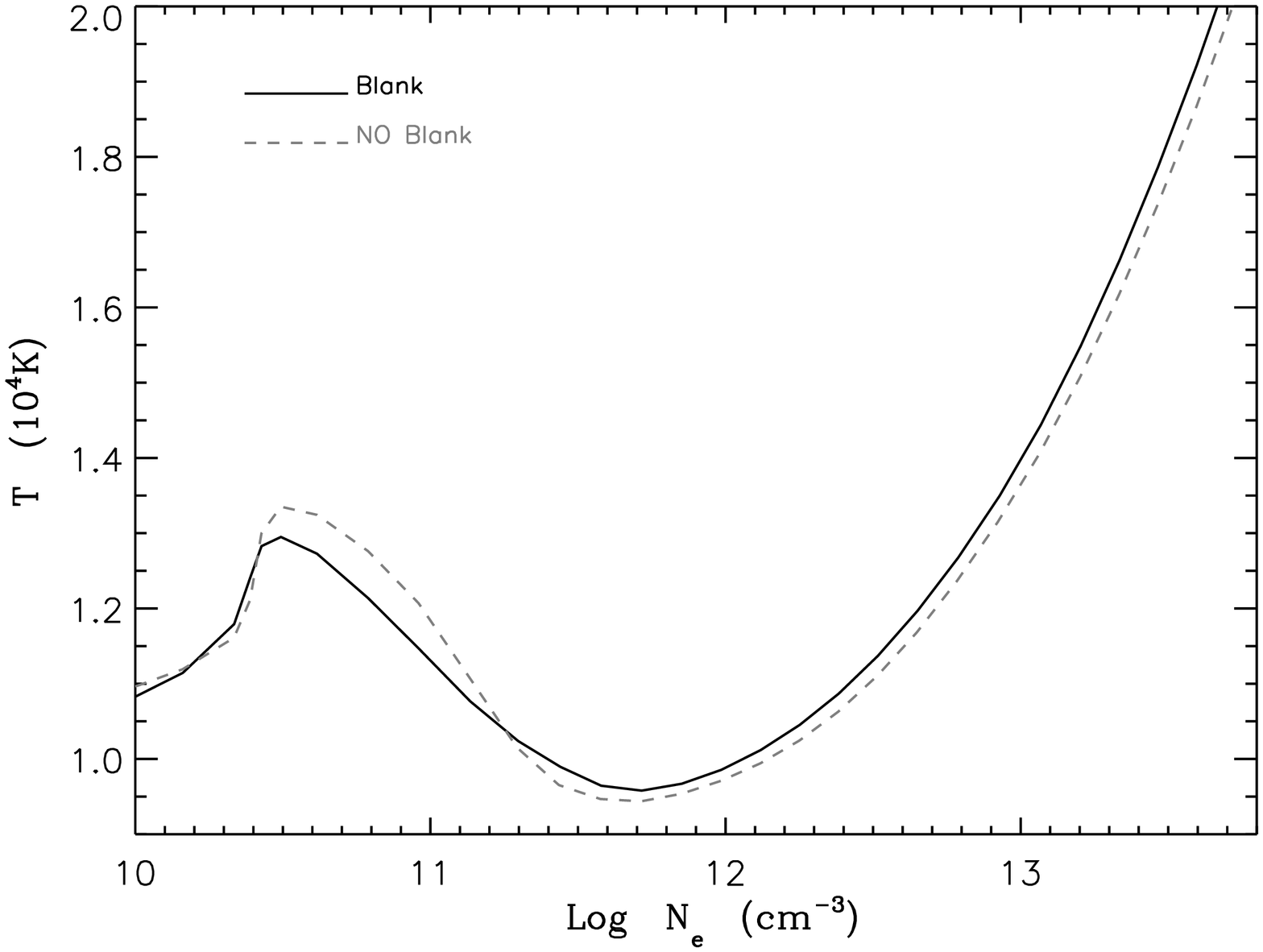}}
\end{minipage}
\caption {Effects of line blanketing (solid) vs. unblanketed models (dashed)
on the flux distribution ($\log F_{\nu}$ (Jansky) vs. $\log \lambda$ (\AA), 
left panel) and temperature structure ($T (10^4$~K) vs. $\log\, \xne$, 
right panel) in the atmosphere of a late B-hypergiant. Blanketing blocks
flux in the UV, redistributes it towards longer wavelengths and causes
back-warming.} 
\label{fig-newmodblanflutem}
\end{figure*}

To illustrate the two main direct effects of line-blan\-ke\-ting, we have
made use of CMFGEN and computed a model for a late B-hypergiant with a
strong wind. Fig.~\ref{fig-newmodblanflutem} shows the comparison of the 
blanketed/non-blanketed cases for the emergent flux distribution and 
temperature structure of the atmosphere. The models confirm the effects
anticipated from early on and discussed above. We clearly see how the
blocked flux in the UV emerges at optical and longer wavelengths. As the
flux is blocked and a certain amount of photons are backscattered, flux
(luminosity) conservation demands a more efficient photon diffusion in the
inner photosphere. Since this effect is controlled by the temperature
gradient, the temperature is increased in this region (back-warming) as
shown in Fig.~\ref{fig-newmodblanflutem}. Back-warming, in combination with
the enhanced radiation field due to backscattering, will lead to enhanced
ionization in the inner parts of the atmosphere. The question arising now is
how the spectral lines will be affected and whether the effect of blanketing
could be considered as a general shift in the ionizing conditions throughout
the whole atmosphere (as found by \citealt{abbotthummer85}) or whether it
will affect differently the inner photosphere and the outer wind. Thus, we
want to know whether this enhanced ionization will be kept throughout the
atmosphere or whether the stellar wind will block enough ionizing radiation
and decrease the ionization degree in the outer wind parts. The outcome of
this competition will depend on the parameter domain in which the star is
located, and will be very sensitive to stellar properties such as effective
temperature, gravity and wind strength. For an illustration, we will compare
the situation for a mid B-supergiant and a very hot early O-supergiant.

\begin{figure*}
\hspace{-0.3cm}
\begin{minipage}{6.0cm}
\resizebox{\hsize}{!}
   {\includegraphics[angle=90]{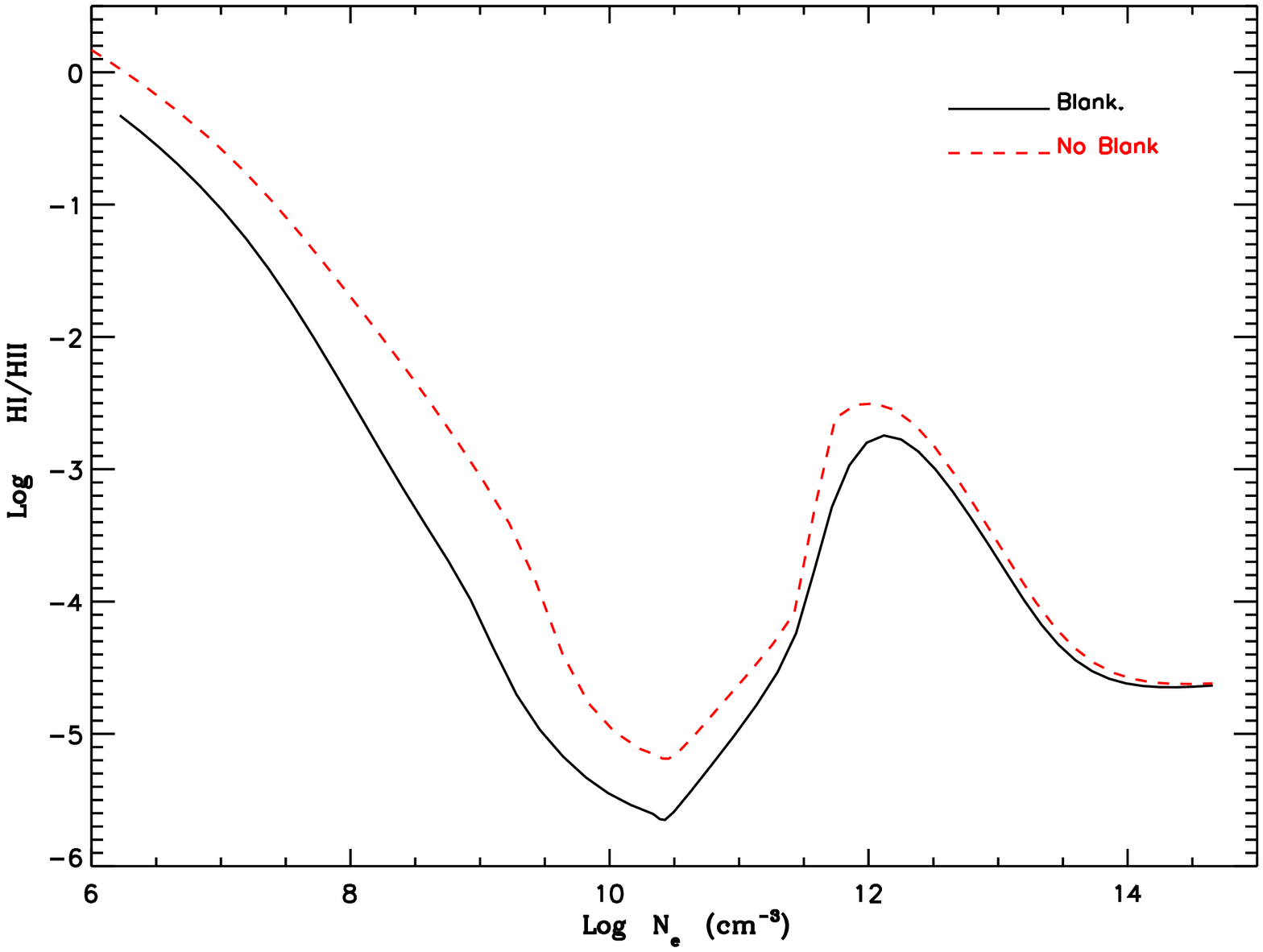}}
\end{minipage}
\hspace{-.3cm}
\begin{minipage}{6.0cm}
   \resizebox{\hsize}{!}
   {\includegraphics[angle=90]{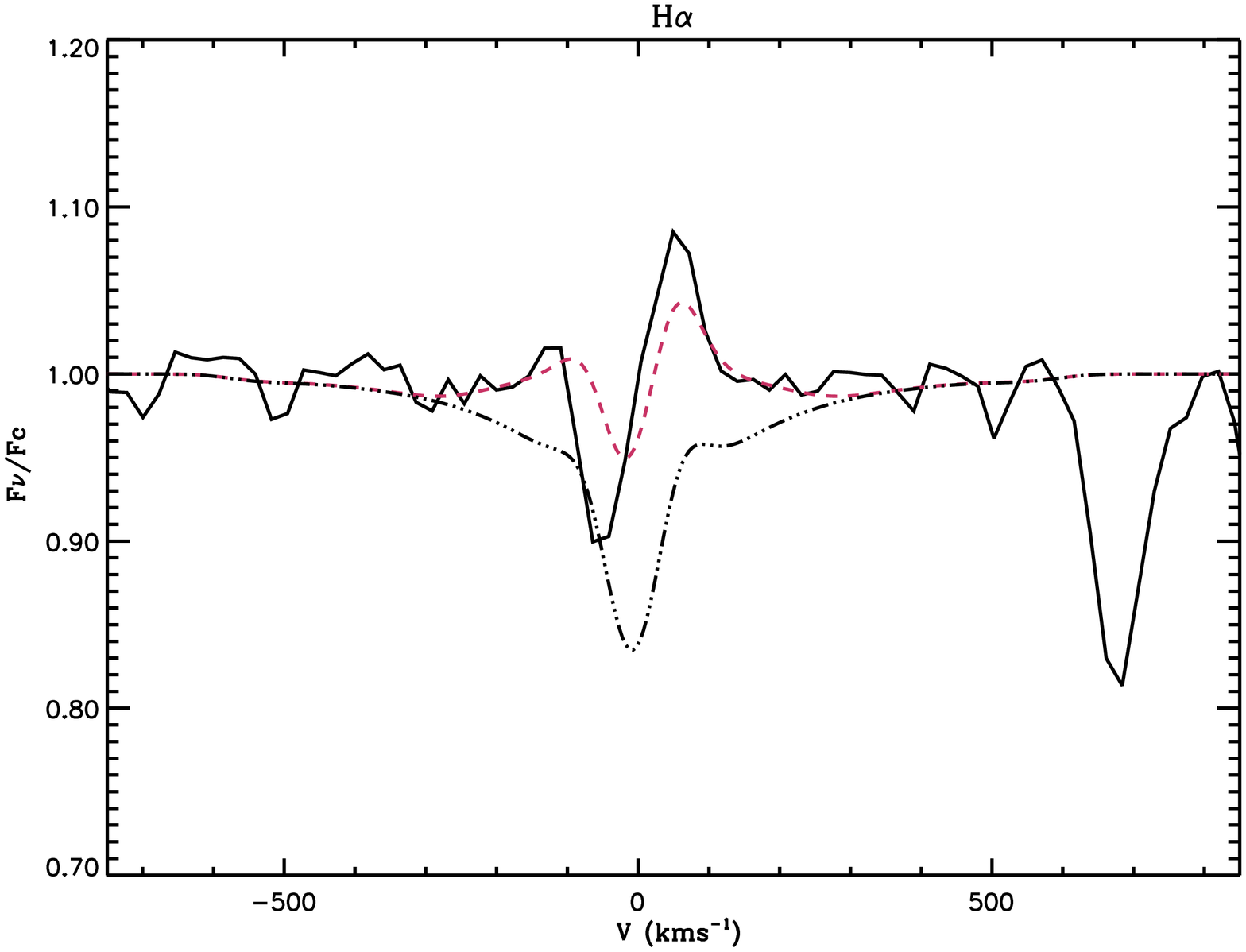}}
\end{minipage}
\caption{Effects of line blanketing on the hydrogen ionization structure
($\log$ \HI\,/\HII\ vs. $\log \xne$, left panel) and line profiles (\Ha,
right panel) of a mid B-supergiant.  The enhanced ionization caused by
blanketing (solid vs. dashed) weakens considerably the \Ha-profile
(dashed-dotted) with respect to the unblanketed model (dashed) and requires
an increase of \mdot\ by up to a factor of three in B-supergiants to balance
the effects of blanketing and to fit the observations (solid). 
\label{fig-newmodbhhalp}}
\end{figure*}

Fig.~\ref{fig-newmodbhhalp} shows the effects of line blanketing on the
hydrogen ionization structure of the mid B-supergiant. Despite the strong
wind, the enhanced ionization caused in the inner photosphere is held
throughout the wind. Further, the increased depletion of \HI\ in the
atmosphere through blanketing requires an enhancement of \mdot\ by up to a
factor of three \label{blanketing_mdot} to compensate the extra ionization
and match the observed profile (see Fig.~\ref{fig-newmodbhhalp}). Thus, for
mid B-supergiants, blanketing affects majorly the derived mass-loss rate,
while milder changes appear on the derived effective temperature. Similar
trends have been found by \citet{urbaneja04}, \citet{Crowther06} and
\citet{MP08}.

\begin{figure*}
\hspace{-0.3cm}
\begin{minipage}{4cm}
\resizebox{\hsize}{!}
   {\includegraphics[angle=90]{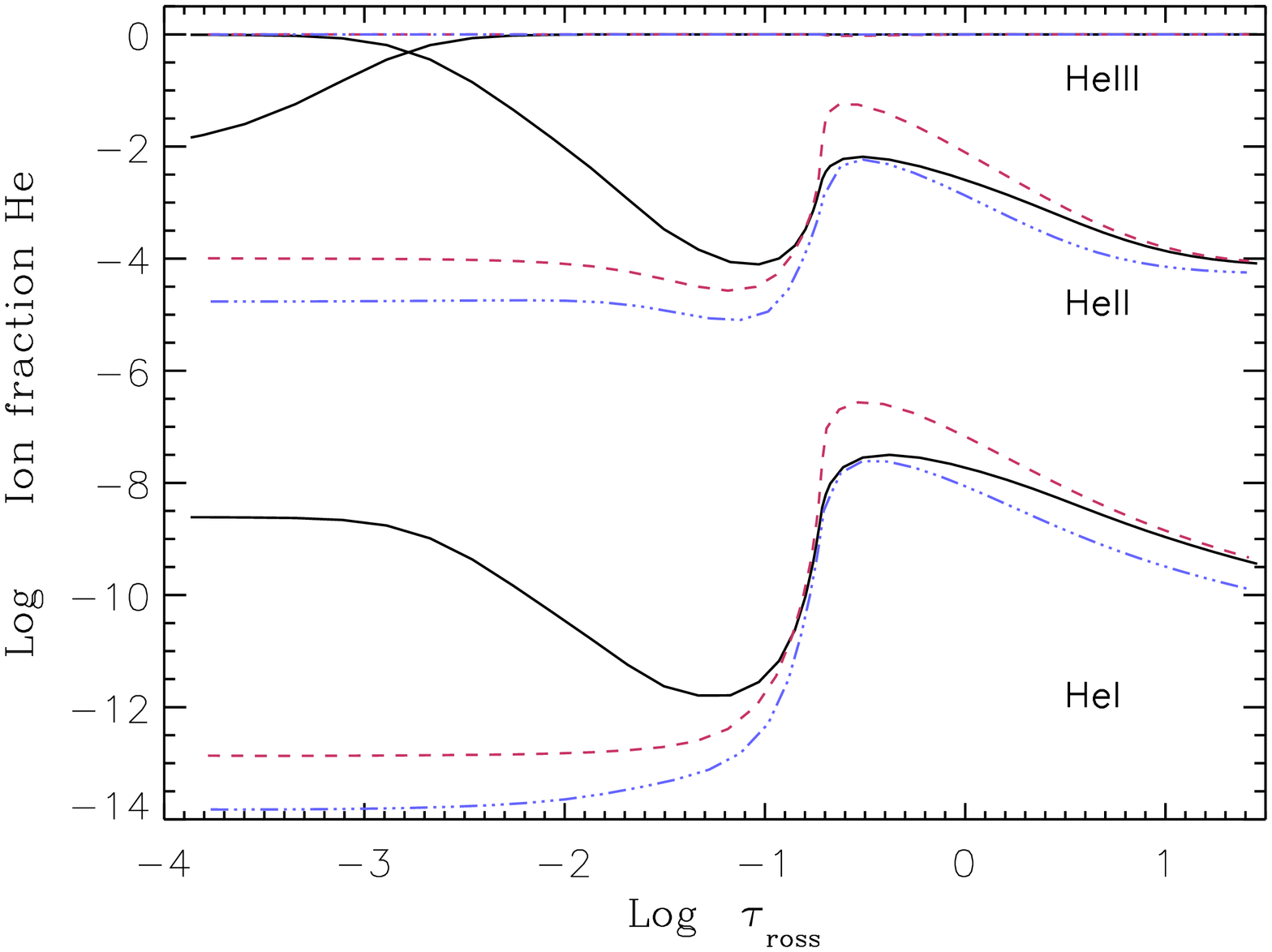}}
\end{minipage}
\hspace{-.3cm}
\begin{minipage}{4cm}
   \resizebox{\hsize}{!}
   {\includegraphics[angle=90]{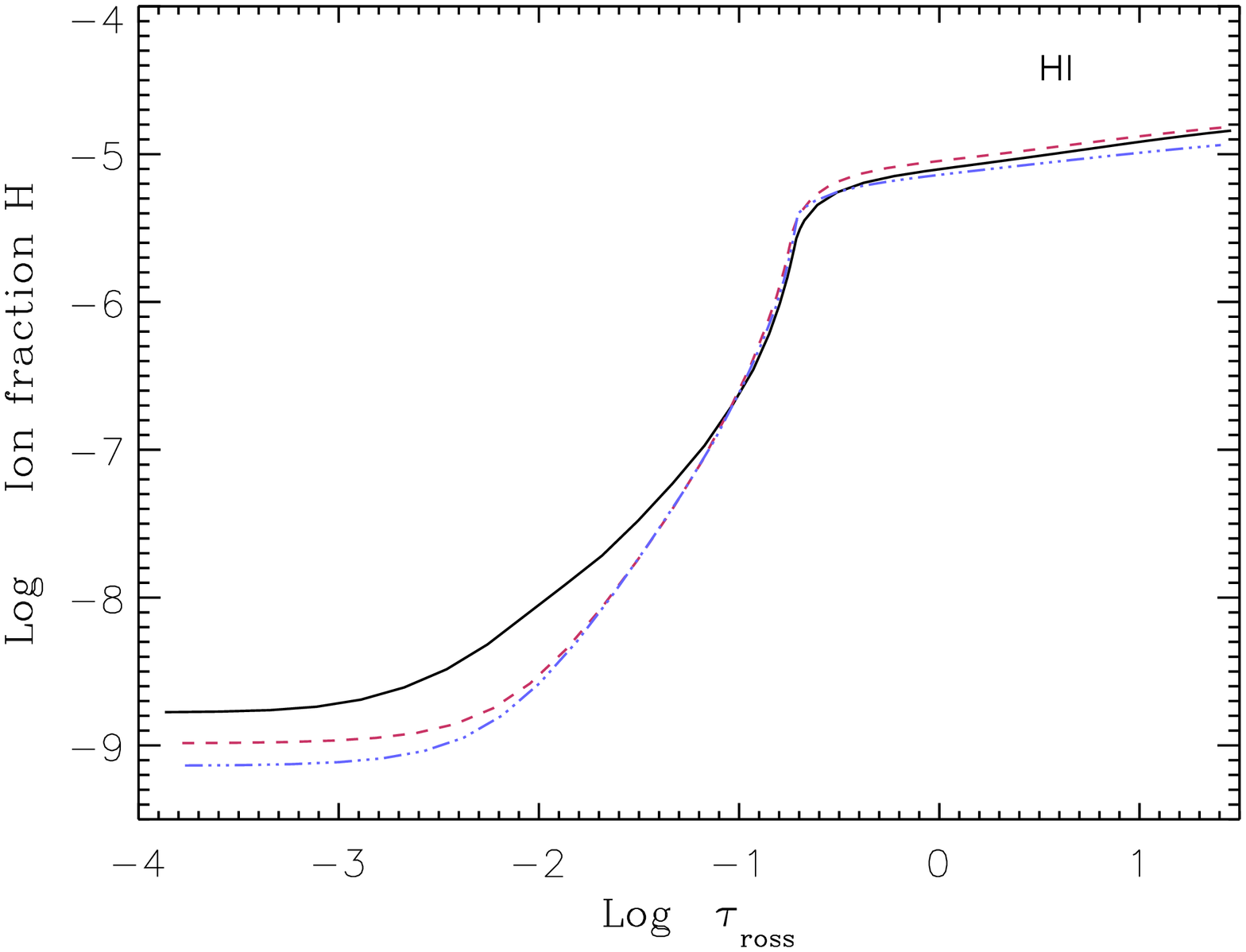}}
\end{minipage}
\hspace{-0.3cm}
\begin{minipage}{4.2cm}
   \resizebox{\hsize}{!}
   {\includegraphics[angle=90]{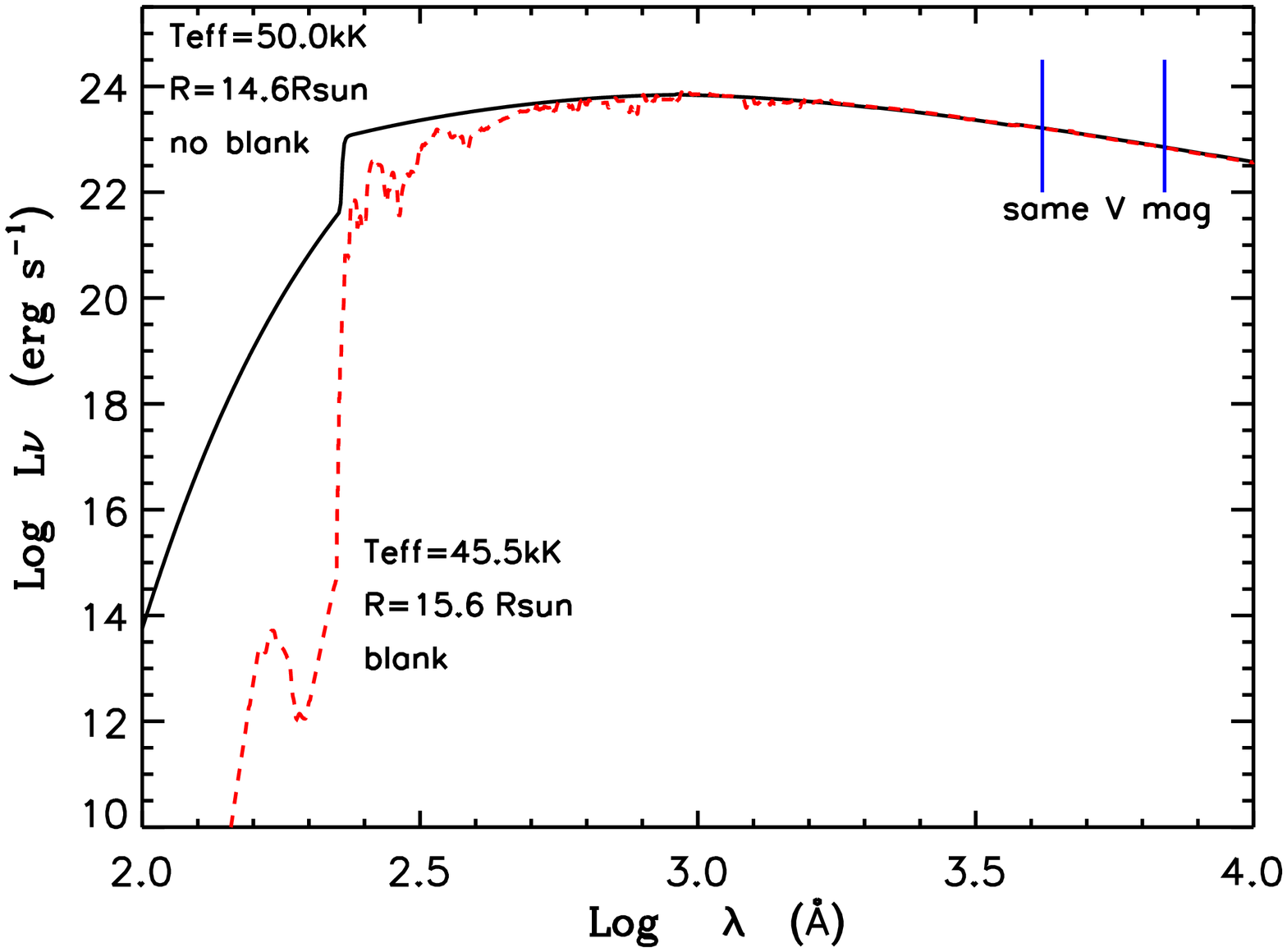}}
\end{minipage}
\caption{
Effects of line blanketing on the helium (left) and hydrogen 
(middle) ionization structure of an O3~If supergiant. Solid: \Teff=
45.5~kK, blanketed; dashed: \Teff= 45.5~kK, not blanketed; dashed-dotted:
\Teff= 50~KK, not blanketed. See text. Right: Flux distribution of models
with (dashed) and without (solid) blanketing matching the stellar line
spectra and photometry. \label{fig-newmodblano3}}
\end{figure*}

Fig.~\ref{fig-newmodblano3} displays the effects of line blanketing on the
He and H ionization structure of an O3~If star. For the case presented in
Fig.~\ref{fig-newmodblano3}, we have selected three different models. The
first one corresponds to a blanketed model with \Teff\ = 45.5~kK, \logg\ =
3.7 and $R$ = 15.6~\rsun. The second model has the same parameters but no
blanketing, while the third one, again not blanketed, is hotter, $\Teff =
50$~kK ($\logg = 3.8$ and $R$ =14.6~\rsun), and reproduces equally well the
observed spectra of the O supergiant {\it and} the observed V magnitude. All
models have the same wind-strength parameter $Q$. From
Fig.~\ref{fig-newmodblano3}-left we see that blanketing (solid) severely
enhances the He ionization in the line forming region with respect to the
cool unblanketed model (dashed), so that a degree of He ionization similar
to the hot unblanketed model (dashed-dotted) is obtained by decreasing
\Teff\ by almost 5000~K. In other words, the basic diagnostic for
temperature determinations in O stars, namely the He ionization balance,
reacts strongly to blanketing and causes the observed reduction on the
effective temperature scale (for further details, see \citealt{repolust04}).

Inspection of the impact of blanketing on the H ionization structure
of O stars (Fig.~\ref{fig-newmodblano3}-middle) tells us that only minor
changes are produced in the ionization degree of H and, therefore, only
small variations should be expected in the inferred mass loss rates
(assuming they are derived from \Ha, as \HeII\ 4686 would
be severely altered). Finally, it should be noted that the relatively strong
wind of the star blocks the \HeII\ ionizing radiation and forces recombination
of \HeIII\ to \HeII, with important implications on the
number of \HeII\ ionizing photons (see also
Fig.\ref{fig-newmodblano3}-right). Thus, this is one case where the strong
wind controls the ionization and outweighs the enhancement produced through
backwarming in the inner regions.

\paragraph{First impact: The new temperature scale for massive OB
stars.~~} \label{newtemp}
\citet{vacca96} presented a compilation of the spectroscopic determinations
of effective temperatures of massive OB stars. They gave preference to the
most recent calculations which {\it at that time} were mostly based on
plane-parallel, hydrostatic, unblanketed NLTE model atmospheres. Interestingly,
in their paper, \citet{vacca96} warned about the significantly lower
temperatures obtained by the few available wind-blanketed analyses 
when compared to unblanketed results.

\begin{figure*}
\hspace{-.5cm}
\begin{minipage}{4cm}
\resizebox{\hsize}{!}
   {\includegraphics{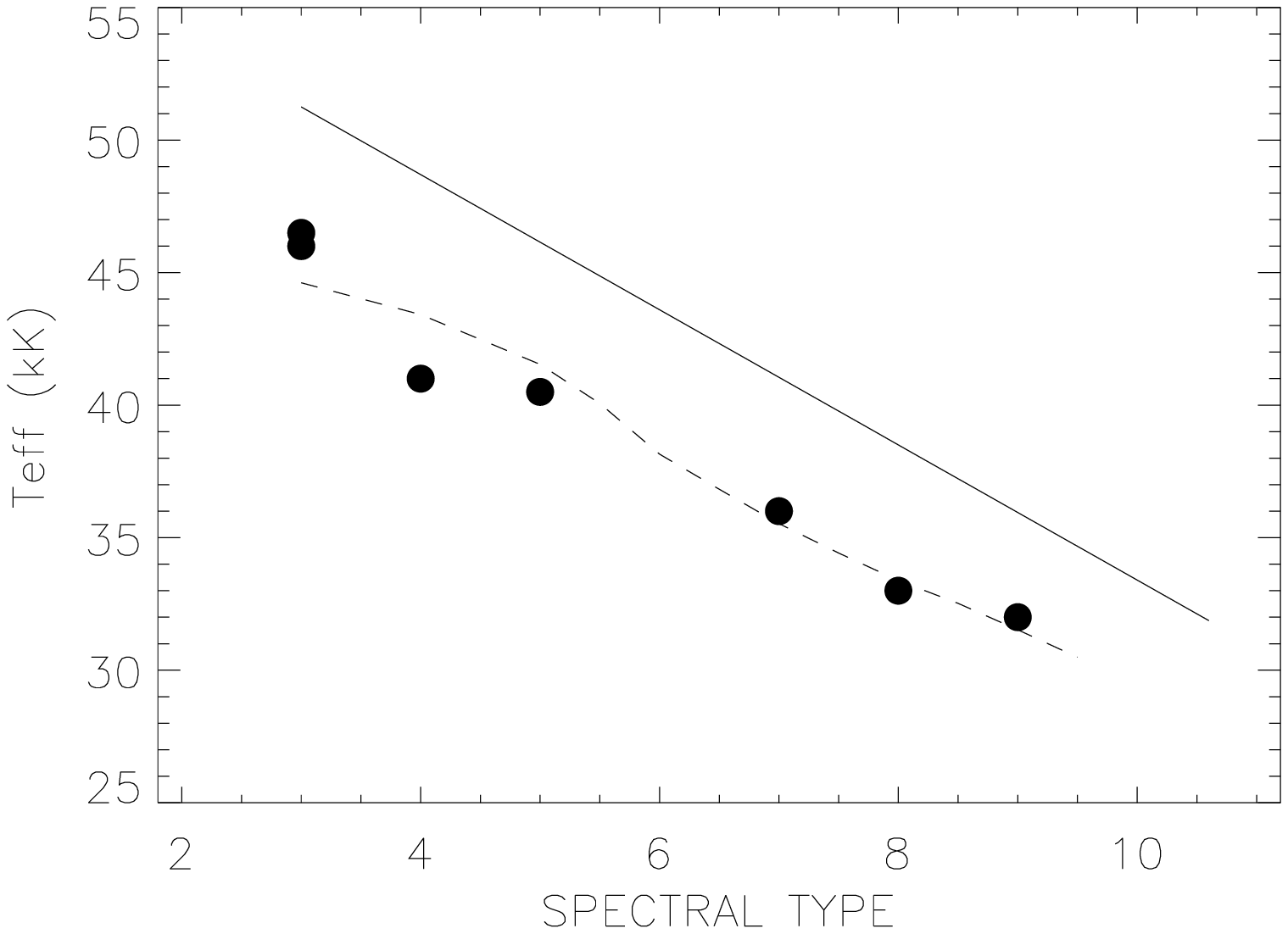}}
\end{minipage}
\hspace{-.5cm}
\begin{minipage}{4cm}
   \resizebox{\hsize}{!}
   {\includegraphics{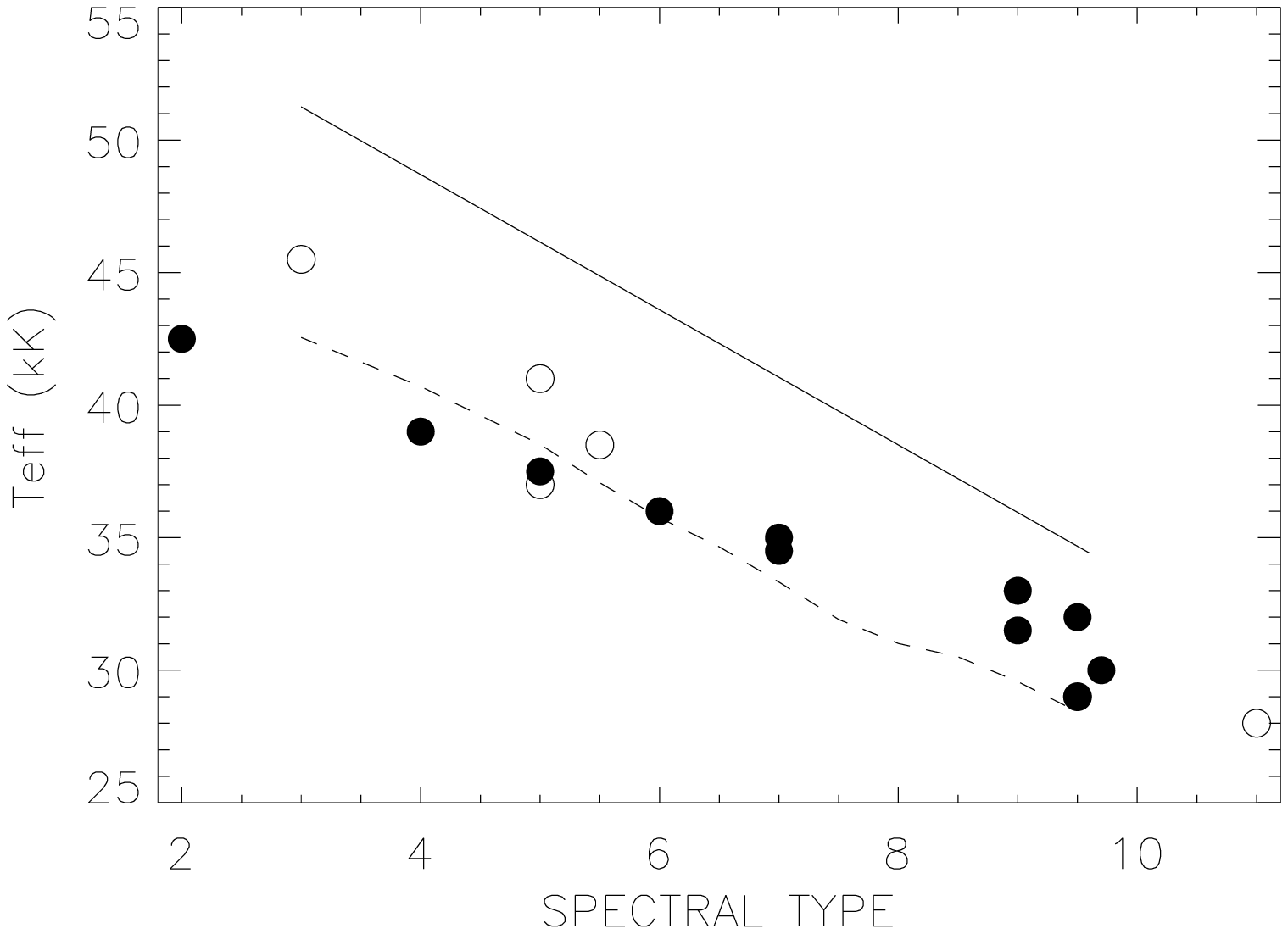}}
\end{minipage}
\begin{minipage}{4.3cm}
   \resizebox{\hsize}{!}
   {\includegraphics[angle=-90]{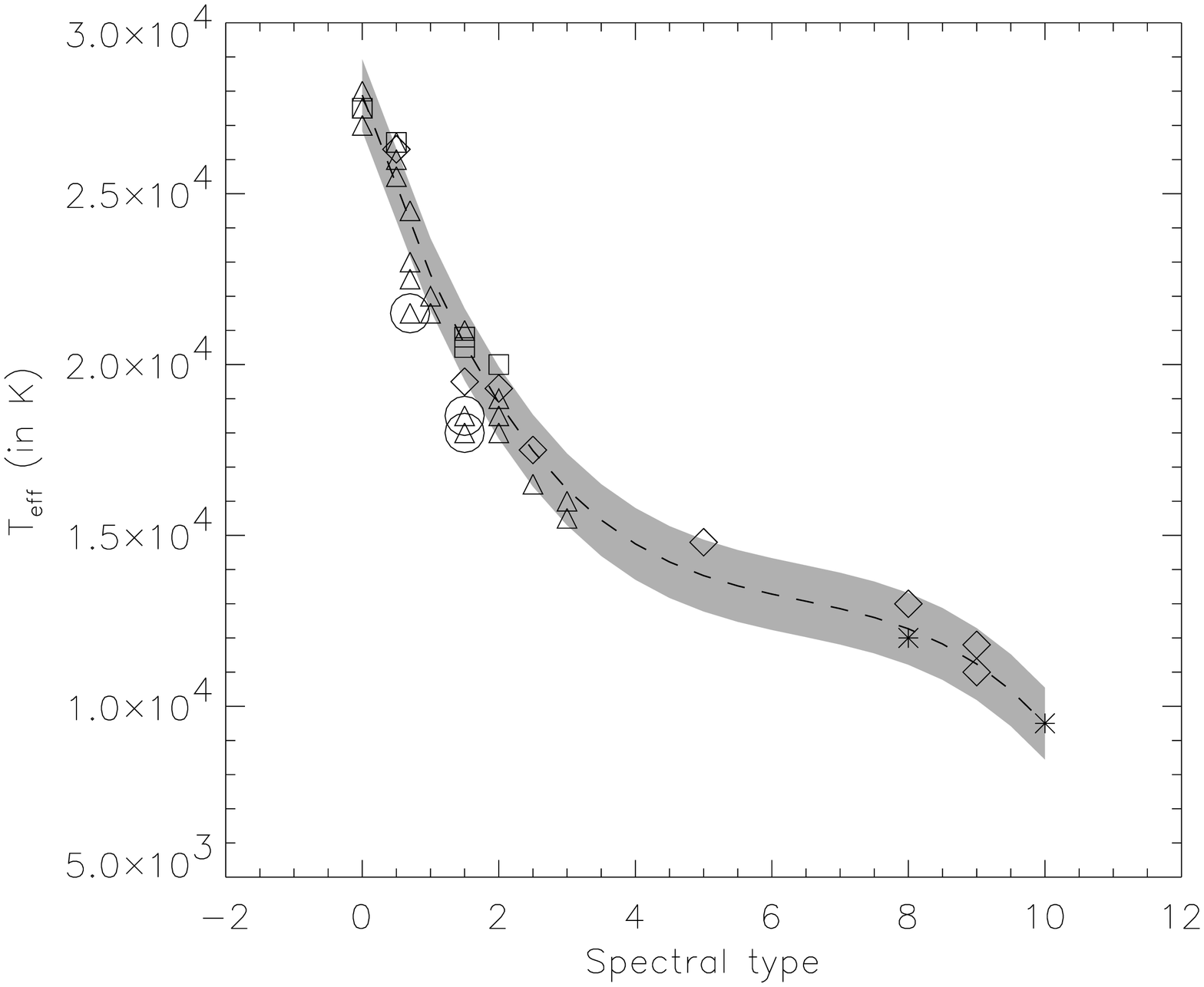}}
\end{minipage}
\caption{The temperature scale for Galactic O-dwarfs (left) and
supergiants (middle). The solid lines are for the \citet{vacca96} scale, the
dashed line for the ``theoretical'' scale defined by \citet{martins05a},
filled symbols are data from \citet{repolust04} and open symbols are data
from \citet{herrero02}. Right: The new temperature scale for B-supergiants
from \citet{MP08}, combining results from
\citet{urbaneja04,Crowther06,przybilla06,lefever07} and their own data. See
text.} 
\label{tempds}
\end{figure*}

Once the codes cited above became available, a series of papers were
published mostly devoted to the issue of temperature scales. The first
``modern'' calculations pointing to a cooler temperature scale were those
from \citet{martins02} using CMFGEN. These authors limited their
calculations to OB-dwarfs, so that the influence of mass-loss effects were
negligible. Therefore, the main differences with \citeauthor{vacca96} were
clearly due to line-blanketing. These differences could reach up to 4\,000~K
for early types, and decreased towards O9 and B0 types, as can be
appreciated in Fig.~\ref{tempds}.

\citet{herrero02} provided a temperature scale for supergiants in Cyg~OB2
using FASTWIND. They found differences of up to 8\,000 K. In this case both 
mass loss and line-blanketing played a role. These authors also showed that
two stars with the same spectral type and luminosity class may have
different effective temperatures if their wind densities are different.
While the results from \citeauthor{herrero02} were based on the analyses of
only seven Cyg OB2 supergiants, \citet{repolust04} presented an analysis of
24 stars (17 supergiants and giants and 7 dwarfs), based on a slightly
improved version of FASTWIND that confirmed the same trends. 

(Almost) identical trends have been found by \citet{martins05a} who 
provided\footnote{among other scaling relations for atmospheric parameters
and ionizing photon numbers.}, by means of CMFGEN models, new temperature
scales for massive OB stars of different luminosity classes. These scales
agree quite well with those from \citeauthor{repolust04}, and confirm that
new models result in effective temperatures that are several thousands
Kelvin cooler for early and intermediate spectral types, decreasing towards
late spectral types. A comparison of all these different temperature scales
is presented in Fig.~\ref{tempds}. 

Likewise detailed studies have been performed for B-supergiants
\citep{urbaneja04,Crowther06,lefever07,MP08} which show that also the
temperature scale of Galactic B-supergiants needs to be revised downwards.
Compared to the unblanketed and ``wind-free'' scale by \citet{McErlean99},
this amounts from 10 up to 20\%, the latter value being appropriate for stronger
winds. The resulting new \Teff-scale (described by a third order polynomial
fit, with gray-shaded 1-$\sigma$ uncertainties) is displayed in
Fig.~\ref{tempds}-right, and agrees perfectly with the results from an 
independent study by \citet{Trundle07}. 

Similar investigations have been performed for OB-stars in the Magellanic
Clouds (MCs), in particular to study metallicity effects on the effective
temperatures and mass-loss rates. \citet{massey04,massey05} investigated a
large sample of MC O-stars by means of FASTWIND, and provided a
spectral-type-\Teff\ calibration for the SMC. For the LMC, the situation
remained unclear, since their sample was concentrated towards the hottest
objects, O2-O4. Overall, it turned out that for a given spectral sub-type,
\Teff(SMC~dw) $>$ \Teff(MW~dw) $\approx$ \Teff(SMC~sg) $>$ \Teff(MW~sg),
where the \Teff-scale for SMC O-stars differs much less from the unblanketed
\citet{vacca96} calibration than the scale for their Galactic counterparts
(`dw' = dwarfs, `sg' = supergiants). This finding has been attributed to
less blanketing and weaker winds because of the lower SMC metallicity (see
below). For {\it dwarfs}, most of the results derived by Massey et al. are
in good agreement with investigations of {\it different} samples performed
with CMFGEN \citep{martins04} and TLUSTY\footnote{augmented by a CMFGEN
analysis of the wind parameters.}\citep{bouret03}, whereas a large number of
MC {\it supergiants} analyzed by means of CMFGEN \citep{Crowther02,
hillier03, evans04b} turned out to be significantly cooler, even cooler than
implied by the Galactic scale. In certain cases, this might be explained by
wind-blanketing effects, since some of the discrepant objects are rather
extreme, but other cases certainly need further inspection. \citet{heap06}
analyzed a sample of SMC O-dwarfs and giants, and compared their results
(using TLUSTY) with other studies. Though they stress the large scatter of
\Teff\ within individual sub-types, they found fair agreement with other
investigations, except for Galactic data derived by \citet{BG02} and
\citet{GB04} by means of a pure UV analysis (using WM-Basic), which seem to
suggest systematically cooler temperatures than all other studies. Also this
discrepancy has to be investigated in the near future.  

\paragraph{The VLT-{\sc flames} survey of massive stars.~~} One of the most
important recent projects on OB-stars was the VLT-{\sc flames} survey of
massive stars ({\sc flames} = {\sc f}ibre {\sc l}arge {\sc a}rray {\sc
m}ulti-{\sc e}lement {\sc s}pectrograph). By means of this campaign, the
massive stellar content of 8 young and old Galactic/MC clusters has been
spectroscopically investigated, in order to answer urgent questions
regarding (i) rotation and abundances (rotational mixing), (ii) stellar
mass loss as a function of metallicity and (iii) fraction and impact of
binarity. In total, 86 O-stars and 615 B-stars have been observed at high
resolution. An overview of objects and objectives has been given by
\citet{Evans05, Evans06}, and a summary of important results can be
found in \citet{Evans08}. 

\citet{mokiem06, mokiem07a}\footnote{by means of a genetic algorithm
combined with FASTWIND, see Sect.~\ref{sec:nltediagnostics}.} studied the
O-/early B-star targets of the {\sc flames} survey in the SMC and LMC,
respectively. They confirmed the basic results from \citet{massey04,
massey05}, but refined the spectral-type-\Teff\ scale, particularly with
respect to the LMC objects. They showed that, at least for O-dwarfs (which
are not ``contaminated'' by additional wind-effects), the effective
temperatures for a given spectral sub-type decrease with increasing
metallicity, i.e., \Teff(SMC) $>$ \Teff(LMC) $>$ \Teff(MW). A similar result
was derived by \citet{Trundle07} for the {\sc flames} B-type dwarfs, in this
case based on TLUSTY model atmospheres. 

\paragraph{Wind properties of OB stars at different metallicities.~~}
\label{winds_metallicity}

Wind-properties of Galactic/MC OB-stars (primarily mass-loss rates, \mdot,
and velocity field parameters, $\beta$) have been determined by numerous
investigations (overlapping with those mentioned above and detailed in
\citealt{Puls08}), from \Ha\ (partly combined with \HeII\ 4686) and the UV.
In most cases, terminal velocities, \vinf, have been adopted from
UV-measurements (see Sect.~\ref{sec:approxdiag}) and/or calibrations
\citep{KP00}. Wind-momentum luminosity relations (WLR) have been inferred
and compared with theoretical predictions, mostly from \citet{Vink00,
Vink01}, see Sect.~\ref{sec:statmodels}. In summary, the following results
have been obtained:

\smallskip \noindent (i) For most {\it O-/early B-stars}, the theoretical
WLR is met. Notable exceptions are O-supergiants with rather dense winds,
where the ``observed'' wind-momenta appear as ``too large'' (see
\citealt{markova04, repolust04}), which has been attributed to wind-clumping, see
Sect.~\ref{sec:clumping}, and low luminosity O-dwarfs, where the
``observed'' wind-momenta are considerably lower than predicted, which has
been denoted as the ``weak wind problem'' and will be covered in
Sect.~\ref{sec:weakwinds}. 

\smallskip \noindent (ii) The bi-stability jump in the B-star domain is
represented by a gradual decrease in \vinf\ over the bi-stability region
\citep{evans04a, Crowther06}, by a factor of roughly 2.5 with respect to the
ratio of \vinf/\vesc. According to \citet{MP08}, the limits of this region
are located at somewhat lower \Teff\ (betweeen 18 to 23~kK) than those
discussed by \citet{Vink00}. Most important, however, is the finding that
\mdot\ changes over the bi-stability region by a factor (in between 0.4 and
2.5, \citealt{MP08}) which is smaller than the predicted factor of 5. Thus 
it seems that the decrease in \vinf\ over the bi-stability region is not
over-compensated by an increase of \mdot\footnote{at least not if the winds
from hotter objects are not {\it substantially} stronger clumped than those
from the cooler ones.}. This finding seems to apply not only to the
bi-stability region itself, but to the complete ``low temperature region''
(i.e., all mid/late B-type stars), where the predicted mass-loss rates are
larger than those found from NLTE analyses based on unified atmospheres
\citep{Vink00,Trundle05,Crowther06,lefever07,Benaglia07,MP08}. This problem
is reminiscent of the findings by \citet{Kud99} who derived wind-momentum
rates of mid B-supergiants which were much smaller (by a factor of 10) than
those of early B-supergiants. Bearing in mind that \citet{Kud99} used
unblanketed atmospheres, even if applying an upwards ``correction'' of their
mid B-supergiant mass-loss rates by a factor of three (accounting for the
arguments from page~\pageref{blanketing_mdot}), their results would still be
in contrast to the predictions.

Nevertheless, a certain effect dividing hotter from cooler winds is present
indeed, evident from the decrease in the ratio \vinf/\vesc\ and likely
related to the principal bi-stability mechanism. Moreover, as shown by
\citet[ see also \citealt{MP08}]{Benaglia07}, a {\it local} increase of the
ratio \mdot/\vinf\ seems to be present at least {\it inside} the transition
zone, which would partly support the predictions by Vink et al., though not
regarding the mass-loss properties of objects below the jump.  Further
effort to clarify the situation is certainly needed.

\smallskip \noindent (iii) Combining the results of the various
investigations of Galactic OB-stars and A-supergiants (the latter from
\citealt{Kud99}), the corresponding WLRs (as a function of spectral type)
extend over significant ranges in \Teff, assuming that clumping is not a
strongly varying function of spectral type. Excluding weak winds, three
rather well-defined relations for O and early B-stars (roughly above 23~kK),
for mid and late B-supergiants (between roughly 18~kK and 10~kK) and for
A-supergiants seem to exist, where the slope of the former two relations is
rather similar, and the slope of the latter is steeper, in accordance with
theoretical predictions. In particular, \citet{Kud99} derived a slope
corresponding to $\alpha' = 0.38 \pm 0.07$ for their sample of
A-supergiants, which is in perfect agreement with corresponding results from
line-statistics (page~\pageref{alpha_asg}). Within the (re-defined)
bi-stability region (18 to 23~kK), on the other hand, the scatter seems to
be much larger, possibly due to ongoing changes in the global ionization
equilibrium \citep{MP08}. 

Translated into the number of driving lines (or force-multiplier parameter
$k$, cf. Eq.~\ref{eq_wlr}), the above results imply that this number remains
rather constant over large ranges in \Teff\ (again in accordance with
theoretical predictions, \citealt{Vink00} and
Sect.~\ref{sec:standardmodel}), which are divided by the transition from
\FeIV\ to \FeIII\ and from \FeIII\ to \FeII\ as the major driving agents.
The ``only'' difference between these principal predictions and the
observations is a quantitative one, regarding the location of the (first)
bi-stability region and the degree to which extent \Neff\ (or $k$) changes
from one region to the other.

\begin{figure}
\centerline{\includegraphics[width=8cm]{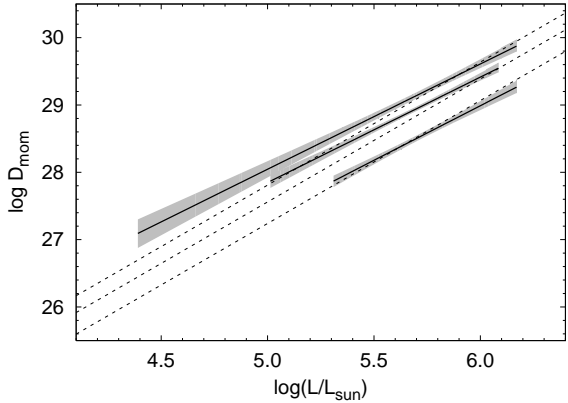}}
\caption{Comparison of observed (solid, with grey shaded 1-$\sigma$
confidence intervals) and theoretically predicted (dashed, \citealt{Vink00,
Vink01}) WLRs for Galactic, LMC and SMC O-/early B-stars (from top to
bottom). Adapted from \citealt{mokiem07b}, see text.} 
\label{wlr_comp}
\end{figure}

\smallskip \noindent (iv) For a given luminosity, the mass-loss rates of
SMC-stars are lower than for their Galactic counterparts, consistent with
theory \citep{massey04, massey05}. A more precise {\it quantification} of
the metallicity dependence of the winds from O-/early B-stars was possible 
within the {\sc flames} survey. From the analysis of the SMC/LMC objects by
\citet{mokiem06} and \citet{mokiem07a}, respectively, and in combination
with data from previous investigations, \citet{mokiem07b} derived the WLRs
for Galactic, LMC and SMC objects, with rather narrow 1-$\sigma$ confidence
intervals, and showed that the wind-momenta strictly increase with
metallicity $Z$ (i.e., the WLR of the LMC lies in between the corresponding
relations for the MW and the SMC, see Fig.~\ref{wlr_comp}). Using $Z$(LMC)
$\approx 0.5 Z_\odot$, $Z$(SMC) $\approx 0.2 Z_\odot$, and allowing for a
modest ``clumping correction'' for dense winds (following the arguments by
\citealt{repolust04}, see also \citealt{markova04} and Sect.~\ref{sec:clumping}), they obtained $\mdot
\propto Z^{0.72 \pm 0.15}$. This result is in very good agreement with
theoretical predictions, both from (stationary) models
(Sect.~\ref{sec:lowzmodels}) and from line statistics
(page~\pageref{sec:linestat}). \citet{mokiem07b} pointed out that the
derived metallicity dependence should remain unaffected from potential
future changes in the absolute values of mass-loss rates due to global
clumping corrections, as long as the clumping properties were similar in
winds of different metallicity. 

\begin{figure}
\centerline{\includegraphics[width=8cm]{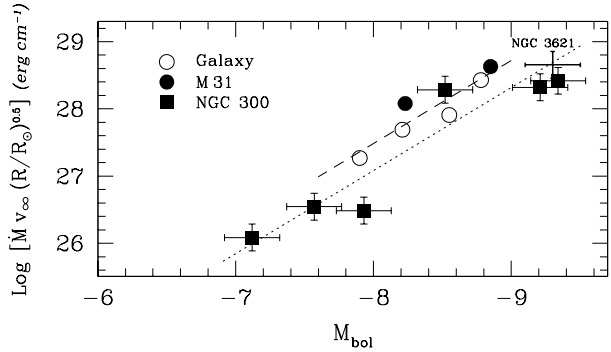}}
\caption{WLR for Galactic and extragalactic A supergiants. Dashed: Linear
regression for Galactic and M31 objects. Dotted: Galactic relation scaled to the
mean abundance found for NGC 300 and NGC 3621, $Z/Z_\odot = 0.4$. Adapted 
from \citealt{BK04}.} 
\label{wlr_a_extra}
\end{figure}

\paragraph{Beyond the Magellanic Clouds.~~}
\label{aextra}

As already noted in our discussion of the (theoretical) concept of the WLR
(Eq.~\ref{wlr}), one of the early aims to use this relationship was 
to derive extragalactic distances by purely spectroscopic means.
Particularly advantageous for this purpose are early A-supergiants, because
they are the visually brightest ``normal'' stellar objects. Also, problems of
multiplicity and crowding are (almost) negligible because of their
brightness, their short life-times, and their relatively old evolutionary
ages (see \citealt{Kud08a}). Moreover, their \Ha-line can be used to
measure \mdot\ and \vinf\ simultaneously (e.g., \citealt{Kud99, KP00}).
Subsequently, a number of these objects have been analyzed in Local Group
galaxies and beyond to compare with Galactic counterparts and to check the
theoretical concept. A summary of the results has been given by
\citet{BK04}, highlighted in Fig.~\ref{wlr_a_extra} which displays the
corresponding WLRs. The relation for ``solar'' metallicity is defined by
Galactic objects plus two M31 stars, and additional wind-momentum rates for
6 A-supergiants in NGC~300 and one object in NGC~3621 are displayed.
Assuming that the latter stars have an abundance similar to the mean found
for their host galaxies, $Z \approx 0.4 Z_\odot$, the dotted line provides
the theoretical expectation when the Galactic WLR is scaled to the lower
metallicity. It is evident that the WLR concept seems to hold. Details
regarding extragalactic samples, observations and analyses can be found in
\citet{BK04} and references therein.

After these promising investigations had been finished, however, no further
attempt was made to derive distances via the WLR. The reason for that 
was that \citet{Kud03} had developed an alternative possibility to use
late B-/early A-supergiants as standard candles, by means of what they
called the {\it F}lux-weighted {\it G}ravity {\it L}uminosity {\it
R}elationship. The authors showed that the flux-weighted gravity, $g/\Teff^4$,
and the absolute bolometric magnitude are strongly correlated for these
objects, since they pass their short evolutionary phase with {\it nearly
constant luminosity and mass}, and because luminosity and mass are related via
$L \propto M^x$ with $x \approx 3$. Thus, the underlying difference in both
methods is that the WLR method exploits the information on radius contained
in the {\it stellar wind}, whereas the FGLR method relies on {\it
evolutionary} facts. 

As shown in the original paper and a follow-up study \citep{Kud08b}, this
new method to derive distances seems to have advantages over the WLR method,
because of the high precision that is possible even for low-resolution
observations (0.1 mag in the distance module when analyzing 10 target stars)
and its simplicity. E.g., the FGLR method requires much less (and only
photospheric\footnote{at least when discarding extreme objects with
wind-affected SEDs.}) parameters to be determined (\Teff, \logg, and metallicity),
and is also much less affected by additional dependencies. Remember,
e.g., that the mass term in Eq.~\ref{eq_mdot} only partly vanishes for
A-supergiants, since $\alpha' \sim 0.4$ and not close to 2/3
as in the OB-star case. Moreover, it is still unclear whether and in how far
clumping affects the WLR method. 

A first impressive demonstration of the power of the FGLR method has been
given by \citet{Urbaneja08} who determined the distance to the Local Group
Galaxy WLM with a distance modulus of 24.99$\pm$0.10 mag (which is only
0.07~mag larger than the most recent Cepheid distance). Further applications
within the {\it Araucaria} project (``Measuring improved distances to nearby
galaxies'', \citealt{Gieren05}) are to be expected soon.

\paragraph{Central Stars of Planetary Nebulae.~~}
\label{cspn}

Although Central Stars of Planetary Nebulae (CSPN) are objects of
significantly lower luminosity and of completely different evolutionary
status, they have winds with a {\it mean density} similar to massive stars,
and their wind-momenta seem to correspond to an extrapolation of the WLR of
O-stars \citep{Kud97, Kud06, Hultzsch07}, though with a larger scatter. This
finding must be regarded as an encouraging success of the interpretation of
winds in terms of radiative driving and of the concept of the WLR. Let us
point out that recent analyses by means of FASTWIND have suggested that some
(but not all) of these winds are {\it strongly} clumped, which could be
quantified because of the different reaction of the \HeII\ 4686 and the \Ha\
line on wind-clumping at temperatures around $\Teff \approx$~30~kK
\citep{Kud06}. 

In an interesting study, \citet{Pauldrach04} analyzed the winds of 9
CSPN by means of WM-Basic and {\it method (i)} as outlined in
Sect.~\ref{sec:nltediagnostics}, i.e., they derived the complete set of 
stellar parameters by comparing {\it self-consistent} wind models with 
UV-observations. The information provided by the wind features (\vinf,
\mdot) permits to derive the stellar radius via such models, since the
corresponding dependencies $\vinf=f(\vesc)$ and $\mdot =
f(\Lstar,\Mstar,\Gamma)$ allow for a unique solution for \Rstar\ {\it
given the observational constraints} (for a discussion of the underlying
principles, see \citealt{Kud92}). By this method, all quantities such as
luminosity, mass, and distance can be constrained {\it without} relying on the
(core)mass-luminosity relation taken from stellar structure theory 
and post-AGB evolution.\footnote{which is usually done when {\it purely
spectroscopic analyses} such as outlined above are performed.} Notably, the
results by \citet{Pauldrach04} showed severe departures from
this generally accepted relation, and masses from 0.4 to 1.4 \msun\ have been
derived, with five out of the nine stars located close to (but not above)
the Chandrasekhar mass limit for white dwarfs. \citet{Pauldrach05} argues
that these objects might belong ``to a not yet understood subgroup of CSPNs
that evolve to white dwarfs which can end up as supernovae of type Ia'' (see
also \citealt{Maoz08}). 
For further progress on this issue, it needs to be clarified in how
far the wind and spectrum synthesis calculations are not biased by
contaminating effects. An important step towards this is the
determination of precise distances to CSPN (to enable a measurement of 
the stellar radius) for which fairly resolved spectra can be obtained as well.

\subsubsection{Luminous Blue Variables}
\label{sec:lbvs} 
\begin{figure}
\begin{center}
   {\includegraphics[width=8cm]{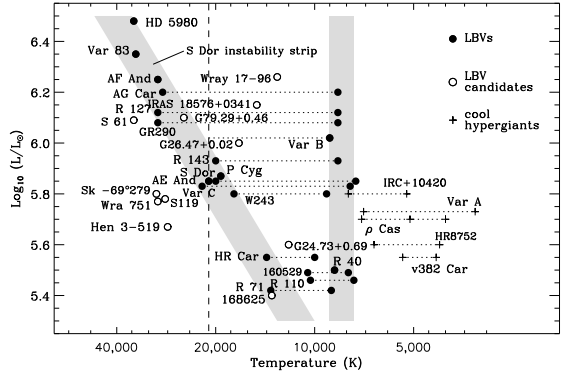}}
\end{center}
\caption{The position of the LBVs in the upper Hertzsprung-Russell diagram.
Note that the general population of blue supergiants that occupy the same
part of the HRD are not plotted.  The slanted band running from 30~kK at
high $L/\lsun$ to 15~kK at lower luminosity is often referred to as the
S\,Dor instability strip. The vertical band at a temperature of $\sim$
$8\,000$ K represents the position of the LBVs in outburst. The vertical
line at 21\,000 K is the position of the observed bi-stability jump (the
center of the observed bi-stability region). Adapted from \citet{Smith04}.} 
\label{hrd}
\end{figure}

\paragraph{What is an LBV?~~} Luminous Blue Variables represent a
short-lived ($\sim 10^{4}-10^{5}$ years) phase of massive star evolution in
which the stars are subject to significant effective temperature changes.
They come in two flavors. The largest population of $\sim$30 LBVs in the
Galaxy and the Magellanic Clouds is that of the S\,Doradus variables with
magnitude changes of 1-2 magnitudes on timescales of years to decades.
These are the characteristic S\,Dor variations, which are represented by the
dotted horizontal lines in Fig.~\ref{hrd}.  The general understanding is
that the S\,Dor cycles occur at approximately constant bolometric luminosity
(which has yet to be proven) -- principally representing temperature
variations. The second type of LBV instability involves objects that show
truly giant eruptions with magnitude changes of order $3-5$ during which the
bolometric luminosity most certainly increases. In the Milky Way it is only
the cases of P~Cygni and Eta Carina which have been witnessed to experience
such extreme behaviour. 

Whether these two types of variability occur in similar or distinct objects
is not yet clear, however in view of the ``unifying'' properties of the LBV
P~Cygni it is highly probable that the S\,Dor variables and giant eruptors
are related, that they are in a similar evolutionary state, and that they
are subject to the same type of instabilities near the Eddington limit (see
\citealt{vink09} for a review). We note that the true nature of the LBV
instability has yet to be revealed. 

\paragraph{LBV parameters and abundances.~~} Reliable determinations of
stellar properties in LBVs became possible only after the advent of unified
model atmospheres. Even then, however, in some extreme objects like
$\eta$~Car, the enormous complexity of the observed ground based spectra
\citep{hillier92}, contaminated by multiple emission regions, hinders a
proper quantitative spectroscopic analysis. In this case, stellar
parameters need to be determined either from ``cleaner'' wavelength regions
(e.g., from the millimeter continuum, \citealt{cox95}) or from ``cleaner'',
resolved observations (e.g., using the {\sc hst}, \citealt{hillier01}, or the
recent, really promising results from long baseline interferometers such as
{\sc amber-vlti}, \citealt{weigelt07}).

The
outstanding wind density (with \mdot$\sim 10^{-3}$\msunyr as estimated for
this object, \citealt{cox95, hillier01}) places $R(\taur=2/3)$ at 80\% of
the terminal velocity, impeding any derivation of the hydrostatic radius.
Nevertheless, CNO abundances could be determined and are consistent with
those found for the surrounding nebula (for a recent review, see
\citealt{najarro08c}).

For less extreme LBVs such as P~Cygni (\mdot$\sim 10^{-5}$\msunyr,
\citealt{najarro01}), more detailed investigations could be carried out. A
major breakthrough was obtained by \cite{langer94} who presented a 
quantitative study of this object, combining evolutionary and hydrodynamical
aspects. Subsequently, \cite{najarro97b} were able to accurately constrain
the main stellar properties, including Helium abundance, by means of
homogeneous, unblanketed atmospheric models and UV to radio observations.
These results were only minorly altered by utilizing blanketed models
including clumping \citep{najarro01}, which confirmed the presence of CNO
processing and provided, for the first time, direct metallicity measurements
for an LBV.

Apart from the problem of determining the (pseudo-hydrostatic) core radius
(see also below), two other major difficulties arise when analyzing the
spectra of LBVs. The first one is the determination of \Teff. Only if both
\HeI\ and \HeII\ are available (as for, e.g., He3-519 and AG~Car,
\citealt{smith94}) a robust estimate may be obtained (although in some cases
the \HeI\ spectra may be controlled by the presence of a binary companion,
e.g., in $\eta$~Car). Otherwise, the effective temperature has to be
obtained from simultaneous fits to lines from species with different
ionization potentials (e.g. \SiII, \MgII, \FeII). This is the case for the
two LBVs in the Quintuplet cluster at the Galactic Center (the Pistol Star
and \#362, \citealt{figer98,geballe00,najarro08a}). The second problem
arises for low excitation LBVs and is related to the Helium abundance. Given
the low excitation, \HeII\ recombines in the inner regions and the weak
\HeI\ lines form very close to $R(\taur=2/3)$. In consequence, a degeneracy
of the helium abundance appears (as for, e.g., HDE\,316285,
\citealt{hillier98b}) as basically for each H/He ratio there is a
\mdot-value that reproduces the observed H and \HeI\ lines equally well.

\begin{figure}
\begin{center}
   {\includegraphics[width=8cm]{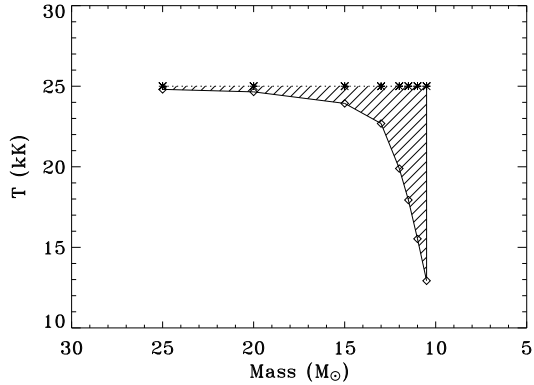}}
\end{center}
\caption{The possible formation of a ``pseudo-photosphere'' for a low mass
LBV. The difference in inner (dashed) and apparent temperature
(representative for the size of the computed pseudo-photosphere) is plotted
against the stellar mass. These computations have been performed for a
constant luminosity of log $L/\lsun$ $=$ $5.7$. The mass is gradually
decreased until the object approaches the Eddington limit, and the apparent
temperature drops as a result of low effective gravity and very high mass loss,
forming a pseudo-photosphere. Adapted from \citet{Smith04}.} \label{app}
\end{figure}

\paragraph{Do LBVs form pseudo-photospheres?~~} It is evident that extreme
objects such as $\eta$\,Car with present-day mass-loss rates as high as
10$^{-3}$ $\msunyr$ form pseudo-photospheres, but it is not generally
accepted whether the mass-loss rates of the more common S\,Dor variables are
large enough to form false photospheres. The issue is of particular interest
for the following reason: whilst it {\it appears} that the photospheric
temperatures of the objects in Fig.~\ref{hrd} change during HRD transits,
there is the alternative possibility that the underlying object does not
change its temperature but changes its mass-loss rate instead -- resulting
in the formation of an optically thick wind (e.g. \citealt{Davidson87}). As
a result of increased mass loss it is hypothetically possible to form a
pseudo-photosphere. Until the late 1980s this was the leading idea to
explain the color changes of S\,Dor variables. Using more advanced NLTE
model atmosphere codes, \citet{Leitherer89} and \citet{deKoter96} showed
that the color changes consistent with measured LBV mass-loss rates are not
large enough to make an LBV appear cooler than the temperature of its
underlying surface.  But is the issue of pseudo-photospheres in LBVs
completely settled?  

In NLTE codes used to investigate pseudo-photosphere formation in LBVs, the
surface layers are based on the physical relation between $L$, $R$, and $T$,
with $L = 4 \pi R_{\rm in}^2 \sigma_{\rm B} T_{\rm in}^4$. As the inner
radius, $R_{\rm in}$, is chosen to lie in the photosphere, the input
temperature $T_{\rm in}$ does not necessarily equal the output apparent
temperature, and one therefore defines the ``effective'' temperature $\Teff$
at a position where the thermalization optical depth equals $1/\sqrt{3}$ in
the optical V band (see \citealt{deKoter96} and references therein). For the
optically thin winds of OB stars, $\Teff$ equals $T_{\rm in}$, but for some
LBVs that lose mass at rates as high as $\mdot \sim10^{-4}\,\msunyr$ cases 
might exist where there are significant differences between $T_{\rm in}$ and
$\Teff$. If the wind is dense enough for the optical photons to emerge from
a layer of rapid wind acceleration, the LBV can be considered to be forming
a pseudo-photosphere. This may be favoured by lower LBV masses, as proximity
to the Eddington limit provides larger mass loss and an increase of the
photospheric scale-height. 

Despite the proximity of LBVs to the Eddington limit, current consensus is
that optically thick winds are generally not present in LBVs. However,
\citet{Smith04} studied a special group of ``missing'' LBVs where the
luminosity is relatively large (with log $L/\lsun$ $=$ 5.7) but the stellar
masses are possibly much lower than those of the classical LBVs (at log
$L/\lsun$ $\sim$ $6$). In other words, the lower-luminosity LBVs may
well be in closer proximity to {\it their} Eddington limit. Figure~\ref{app}
shows the potential formation of an optically thick wind for a relatively
low-mass LBV. The size of the temperature difference (dashed vs. solid) is a
proxy for the extent of the pseudo-photosphere. The figure demonstrates that
for masses in the range 15-25 $\msun$, the winds remain optically thin, but
when the stellar mass approaches 10 $\msun$ and the star enters the regime
near the Eddington limit, the photospheric scale-height blows up --
resulting in the formation of a pseudo-photosphere.  It should be noted that
the reality of this scenario hinges critically on the input stellar
mass, and LBV masses are poorly constrained from observations.

\paragraph{The line-driven winds during S Doradus variations.~~} Whilst most
Galactic and Magellanic Cloud LBVs have been subject to photometric
monitoring, only few have been analyzed spectroscopically in sufficient
detail to understand the driving mechanism of their winds.  As reported
above, mass-loss rates are of the order of $10^{-3}$ - $10^{-5}$ \msunyr,
whilst terminal wind velocities are in the range $\sim100-500\,\kms$. Of
course, these values vary with $L$ and $M$, but there are also indications
that the mass loss varies as a function of $\Teff$ when the S\,Dor variables
transit the upper HRD on timescales of years. It is this aspect that
provides us an ideal laboratory for testing the theory of radiation-driven
winds.

The Galactic LBV AG\,Car is one of the more comprehensively monitored S\,Dor
variables. \citet{Stahl01} investigated its mass-loss behaviour during the
1990s -- modeling the \Ha\ line profile in detail. \citet{Vink02} showed
that the Stahl et al. mass-loss rates rise, drop, and rise, in line with
radiation-driven wind models in which the variations in mass-loss rate and
wind velocity are attributable to ionization shifts of Fe -- the dominant
element driving the wind. 

It is relevant to mention here that this variable wind concept (basically
wind bi-stability) has been put forward as a potential explanation for the
circumstellar density variations inferred from quasi-periodic modulations in
the radio light-curves of some transitional supernovae \citep{Kotak06}.
Additionally, the same mechanism is able to account for wind-velocity
variations in the P~Cygni absorption line spectra of supernovae such as
SN2005gj \citep{Trundle08}. As current stellar models predict massive stars
with $M$ $\ge$ 30 \msun\ to explode at the end of the WR phase, rather than
during or after the LBV phase, the ultimate implications could be
far-reaching -- impacting even our most basic understanding of massive-star
evolution.

\paragraph{Continuum driven winds during giant eruptions.~~}

Whilst, during ``quiet'' phases, LBVs lose mass most likely via ordinary
line-driving, they also appear to be subject to one or more phases of much
stronger mass loss. For instance, the giant eruption of $\eta$ Car with a
cumulative loss of $\sim$10 \msun\ between 1840 and 1860 \citep{Smith03}
which resulted in the Homunculus nebula corresponds to $\mdot
\approx$~0.1-0.5~\msunyr, which is a factor of 1000 larger than that 
expected from a line-driven wind at that luminosity. Such a strong mass loss
has been frequently attributed to the star approaching or even exceeding
the Eddington limit\footnote{here and in the following, we will consider a
generalized Eddington limit, accounting for Thomson scattering and bf/ff
continuum processes.}, but line-driven mass loss is difficult to invoke,
since the formal divergence of \mdot\ for $\Gamma \rarrow 1$
(Eq.~\ref{eq_mdot}) is accompanied by a vanishing terminal speed, \vinf\
$\propto$ \vesc\ $\propto (1-\Gamma)^{0.5} \rarrow 0$. (Remember that the
wind-momentum rate is almost independent of $\Gamma$). 

Building upon pioneering work by \citet{Shaviv98, Shaviv01}, \citet{OGS04}
investigated this problem in some detail and developed a simple theory of
``porosity-moderated'' continuum driving in stars that exceed the Eddington
limit. At first, they showed that the divergence of {\it line-driven} mass
loss for $\Gamma \approx 1$ is actually limited by the {\it photon tiring}
\label{tiring} associated with the work needed to lift material out of the
star's gravitational potential. The authors argue that {\it
continuum-driven} winds in super-Eddington stars with a mass loss close to
this {\it tiring limit}, $\mdot_{\rm tir} = \Lstar$/ $(G \Mstar/\Rstar)$, will
result in a stagnating flow at a finite radius, which should lead to
extensive variability and spatial structure. This should also be true for
other instabilities (such as a convective one), which are expected to be
present in the envelopes and atmospheres of stars close to or beyond the
Eddington limit. 

As first noted by \citet{Shaviv98}, the {\it porosity} of such a structured
medium can reduce the effective coupling between the matter and radiation 
by lowering the {\it effective} opacity in deeper layers (i.e., $\Gamma_{\rm
eff} < 1 $ for $\Gamma >1$), thus enabling a quasi-hydrostatic photosphere,
but allowing for a transition to a supersonic outflow when the structures
become optically thin, i.e., $\Gamma_{\rm eff}$ $\rarrow \Gamma$ for
decreasing optical depth. As detailed in Sect.~\ref{sec:clumpquant}, the
effective opacity can be approximated by 
\beq
\kappa_{\rm eff} \approx \frac{\bar \kappa}{\tau_{\rm c}} = \frac{1}{h} 
\quad(\tau_{\rm c} > 1),\qquad 
\kappa_{\rm eff} \approx \bar \kappa \quad(\tau_{\rm c} < 1), \qquad 
\tau_{\rm c} = \int \kappa_{\rm eff} \dd r
\label{porositylength}
\eeq
for the optically thick and thin case, respectively, where $\bar \kappa$ is
the mean opacity allowing for micro-clumping, $\tau_{\rm c}$ is the
continuum optical depth and $h \equiv L^3/l^2$ the {\it porosity length}
introduced by \citet{OGS04} with $l$ the size and $L$ the separation of
individual clumps. In other words, the porosity length is the photon's mean
free path for a medium consisting of optically thick clumps.

Based on this porosity length and the Ansatz that it should scale with the
gravitational scale height $H$ (in analogy to the mixing length theory), the
authors derived scaling laws for a porosity-moderated, continuum-driven
mass-loss rate from stars with $\Gamma > 1$. For a super-Eddington model
with a {\it single} porosity length $h \approx H$ and $\Gamma >> 1$, $\mdot
\approx \Lstar/(a c)$, consistent with the earlier findings by
\citet{Shaviv01}. This is much higher than typical for line-driven winds but
still only a few percent of the tiring limit. Even higher \mdot\ as
implied from the ejecta of $\eta$~Car could be obtained by a power-law
porosity model (in analogy to the line-strength distribution) which should
mimic a broad range of clump scales. For power law exponents $\alpha_p < 1$,
\mdot\ becomes enhanced over the single-scale model by a factor increasing
with $\Gamma^{1/\alpha_p -1}$, approaching the tiring limit under certain
circumstances. Together with quite fast outflow speeds, \vinf\ $\approx
\mathcal{O}(\vesc)$\footnote{calculated here without the correction for
$\Gamma$.} and a velocity law corresponding to $\beta
=1$, the derived wind structure can actually explain the observational
constraints of giant outbursts in $\eta$~Car and other LBVs. Moreover, the
porosity model retains the essential scalings with gravity and radiative
flux (the von Zeipel theorem, cf. Sect.~\ref{sec:rot}) that would give a
rapidly rotating, gravity-darkened star an enhanced polar mass loss and flow
speed, similar to the bipolar Homunculus nebula. As stressed by
\citet{vanMarle08}, continuum driving does {\it not} require the presence of
metals in the stellar atmosphere. Thus, it is well-suited as a driving agent
in the winds of low-metallicity and First Stars and may play a crucial role
in their evolution.

\subsubsection{WR-stars}
\label{sec:wrstars}

The subject of WR stars was recently reviewed by \citet{Crowther07} to
which we refer for details. Here we briefly discuss some of the recent 
progress in WR wind research.

WR stars can broadly be divided into nitrogen-rich WN stars and carbon (and
oxygen) rich WC (and WO) stars. The principal difference between the two
subtypes is believed to be that the N-enrichment in WN stars is merely a
by-product of H-burning, whilst the C in WC/WO stars is a sign of the fact
that He-burning products have reached the stellar surface. As a result, WC
stars are thought to be more evolved than WN stars. 

We re-iterate that a classification as a WR star reflects purely
spectroscopic terminology -- signaling the presence of strong and broad
emission lines in the spectra which are produced by strong winds. Such
spectra can originate in evolved stars that have lost a significant amount
of their initial mass, or alternatively from an object that has formed with a
large initial stellar mass and luminosity. This latter group of WR stars may
thus include objects still in their core H-burning phase of evolution.

During the last decade, we have witnessed at least two important
developments in the study of WR stars. One advancement concerned the
inclusion of line-blanketing in NLTE models which, contrary to the 
situation for the O stars (see Sect.~\ref{sec:obastars}), led to an {\it
increase} in $\Teff$ and $L$ for WR stars \citep{schmu97,hilliermiller99}:
The strong dependence of the ionization equilibrium of the WR wind on the
interaction of the \HeII\ \Lya\ line with metal lines close in wavelength
\citep{schmu97} and the correct treatment of the \HeII\ n$\rightarrow$1
series near the \HeII\ Lyman limit \citep{hilliermiller98} lead to a lower
ionization in the wind. Thus, a higher temperature is required to reproduce
the observed WR lines.

\begin{figure}
\begin{center}
   {\includegraphics[width=6cm]{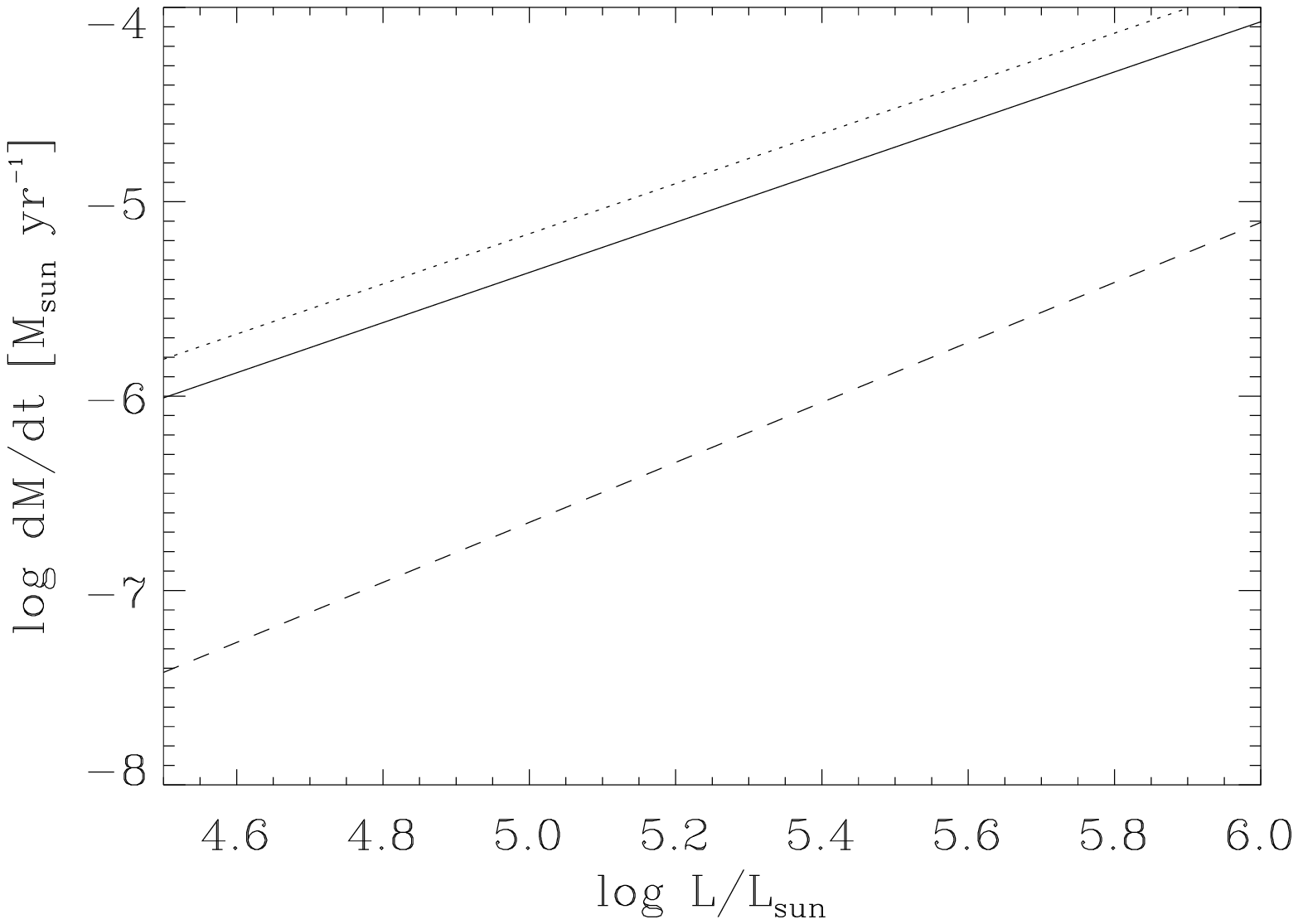}}
\end{center}
\caption{Comparison of mass-loss rates from WR and Galactic OB-supergiants. 
Solid and dotted lines represent mean relations for H-poor WN (solid) and WC
stars (dotted) as provided by \citet{NL00}, with \{$Y=0.98,Z=0.02$\} and
\{$Y=0.55,Z=0.45$\}, respectively. See also Fig.~7 in \citet{Crowther07}.
The dashed line corresponds to Galactic O-supergiants, and has been taken
from the WLR derived by \citet[see Fig~\ref{wlr_comp}]{mokiem07b}, assuming
prototypical values for \vinf (2000~\kms) and \Rstar (15 \rsun).}
\label{mdot_wr}
\end{figure}

A second
development was the result of the realization that WR winds are clumped
(e.g. Moffat et al. 1988) leading to a decrease in the empirical WR
mass-loss rate by a factor $2-4$ (e.g., \citealt{Hamann98}). Because on the
one hand mass-loss rates have been down-revised, whilst on the other hand
luminosities have been revised in an upwards direction, the burden for
radiative acceleration to drive WR mass loss to values exceeding the
single-scattering limit has become less severe. Nevertheless, compared to
O-stars, WRs display significantly larger wind densities and mass-loss rates
(factors $\ga 10$ for the latter, see Fig.~\ref{mdot_wr}). In contrast, the
terminal velocities of WR stars are comparable to those from O-stars, of the
order of 1000~\kms (late WNs and WCs) to 2500 \kms and higher (early WNs and
WCs), see Table~2 in \citet{Crowther07}.

It is relevant to note that despite the fact that WR temperatures are now
calculated to be larger than a decade ago thanks to the inclusion of
metal-line blanketing, the inner core radii (which are a factor of several
smaller than the ``photospheric'' radii) computed from the most 
sophisticated NLTE models are still a factor of several ($\sim$3) larger
than those predicted for the He-main sequence by stellar evolution models.
In other words, the problem of how WR winds are connected to their stellar
core is still lingering, despite the impressive progress recently reported 
\citep{Crowther07}. 

One might now question why there is a difference between the winds from WR and
O-stars {\it if} also WR winds were accelerated by radiative line-driving,
as supported by a number of investigations (for references, see
Sect.~\ref{sec:selfenrichment}). This problem has been extensively discussed
in the literature, but cannot yet be conclusively answered. However, there
are some facts and findings which might serve as a guideline.

({\it i}) Since for a given luminosity the initiated mass-loss rate is
considerably larger than in O-stars, whereas the terminal velocities are
quite similar, the momentum transfer in WR winds must be more efficient.
This efficiency is supposed to be related to very efficient multi-line
scattering events, possible because of {\it photon trapping} \citep{LA93,
SpringmannPuls98}: In O-star winds the ionization equilibrium is almost
frozen-in, leading to gaps in the frequency distribution of the lines and
thus to a ``premature'' photon escape when such gaps are encountered. In WR
stars with a high core temperature and a larger wind-density, the ionization
equilibrium is stratified, and the previous gaps are closed due to the
emergence of new lines from a new ionization stage. Thus, photons become
trapped instead of escaping, and can transfer more momentum.  ({\it ii})
Moreover, again due to the larger wind-densities, the ionization equilibrium
in the outer WR winds is shifted towards lower stages than present in the
corresponding O-star winds, and the number of lines is larger, leading to an
efficient acceleration of the outer wind.  ({\it iii}) Following the
ionization stratification inwards, there is a (hot) iron opacity peak around
160~kK in the deeper atmospheres of WR stars\footnote{Additionally, there is
also a weaker, (cool) opacity peak around 40~kK.} \citep{NL02} which
according to \citet{Graf05} provides the required opacity for {\it
initiating} WR winds (as shown for a WC5-type object).  (iv)~Finally, it
should be mentioned that theoretical models of WR stars display the
strongest {\it strange-mode} instabilities (\citealt{Gautschy90}; for a
review, see \citealt{Glatzel01}) found so far \citep{Glatzel93}. 
\label{strangemode} Recent non-linear simulations \citep{Wende08} have shown
that Wolf-Rayet models can reach radial velocity amplitudes which amount to
up to 30\% of their escape velocity.  The corresponding mechanical energy
transferred by shock waves to the stellar atmosphere might easily help or
even dominate the initiation of mass-loss and the acceleration of the lower
WR wind. But note also that \citet{Moffat08} found no coherent oscillations
in the hot WC-star WR 111 after photometrically monitoring this object with
the microsatellite {\sc most} for three weeks. The derived upper limit for
coherent Fourier components turned out to be almost two orders of magnitude
below corresponding predictions from strange-mode pulsation simulations.

Returning to the (more conventional) WR wind model by \citet{Graf05}, let us
stress that there are two crucial aspects which had to be adopted in order
to enable a {\it consistent} wind driving: a rather large clumping in the
outer wind (\fcl = 50, but see Sect.~\ref{sec:clumpquant}), and the
proximity of WRs to the Eddington-limit. The latter was suggested by these
authors as the most critical parameter for initiating a strong mass flux
(see Eq.~\ref{eq_mdot} and Sect.~\ref{sec:modelsclosetoedd}). Note that
there are no present self-consistent models of WR winds with the underlying
star being (far) away from this limit!

\smallskip
We now turn to the empirical dependence of WR winds on initial metallicity. 
Until a few years ago, there were both indications for as well as against
the existence of a mass-loss-$Z$ dependence for WR winds, however
quantitative numbers on the exponent $m$ of the mass-loss $Z$ exponent were
not provided until theoretical results of a $\dot{M}$-$Z$ dependence for WR
stars became available (\citealt{Vink05} and Sect.~\ref{sec:selfenrichment}),
and empirical exponents have been quoted for both WN ($m$ $\sim 0.8$) and WC
($m$ $\sim 0.6$) stars -- similar to those predicted.

An oft-used formula in stellar evolution computation for WR winds is the one
provided by \citet[ see also Fig.~\ref{mdot_wr}]{NL00} in which the mass-loss rate
depends on both the He (Y) and the total metallicity (Z). We note that the
$Z$ in this formula represents the total $Z$, whilst, theoretically, Fe is
thought to be the dominant wind driving ion for WR stars \citep{Vink05,
Graf08}. For this reason we should not attach too much physical significance
to the $Z$ dependencies quoted by \citet{NL00}.  Nevertheless, the formula
may well provide a proper representation of empirical mass-loss rates (cf.
Fig.~7 in \citealt{Crowther07}), and since a {\it general} theoretical
description of mass-loss rates from WR stars is no yet available, they
probably present the most reliable formulae for use in stellar evolution
calculations of He stars.

Particularly relevant for the issue of LGRBs are the rotational
properties of their progenitor stars.  Owing to the broad emission lines in
their spectra, traditional \vsini\ studies are not feasible for WR stars.
Luckily, there is a powerful alternative in the form of linear
spectropolarimetry. Apart from its role in constraining wind clumping
properties (Sect.~\ref{sec:clumpquant}), large scale asymmetries resulting from
rotationally-flattened winds may also be probed with the technique. 
\citet{Harries98} and \citet{Vink07} studied samples of resp. Galactic and
LMC WR stars and found equal incidences of polarization ``line effects'' in
WR stars at $\Zsun$ and $\sim$0.4 $\Zsun$.  As there exists no evidence for
significant differences in rotational properties of the presumed 
progenitors of LGRBs, these results suggest that one needs to study lower
than LMC $Z$ environments to detect a significant population of rapid
rotators at the end of their lives, if there is a low $Z$ bias for the
creation of LGRBs via the collapsar model.

\section{The weak wind problem}
\label{sec:weakwinds}

First statistically relevant evidence for the presence of O-stars with
theoretically unexpected thin, weak winds was provided already by
\citet{chlebo91}. From studies of the wind and X-ray properties of O-stars,
they found that the mass-loss rates derived for the {\it O-dwarfs of later
spectral type} were significantly lower (up to a factor of 10) than expected
from the $\mdot \propto L^x$ power-law relation
(Sects.~\ref{sec:standardmodel} and ~\ref{sec:statmodels}) known for
brighter O-stars. In the same year, \cite{kudEUV91} analyzed the UV spectrum
of the B-type giant $\beta$\,CMa (B1\,II) and concluded that its
theoretically predicted mass-loss rate ($3 \cdot 10^{-8}$\msunyr) had to be
reduced by a factor of 5 to reproduce the \siivuv, \ciiiuv, and \civuv\
profiles. This result was confirmed by \cite{drew94} from an analysis of the
{\sc rosat} X-ray emission of $\beta$\,CMa, who claimed a similar reduction
for the B2\,II star $\epsilon$~CMa and presented two alternative
explanations for this behaviour. The first one invoked the X-ray emission
from an ``outer zone of hot shocked gas'' to penetrate far enough inwards to
increase the ionization around the critical point, thus reducing the
radiative acceleration and the mass-loss rate. 
Alternatively, the source of wind heating could be due to ion-drag
effects\footnote{frictional heating accompanying the collisionial momentum
transfer between radiatively accelerated metals and the bulk wind matter,
see page \pageref{coulomb_coupling}.} \citep{SP92}. 
Using corresponding scaling relations from \cite{gayley94},
\citeauthor{drew94} argued that the winds of these two B-giants were not
significantly affected, leaving wind-shocked X-rays as the most plausible
heat source.

\begin{figure}
\hspace{-.3cm}
\begin{minipage}{7.cm}
\resizebox{\hsize}{!}
   {\includegraphics{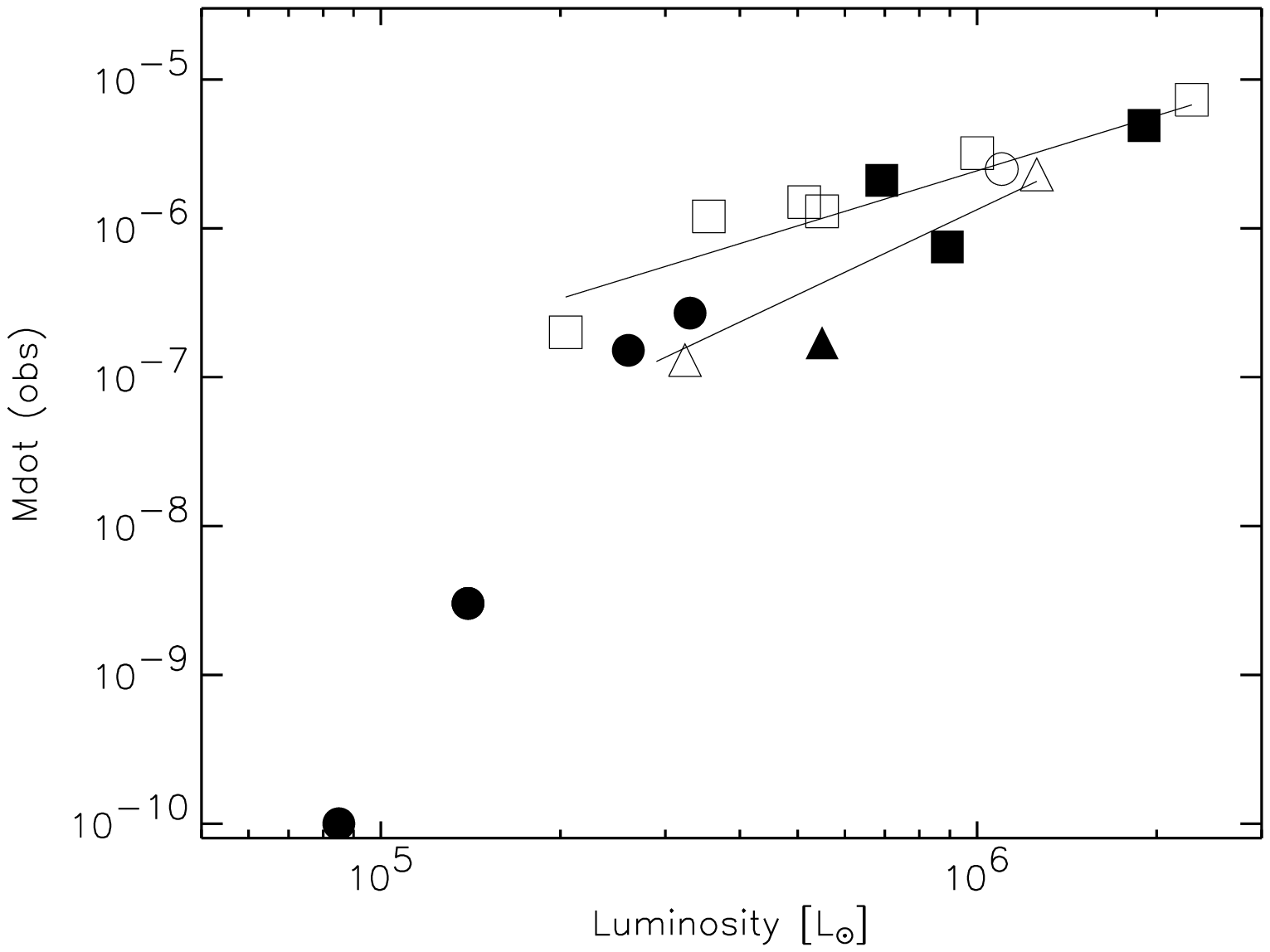}}
\end{minipage}
\hspace{-.3cm}
\begin{minipage}{5.cm}
   \resizebox{\hsize}{!}
   {\includegraphics{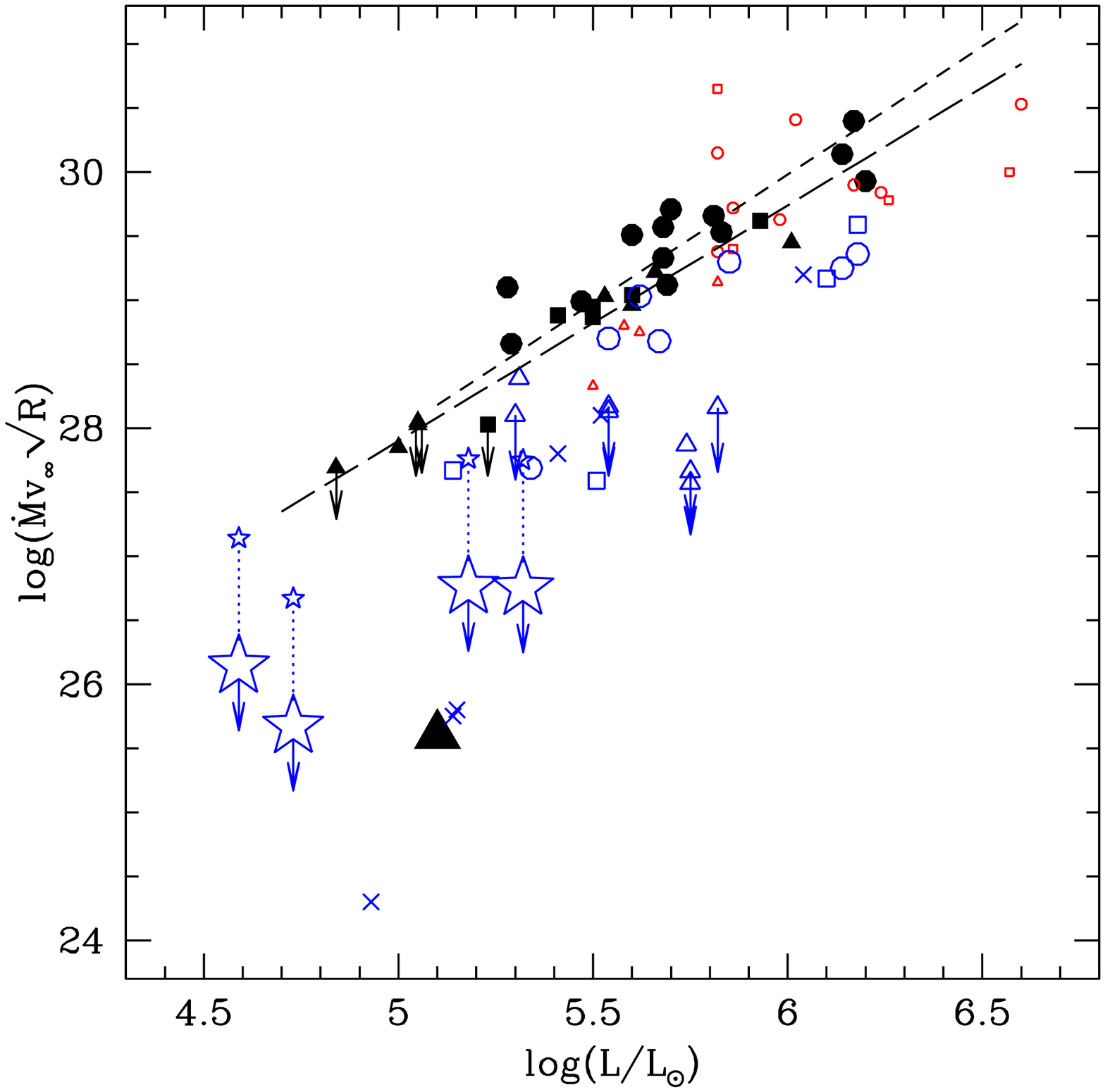}}
\end{minipage}
\caption{The weak wind problem. Left: Mass-loss rates for O-dwarfs in
NGC~346 \citep[][ LMC, filled circles]{bouret03}, compared to ``normal''
winds from Galactic (squares) and other SMC stars (triangles) as analyzed by
\citet{Puls96}. Right: Modified wind momenta as a function of stellar
luminosity for O-stars (adapted from \citealt{martins04}). 
Filled and open symbols correspond to Galactic and LMC (small red)/SMC
(blue) stars, respectively.
Luminosity class V, III and I are represented by asterisks, squares and
circles. Note the low wind-momenta of the SMC objects and the Galactic star 
10~Lac (large triangle). \label{fig-thinwind}}
\end{figure}

Using \Ha\ as a more reliable mass-loss indicator, \citet{Puls96} showed
that the observed WLR for O-type dwarfs exhibited a severe curvature towards
very low wind-momenta for luminosities lower than $10^{5.3}$ \lsun,
indicating a break-down of the WLR in that domain. It should be stressed,
however, that their sample was biased towards higher luminosity stars for
which \Ha\ could be optimally used. In fact, their \Ha-diagnostics could
provide only upper \mdot-limits for the lower luminosity objects, and
subsequently the \mdot-determination in weak-winded stars was deferred 
until unified atmosphere models were able to do so.

From a theoretical point of view, physical possibilities to explain this
behaviour continued to be explored
\citep{puls98,OP99,babel95,babel96,krticka00}, where three effects were
considered as particularly promising and have been reviewed by \citet{KP00}.
The first one relates to the potential decoupling of metal ions from the
bulk plasma, which should happen below a critical density when the
Coulomb-collisions become ineffective (Sect.~\ref{sec:standardmodel} and
references therein). The second one refers to the shadowing of the
wind-driving lines by the photospheric ones, which considerably lowers the
line force, resulting in a reduction of \mdot, especially in the case of
B-dwarfs \citep{babel96}. Finally, for low mass-loss rates where the
continuum is thin throughout the transonic region, curvature terms of the
velocity field leading to source function gradients may reduce the
line-accelerations to values much smaller than in the standard computations,
resulting again in reduced mass-loss rates \citep{puls98, OP99}.
Unfortunately, due to the lack of reliable \mdot\ estimates it remained open
to which extent the above processes could account for the weak wind problem.

\begin{figure}
\begin{center}
\includegraphics[width=6cm,angle=90]{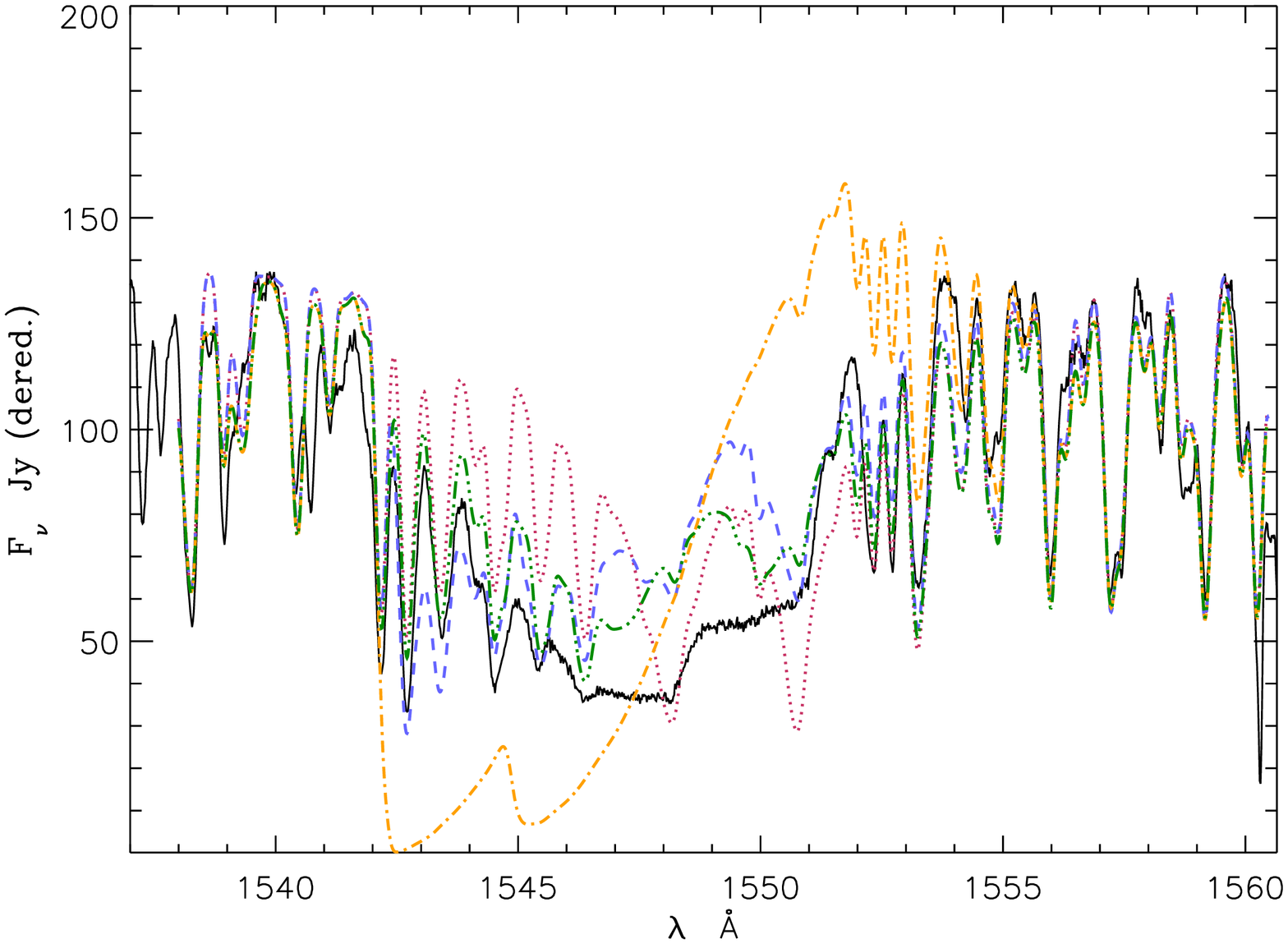}
\end{center}
\vspace{-0.5cm}
\caption{The role of X-rays in the O-dwarf weak wind problem. Model fits to
\civuv\ from the O9V star 10~Lac, using different mass-loss rates. Without
X-rays, an \mdot-value as low as 2.5$\cdot 10^{-10}$ (dashed blue, \mdot\ in
\msunyr) is required, whilst a value of $10^{-10}$ (dotted red) produces a
much too weak profile, and a value of $10^{-9}$ (dashed-dotted orange)
results in a too strong, P~Cygni type profile. If X-rays are included 
($L_x$/\Lstar$\sim 10^{-6.7}$, dashed-dotted-dotted green curve), however,
the observations can be reproduced with \mdot\ = $10^{-9}$ \msunyr. 
Adapted from Najarro, Puls \& Herrero (in prep. for A\&A).
\label{fig-thinwind-10lac}}
\end{figure}

First results with modern unified codes using both UV and optical lines were
obtained by \citet{herrero02}. For the Galactic O9V star 10~Lac they derived
an upper limit of \mdot $\le 10^{-8}$\msunyr\ from the optical and \mdot $\sim
10^{-10}$\msunyr\ by including additional constraints from the UV, which is 
more than one dex below the theoretical predictions (see also
Fig.\ref{fig-thinwind-10lac}). Subsequently, \cite{bouret03} and
\cite{martins04} obtained similar results for a larger sample of SMC
O-dwarfs (see Fig.~\ref{fig-thinwind}). The latter authors studied in detail
the reliability of their mass-loss rate estimates derived from UV spectra and
concluded that, at most, the true \mdot\ could be underestimated by a factor
of six. With these determined mass-loss rates \citeauthor{martins04}
revisited in detail the physical possibilities discussed above and concluded
that none of them could account for such a strong mismatch between observed
and theoretically expected \mdot. They speculated that although the physical
mechanism leading to such weak winds remained unknown, the low mass-loss
rates were intrinsically related to the youth of the stars, possibly
testifying the phase of wind onset in young O-stars shortly after their
formation. 

\citet{martins05b} extended this study to Galactic objects by performing a
combined UV and optical analysis of a sample of 12 O-dwarfs. Again, a
mismatch as high as a factor of 100 was found for later spectral types with
$\logLL < 5.2$. A major step was the investigation of the effects of X-rays,
magnetic fields, advection and adiabatic cooling in the determination of
\mdot\ from UV spectra. This lead \citeauthor{martins05b} to conclude that
the inclusion of X-ray emission (likely related to magnetic mechanisms, see
Sects.~\ref{sec:mag} and \ref{sec:xraylinesmagnetic}) in models with low
density is crucial to derive accurate mass-loss rates from UV lines, while
it was found to be unimportant for high density winds. This is shown in
Fig.\ref{fig-thinwind-10lac}, where models with and without X-rays and
different mass-loss rates are compared for \civuv\ from 10~Lac. 

One important conclusion from the above studies is that the discrepancy in
the derived mass-loss rates seems not to be related to metallicity, as it is
present in both the Galactic and SMC stars. Interestingly, however, the break-down of
the WLR starts at earlier spectral types (higher luminosity) for the lower
metallicity environment of the SMC \citep[around O6-6.5V,][]{bouret03,martins04} 
than for the Galaxy \citep[O9V, ][]{martins05b}. Thus, although the principal
problem appears to be unrelated to metallicity, its onset (with respect to
luminosity) could be linked to the metal content.

\begin{figure}
\hspace{-0.3cm}
\begin{center}
\includegraphics[width=11.8cm,angle=0]{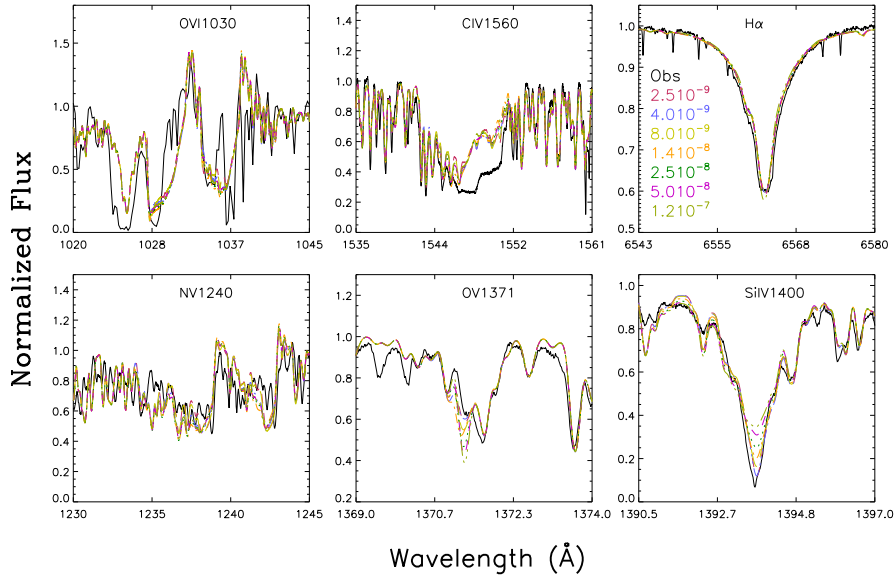}
\end{center}
\vspace{-0.5cm}
\caption{The \mdot-X-ray degeneracy. Model profiles computed for different
mass-loss rates and scaled X-ray fluxes for strategic UV lines from 10~Lac,
with \mdot-values in units of \msunyr. 
Adapted from Najarro, Puls \& Herrero (in prep. for A\&A).
\label{fig-thinxrays-10lac}}
\end{figure}

Not only the mass-loss rates, but also the terminal velocities seem to
behave unexpectedly in weak winds. The values derived by \cite{martins05b}
lie far below the expected and observed average value of \vinf/\vesc
$\approx 3$ for stars above the bi-stability jump, rather close to a ratio
of unity. Given the low wind densities, only upper limits could be obtained
from \civuv, which constitutes the line with the highest sensitivity to
\mdot\ and \vinf\ for these late spectral types. Inverting the scaling
relations for radiation driven winds (Eq.~\ref{eq_vinf}),
\citeauthor{martins05b} argued that a value of $\alpha \sim 0.3$ was
required to reproduce the observed ratios, again quite different from the
expected $\alpha \sim 0.6$ value. Interestingly, the mechanism related to
X-ray emission proposed by \citet[ see above]{drew94} was recalled to
explain such a low value. 

A crucial point regarding the reliability of UV \mdot-determinations for
weak-winded stars is posed in Fig.~\ref{fig-thinxrays-10lac}. Here, we show
the diagnostic UV and \Ha\ profiles for a set of models where the mass-loss
rate is varied by almost two orders of magnitude whilst the X-ray luminosity
is increased in parallel to keep the ionization structure and thus the UV
lines at the observed level as far as possible. Note that the changes
imposed on the wind do {\it not} reach the photospheric levels, which is
obvious from the fact that the underlying UV iron forest does not change. 
Apart from the cores of \siivuv\ and \ouv, the observations can be equally
well reproduced with {\it any} \mdot\ combined with an appropriate $L_x$.
Thus, no independent UV ``measurement'' of mass-loss rates may be obtained
unless the X-ray properties of the star are accurately known.

\begin{figure}
\hspace{-.3cm}
\begin{minipage}{6.cm}
\resizebox{\hsize}{!}
   {\includegraphics{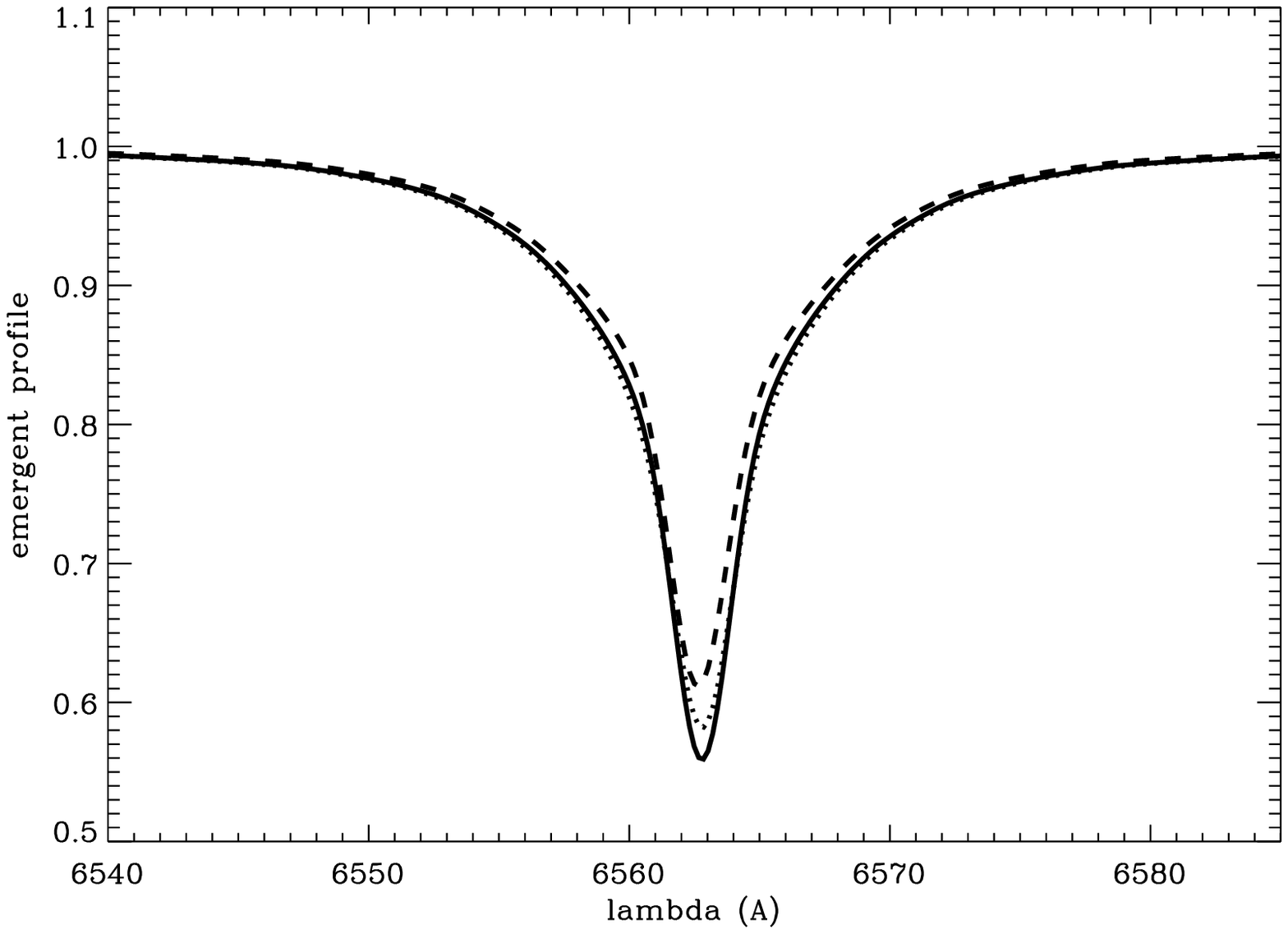}}
\end{minipage}
\hspace{-.3cm}
\begin{minipage}{6.cm}
   \resizebox{\hsize}{!}
   {\includegraphics{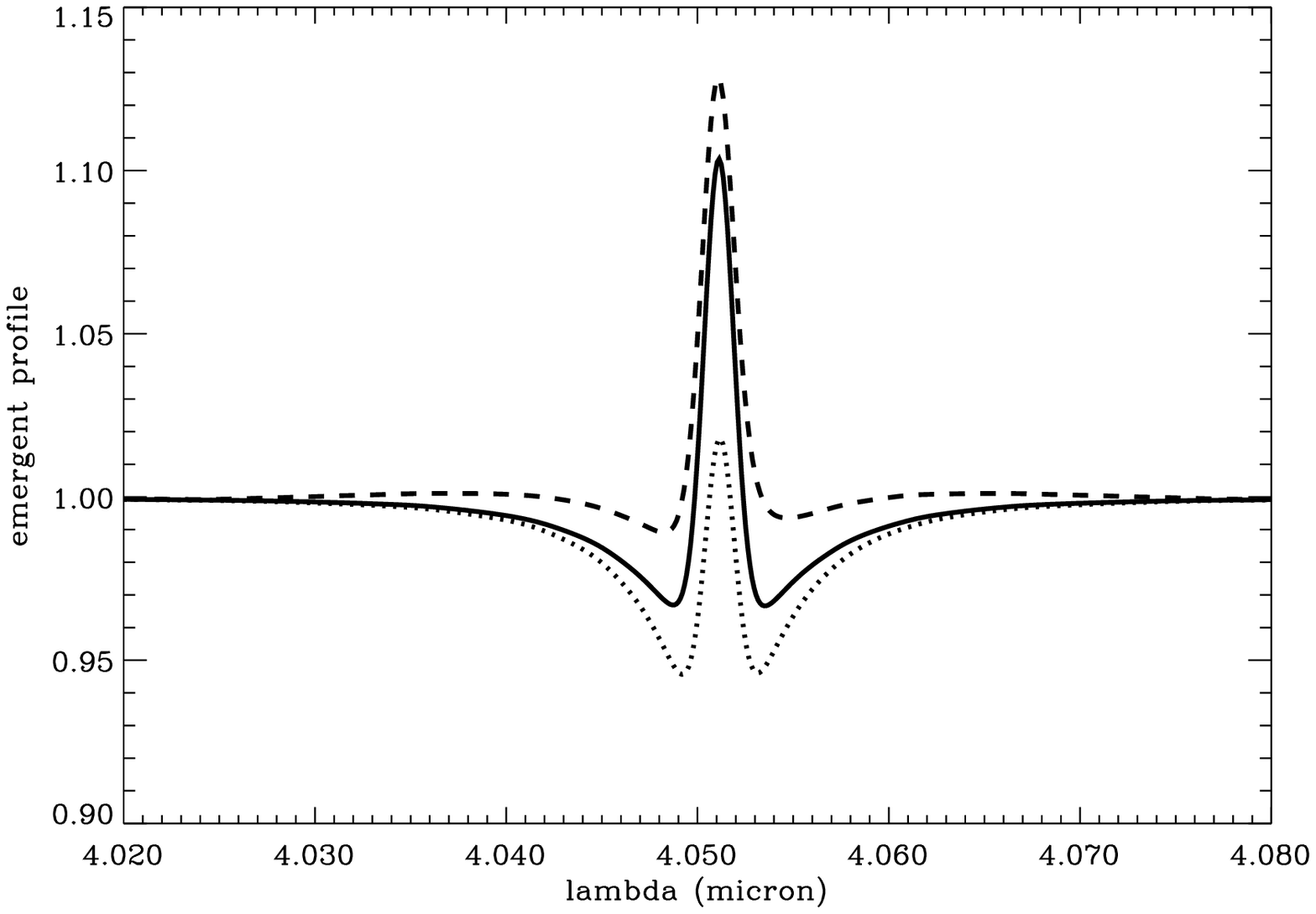}}
\end{minipage}
\caption{Left: Synthetic \Ha\ profiles for $\tau$ Sco, at different mass-loss
rates \mdot\ = $2\cdot10^{-9}, 2\cdot10^{-8}$ and $6\cdot10^{-8}$ \msunyr 
(dotted, bold, and dashed, respectively). Note the insensitivity of \Ha\ 
at lowest mass-loss rates. Right: Synthetic
\Bra\ profiles, for the same models. The line is a sensitive diagnostics for
\mdot\ even at lowest values. \label{fig-bralpha}}
\end{figure}

\begin{figure}
\begin{center}
\includegraphics[width=8cm,angle=0]{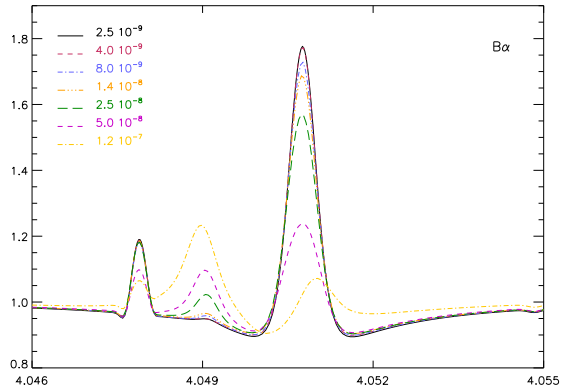}
\end{center}
\vspace{-1cm}
\caption{Synthetic \Bra\ profiles for 10~Lac for the model sequence
displayed in Figure\ref{fig-thinxrays-10lac}. Note the sensitivity of \Bra\
on \mdot\ even at lowest values. At larger \mdot, the \HeII-component becomes
sensitive on the amount of X-ray emission  as well (see text).
\label{fig-bralpha-xrays}}
\end{figure}

Consequently, a reliable \mdot-diagnostics is required that does not suffer
from effects such as X-rays, advection, or adiabatic cooling. This is
provided by the \Bra\ line in the infrared. The relevance of \Bra\ has been
pointed out already by \citet{auer69}, who predicted that even for
hydrostatic atmospheres the (narrow) Doppler-cores should be in emission,
superimposed on rather shallow Stark-wings. The origin of this emission is a
combination of various effects in the upper photosphere and the transonic
region that depopulate the lower level of \Bra, $n=4$, stronger than the
upper one, $n=5$. As detailed by Najarro, Hanson \& Puls (in prep. for
A\&A), this is the consequence of a typical nebula-like situation, due to
the competition between recombinations and downwards transitions, where the
strong decay channel $(4 \rarrow 3)$ (stronger than $5 \rarrow 4$ itself) is
of particular importance. 

As shown in Fig.~\ref{fig-bralpha} (right), the emission component of \Bra\
(and the line wings!) react strongly on \mdot, particularly for very weak
winds \citep[see also][]{Najarro98}. In fact, if we now investigate the same
\mdot-$L_x$ combinations as in Fig.~\ref{fig-thinxrays-10lac} using \Bra, we
obtain a really promising result. From Fig.~\ref{fig-bralpha-xrays} we note
that \Bra\ reacts to changes in \mdot\ even at lowest values. Evidently, for
such low values (with a deep-seated line formation region), the line profile
is basically not affected by X-rays. With increasing mass-loss rate,
however, the line starts to react on changes both in \mdot\ and X-rays, due
to an outwards moving line formation zone, reaching the layers to which
X-rays can penetrate. This is clearly seen in Fig.~\ref{fig-bralpha-xrays}
from the strong reaction of the \HeII\ emission component of \Bra,
which is caused by the large sensitivity of the \HeII/\HeIII\ ionization 
equilibrium to X-rays in the region where the emission arises. Our models
indicate that under ``normal'' X-ray luminosities ($L_{\rm x}/L_{\rm bol} <
10^{-4})$ the (hydrogen) core of \Bra\ is not affected though.

We thus expect that \Bra\ will become {\it the} primary diagnostic tool to
measure very low mass-loss rates at unprecedented accuracy, thus allowing to
clarify whether they actually lie well below the theoretical predictions and
to which degree.

\section{Inhomogeneous winds}
\label{sec:inhomowinds}

In our discussion of diagnostical methods (Sect.~\ref{sec:diagnostics}) and
recent results (Sect. ~\ref{sec:results}) we have encountered a multitude of
findings which are difficult or even impossible to reconcile with the
standard model of a smooth and stationary wind. Even the inclusion of
rotation and magnetic fields does not help to explain additional features,
at least not at first glance. In the following, we will summarize various
observational findings (ordered from low to high energies) which point
towards a structured and variable wind. Further evidence will be outlined in
Sect.~\ref{sec:clumping} where we will concentrate on wind-clumping, maybe 
the most important aspect with respect to deviations from the standard
model.

\subsection{Observational background}
\label{sec:inhomoobs}

\noindent
{\it Non-thermal radio emission} has been discussed on
page~\pageref{nonthermal}. According to recent investigations, the chances are
high that this emission might be related to colliding winds alone. Insofar, we
refrain here from considering non-thermal emission as an indication for
shocks being present {\it in the wind}, in contrast to previous reasoning by
various authors (including ourselves). 

\smallskip
\noindent
{\it Electron-scattering wings} of recombination lines in Wolf-Rayet
stars are weaker than predicted by smooth models, indicating the presence of
clumping \citep{Hillier91}. Due to a routine inclusion of the latter process
into recent spectroscopic analyses of WR-stars, perfect fits have been
obtained for (almost) all lines, and the derived mass-loss rates are a
factor of two to three lower than if derived by means of unclumped models.
Note that this diagnostics cannot be used in OB-stars, due to the much lower
wind-density and the corresponding lower optical depth in
electron-scattering.

\smallskip
\noindent
{\it Optical line profile variability} has been detected in the wind
lines of OB-stars (\Ha: \citealt{Markova05} and references therein;
\HeII\ 4686: \citealt{Grady83, Henrichs91}) and BA-supergiants
\citep{Kaufer96}. Typical time-scales range from few hours/days (wind-flow
times) to several months. Direct evidence for outward accelerating wind
inhomogeneities (clumps) is implied by the ``moving  bumps'' superimposed on
the emission lines of Wolf-Rayet stars (e.g., \citealt{Robert94}), whilst
similar evidence for O-stars has been reported for only few objects (see
Sect.~\ref{sec:clumping}).

Not only wind lines, but also lines which form to a significant or major
part in the photosphere (e.g., \HeI\,$\lambda$5876) are variable. \citet[ and
references therein]{Fullerton96} studied a large sample of O-stars and found
that almost 80 \% of the investigated stars (all supergiants and a few
late-type dwarfs) displayed photospheric line profile variability, with an
amplitude increasing with stellar radius and luminosity. These findings were
interpreted as being likely related to strange-mode oscillations (page
\pageref{strangemode}), linking photospheric variability and
structure formation in the wind.

\smallskip
\noindent
{\it Spectropolarimetric variations} were detected by
\citet{Lupie87} in a sample of 10 OB supergiants which pointed to the
presence of clumps.  Given that the position angles of the polarization, and
thus the orientations of the electron scattering plane, were found to vary
randomly, this implied that the polarization changes were not attributable
to an orderly interaction with a companion, but the data provided direct
evidence for the presence of an inhomogeneous wind structure. 

\smallskip
\noindent
{\it Black troughs in saturated UV P~Cygni profiles} strongly point
towards the presence of multiply non-monotonic flows, and can be simulated
in stationary models only if allowing for a {\it supersonic} velocity
dispersion (increasing with distance from the star) as the dominating
line-broadening agent (page~\pageref{veldisp}).

\smallskip
\noindent
{\it UV line profile variability} is evident in UV P~Cygni
lines, with large variations of the blue edge, contrasted to an almost
stationary red emission part \citep{Henrichs91, Prinja92}, implying a
significant fraction of small-scale structure.

\smallskip
\noindent
{\it Discrete absorption components (DACs)} detected in the
absorption troughs of unsaturated P-Cygni profiles are the most intensively
studied indicators of wind structure and variability. To date, they are
believed to be formed in co-rotating interaction regions (CIRs) and will be
discussed in Sect.~\ref{sec:cirs}, together with the so-called ``modulation
features''. 

\smallskip
\noindent
{\it X-ray emission} from hot stars has been measured mostly at soft
energies (0.1 to $\ga$2~keV), either at low resolution in the form of a
quasi-continuum or at high resolution allowing to see individual lines.  The
latter will be covered in the last part of this section
(\ref{sec:xraylines}). 

Already the first X-ray satellite observatory, {\sc einstein}, revealed that
O-stars are soft X-ray sources. Follow-up investigations, mostly by {\sc
rosat}, have confirmed this notion and also allowed for quantifying the
X-ray properties for a multitude of OB-stars (see \citealt{KP00} and
references therein). Roughly, the (soft) X-ray luminosity scales with
$L_{\rm x} \approx 10^{-7} L_{\rm bol}$.  First there appeared to be a
rather large scatter indicating the influence of additional parameters,
however a more recent {\sc xmm-Newton} study by \citealt{Sana06} of the
young open cluster NGC 6231 showed that the majority of O-stars display a very
limited dispersion (of 40 or only 20 per cent depending on whether cool
dwarfs are included or not). This suggests that the intrinsic X-ray emission
of normal O-stars is rather constant for any given object and that the level
of X-ray emission is accurately related to the basic stellar and wind
parameters. 

The source of the X-ray emission is widely believed to be shocks embedded in
the stellar wind and related (though not necessarily directly, see
Sect.~\ref{sec:lineinstab}) to the line-driven instability. In terms of a
stationary description\footnote{Generally, the X-ray luminosity of O and
early B-stars is constant to within 10 to 20\%, see
\citealt{BerghoeferSchmitt94}.}, a simple model (e.g., \citealt{Hillier93,
Cassinelli94}) assumes randomly distributed shocks where the hot shocked gas
(with temperatures of a few million Kelvin and a volume filling factor of
the order of $10^{-3}{\ldots} 10^{-2}$ in the case of denser O-type winds)
is collisionally ionized/excited and emits X-ray photons due to spontaneous
decay, radiative recombinations and bremsstrahlung. The ambient ``cool''
stellar wind (with a kinetic temperature of the order of \Teff) can
re-absorb part of the emission due to K- and L-shell processes if the
corresponding optical depths are large. This simple model (sometimes
extended to account for the different post-shock cooling zones of radiative
and adiabatic shocks, see \citealt{Feldmeier97a}) has been used to analyze
large samples of O-stars observed with {\sc rosat} \citep{Kudritzki96}, and
is still used in the context of NLTE diagnostics to account for the
influence of X-ray/EUV emission on the photo-ionization rates (see below). A
simple scaling analysis of the involved X-ray emission and absorption
processes by \citet{OwockiCohen99} showed that the ``natural'' scaling for
optically thin (with respect to X-ray absorption) winds is given by $L_{\rm
x} \propto (\mdot/\vinf)^2$ and for optically thick winds by $L_{\rm x}
\propto (\mdot/\vinf)^{1+s}$, if the volume filling factor follows a radial
stratification $f \propto r^s$. This would imply, for O-stars, $s \approx
-0.4$ to explain $L_{\rm x} \propto L_{\rm bol}$ by means of the scaling
relations for line-driven winds (Eqs.~\ref{eq_vinf}, \ref{eq_mdot}).

Whereas {\sc rosat} observations of OB-star winds showed correlations between 
$L_{\rm x}$ and $L_{\rm bol}$, 
WC stars were not found to emit in the X-ray
regime, unless the star is part of a binary system and the X-ray emission
can be attributed to colliding winds.  The fact that single WC stars 
(for WN stars the situation is less clear)
are no significant X-ray emitters does not imply that shocks do not develop in
their winds.  In fact, \citet{Ignace99} could explain the lack of X-ray
emission in single WR winds by realizing that the wind is optically thick to
the hot X-ray gas. The absence of X-ray detections in single WC stars was 
reinforced by more sensitive {\sc xmm-Newton} and {\sc chandra} data, where
the lack of X-ray emission was attributed to photoelectric absorption by 
the wind and where the large opacity of the WR winds was thought to put the
radius of optical depth unity at hundreds or thousands of stellar radii for
much of the X-ray band \citep{Oskinova03}. 

Since the X-ray and associated EUV luminosity emitted by the shocks is quite
strong, it can severely affect the degree of ionization of highly ionized
species such as C{\sc v}, N{\sc v} and O{\sc vi}, by Auger-ionization
\citep{Macfarlane93} and even more by direct ionization in the EUV
\citep{Pauldrach94}. In fact, the so-called superionization 
\label{superion} in the winds of
O supergiants (exemplarily, the strength of the O{\sc vi} resonance line in
$\zeta$~Pup, see Sect.~\ref{sec:diagnostics}) can be explained only by means
of this additional source of photons. A systematic investigation of these
effects on the complete FUV spectrum, as a function of stellar parameters,
mass loss and X-ray luminosity has been performed by \citet{Garcia05}.

\smallskip
\noindent
{\it Correlated variability.} Finally, let us note that both \Ha\
and UV variability \citep{Kaper97} and \Ha\ and X-ray variability
\citep{Berghoefer96} have been found to be correlated, which indicates the
propagation of disturbances throughout the {\it entire} wind.

\subsection{Theoretical background: The line-driven instability and
time-dependent models}
\label{sec:lineinstab}

Historically, most of the theoretical effort to explain the above
observational features has concentrated on the line-acceleration itself. 
Later on, however, it turned out that at least one of the primary indicators
of time-dependence and structure, the DACs, can be understood within the
concept of CIRs, rather (but not completely) independent on the specific
physics of the wind acceleration (for details, see Sect.~\ref{sec:cirs}).

\paragraph{Linear stability analysis.~~} Already in their pioneering paper
on line-driven winds, \citet{LS70} pointed out that the accelerating
mechanism of these winds should be subject to a strong instability, and
subsequent investigations tried to quantify its properties, at first based on
linear stability analyses. By linearizing the time-dependent equations for
density and momentum in a frame comoving with the  {\it mean} fluid and
using a planar approximation with height co-ordinate $z$, it is easy to show
that the conventional ansatz 
\beq
\left. \begin{array}{r} \rho(z,t) \\ v(z,t) \\ \grad(z,t) \end{array} \right\}
= \left. \begin{array}{rl} \rho_0(z) + &\delta \rho(z,t) \\ &\delta v(z,t) \\ 
g_{{\rm rad},0}(z) + &\delta \grad(z,t) \end{array} \right\} \cdot 
\eu^{\displaystyle{\ii (kz - \omega t)}}
\eeq
for the density, velocity and line force (quantities with subscript `0'
refer to the mean flow, $v_0=0$, and $\delta \rho, \delta v, \delta \grad$
are the corresponding amplitudes of a sinusoidal disturbance with wavenumber
$k$ and frequency $\omega$) results in 
\beq
\left. \begin{array}{rllclcr}
- \ii \omega \delta \rho + \rho_0 \ii k \delta v &=& 0 &
\quad  \Rightarrow \quad &  
\displaystyle{\frac{\delta v}{\delta \rho}} & = &
\displaystyle{\frac{1}{\rho_0} \frac{\omega}{k}}  \\ 
- \ii \omega \delta v &=&\delta \grad & \quad  \Rightarrow \quad &
\omega & = &\ii \displaystyle{\frac{\delta \grad}{\delta v}} 
\end{array} \right\} \quad \frac{\delta v}{\delta \rho} = \frac{\ii}{\rho_0 k}
\frac{\delta \grad}{\delta v}.
\eeq
In the latter equation, we have neglected the pressure terms (for a complete
analysis, see \citealt{ORI}). Thus, the phase angle between density
and velocity disturbance, $\cos \varphi$, the growth-rate of the instability,
$\Omega$, and its phase speed with respect to the mean flow, $v_\varphi$,
\beq
\cos \varphi = \frac{-\Im(\delta \grad/\delta v)}{|\delta \grad/\delta v)|},
\qquad \Omega = \Im(\omega), \qquad v_\varphi = \Re(\omega)/k
\eeq
are {\it completely determined by the response of the line-acceleration on
the velocity disturbance}, at least in the supersonic regime where the
neglect of the pressure force is justified ($\Re(x)$ and $\Im(x)$ are the
real and the imaginary part of the $x$ quantity, respectively). Thus, a
correct description of the line force is of highest importance for
time-dependent simulations (see below).

Within the Sobolev approximation\footnote{Remember that most stationary wind
models are based on this approach.} (page~\pageref{sobo}) it turns out that 
$\delta g_{\rm rad}^{\rm Sob}/\delta v \propto \ii k$, i.e., it is purely
imaginary. Consequently $\cos \varphi = -1$ and density and velocity are
completely anti-correlated. There is no growth of the instability, 
$\Omega^{\rm Sob} = 0$, but only an oscillatory behaviour, and the phase
speed (independent of $k$, i.e., the medium is non-dispersive in this
approximation) is inwards
directed. The inclusion of pressure terms leads to slight modifications: 
the inwards directed phase-speed becomes marginally larger
(roughly by $a^2/v_\varphi$), and a slow, outwards
propagating mode becomes possible, at a speed $\approx a^2/v_\varphi < a$
\citep{Abbott80}. 

The (absolute) value of $v_\varphi$ (inwards mode) at the critical point
(page~\pageref{critpoint}) of stationary models based on a Sobolev
line-force turns out to be identical with the corresponding flow speed, much
larger than the speed of sound. Insofar, this critical point may be
considered as a critical point in the ``usual'' sense \citep{Holzer77}: It
is the outermost point in the flow where a communication with lower layers
is possible, in this case via the above radiative-acoustic waves which have
been named as {\it Abbott-waves} according to their ``inventor''. Following
Holzer's ``wind-laws'', \mdot\ then depends solely on processes in the
sub-critical region, whilst \vinf\ is controlled by the acceleration in the
super-critical part. For a detailed discussion of Abbott-waves and the
corresponding modifications of the above wind-laws for line-driven wind we
refer the reader to \citet{FeldmeierShlosman02}. In any case, however,
remember that the use of the Sobolev approximation is justified only for
wavelengths which are large compared with the Sobolev length (page
\pageref{sobolength}), $L^{\rm Sob} = \vth/(\dvdr)$, i.e., for small $k$. A
discussion of critical points in models based on non-Sobolev line-forces is
beyond the scope of this review.

A somewhat opposite approach \citep{MacGregor79, Carlberg80} results in a
purely real response, $\delta g_{\rm rad} =A \delta v$, with $\Im(A) =0$ and
$A>0$, which is valid as long as the disturbances remain optically thin. In
this case, the finite sound-speed is of importance and disturbances can
propagate only for large $k$, in {\it both directions}  and becoming 
strongly amplified, with a phase-speed of the order of $a$.

A combination and generalization of these two alternative approaches has
been obtained by \citet{ORI} in the form of a ``bridging-law'', which
results from a perturbation analysis of the line-force. In this case,
$\delta g_{\rm rad}/\delta v$ has both a real and an imaginary component,
and includes the two limiting cases for small and large $k$ from
above.\footnote{The critical length scale is here the ``bridging length''
$\chi_{\rm B}^{-1}$, with $\chi_{\rm B}$ the line opacity required to reach
unit optical depth.} The resulting phase relation between density and
velocity is predominantly anti-correlated for all $k$, and there is a fast,
exponential growth of the inwards propagating modes, within a {\it
dispersive} medium. Important additions to this analysis have been provided
by \citet{Lucy84a} and \citet{ORII}, by considering the effects of the
disturbed diffuse radiation field and associated line-force 
(the so-called 
``line drag''\footnote{The disturbed diffuse radiation field in resonance
lines leads to a damping effect which, at the foot-point of the wind,
exactly cancels and otherwise diminishes the instability resulting from pure
absorption lines.}) which had been neglected in the previous approaches.

\paragraph{Time-dependent models.~~} After these initial studies, effort has
concentrated on a direct, hydrodynamical modeling of the time-dependent
structure of line-driven winds (for reviews, see \citealt{Owocki_isle94} and
\citealt{Feldmeier99}). As already pointed out, the key aspect concerns a
consistent description of the line force which follows the local and {\it
non-local} conditions (due to the coupling with the radiation field) at {\it
all} scales of structure formation. In order to keep the problem
computationally feasible, an ``exact'' calculation of the line-force is
prohibitive, and complex integral forms of various degrees of approximation
have been developed \citep{OP96, OP99} to allow for this objective.

After incorporating the stabilizing effect of the line drag by the
diffuse, scattered radiation field (see above), hydrodynamical
models have proven to be significantly more stable in the lower wind than
the first ones based on pure absorption \citep{OCR88}. The outer wind ($r
\ga 1.3 \Rstar$), however, still develops extensive structure that consists
of strong {\it reverse}\footnote{Reverse shocks travel backwards in the
comoving frame (though outwards in the stellar frame). In the pre-shock
region (starwards from the shock), the velocity is high and the density low,
and vice versa in the post-shock region.} shocks separating slower, dense
material from high-speed rarefied regions in between. This is a direct
consequence of the strong amplification of the inwards propagating mode with
a predominant anti-correlation of density and velocity, as derived from the
linear analysis. Such structure is the most prominent and robust result from
time-dependent modeling, and {\it the basis for our interpretation and
description of wind-clumping}.  Moreover, the multiply non-monotonic nature
of the velocity field explains the black absorption troughs observed in
saturated UV P~Cygni lines in a natural way \citep{Lucy82, Lucy83, POF93,
Owocki_isle94}.

\citet{Feldmeier95} extended these models by accounting for the energy
transport including radiative cooling. By investigating the impact of
various photospheric disturbances which can affect the on-set of structure
formation (via exciting the line-force instability), \citet{Feldmeier97a}
concluded that clump-clump collisions (and not the cooling of the reverse
shocks themselves, as previously suspected, since this results in much too
low filling factors) are the origin of the observed X-ray emission. 

Recent effort has focused on more-dimensional simulations and on the study
of the outermost wind. By means of an efficient 1-D ``pseudo-periodic box
formalism'' to investigate the evolution at distances very far from the central
star, \citet[ see also \citealt{RO02}]{RO05} showed that structure can
survive out to distances of more than 1000 stellar radii, due to supersonic
clump-clump collisions which counteract the pressure expansion.

\label{linedrag}
Initial results of 2-D simulations have been obtained using different 
approximations for the calculation of the line-force. In the simplest
case (radiatively isolated azimuthal zones and neglect of {\it lateral}
line-drag\footnote{see \citet{ROC90}.}), the shells arising in 1-D models are
broken up by Rayleigh-Taylor or thin-shell instabilities, resulting in a
completely incoherent lateral structure \citep{DO03}. A follow-up
investigation \citep{DO05} showed that by accounting for the lateral
line-drag and the lateral mixing in a more self-consistent manner (three
rays \label{3rays} at different impact parameters, as already used by \citealt{Owocki99}),
such models - at least at highest resolution - show a much larger lateral
coherence than corresponding one-ray models. Quantitative results are
still lacking though. Further effort is needed to answer the important question
about the lateral and radial length scale of clumps, required to check and
improve present assumptions when modeling the spectra from the IR to the X-ray
regime. 

\paragraph{Time-dependent versus stationary approach.~~}

Obviously, time-dependent models based on a non-local, unstable line-force 
are able to explain a large number of observational findings, at least
qualitatively. On the other hand, they appear to be in stark contrast with
our assumptions within the standard model (Sect.~\ref{sec:standardmodel}),
both with respect to modeling and diagnostics. However, when viewed with
respect to the {\it mass} distribution of density and velocity, the
stationary and time-dependent models are quite similar (e.g.,
Fig.~\ref{fig:owockivoro}, left panel). Given the intrinsic
mass-weighting of spectral formation, this explains why detailed
line-synthesis calculations assuming line-opacities proportional to the
local density are able to generate (time-dependent) line profiles quite
similar to observations \citep{POF93}. 

Let us also emphasize that the gross wind properties \label{grosswind}
derived from time-de\-pen\-dent models like the {\it terminal} flow speed
and {\it time-averaged} mass loss rate agree well with those following from
a stationary approach. Deviations from this rule due to source function
gradients in the transonic region (\citealt{OP99}, see also
Sect.~\ref{sec:weakwinds}) are not a consequence of the line-driven
instability. Instead, they originate from an incorrect use of the local,
Sobolev approach in the stationary models in a region where this approach is
no longer justified, mainly because of the strong curvature of the velocity
field.\footnote{Remember that {\it standard} Sobolev theory requires a
vanishing curvature over the corresponding Sobolev length, which is small in
the transonic region.} 

\subsection{Wind clumping}
\label{sec:clumping}

In our review of ``observed'' wind parameters (Sect.~\ref{sec:obswindpara}),
we have frequently referred to the process of {\it wind-clumping}, and
particularly to its impact on the derived mass-loss rates, without providing
further details. Indeed, the investigation of this process and its
consequences constitutes a major fraction of studies performed in the last
decade, and various contributions regarding the present status-quo can be
found in the proceedings of a recent workshop dedicated to this topic
\citep{Hamann08}. 

In the previous two sections, we have summarized important observational and
theoretical findings which unambiguously indicate that radiation driven winds are
structured and time-variable, and we have outlined that the line-driven
instability is a primary candidate for explaining at least part of these
findings. Thus, we are now in a position to consider wind-clumping in more
detail, starting with a discussion of more specific evidence.

\subsubsection{Observational evidence}
\label{sec:clumpobs}

\paragraph{Half a century of clumping.}
The presence of clumping as ``extreme density fluctuations'' around hot
stars was first invoked by \citet{oste59} to explain the discrepancy between
observed and predicted fluxes in the \HII\ region in the Orion nebula. Two
decades later, \citet{luwhi80} proposed a phenomenological model to explain
the observed X-ray luminosity of $\zeta$~Pup, in which the stellar wind
would break up into a population of radiatively driven blobs and where the
radiation should originate from the bow shocks preceding the blobs (see also
Sect.~\ref{sec:xraylinesnonmagnetic}). Applying the same formalism as 
\citet{oste59}, \citet{abbott81} investigated the effects of dense clumps on
the thermal free-free radio emission of stellar winds as a function of both
the volume filling factor (\fv$\le$1) and the density ratio between
clumped and inter-clump matter. For the standard assumption of vanishing
inter-clump density (see below), they showed that the radio flux is a factor
of \fv$^{-2/3}$ larger than that from a homogeneous wind with the same
average mass-loss rate. Thus, using Eq.~\ref{fnuradio}, radio mass-loss
rates derived from clumped winds are lower by a factor of \fv$^{1/2}$
compared to unclumped winds.

\paragraph{Wind-clumping in dense winds.~~} In the mid-80s, two more 
observational diagnostics suggested that the standard, and convenient, 
assumption of homogeneous stellar winds might fail due to clumping. The
first line of evidence came from studies of the light-curve eclipses as a
function of wavelength in the WR + O binary V444~Cyg \citep{cherepa84}.
The differences in shape and depth of eclipses from the UV to the IR could
not be simultaneously interpreted using a smooth electron density
distribution in the wind. The authors required the presence of individual
density condensations becoming optically thick to reconcile the electron
scattering ($\propto \rho$) dominated UV and optical fluxes with the
free-free and bound-free ($\propto \rho^2$) infrared values.

The second line of evidence relied on studies of {\it continuum and line
variability}. For continuum investigations, linear polarization variability
is particularly advantageous. This process, being directly related to the
electron density through electron scattering processes, provides information
on the presence of localized inhomogeneities and is more reliable than pure
continuum photometric variability studies, where other processes such as
rotational modulation or pulsations may come into play. In addition to the
OB-star work by \citet{Lupie87}, mentioned already in
Sect.~\ref{sec:inhomoobs}, extensive and systematic monitoring of linear
polarization in WR-stars \citep{stlouis87} and LBVs \citep{taylor91,
Davies05} revealed stochastic variability on short time scales, indicating
that the material had to be confined in clumps. More recent results helping
to {\it quantify} the clumping properties are discussed below.

In the case of lines, time-resolved spectroscopic monitoring of WR-stars 
\citep{Moffat88, Robert94} revealed the presence of narrow emission features
on top of the broad emission lines that appeared to propagate from the line
centers to the line wings on timescales of the wind flow-time. These moving
sub-peaks were interpreted as evidence for blob ejections into the wind and
are now considered to be one of the most clear-cut evidences for clumping in
WR winds.

\citet{Lepine99} monitored the line-profile variations in the \HeII\
emission lines of WN stars and in \CIII\ emission lines of WC stars, and
studied these variations by means of a detailed wavelet analysis. As well,
these authors performed simulations of line-profile variability (lpv)
assuming a wind consisting of a large number ($\sim 10^4$) of discrete
wind-emission elements (DWEEs). Observations were best reproduced by
assuming a large ratio of the radial to the lateral velocity dispersion for
the DWEEs, consistent with the nature of the line-driven instability and the
corresponding lateral line-drag (page~\pageref{linedrag}). Physical models
of such lpv interpreted in terms of clumps generated by the line-driven
instability have been provided by \citet{DO02a, DO02b}, where the latter study
extends the \citet{Lepine99} wavelet formalism to modeling lpv from
hydrodynamic simulations. A key result is that one can fit most of the
characteristics if one assumes a lateral coherence scale of $\sim$3 degrees.

\paragraph{Wind-clumping in OB-star winds.~~}

For O-stars, observational lpv-studies are more difficult due to their
roughly factor of ten thinner winds. First investigations were carried out
by \cite{Eversberg98} on the \HeII\ 4686 line in $\zeta$~Pup, which revealed
outward-moving inhomogeneities in its wind {\it that started near the
stellar surface} and became invisible at $\sim 2 \Rstar$. From the wind
acceleration, the authors derived a value of $\beta=1$, consistent with
results from independent diagnostics assuming unclumped winds. Thus, it
seems that the clumping properties of O- and WR stars are similar, and this
``universality'' has been recently confirmed by \cite{lepine08} who extended
the study to five hot stars in different evolutionary stages. They show that
the moving sub-peaks identified on top of the broad emission lines in
WR-stars appear also in the Of stars $\zeta$~Pup and HD\,93129A, in the more
evolved hydrogen-rich and luminous Of-like WN stars HD\,93131 and HD\,93162,
and in the more mass-depleted WC star in $\gamma^2$~Vel. The authors
conclude that stochastic {\it wind clumping is a universal phenomenon in the
radiation-driven, hot winds from all massive stars.} 

\cite{Markova05} presented a study on the lpv of \Ha\ for a sample of 15
O-type supergiants, ranging from O4 to O9.7. {\it All} stars showed evidence
of significant lpv, mostly dominated by processes in the wind, where the
variations were found to occur from the wind base on out to $\sim$ 0.3
\vinf. From profile simulations, they concluded that the mass-loss rate is
only marginally variable, resulting in insignificant variations of the wind
momenta and hence of the WLR.

Of course, best {\it direct} evidence for the presence and behaviour of
clumps in stellar winds would be reached using imaging devices with enough
spatial resolution to ``see'' the individual clumps close to the star and
track them as they expand with the wind flow. By such imaging, also
geometric constraints on, for instance, the sizes of the clumps might be
obtained (e.g. \citealt{Chesneau00}). Unfortunately, micro-arcsec resolution
would be required to carry out such studies, even for the nearby O-stars.
Present observing facilities allow to see such clumps only at much larger
scales, such as the outer wind regions where the thermal radio emission
originates from \citep{williams97}. However, although these observations
indicate that clumps seem to survive over the entire stellar wind \citep[see
also][]{RO05}, no information is provided on their formation region nor how
they evolve throughout the wind. It should be noted that even after the
possible commissioning of 100-m class telescopes, one would still not be
able to spatially resolve the innermost wind regions where the clumps
originate, and it is thus pivotal that alternative methods to constrain
clumping properties are further developed, such as linear polarimetry (see
below). 

\label{clumprepo} At least for O-stars, however, we presently have to rely
mostly on {\it indirect} methods to address the behaviour of clumping
throughout the stellar wind. Such indirect evidence for the presence of
clumping was found 
in parallel by \citet{markova04} and \citet{repolust04}.
By means of a comparison between the observed and predicted WLR for large
samples of Galactic O-stars, these authors showed that if the stellar sample
was plotted as a function of \Ha\ profile type (stars with \Ha\ in emission
and those with \Ha\ in absorption, partly filled in by wind emission), two
different WLRs appeared, inconsistent with the basic theoretical prediction
that the wind-momentum rates of O-stars should depend on luminosity alone
(Sect.~\ref{sec:ldw} and Fig.~\ref{wlr_vink}). It was
suggested\footnote{following a previous idea by \citet{puls03lanzarote}.}
that this difference is related to wind-clumping (being ``visible'' only in
denser winds, i.e., in those with \Ha\ in emission) and shown that a
unification of both ``observed'' WLRs with the theoretical one is possible
when the average clumping factor, \fcl (=\fv$^{-1}$, see below) of the
clumped matter in stars with \Ha\ in emission is of the order of $\fcl
\approx 5$, implying a net reduction of \mdot\ by a factor of 2.3. Note that
these values differ significantly from those derived by means of UV studies
with net \mdot\ reductions up to factors of 10 (e.g., \citealt{bouret03,
fullerton06}, see below).

\smallskip
\noindent
Therefore, at this stage, several fundamental questions need to be addressed:
\begin{itemize}
\item Which are the upper and lower limits for the clumping factor obtained 
from {\it observationally driven} studies and how reliable are they?
\item What is the stratification of the clumping factor throughout the wind,
and what is the shape of the clumps?
\item How realistic is the treatment of clumping assumed so far and 
how well can modern, sophisticated atmospheric codes account for it?
\item How much mass-loss reduction can evolutionary models of 
massive stars cope with to still be consistent with independent constraints
such as the observed numbers of WR-stars?
\end{itemize}
Before we can consider these points in detail, we have to 
describe the basic concept underlying most of the following analyses.

\paragraph{``Micro-clumping''.~~} \label{microclumping} This concept bases
on the hypothesis that the wind consists of small-scale density
inhomogeneities, which redistribute matter into overdense clumps and
underdense inter-clump matter. Motivated by the principal results from
hydrodynamical simulations including the line-driven instability, the
inter-clump matter is assumed to be void.\footnote{Alternatives from this
assumption are rare, and not used in standard diagnostical tools until
recently when \citet{Zsargo08b} re-investigated the ``superionization'' of
\OVI\ (page~\pageref{superion}) in $\zeta$~Pup.  Their simulations show that clumped wind models that
assume a void interclump medium cannot reproduce the observed \OVI\
profiles, though enough \OVI\ can be produced if the voids are filled by a
low-density gas.} In this case then, the {\it average}
density \rhob =\mdot/(4 $\pi r^2 v$) can be described by 
\beq 
\rhob = \fv \rho^+,\quad \rhobtwo = \fv (\rho^+)^2 
\eeq 
where $\rho^+$ is the density inside the overdense clumps, and \rhobtwo\ is
the mean of the squared density. Thus, the clumping factor as introduced by
\citet{OCR88}, 
\beq 
\fcl
= \rhobtwo/\rhob^2 \quad \Rightarrow \quad \fcl=\fv^{-1} \quad \mbox{and}
\quad \rho^+ = \fcl \rhob, 
\eeq 
corresponds to the inverse of the volume filling factor and measures the
overdensity of the clumps. Since we assume void inter-clump material, matter
is present only inside the clumps, with density $\rho^+$, and the
corresponding opacity is given by $\kappa=\kappa_C(\fcl \rhob)$, where $C$
denotes the process evaluated inside the clump. Optical depths have to be
calculated via $\tau=\int \kappa_C(\fcl \rhob) \fv \dd r$, with a reduced
path length $(\fv \dd r)$ to correct for the volume where a clump is
``seen''. Note that this expression is valid only if all clumps are
optically thin. Extensions for optically thick clumps are given in
Sect.~\ref{sec:clumptreat}. Combining these results, optical depths for a
clumped medium can be expressed by a mean opacity, $\bar \kappa$,
\beq 
\tau=\int \bar \kappa \dd r \quad \mbox{with} \quad \bar \kappa =
\kappa_C(\fcl \rhob)\fv = \frac{1}{\fcl} \kappa_C(\fcl \rhob).  
\label{kappabar}
\eeq 
Thus, for processes that are linear in density, the mean opacity of a
clumped medium is the same as in the homogeneous case, whereas for processes
scaling with the square of density, mean opacities (and emissivities) are
effectively enhanced by a factor \fcl. Moreover, radiative processes 
described by the micro-clumping approach do not
depend on the size, the geometry or the distribution of the clumps, but on
the radial stratification of the clumping factor alone. The enhanced opacity
for $\rho^2$ dependent processes has the consequence that mass-loss rates
derived by such diagnostics are lower by a factor of $\sqrt{\fcl}$ than
corresponding mass-loss rates assuming no clumping. Consequently, the
optical depth invariant, $Q$ (see Eq.~\ref{qq}), for such processes might be
extended according to 
\beq
Q = \frac{\mdot \sqrt{\fcl}}{(\Rstar \vinf)^{1.5}} = 
\frac{\mdot}{\sqrt{\fv}(\Rstar \vinf)^{1.5}} 
\label{qclump}
\eeq
as long as there is no strong radial variation of \fcl\ or \fv.

\subsubsection{Quantification of wind-clumping} 
\label{sec:clumpquant}

Let us now review how observations and models have evolved during the last
decade and summarize the most important results concerning the ``chase'' of
the clumping factor.

One of the first (and still most important) attempts to quantify
wind-clumping was carried out by \cite{Hillier91} who analyzed the
corresponding effects on the $\rho^2$-dependent emission components and the
$\rho$-dependent electron scattering wings in WR emission lines. Noting that
standard, homogeneous models severely overestimated the electron scattering 
wings of the observed profiles, \cite{Hillier91} showed how both components
could be simultaneously reproduced when clumping is accounted for and the
mass-loss rate is reduced by a factor of $\la 2$. 

For the method to work, the electron scattering wings have to be strong and
clearly defined, i.e., a good continuum rectification is required. The
method also assumes that the strength of the $\rho^2$-dependent emission can
be described by micro-clumping alone. Thus, this method may also be applied
to ``well behaved'' LBVs like P~Cygni, for which \cite{najarro01} found $\fv
\approx 0.5$ by analyzing all the Balmer lines. 

\begin{figure}
\begin{center}
\centerline{\hspace{-0.2cm}
\includegraphics[width=11.8cm]{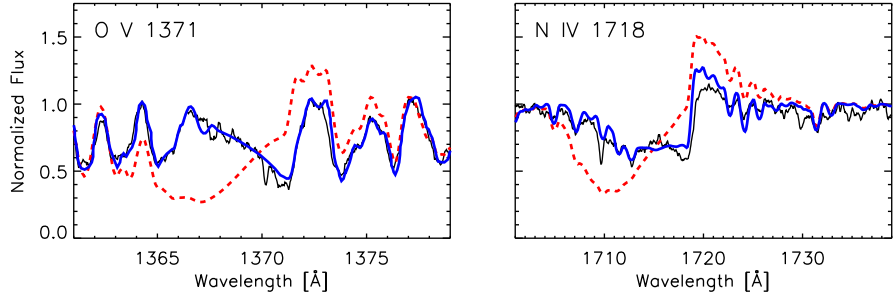}}
\vspace{-0.1cm}
\caption
{Effects of clumping and a correspondingly reduced mass-loss rate in the UV
\ouv\ and \nivuv\ lines of HD\,96715 (O4((f))).  The derived \fv=0.02 value
implies a factor of 7 reduction in \mdot. Dotted: Profiles from an unclumped
model which reproduces other, clumping-insensitive lines equally well.
Adapted from \citet{bouret05}.\label{fig:bouretclump}}
\end{center}
\end{figure}

Results using the new generation of unified model atmospheres accounting for
blanketing and clumping started to appear with the new millennium (see
Sect. \ref{sec:obastars}). From quantitative IR studies in the Galactic
Center, \cite{figer02} found that \fv-values around 0.1 (\mdot-reductions
of $\sim$3 with respect to homogeneous models) were required to best reproduce the
spectra of WNL and Of stars in the Arches Cluster. Analyses carried out for
O supergiants in the Magellanic Clouds \citep[ using UV and optical
data]{Crowther02} revealed that \fv = 0.1 could reproduce the observed \puv\
doublet without requiring a reduced phosphorus abundance as suggested by 
\citet{Pauldrach94, pauldrach01}. Further analyses in the SMC found likewise \fv\ values around
0.1 \citep{hillier03} and lower (\fv = 0.1 - 0.01 in four O stars of
the SMC cluster NGC346, \citealt{bouret03}). A similar study was carried out using
UV (FUSE + IUE) spectra of two Galactic O4 stars by \cite{bouret05}. In this
case they found that values as low as \fv = 0.02 and 0.04 were required to
satisfactorily reproduce not only the \puv\ and the \ouv\ lines, but also
\nivuv, as shown in Fig.~\ref{fig:bouretclump}. Though indicating an 
enormous degree of clumping (overdensities of factors between 25 to 50!),
these numbers are consistent with the alternative explanation in terms of a
strong oxygen depletion (factor 50) in early O stars, suggested by
\cite{pauldrach01} as a result of analyses using {\it homogeneous} models.

Interestingly, the lower \fv-values found for the SMC objects seem to
correspond to the Galactic ones down-scaled by the abundance ratio.  Major
uncertainties in these UV-based methods are due to the role of X-rays and
the validity of micro-clumping (see below) assumed for the models. Other
effects such as changes in the ionization structure driven by clumping are
included, however.

\paragraph{The \PV\ problem.~~} Although the problem with the \ouv\ line had
been extensively debated in the literature
\citep{haser98,pauldrach01,bouret03}, it was the advent of the FUSE mission
providing the cleaner \puv\ mass-loss diagnostics that gave rise to the
so called \PV\ problem (see \citealt{fullerton08} for a comprehensive review
of this issue).  Due to the very low cosmic abundance of phosphorus, the
\puv\ doublet basically never becomes saturated, not even when P$^{+4}$ is
the dominant ion. This enables to obtain a direct estimate of the product
\mdot\qb, where \qb\ is a spatial average of the ion fraction of the species
(cf. page~\pageref{mdotq}). Unfortunately, estimates of \qb\ for any given
resonance line is problematic unless detailed NLTE modeling is performed,
and even then, problems such as X-rays or subtle blanketing/blocking effects
may lead to incorrect predictions. On the other hand, an empirical
determination of ionization fractions is not feasible, since access to
resonance lines from consecutive ionization stages are not available in the
far-UV and UV regions. In the case of \PV, some insight is gained from the
available FUSE atlases of Galactic and Magellanic OB spectra
\citep{pellerin02,walborn02}. These show that the doublet is very weak for
the earliest O-subtypes, exhibits a stable maximum over a wide range of
spectral types, and then starts to disappear rather quickly for stars
between O9.5 and B0. Such morphology seems to secure a maximum of the
P$^{+4}$ ionization fraction for mid O-type stars, a behaviour confirmed by
model atmosphere calculations. Therefore, for those O-stars within a safe
$\qb \approx 1$ region, the \puv\ line should provide a robust and
independent estimate of \mdot\ (pure $\rho$-diagnostic).

\begin{figure}
\begin{minipage}{5.9cm}
   \resizebox{\hsize}{!}
      {\includegraphics{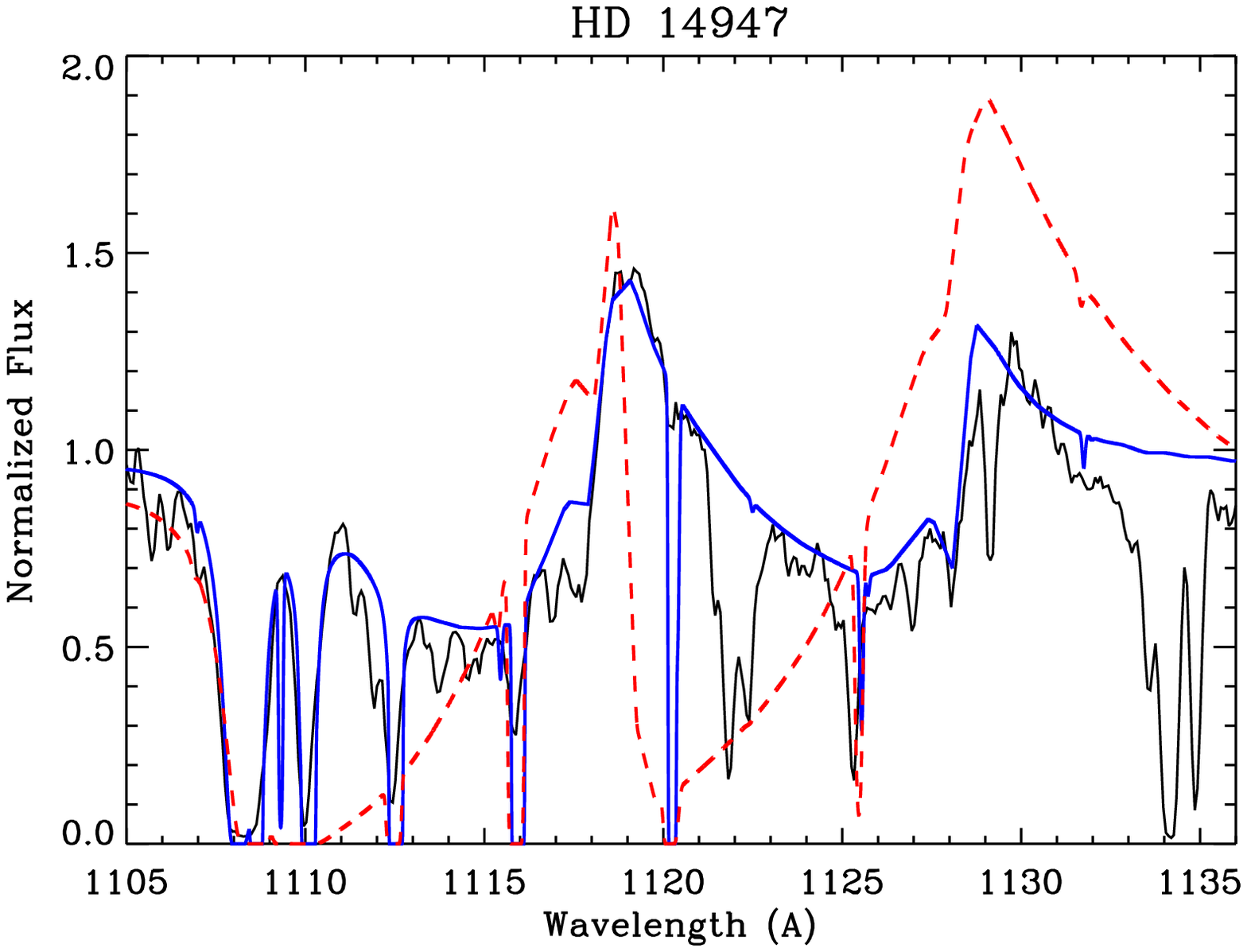}}
\end{minipage}
\begin{minipage}{5.9cm}
   \resizebox{\hsize}{!}
      {\includegraphics{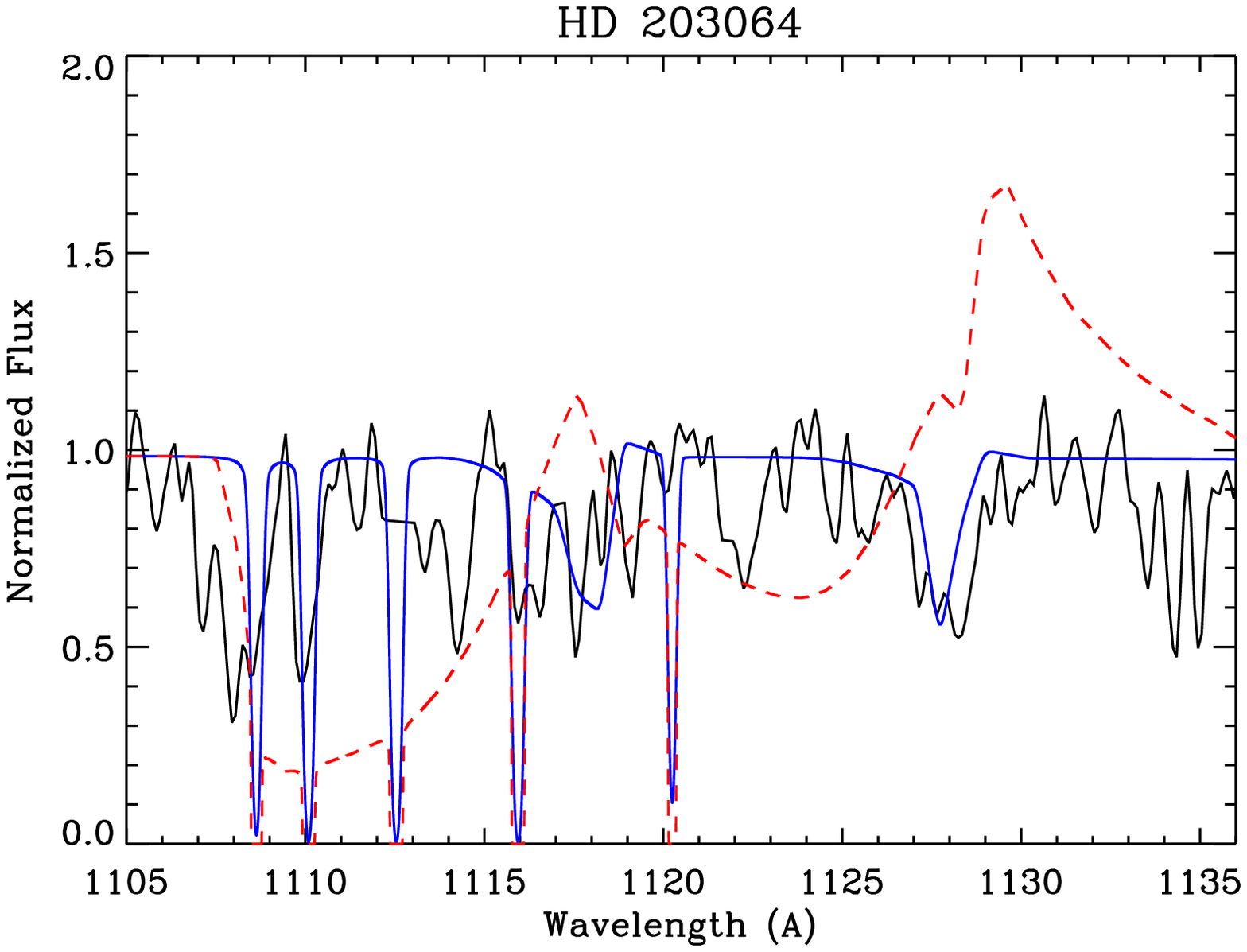}}
\end{minipage}
\caption{FUSE spectra of HD\,14947 (O5\,If$^+$) and HD\,203064 (O7.5\,III)
for the \puv\ doublet, together with the best fits from the simulations by
\citet{fullerton06} with a low \mdot\ (solid) and the \mdot\ derived from
\Ha\ using homogeneous models (dashed). To unify both analyses and assuming
\qb(P$^{+4}$) = 1, \fv\ values as low as $\fv \approx 0.03$ (HD\,14947) and
$\fv \approx 0.003$ (HD\,203064) are required, resulting in a down-scaling of
the ``unclumped'' mass-loss rates by factors of 6 and 18, respectively.
\label{fig:fullpv}}
\end{figure}

\begin{figure}
\begin{center}
\includegraphics[width=11.cm,angle=0]{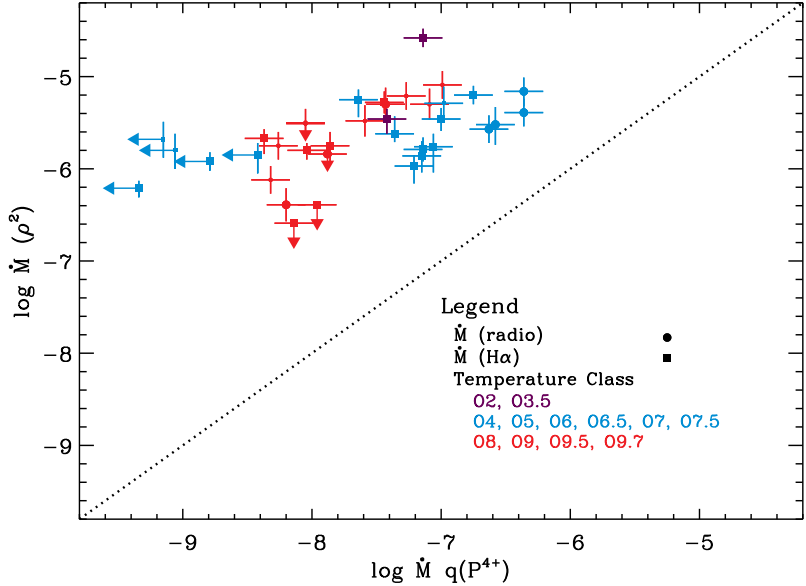}
\vspace{-0.1cm} \caption{Comparison of \mdot\ (derived from
$\rho^2$-diagnostics) with \mdot \qb($P^{+4}$) for a large sample of
Galactic O-stars. Adapted from \citet{fullerton06}, see text.
\label{fig:fullmdotpv}}
\end{center}
\end{figure}

\cite{fullerton06} carefully selected a sample of 40 Galactic O-stars, for
which also reliable $\rho^2$ \mdot-estimates (\hap, radio) were available,
and carried out a detailed comparison between both methods. Surprisingly
(see Figs. \ref{fig:fullpv} and \ref{fig:fullmdotpv}) they found a severe
discrepancy by a factor of $\sim$10 in \mdot\ for mid-range O-stars and by
substantially larger factors for earlier and later spectral types. A more
restrictive consideration, taking into account only those stars for which
\qb(P$^{+4}$) is expected to be 1, \ie\ the ``safer'' mid-range O
supergiants, resulted in a median discrepancy
\mdot($\rho^2$)/(\mdot(\PV)\qb) = 20, which would imply \fcl $\approx$ 400
if interpreted in terms of micro-clumping, i.e., {\it extreme} clumping,
much larger than derived by other methods. This discrepancy is the \PV\
problem and is nicely illustrated in Fig.~\ref{fig:fullmdotpv}. At the end
of this section we present a possible solution of this problem.

\paragraph{Polarimetric studies.~~} Soon after the detection of moving
subpeaks on top of the broad emission lines, other indications for clumping
of WR winds were found. Besides the diagnostics provided by the electron
scattering wings, a straightforward comparison of mass-loss rates based on
$\rho^2$-diagnostics with rates determined using other methods, such as
those from polarization variations in WR+O binaries\footnote{The basic
principle is that the light from the O-star acts as a probe of the WR-wind
via electron scattering. O-star photons scatter off the abundant free
electrons in the WR wind, which polarizes the light received by the
observer, where the level of polarization depends on the orbital-phase.}
(e.g. \citealt{StLouis93}), suggested that the mass-loss rates determined
from emission-lines are overestimated by about a factor three
\citep{Robert94}, which agrees with the factors of 2-4 derived from
NLTE analyses using clumped models (e.g. \citealt{Hamann98}).

Extending the above technique, \cite{stlouis08} studied the polarimetric
variability of O-stars winds in six massive O+O binary systems. They
reported clumping independent mass-loss rate estimates that suggest only
small clumping corrections. However, the O-stars analyzed by this method had
relatively weak winds and, as a consequence, the electron scattering optical
depth of the wind is low. Thus, a significant fraction of the
polarization signal originates from scattering off the photosphere, implying 
that the derived mass-loss rates should be lowered.

In their investigation of the polarization variability in WR-winds, 
\citet{Robert89} found an anti-correlation between the terminal wind 
velocity and the degree of random, intrinsic scatter in polarization, and
interpreted this finding as due to the presence of blobs that survive or
grow more effectively in slow winds. \citet{Davies05} found this trend to
continue into the regime of the LBVs, which have much lower $\vinf$ than
WR-winds and are an ideal testbed for constraining clumping properties, due
to their long wind-flow times.

\begin{figure}
\begin{center}
   {\includegraphics[width=8cm]{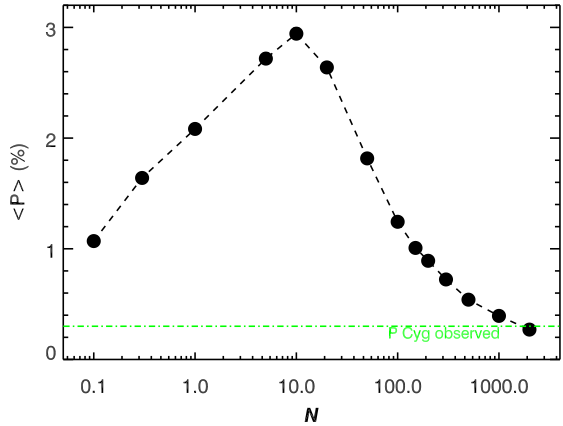}}
\end{center}
\caption{Time-averaged polarization over a broad range of ejection rates per
wind flow-time, $\mathcal{N}$. At $\mathcal{N} \sim 20$, the optical depth
per clump exceeds unity and the overall polarization falls off (see
\citealt{Davies07} for details).  The observed level of polarization for
P~Cygni is marked by the dash-dotted line. Note that there are two
ejection-rate regimes where the observed polarization level can be
achieved.}
\label{fig:pcyg_thick}
\end{figure}

\citet{Davies05} performed a spectropolarimetric survey of LBVs in the 
Galaxy and the Magellanic Clouds and found more than 50\% of them to be 
intrinsically polarized. As the polarization angle was found to vary 
irregularly with time, the polarization line effects were attributed to 
wind clumping. Monte Carlo models for scattering off wind clumps were
developed by \citet{Code95, Rodrigues00, Harries00}, whilst analytical
models to produce the variability of the linear polarization were presented
by \citet{Davies07}. 

An example of an analytical model that predicts the time-averaged
polarization for the LBV P\,Cygni is presented in 
Fig.~\ref{fig:pcyg_thick}. The clump ejection rate per wind flow-time
$\mathcal{N}$ is defined as ${\mathcal N} = \dot{N} t_{\rm fl} = \dot{N}
R_{\star}/v_{\infty}$, where the clump ejection rate, $\dot{N}$, is related
to the mass-loss rate via $\dot{M} = \dot{N} N_{e} \mu_{e} m_{H}$, with
$N_{e}$ the number of electrons in each clump and $\mu_{e}$ the mean mass
per electron. There are two
ejection-rate regimes where the observed polarization level can be achieved.
One regime is where the ejection rate is low and a few optically thick
clumps are expelled, and the other regime is that of optically thin clumps
where the number of clumps is very large. These two models can be
distinguished via time resolved polarimetry. Given the relatively short
timescale of the observed polarization variability, \citealt{Davies07} favor
the latter scenario, suggesting that LBV winds consist of thousands of
optically thin clumps in close proximity to the photosphere. 

\paragraph{Stratification of the clumping factor.~~} Most of the studies
considered so far assume that clumping starts just above the sonic point 
and remains constant throughout the wind. We now address whether this
approximation is feasible or needs to be relaxed by requiring stratified
clumping. Obviously, this can be achieved by identifying reliable clumping
diagnostics at different distances from the star. First indications that
\fv\ (or \fcl) vary as a function of radial distance from the star were
obtained by \cite{Nugis98} by means of IR to mm and radio continuum studies
of WRs.  To explain a spectral index from mm to cm wavelengths of
the order of $\sim$0.76 (steeper than a value of 0.6 expected for smooth
winds, see Eq.~\ref{specindradio}), \cite{Nugis98} proposed a stratification
of the clumping factor such that \fcl\ is unity at the wind base, reaches a
maximum at 5-10 \Rstar, and then recovers unity in the outer wind parts.
With respect to mass-loss rates derived from homogeneous models,
\citet{Nugis98} concluded that these were overestimated by $\sim$0.2 dex for
WN stars and by $\sim$0.6 dex for WC stars\footnote{In a subsequent study,
\citet{NL00} derived an empirical mass loss formula for Galactic WR stars
accounting for these reduced rates, see Sect.~\ref{sec:wrstars}.}

\begin{figure}
\begin{center}
   {\includegraphics[width=8cm]{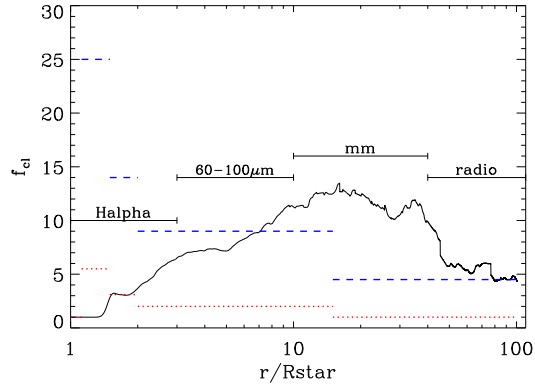}}
\end{center}
\vspace{-0.6cm}
\caption{Radial stratification of the clumping factor \fcl\ for an 
O-supergiant wind. Solid: Theoretical predictions by \citet{RO02} from
hydrodynamical models, with self-excited line-driven instability. The
formation regions of different diagnostics are indicated. Dotted: Average
clumping factors, as derived for the wind from $\zeta$~Pup, assuming an
unclumped outer wind \citep{Puls06}. Dashed: as dotted, but assuming an
outer wind with clumping as theoretically predicted. The corresponding
mass-loss rate is a factor of $(f_{\rm cl}^{\rm out})^{0.5}$ smaller than in
the dotted case. See text.} 
\label{fcl}
\end{figure}

A similar parameterization of the clumping factor was used by
\cite{figer02} in their analysis of WNL and OIf$^+$ stars in the Arches
Cluster. Only a vanishing clumping (\fcl = 1) in the outer wind zones 
of these objects could account for the observed \pap\
equivalent widths and radio fluxes of the stars. These authors used \pap,
\bgam, and the radio fluxes as diagnostics for the radial dependence of \fcl.

A further step towards the characterization of wind clumping was taken by
\citet{Puls06}, who were able to derive constraints on the radial
stratification of \fcl\ by simultaneously modeling \hap\ and the IR/mm/radio
emission from a sample of 19 O-stars with well-known parameters. Actual
constraints could be obtained in the (1 - 5 \Rstar) lower and intermediate
wind region where \hap\ and the IR form, and in the outer (10 - 50 \Rstar)
wind region where the mm/radio emission arises. Particularly for the best
constrained object, $\zeta$~Pup, the derived stratification is in strong
contrast to theoretical predictions (see Fig.~\ref{fcl}). Actually, 
\citet{Puls06} found considerable clumping already close to the stellar
surface for all denser winds, consistent with the findings from other diagnostics
covered in previous sections.  Clumping would then remain rather constant
over a large volume before decreasing in the outer wind, with ratios of
clumping factors between the inner and outer regions ranging from 3 to 6
(\mdot(\hap) = 1.7 - 2.5 \mdot$_{\rm radio}$), which is very similar to the
values suggested by \citet{markova04} and \citet{repolust04} on a completely different basis
(page~\pageref{clumprepo}).
For weaker winds, on the other hand, \citet{Puls06} found similar
clumping factors for the inner and outermost regions, i.e.,
\mdot(\hap) $\approx$ \mdot$_{\rm radio}$. 

A major shortcoming of this study is caused by the $\rho^2$ dependence of
{\it all} considered diagnostics, which lead the authors to derive only {\it
relative} clumping factors. Note, however, that mass-loss rates based on the
{\it assumption} that \fcl(radio) = 1, i.e., an unclumped outer wind, would
give rise to a unique WLR (dependent only on $L$), being in very good
agreement with the theoretical predictions by \citet{Vink00}.

The next natural step is then to perform multi-wavelength analyses including
both $\rho$ and $\rho^2$ diagnostics. These have been recently presented by
\citet{najarro08a} and \citet{najarro08b} by means of UV to radio
observations of OB stars and IR observations of LBVs. These authors identify
key diagnostics lines to obtain the clumping structure throughout the
stellar wind, and provide guidelines to constrain the degree of clumping in
stellar winds for different stellar types. 
 
\begin{figure}
\begin{center}
\includegraphics[width=8cm,angle=0]{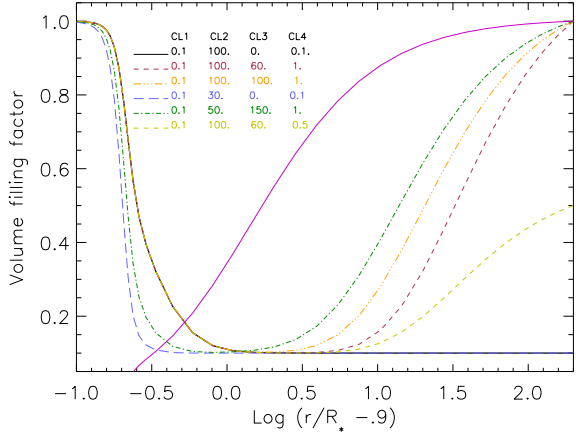}
\vspace{-.1cm}
\caption{Examples for the radial structure of the volume filling factor,
\fv(r), for different values of CL$_{2,3,4}$ according to
Eq.~\ref{eq:clump}, and CL$_1$ = 0.1. For models with an unclumped outer
wind, CL$_4$ = 1, whilst for models with constant clumping in the outer
part, CL$_4$ = CL$_1$. The velocity structure (solid magenta, in units of \vinf) is
displayed to illustrate the onset of the clumping variations.
\label{fig:runclump}}
\end{center}
\end{figure}

To investigate the clumping stratification in detail, \cite{najarro08b} 
utilized the O3If$^+$ star CyOB2\#7 as a work bench and suggested the
following clumping law (see Fig.~\ref{fig:runclump}):
\begin{equation}
\label{eq:clump}
  \fv(r) = CL_1 + ( 1 -CL_1 )\, {\rm e}^{\frac{-v(r)}{CL_2}} + 
  ( CL_4 - CL_1)\,{\rm e}^{\frac{-(v_\infty-v(r))}{CL_3}},
\end{equation}
where CL$_1$ and CL$_4$ are volume filling factors and CL$_2$ and CL$_3$ are
velocity terms defining locations in the stellar wind where the clumping
stratification changes. CL$_1$ sets the maximum degree of clumping reached
in the stellar wind (provided CL$_4$ $>$ CL$_1$) whilst CL$_2$ determines
the velocity of the onset of clumping. CL$_3$ and CL$_4$ control the
clumping structure in the outer wind. This is illustrated in
Fig.~\ref{fig:runclump}, which displays the behavior of clumping in a
stellar wind for different values of CL$_2$, CL$_3$ and CL$_4$. The clumping
law Eq.~\ref{eq:clump} has been constructed in such a way that as the wind
velocity approaches \vinf, the volume filling factor starts to change from
CL$_1$ towards CL$_4$. If CL$_4$ is set to unity, the wind will be unclumped
in the outermost region (as suggested by \citealt{Nugis98} and adopted by
\citealt{figer02} and \citealt{najarro04} for the case of WR-winds, see
above). From Eq.~\ref{eq:clump} we note that for CL$_3 \rightarrow$ 0 we
recover the law as proposed by \cite{hilliermiller99}. 

\begin{figure}
\begin{center}
\includegraphics[width=8cm,angle=0]{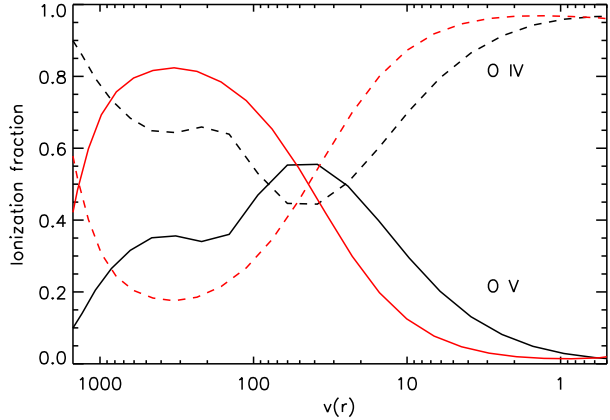}
\vspace{-0.1cm} \caption{Run of the \OIV\ (dashed) and \OV\ (solid)
ionization fractions of an O supergiant wind for a clumped (black) and
homogeneous (light color) model. Adapted from \citet{bouret05}. 
\label{fig:bouret05ion}}
\end{center}
\end{figure}

All clumping effects considered so far refer to a situation where the
ionization fraction of the ion responsible for the considered process is not
or only weakly affected by the clumping itself (at least beyond the
conventional \mdot$\sim \sqrt{\fv}$-scaling Eq.~\ref{qclump}). Thus, for a
resonance line (e.g., \PV) it must be secured that the ionization stage the
line belongs to (P$^{+4}$) clearly remains as the dominant stage, so that by
varying the clumping alone no changes in the line profile are produced. On
the other hand, for a recombination line, the ionization stage the line
belongs to must be clearly less populated than the next higher ionization
stage (i.e., \HII$\gg$\HI\ for the hydrogen lines), so that an \mdot$\sim
\sqrt{\fv}$-scaling preserves the number of recombinations. We consider the
corresponding regime of the ionization equilibrium to be on the ``safe''
side. 

\begin{figure}
\begin{center}
\includegraphics[width=0.9\textwidth]{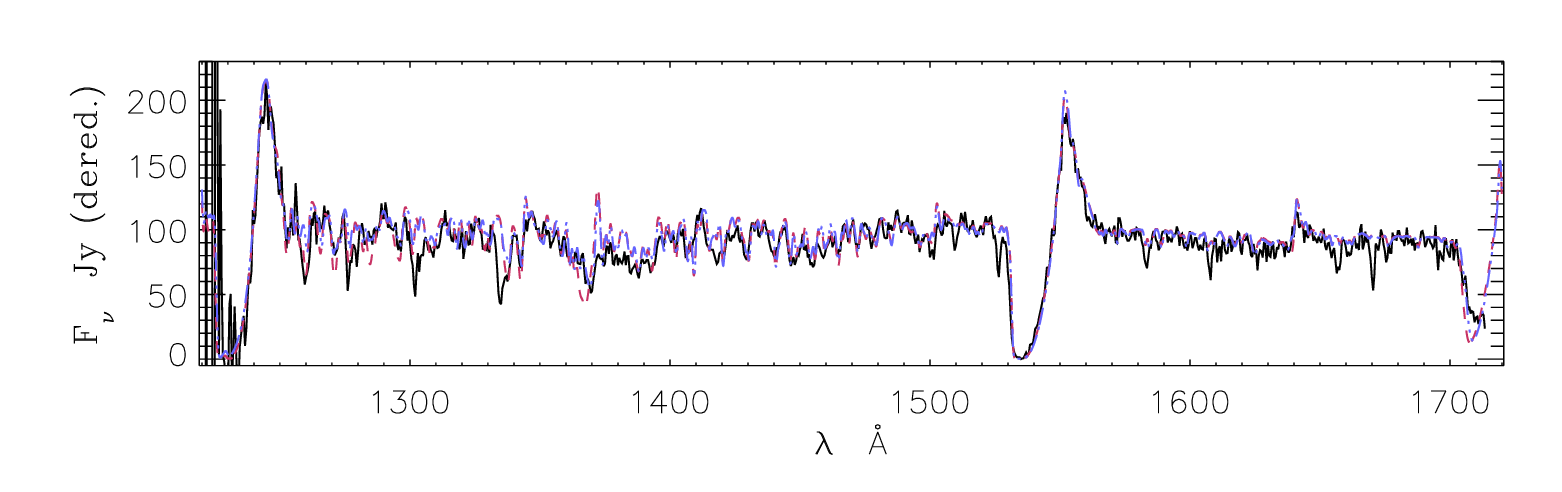}
\vspace{-0.2cm}
\includegraphics[width=0.9\textwidth]{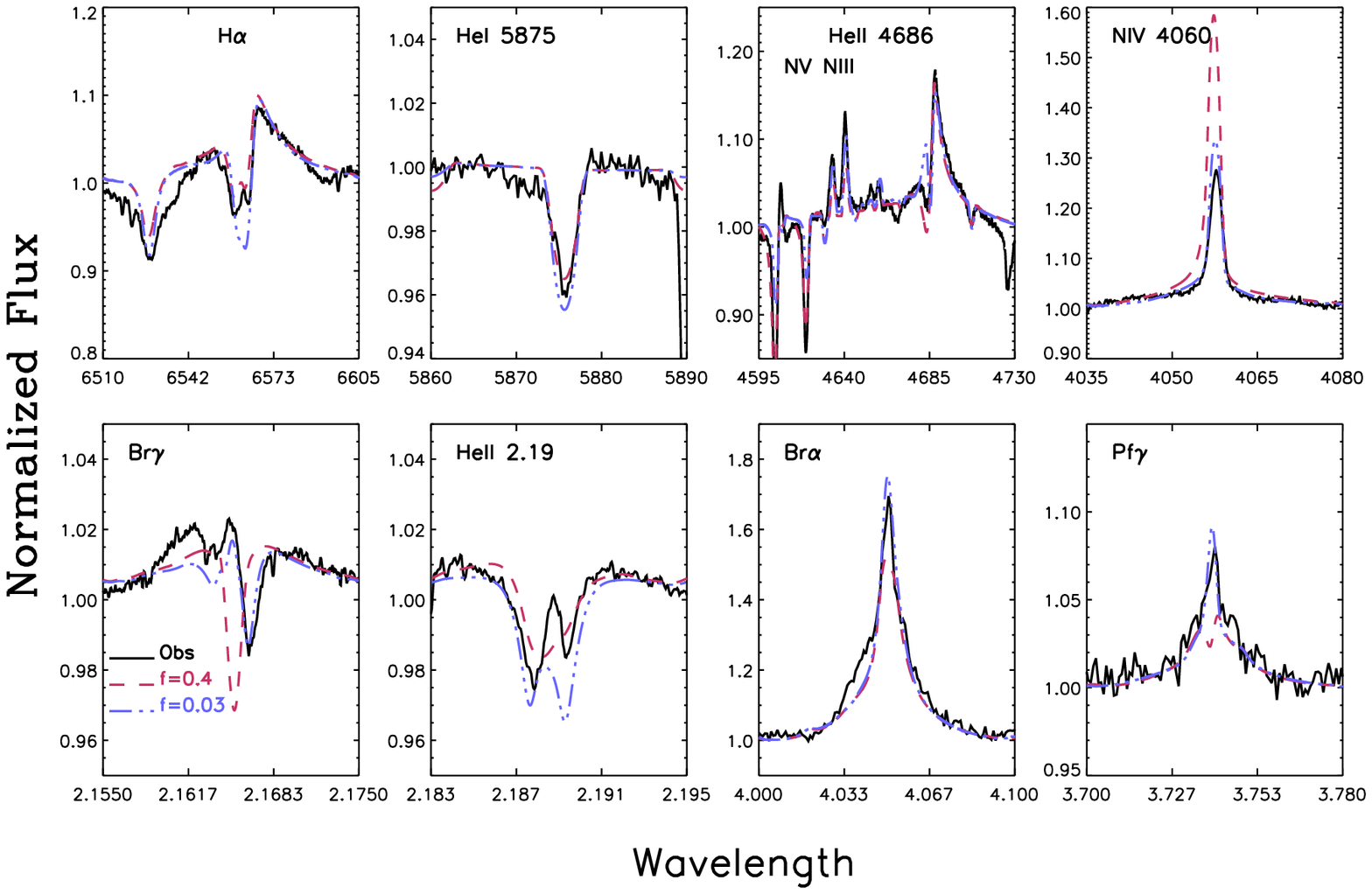}
\vspace{-0.1cm} \caption{Best model fits to UV, optical, and infrared
profiles of CygOB2~\#7, for two values of CL$_1$, denoted in the figure by
$f$ and corresponding to the maximum degree of clumping in the wind (=
minimum volume filling factor). The weakly clumped model (dashed) with CL$_{1,2,3,4}$
= (0.4, 350~\kms, 25~\kms, 1.) has a mass-loss rate of 6\Mdu, whereas the
strongly clumped model (dashed-dotted) with a clumping law according to
\citet{hilliermiller99}, CL$_{1,2}$ = (0.025, 120~\kms), has \mdot\ = 1.4\Mdu.
\label{fig:line7}}
\end{center}
\end{figure}

Most worrying is the case where two adjacent ionization stages have similar
populations. Then clumping will cause a net reduction of the mean
ionization, due to an enhanced number of recombinations ($\propto \rho^2$)
compared to ionizations (roughly $\propto \rho$). This will result in a
line-profile behaviour similar to a reduction of the stellar temperature. We
denote such a regime of the ionization equilibrium by the term ``changing''.
This indirect effect of clumping has been warned about from early times on
(e.g., \citealt{Puls93}), and \cite{bouret05} have investigated this effect
in detail. As an example, Fig.~\ref{fig:bouret05ion} shows how the oxygen
ionization equilibrium is significantly altered due to wind clumping.

One may, therefore, distinguish between lines formed in a ``safe'' region
and those arising from a ``changing'' region. Figure~\ref{fig:line7} shows
that the \NIV\,$\lambda$4058 and \NV\,$\lambda$4600 lines in CygOB2~\#7 are
clearly formed in such a ``changing'' region, whilst the rest of the optical
lines are less affected by the large change of clumping and seem to follow
the ``safe'' region behaviour. On the other hand, most of the infrared lines
react strongly to changes in the clumping law. Since, as outlined on
page~\pageref{nirspec}, the continuum adjacent to NIR-lines is already
formed in the wind, not only the lines, but also the continuum will depend
on clumping (via bound-free and free-free processes $\propto \rho^2$),
resulting in a high sensitivity of the {\it continuum-rectified} line
profiles on the clumping {\it stratification}, particularly on CL$_1$ and
CL$_2$. \cite{najarro08b} found that if clumping starts with large clumping
factors relatively close to the photosphere of CygOB2~\#7 (with \fcl\ = 5
reached at $\sim$200~\kms, corresponding to $\sim$1.1~\Rstar, similar to the
results by \citealt{Puls06} for this object), {\it consistent} simultaneous
fits to the UV, optical, and IR observations are possible (see
Fig.~\ref{fig:line7}).

Whilst the optical and IR spectra of CygOB2~\#7 provide strong constraints on
CL$_1$ and CL$_2$, the UV, submillimeter, and radio observations constitute
crucial diagnostics to determine CL$_1$ and CL$_3$. Indeed, UV and
submillimeter data support the presence of constant
clumping \citep{najarro08b}, at least up to mid/outer wind regions where the
millimeter continua of CygOB2~\#7 are formed. However, radio observations by
\cite{Puls06} with a formation region at much larger radii showed that
clumping may begin to vanish in the outermost wind regions, which is
consistent with the fact that the radio emission from such models with
constant clumping severely overestimates the upper limits provided by the
observations by \cite{Puls06} of CygOB2~\#7.  This demonstrates the need of
multi-wavelength observations to constrain the run of the complete clumping
stratification.

Finally, for quantitative analyses one would wish to have a nice toolkit
providing diagnostic lines that, depending on the stellar type of the
object, could be used to constrain the {\it absolute} degree of wind
clumping. At temperatures around 30~kK, there is the possibility to use
\HeII\ 4686 and \Ha\ in parallel, as outlined on page~\pageref{cspn}.
Otherwise, one has particularly to rely on the ``changing'' region situation
discussed above. No strong UV lines should be included as they may be
significantly affected by X-rays and corresponding EUV emission. Note, however,
that even UV-lines formed in the lower wind could be affected by very hot
material potentially located close to the wind base (see
Sect.~\ref{sec:xraylines}). For early O supergiants, the \NIV\,$\lambda$4058
and \NV\,$\lambda$4600 lines, and for late O supergiants, the \HeII, \NIII\ and
\CIII\ lines, turn into important clumping diagnostic lines. Valuable
clumping information may be obtained from cool LBVs for which the \HeI\ and
\FeII] \citep{najarro08a} lines will arise from ``changing'' regions. 

\subsubsection{Treatment of clumping - Micro- and macro-clumping} 
\label{sec:clumptreat}

With studies yielding clumping factors ranging from $\fcl$ = 2 to $\ga 100$
(corresponding to volume filling factors $\fv$ = 0.5 to $\la 0.01$), one may
question whether our present treatment of clumping is physically correct. If
related to a certain type of instability (e.g., the line-driven one), we may
expect this treatment to be inadequate in those regions where the
instability is not fully grown but still in its linear or only weakly
non-linear phase. In these (lowermost) wind regions, the assumption of a
void inter-clump medium is certainly questionable. Most of the atmospheric
codes take into account only variations in the density, but the hydrodynamic
simulations also reveal strong changes of the velocity field inside the
clumps, sometimes even changing the sign of the velocity gradient.  Most
important, however, is the recent finding that the prerequisite of the
standard micro-clumping approach, namely that the clumps are optically thin,
needs to be relaxed. 

Within the micro-clumping approach (page~\pageref{microclumping}), an
optically thin clump is identified as a clump with a size smaller than the
photon mean free path of the relevant matter-light interaction. In an
optically thick clump (``macro-clumping''), photons will interact with
matter many times before eventually being destroyed or scattered off the
clump and escaping through the inter-clump matter. Of course, whether a
clump is optically thin or thick will depend on the abundance, ionization
fraction, and cross-section of the involved transition. Thus, we will have
clumps that are optically thick for some processes while being optically
thin for others. For a recent review of micro- vs. macro-clumping, we refer
the reader to \citet[ see also \citealt{oskinova07}]{Hamann08a}.

Micro-clumping has been firstly implemented into NLTE model atmospheres by
\citet{Schmutz95}, and the corresponding formalism was outlined on 
page~\pageref{microclumping}. For our discussion of {\it macro-clumping} we
follow the same notation (see also \citealt {OGS04} and the review by
\citealt{Hamann08a}). 

For optically thick clumps, photons {\it do} care about the distribution,
the size and the geometry of the clump (see Sect.~\ref{sec:xraylines}) they
will encounter and multi-interact with. 
The conventional description of macro-clumping bases on a uniform 
clump size, $l(r)$, and an average separation among the statistically
distributed clumps, $L(r)$, which are related by the volume filling factor,
\beq
\fv = \bigl(\frac{l}{L}\bigr)^3 = \frac{1}{\fcl}.
\label{eq:fmacroclu}
\eeq
Following the nomenclature of Eq.~\ref{kappabar}, the optical depth across
a clump of size $l$ and constant opacity $\kappa_C$ is given by
\beq
\tau_C = \kappa_C l = \bar \kappa \fcl l = \bar
\kappa \frac{L^3}{l^2} = \bar \kappa h,
\label{eq:taumacroclu}
\eeq
with mean opacity $\bar \kappa$ (Eq.~\ref{kappabar}) and porosity length
$h=L^3/l^2$, as introduced in Eq.~\ref{porositylength}. Remember that the
porosity length constitutes a key parameter to quantify the characteristics
of a structured medium and corresponds to the photon mean free path in a
medium consisting of optically thick clumps (page~\pageref{porositylength}).

The {\it effective} cross section of the clumps, i.e., their geometrical
cross section (surface) diminished by the fraction of transmitted radiation,
is given by
\beq
\sigma_C = l^2 \, (1 - \eu^{-\tau_C})
\label{eq:croscluthick}
\eeq
such that the {\it effective opacity} that needs to be used in the models 
results in
\beq
\kappa_{\rm eff} = n_C \sigma_C = \frac{l^2\,(1 - \eu^{-\tau_C})}{L^3} =
\frac{(1 - \eu^{-\tau_C})}{h} = \bar \kappa \frac{(1 -
\eu^{-\tau_C})}{\tau_C},
\label{eq:chicluthick}
\eeq
when $n_C$ is the number density of the clumps. Note that this equation
holds for clumps of arbitrary optical thickness. When taking the optically 
thin limit, we immediately recover the micro-clumping approximation, 
$\kappa_{\rm eff} = \bar \kappa$, which depends on the volume filling factor
alone but not on clump sizes and distribution, whereas in the optically
thick case, the {\it effective opacity becomes strongly reduced},
$\kappa_{\rm eff} = \bar \kappa/\tau_C$ = $h^{-1}$ and depends only on the
porosity length, i.e., it is grey. Further comments on effects resulting
from specific clump geometries are given in Sect.~\ref{sec:xraylines}.

\begin{figure}
\begin{center}
\centerline{ \hspace{-0.2cm}
\includegraphics[width=11.5cm,angle=0]{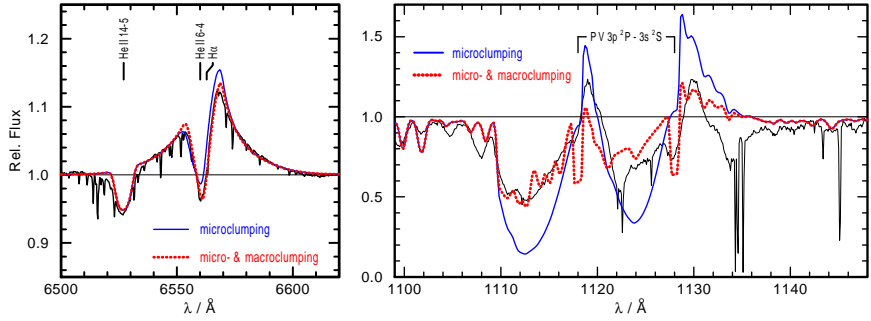}}
\vspace{-0.1cm} \caption{Porosity as a possible solution for the PV problem.
Adapted from \citet {oskinova07}, see text. \label{fig:oskiclump}}
\end{center}
\end{figure}

\paragraph{Effects on wind structure.~~}
\cite{krticka08} have studied the influence of micro- and macro-clumping on
the predicted wind structure of O-type stars, by artificially including
clumping into their stationary wind models which have been 
described in Sect.~\ref{sec:mcakmodels}.

If they assumed all clumps to be optically thin, the radiative line force
increased compared to corresponding unclumped models, due to an increase of
the electron density dependent $\delta$-term in the force-multiplier
(page~\pageref{deltaterm}), representing a changed ionization equilibrium
with more lines from lower stages. This increase of the line-force leads to
an {\it increase} of either the mass-loss rate or the terminal velocity,
depending on the onset of wind-clumping (below or above the critical point).
Similar results have been found by \citet{dekoter08}.

If, on the other hand, the clumps were considered as optically thick, 
\cite{krticka08} showed that a net reduction of the line-force may occur. 
Note, however, that this finding is in contrast to hydrodynamical
simulations\footnote{which do {\it not} find changes of the gross wind
properties compared to a stationary solution, though the medium {\it is}
clumpy (page~\pageref{grosswind}).}, and originates from assuming that the
velocity law remains unchanged, in contrast to what would happen in a
consistent approach.  Finally, an inclusion of wind porosity into the
continuum opacity (with the onset of porosity below the critical point)
might also lead to a reduced \mdot, due to an increase of the
wind-ionization.

In conclusion, a further, consistent investigation of the impact of
wind-clumping (with its various facets) on the wind structure of massive O
stars is crucial, since this might influence the {\it predicted} wind
parameters, to which the observed ones have to be compared.

\paragraph{Effects on line profiles.~~} \cite{oskinova07} investigated the
effects of porosity on the line profiles of $\zeta$~Pup, by means of the
above formalism (Eq.~\ref{eq:chicluthick}) and assuming a ``high'' value for
the mass-loss rate, \mdot = 2.5 \mdu, which is only mildly lower than the
upper limit derived by \citet{Puls06}. Using over-densities, \fcl, inferred
from diagnostic lines, and the equation of continuity to parameterize the
radial dependence of the clump separation, $L(r)$, they were able to account
for (part of\footnote{Note that the original porosity concept had been
developed for continuum processes; line processes are additionally affected
by changes in the velocity field, which are simulated here by adopting a 
velocity dispersion inside the clumps (cf. page~\pageref{veldisp}), but see
below.}) the radiative transfer effects of clumping. 

In particular, only changes in the ``formal integral'' were considered (by
using the effective opacity), whereas the feedback of macro-clumping on the
NLTE-populations was neglected by arguing that most transition rates are not
affected. From Fig.~\ref{fig:oskiclump}, it becomes clear that the most
important effect regards strong resonance lines like \puv, which can be
reproduced by this approach without relying on (very) low mass-loss rates,
due to the reduction of the effective opacity when the clumps become
optically thick. On the other hand, \hap\ is not affected by wind porosity,
remaining optically thin inside the clumps and being reproduced in parallel
with \PV. Insofar, these results might open a way out to the ``\PV\
problem''.

\begin{figure}
\begin{center}
\centerline{
\hspace{-0.2cm}
\includegraphics[width=11.5cm,angle=0]{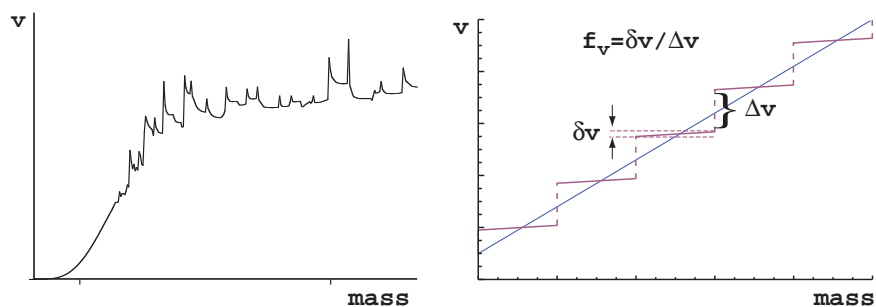}}
\vspace{-0.1cm}
\caption{Vorosity. Multiply non-monotonic velocity structure arising in a
1-D hydrodynamic simulation with self-excited line-driven instability,
plotted versus a mass-coordinate (left), and schematic simplification
(right). The smooth wind flow is represented by the straight line. Adapted
from \citet{owocki08}, see text.\label{fig:owockivoro}}
\end{center}
\end{figure}

\paragraph{Clumping from the other side: ``Vorosity''.~~} As indicated
above, line transitions should be strongly affected by a strongly
non-monotonic velocity field as resulting from the line-driven instability.
Aiming to examine the potential role of such a flow structure in reducing
the observed strength of wind absorption lines, \cite{owocki08} performed
dynamical simulations of the line-driven instability in which the line
strength can be described in terms of a ``velocity clumping'' factor.
Interestingly, this velocity clumping factor is insensitive to spatial
scales. 

The left panel of Fig.~\ref{fig:owockivoro} shows the velocity structure
arising in a 1-D simulation. The intrinsic instability of line-driving leads
to substantial velocity structure, with narrow peaks corresponding to
spatially extended but tenuous regions of high-speed flow. Such regions are
followed by dense, spatially narrow clumps/ shells which, when plotted as a
function of mass-coordinate, give rise to extended velocity plateaus. The
right panel illustrates the situation by means of a simplified model, where
the velocity clumping is represented by a simple {\it staircase} structure,
quantified by a velocity clumping factor. This value is set by the ratio
between the internal velocity width $\delta v$ to the velocity separation
$\Delta v$ of the clumps. 

Using this model, \cite{owocki08} computed dynamic spectra directly from
instability simulations, which indeed exhibited a net reduction in
absorption, due to the porosity in velocity space, an effect he denoted by
``vorosity''. The key problem in explaining a ``reduced'' \PV\ line-strength
in terms of vorosity is that one needs to have a large number of {\it big}
velocity gaps, which is not what comes ``naturally'' from the instability
simulations (Fig.~\ref{fig:owockivoro} - left). Thus, only a modest
reduction at a 10-20\% level was found, which is well short of the factor 10
(or more) required in terms of the \PV\ problem.

Obviously, it remains a future challenge to investigate scenarios including 
both porosity and vorosity and their possible inter-relation.

\subsubsection{Evolutionary limits} 

If mass-loss rates would need to be revised drastically (e.g., by factors of
$\sim$ 10 or more), this would have dramatic consequences for the evolution
of and feedback from massive stars. Recalling a corollary statement in
massive star evolution, that {\it a change of the mass-loss rates of massive
stars by even a factor of two has a dramatic effect on their evolution}
\citep{meynet94}, one immediately realizes the importance of quantifying the
wind clumping phenomenon. This statement illuminates a potential alternative
to constrain the ``allowed'' range of mass-loss reductions, namely by
evolutionary considerations themselves. 

\cite{hirschi08} has recently explored the evolutionary {\it implications}.
Taking for example a 120~\msun\ star with a lifetime of 2.5 million years
and an average \mdot$\sim 2.5 \cdot 10^{-5}$\msunyr, the star will have lost
around 50~\msun\ at the end of the main sequence using the theoretical
prescriptions of \citealt{Vink00,Vink01}, whilst the star would lose only
5~\msun\ if the mass-loss rate was reduced by a factor of 10. The two big
questions regarding evolution are then how to produce WR-stars with such low
mass-loss rates and whether one would end up with too many critically
rotating stars.

Concerning the WR issue, one way out would be to shift the bulk of mass loss
on other evolutionary phases such as the LBV (e.g. \citealt{NL00,Smith06})
and the red supergiant (RSG) stages. However, although mass-loss rates are
more challenging to determine in these two stages, it seems unlikely that
they could cope with the bulk of it. 

A second, provocative, possibility is to place all massive stars in close
binary systems \citep{kobul07}. This scenario, however, would not only fail
to produce the many RSG stars observed, but there is also observational
evidence against {\it all} WR-stars being in close binary systems
\citep[only 30-40\% in the Magellanic Clouds,][]{foellmi03a,foellmi03b}.
Another possibility is to assume that WRs are produced by strong rotational
mixing rather than by mass loss. Since fast rotators constitute only a small
fraction of the whole population, this scenario seems very unlikely though.

Combining the above points, \cite{hirschi08} concluded that evolutionary
models could survive with \mdot\ reductions of at most a factor of
$\sim$2 in comparison to the rates from \citet{Vink00}, whilst factors
around 10 are strongly disfavored. We note that a reduction of the
\citet{Vink00,Vink01} rates by a factor of two may correspond to an
``allowed'' reduction of the empirical mass-loss rates of a factor of about
four - at most.

\subsection{Co-rotating interaction regions (CIRs)}
\label{sec:cirs}

In the previous section, we have extensively discussed features attributed
to {\it micro-}structure in the wind, expressed in terms of micro- and
macro-clumping, and consequences of this. However, there is also large-scale
structure present, inferred most directly from the so-called {\it discrete
absorption components (DACs)}. 

\paragraph{Discrete absorption components.~~} These are optical depth
enhancements observed in the absorption troughs of unsaturated UV P~Cygni
profiles and are present in most O-/early B-star winds \citep{HP89}, but
also in late B-supergiants \citep{BatesGilheany90} and at least one WN7 star
\citep{PrinjaSmith92}\footnote{Absorption components have been also 
observed in the Balmer lines of the LBV P~Cygni \citep{Markova86}.}. 

Typically, these DACs accelerate to the blue wing of the profile on
timescales of a few days, becoming narrower as they approach an asymptotic
velocity. \citet{HP89} and \citet{Prinjaetal90} have used this asymptotic
value to determine $\vinf$ for objects with no black trough in C{\sc iv}
(see Sect.~\ref{sec:diagnostics}). The acceleration and recurrence time
scales of DACs were found to be correlated with \vsini, i.e., with the
rotational period \citep{Prinja88, Henrichs88, Kaper99}. The acceleration of
most strong DACs is much slower than the mean wind acceleration
\citep{Prinjaetal92, Prinja94}, suggesting that DACs might arise from a
slowly evolving perturbation through which the wind material flows. Such a
perturbation could consist of a higher density or a lower velocity gradient
(e.g., a velocity plateau, as speculated already by \citealt{Lamersetal82}),
or result from a combination of both \citep{FO92}.

Though the lack of variability in the emission component of P~Cygni lines
\citep{Prinja92} and the lack of significant infrared variability
\citep{Howarth92} seems to rule out a spherically-symmetric disturbance, the
structure must be large enough to cover a substantial fraction of the
stellar disk in order to produce the observed strong absorption features.
The latter constraint strongly argues against any micro-structure related to
the wind instability as the source of the DACs, because (i) micro-structure
would be averaged out and (ii) the presence of the {\it lateral} line-drag
\citep{ROC90} should suppress any lateral velocity disturbances and
maintain lateral length scales as set by a potential base perturbation.

\paragraph{Dynamical models of CIRs.~~} The above arguments suggest that DACs
should originate from coherent structures of significant lateral extent,
with a clock related to stellar rotation and an increased optical line depth
due to density and/or velocity field effects. Already in 1984, i.e., well before these
constraints were known, Mullan \nocite{Mullan84} had suggested co-rotating
interaction regions\footnote{well known and studied in the solar wind.} (CIRs)
as a potential candidate (see also \citealt{Mullan86}). Over the
following years, this suggestion became more and more likely, due to
an amount of empirical knowledge and since CIRs are compatible with
all constraints as outlined above. 

\citet{CO96} were the first to investigate the CIR scenario in detail, by
means of 2-D time-dependent hydrodynamic modeling of the wind of a rotating
O-star. Photospheric disturbances in the form of {\it azimuthal} variations 
were induced by a local increase of the line force\footnote{Actually, the
authors investigated also the consequences of a local decrease in the line
force due to dark spots. Such a model, however, could be ruled out, since
the resulting synthetic profiles behave quite different from the
observed variability pattern.} due to a bright stellar spot in
the equatorial plane. Such a stellar spot should be regarded as a
representative for various kinds of photospheric disturbances (e.g.,
localized magnetic fields or non-radial pulsations), since the induced wind
structure turned out to be rather insensitive to the particular way by which
the photospheric conditions were disturbed. 

A locally enhanced line acceleration generates streams of higher density and
lower velocity (compared to the undisturbed mean flow), and CIRs of enhanced
density form where the faster mean flow collides supersonically with the
slow material at the trailing edge of the stream. When the unperturbed wind
collides with the CIR, a non-linear signal is sent back towards the star
which forms a sharp, propagating discontinuity (``Abbott-kink'', see also
\citealt{FeldmeierShlosman00, FeldmeierShlosman02}) in the radial velocity
gradient. The inwards directed propagation velocity (with respect to a
comoving frame) of this kink is just the characteristic speed of the
radiative-acoustic Abbott waves considered in Sect.~\ref{sec:lineinstab}.
In the stellar frame, this feature travels slowly outwards and a velocity
plateau-like structure is formed between the trailing Abbott-kink and the
CIR compression.

By calculating the corresponding line optical depths and time-dependent
(``dynamic'') synthetic profiles, \citet{CO96} showed that these slowly
moving kinks, together with their low velocity gradients ($\tau \propto
(\dvdr)^{-1}$!), are the likely origin of the observed DACs rather than the
CIRs themselves (but see below). A fit to the temporal wavelength drift of
the synthesized DACs resulted in a velocity field parameter $\beta \approx 2
-4$, i.e., in a slow {\it apparent} acceleration, in accordance with
observations. Though the recurrence time scale of the DACs is strictly
correlated with $\vrot$, due to the link of CIRs and the rotating
photosphere, no correlation between acceleration time scale and rotational
speed was found in the synthetic profiles, in contrast to the observational
indications as outlined above.  

\smallskip
This ``missing'' correlation is the central topic of an interesting
kinematical investigation by \citet{Hamann01}, who stressed the fact that
the low drift rate is not a consequence of the CIRs themselves. Features
formed within the CIR would display a drift even faster than features formed
in a non-rotating wind. Instead, the slow drift is a consequence of the
difference between mean flow and the velocity field of the pattern in which
the features form. A pattern with upstream propagation (as the kink in the
model by \citealt{CO96}) inevitably results in a wavelength drift with a
slower apparent acceleration than displayed by features formed in the bulk
or within the CIR itself. Any of these drifts, however, were shown to be
independent on the rotation {\it rate}, leaving the observed correlation 
(if actually present) still unexplained. 

\smallskip An implicit assumption in the work by \citet{Hamann01} is that
the CIRs are induced by disturbances {\it locked} to the stellar surface, in
accordance with the model by \citet{CO96}. In a detailed study,
\citet{LobelBlomme08} relaxed this assumption in order to enable a {\it
quantitative} analysis of the time evolution of DACs for the fast-rotating
B0.5 Ib supergiant HD\,64760, one of the best observed objects (see below).
The authors use their own 3-D radiative transfer code, coupled with 3-D
hydrodynamics restricted to the equatorial plane. A large grid of models and
dynamic spectra for different spot parameters (brightness, opening angle and
velocity) has been computed, and the effects of these parameters on the wind
structure and DAC evolution studied. A best fit could be obtained with a
model with two spots of unequal brightness and size on opposite sides of the
equator, with spot velocities being five times slower than \vrot. All basic
conclusions of \citet{CO96} could be confirmed, particularly the importance
of velocity plateaus in between kinks and CIRs. For their models of spots
which are not locked to the stellar surface, however, the independence of
the DAC acceleration on \vrot\ can no longer be warranted. In this case, the
internal clock is given by the rotational speed of the spots, $v_{\rm sp}$,
whereas a change in the stellar rotational speed modifies the underlying
bulk flow (densities and velocities) via a different centrifugal
acceleration (Eqs.~\ref{eq_mdot_eq}, \ref{eq_vinf_eq}) such that the
time-scales are no longer conserved when \vrot\ is changed. In other
words, two models with $v_{\rm sp} = \vrot$ (locked) and $v_{\rm sp} <
\vrot$ (non-locked) which {\it both} reproduce the DAC recurrence time
behave differently with respect to the acceleration time scale. 

The physical scenario of different spot and stellar rotation velocities
follows a suggestion by \citet{Kaufer06}. These authors detected non radial
pulsations in the photospheric lines of HD\,64760, with closely spaced
periods. In order to explain the observed \Ha\ variability, they invoked the
beat period between two of these periods (lower than the rotational one!)
to be responsible for the CIRs in the wind. Interestingly, however, this
beat period is {\it not} consistent with the spot velocity derived from the
UV by \citet{LobelBlomme08}. For further discussion we refer to
the latter publication.

\smallskip
\noindent
{\it Reality or artefact?~~} As should be clear from our
discussion of the onset of Abbott-waves, the dynamic models discussed above
rely on a Sobolev line-force, because of computational feasibility. Thus,
two questions need to be answered.\\ 
(i) In how far do the kinks still occur
when the Sobolev approximation is relaxed?  (ii) In how far can the
line-driven instability disrupt the CIR structure? A preliminary answer
might be found in Fig.~2c from \citet{Owocki99}, which displays the radial
velocity calculated for a CIR model based on a {\it non-local} line force
calculated from a three-ray treatment (cf. page~\pageref{3rays}). Also here,
somewhat "rounded" kinks followed by a plateau are present! Unfortunately,
however, the line-driven instability effectively destroys all
macro-structure, but this might be related to a still inadequate three-ray
treatment and/or insufficient resolution, leading to too low a lateral
damping. To cite Stan Owocki (2008, priv. comm.), ``there is a strong potential
that such kinks and plateaus are real physical effects in a line-driven
wind, and not just an artefact of the Sobolev approximation.''

Future work using computationally expensive methods to calculate a
consistent more-D line force will hopefully prove this expectation. 

\paragraph{Modulation features.~~}

The observed correlations of the DAC properties with \vsini\ inspired the
{\it ``IUE MEGA Campaign''} \citep{Massaetal95}, during which three
prototypical massive stars ($\zeta$~Pup (O4If), HD~64760 (B0.5Ib) and
HD~50896 (WN5)) were monitored almost continuously over 16 days. Results
of this campaign have been summarized by \citet{Fullerton99}. The most
important finding was the detection of a new type of variability, namely
{\it periodic modulations} in the UV wind lines, derived from the dynamic
spectra of HD\,64760 \citep{Prinjaetal95} which most likely is observed
equator-on. As shown by \citet{Fullerton97},
these modulations result from two quasi-sinusoidal fluctuations with periods
of 1.2 and 2.4 days, which propagate, as a function of phase and starting at
roughly 750~\kms, {\it simultaneously} towards lower and higher line-of-sight 
velocities, until a certain minimum and maximum is reached and the pattern begins to
repeat itself. Because of its peculiar shape, this effect has been called
``phase bowing''.  Similar features (though not as pronounced) have been
detected in only two other stars, $\zeta$~Pup \citep{Massaetal95, Howarth95}
and $\xi$~Per (O7.5 III (n)((f)), \citealt{Kaper99}). Note that these
features occur in parallel with the conventional DACs.

The peculiar phase properties of the modulation features indicate the 
presence of absorbing material with the same phase at two different
projected velocities. By means of a kinematical model and corresponding
synthetic spectra, \citet{OCF95} and \citet{Fullerton97} suggested that
these features are formed in azimuthally extended, co-rotating and
spirally-shaped wind structures linked to a surface density which is
modulated by non-radial pulsations. Due to rotation, a certain spiral exits
from the lines of sight in front of the stellar disk first at an 
intermediate projected velocity (750~\kms in case of HD\,64760). At later
times and because of its curvature, this structure exits from the absorption
region simultaneously at lower and higher velocities. Since the spirals are
assumed to be linked to the surface density modulation, different locations
along the spiral have the same phase, because they arise from the same
``spot'' on the surface. In concert with the kinematics of the spirals, this
explains the modulation features and their phase-bowing, verified in the
aforementioned synthetic spectra. Consistent hydro-simulations of this
scenario are still missing, but these features are a strong hint on the
presence of azimuthally extended macro-structure and observational
``evidence for co-rotating wind streams rooted in surface variations''
\citep{OCF95}.

\subsection{X-ray line emission}
\label{sec:xraylines}

After the advent of the X-ray satellites {\sc xmm-Newton} and {\sc chandra}, 
with their capability to perform high-resolution spectroscopy, detailed
knowledge about the X-ray line emission from hot stars has been accumulated
(for recent reviews, see \citealt{Oskinova08} and \citealt{Cohen08}).
Two ``classes'' of spectra have been found. On the one side, the
prototypical spectrum of $\theta^1$ Ori C, one of few massive stars with
a strong magnetic field (Sect.~\ref{sec:mag}), is rather hard (implying high
plasma temperatures) and shows lines with small widths. On the other side,
the spectrum of the prototypical ``normal'' supergiant $\zeta$~Pup is
softer and displays much broader, asymmetric lines. Moreover, the lines
from hydrogen-like ions are stronger in the former class, and vice versa for
the lines from helium-like ions, which immediately points to differences in
the ionization balance and thus again to different plasma temperatures.
\citet{WojdowskiSchulz05} derive a peak value $T=30 \cdot 10^6$~K for
$\theta^1$ Ori C and 1-2$\cdot 10^6$~K for six other O-stars (incl.
$\zeta$~Pup), where the latter values are of the same order as those
derived from earlier, low resolution data (e.g., by {\sc rosat},
\citealt{Kudritzki96}, see Sect.~\ref{sec:inhomoobs}). 

\subsubsection{Magnetically confined winds} 
\label{sec:xraylinesmagnetic}

By means of a tailored MHD model\footnote{extending the work by
\citet{udDoula02} with respect to the energy equation.} for the wind of
$\theta^1$ Ori C, \citet{Gagne05} showed that the magnetically confined 
wind shock (MCWS) model (discussed in Sect.~\ref{sec:mag}) fits quite well
all the data, predicting the temperature, luminosity, and occultation of the
X-ray emitting plasma with rotation phase. Particularly, the bulk of the
shock-heated plasma should be located in the magnetically confined region
close to $r \approx 2\, \Rstar$, where the speed and line-of-sight velocity of
this material is low, giving rise to rather narrow X-ray emission lines.
Similar arguments might hold for the X-ray spectrum of $\tau$~Sco which
also shows narrow lines, consistent with the finding that also $\tau$~Sco
has a strong magnetic field.

\subsubsection{Non-magnetic stars} 
\label{sec:xraylinesnonmagnetic}

In accordance with previous results derived from lower resolution
observations (Sect.~\ref{sec:inhomoobs}), also the gross properties of
highly resolved X-ray spectra from ``normal'' O-stars may be well explained
by the line-driven instability scenario (either self-excited or triggered,
with dominating clump-clump collisions). This refers particularly to the
softness of the spectra and the large line widths, resulting from the high
velocity of the shock-heated wind. Remember, e.g., that the line-driven
instability should be fully grown above 1.3-1.5 \Rstar and that the
predicted shock temperatures are of the order of ``only'' a few million
Kelvin, due to jump velocities of a few hundred \kms.

\paragraph{fir-diagnostics of Helium-like ions.~~} 

\begin{figure}
\vspace{-1cm}
\begin{center}
   {\includegraphics[width=10cm]{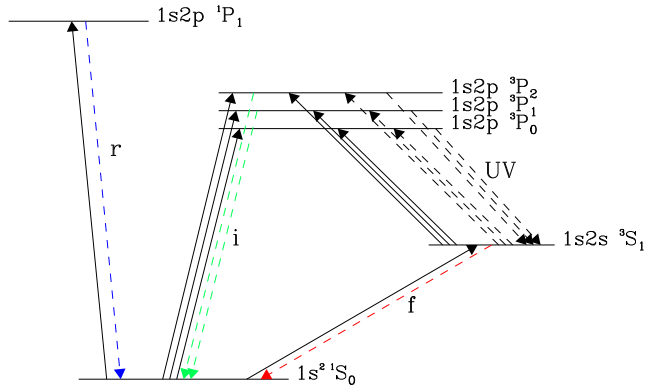}}
\end{center}
\vspace{-1.5cm}
\caption{Schematic Grotrian diagram for Helium-like ions, showing levels
and transitions involved in the formation of the {\it fir-complex} in X-ray
spectra of hot stars. Solid arrows: collisional excitations. Dashed arrows:
radiative transitions. The f,i,r transitions are denoted by corresponding
letters. The transitions between the $^3$S and the $^3$P levels are located
in the (F)UV. Adapted from \citet{GabrielJordan69}.}
\label{fir}
\end{figure}

A key diagnostics to {\it observationally} constrain the onset of the X-ray
emission is provided by the ratio of line fluxes from Helium-like ions such
as O{\sc vii}, Ne{\sc ix}, Mg{\sc xi} and Si{\sc xiii}. These ions show
characteristic {\it fir triplets} \citep{GabrielJordan69} of a forbidden
(f), an intercombination (i) and a resonance (r) component (cf.
Fig.~\ref{fir}).

In hot stars, the ${\cal R} \equiv f/i$ line ratio is sensitive to the local
mean intensity of the UV radiative field, since UV photons can excite the
metastable $^3$S$_1$ level to the $^3$P$_{0,1,2}$ levels. To a good
approximation, this mean intensity is given by $W(r) H_\nu^{\rm phot}$, with
dilution factor $W(r)$ and photospheric flux $H_\nu^{\rm phot}$. If the
latter is known (from observations\footnote{if the transition is located in
the observable part of the UV spectrum, and not contaminated by wind-lines.}
or model atmospheres), the dilution factor $W(r)$ may be inferred from
${\cal R}$, which constrains the formation radius $R_{\rm fir}$ of the X-ray
emitting plasma. Applying this method, \citet{Leutenegger06}\footnote{see
also \citet{Leutenegger07} for the effects of resonance scattering in the
r-component, which give rise to more symmetric fir-complexes.} studied the
X-ray lines of Helium-like ions in four evolved O-stars (incl. $\zeta$~Pup)
and derived minimum radii of X-ray line formation in the range of $1.25
\Rstar < R_{\rm fir} < 1.67 \Rstar$, consistent with the theoretical
expectation. 

On the other hand, \citet{WC07} published rather puzzling results for a
larger sample of OB-stars. Based on similar {\it fir} diagnostics, they
concluded that high-Z ions were predominantly located closer to the
photosphere, a problem which they named the ``near-star high-ion'' problem
(see also \citealt{WC01}). The derived conditions (temperatures
$\sim$1-2$\cdot 10^7$~K close to the star, with $R_{\rm fir}$ as low as 1.1
\Rstar) would be in stark contrast to the standard shocked-wind scenario,
since, e.g., the implied jump velocities would be higher than the velocity
of the underlying wind.  Moreover, and assuming that the average wind
velocity increases with the radius, lines from lower Z ions should be
broader than from ions with higher Z, at least in terms of the standard
model of X-ray emission (see below).\footnote{\citet{Cassinelli08} speculate
that the ``near-star high-ion'' problem might be explained by infalling
clumps and corresponding {\it bow shocks}.} Contradictory to this
expectation, the observed widths of all X-ray lines are similar (e.g.,
\citealt{WC01, Kramer03, WC07}). More investigations are certainly needed to clarify the
situation.

\begin{figure}
\begin{center}
   {\includegraphics[width=6cm,angle=90]{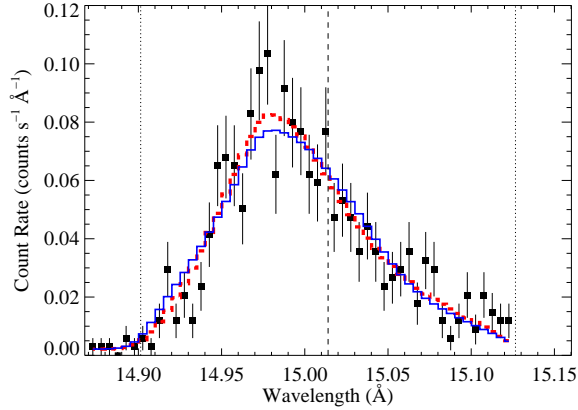}}
\end{center}
\caption {Two different models that fit the Fe{\sc xvii} line in the {\sc
chandra} spectrum of $\zeta$~Pup. Dashed histogram: non-porous model with
$\tau_\ast$ = 2. Solid histogram: porous model with $\tau_{\ast} = 8$. 
$h_{\infty} = 3.3$. The dashed vertical line refers to the transition
frequency at rest and the dotted lines to frequencies corresponding to $\pm
\vinf$. See text. (From D. Cohen, priv. comm.)} \label{zpup_fexvii}
\end{figure}

\paragraph{Analysis of emission line profiles.~~} Further clues on the X-ray
emission from hot stars are provided from the profile {\it shape} of strong,
{\it single} lines (mostly hydrogen-like, e.g., O{\sc viii}, Ne{\sc x},
Mg{\sc xii}, Si{\sc xiv}, but also Fe{\sc xvii}), which is asymmetric and
skewed. Note first that such a line shape is consistent with the ``standard
model'' (e.g., \citealt{Hillier93}) which
assumes the presence of hot, line-emitting material mixed with a ``warm'',
continuum-absorbing mean wind flow.  If the wind is optically thick in the
continuum, the emission from the receding hemisphere will be significantly
attenuated, i.e., a deficit of red-shifted photons will be observed, whereas
from the front, less attenuated side there will be comparatively more
emission, contributing to the blue-shifted part. Taken together, an
asymmetric profile should be created, already predicted by
\citet{Macfarlane91} and indeed observed, as shown in
Fig.~\ref{zpup_fexvii}.

Unfortunately, however, this principal agreement was severely challenged
after {\it quantitative} investigations had been performed. Already from the
first analyses on \citep{OwockiCohen01, Kramer03} it turned out that the red
parts of the profiles were much less attenuated by the underlying wind than
expected. Even worse, the continuum opacity, $\kappa_\nu$, increases with
wavelength, and consequently the lines of lower Z ions are expected to be
more skewed compared to the higher Z ones. Such a dependence, however, has
not been found, and the wavelength shifts of line centroids are similar for
lines of all ions \citep{WC07}.

\paragraph{Lower mass-loss rates or porosity/macro-clumping?~~} Different
hypotheses to explain these problems have been proposed and are summarized
by \citet{Oskinova08}. In the following, we will concentrate on two
promising alternatives that at least allow for reasonable profile fits, but
are still subject to a lively debate (see, e.g., the discussions recorded in
\citealt{Hamann08}). Both approaches rely on the notion that a reduced
continuum opacity can lead to a better agreement between modeled and
observed line shapes. This can be achieved either by reducing the adopted
mass-loss rate or by accounting for porosity effects. The latter have the
additional advantage that the {\it effective} opacity, in the optically thick
limit, becomes frequency-independent, i.e., grey (Eq.~\ref{porositylength}),
thus potentially explaining the similarity\footnote{The issue of the actual degree of
grayness has not been settled though.} of the line-shapes from different
ionization stages.

As a typical example, Fig.~\ref{zpup_fexvii} displays the observed Fe{\sc
xvii} profile from $\zeta$~Pup and two fits. The dashed histogram refers to
the best fit obtained by \citet{Cohen08}, based on the empirical model by
\citet{OwockiCohen01}, which optimizes three parameters: the
normalization, the inner radius below which there shall be no emission
(R$_{\rm o}$), and the continuum optical depth of the ``warm'' wind,
parameterized by the quantity $\tau_\ast=\kappa_\nu \mdot/(4 \pi \Rstar
\vinf)$. The fit is formally good, and the best fit values (within 68\%
confidence limits) are R$_{\rm o}$ = 1.53$^{+0.12}_{-0.15}$~\Rstar\ and
$\tau_\ast = 2.0 \pm 0.4$.

Thus, the R$_{\rm o}$ derived from the profile is consistent with the
expected onset of shock formation and the optical depth is quite small
indeed. Translated to a mass-loss rate, this would imply \mdot\ = 1.5 \Mdu,
which is (much) lower than recent estimates from NLTE analyses based on {\it
unclumped} models (\mdot $\approx$~8.8~\Mdu, \citealt{repolust04}), and a
factor of 2.8 lower than the {\it maximum} mass-loss rate (\mdot\ = 4.2 \Mdu)
constrained by independent $\rho^2$-diagnostics when the {\it outer} wind is
assumed to be homogeneous (\citealt{Puls06}, see Sect.~\ref{sec:clumping}).

Alternative models were simultaneously presented by \citet{Oskinova06} and
\citet{OwockiCohen06}. Both publications (which differ with respect to
important details) suggested that also porosity associated with
macro-clumping (Sect.~\ref{sec:clumptreat}) can account for the low degree
of asymmetry, without any need to worry about reduced mass-loss rates. The
solid histogram in Fig.~\ref{zpup_fexvii} bases on the approach by
\citet{OwockiCohen06} and displays the best fit for such a porous model,
assuming {\it spherical} clumps. By {\it fixing} the mass-loss rate at \mdot
= 6~\Mdu (corresponding to $\tau_\ast = 8$), the porosity length, $h$
(Eq.~\ref{porositylength}) is a free fit-parameter in this model. It is
evident that the best fitting porous model with the ``high'' mass-loss rate
is almost indistinguishable from the best fitting non-porous model with the
low \mdot. The optimum terminal porosity length (the value of $h$ in the
outer wind region) is rather large, $2.5 \Rstar < h_\infty < 4 \Rstar$ at a
68\% confidence level. \citet{Cohen08} argued that this is much larger than
any porosity length seen in state-of-the-art hydrodynamical simulations, and
concluded that ``there is no compelling evidence'' for invoking porosity
rather than a reduced mass-loss rate. 

\citet{Oskinova06} pointed out that by combining the fits for $\tau_\ast$
from $\zeta$~Pup, based on non-porous models (from \citealt{Kramer03}) with
appropriate absorption coefficients, different lines would imply
significantly different mass-loss rates. Based on previous work by
\citet{Feldmeier03} and \citet{Oskinova04}, they describe the wind by a flow
of clumps obeying the equation of continuity.  As in the porosity length
formalism, the effective opacity of the clumps becomes grey when the clumps
become optically thick. They assume that the emission originates between two
radii $r_1$ and $r_2$ (derived from {\it fir}-lines), and investigate the
influence of the {\it clump geometry}. Comparing spherical clumps with
(infinitesimally) thin shell fragments (pancakes) oriented perpendicular to
the radial direction, they argue that the latter geometry leads to more
symmetric profiles than the former one, as long as the clumps subtend a
constant solid angle as they propagate outwards. In their subsequent
simulations of synthetic X-ray lines, the shape of the clumps is simulated
by an anisotropic opacity (per default assumed to be pancake-like), and the
only parameter to be specified (in addition to the wind parameters, \mdot,
\vinf, and $\beta$, adopted from the literature) is the quantity $n_0 =
n(r)v(r)$. This quantity defines the number of fragments passing through
some reference radius per unit time (constant due to mass conservation), and
was set to $n_0 \equiv \vinf/\Rstar$, in order to obtain average clump
separations $L(r)$ which are compatible with predictions from hydrodynamic
simulations, $L(r)= \Rstar v(r)/\vinf \le \Rstar$. This model has no free
parameter, but nevertheless provides a satisfactory agreement between a
variety of synthetic and observed lines from different ions (thus
supporting the grey character of the opacity). \citet{Oskinova08} conclude
that these results provide evidence that the wind clumps are not optically
thin, that they are compressed in the radial direction, and that they are
separated by a few tenths of the stellar radius in the wind acceleration
zone. 

Let us finish this section with {\it our} conclusions. Whilst the X-ray
spectra from magnetic winds seem to confirm the magnetically channeled wind
shock scenario, the situation for non-magnetic winds remains unclear. Both
non-porous models with reduced mass-loss rates and porous winds with
``conventional'' \mdot\ are able to explain the observed spectra, where the
role of non-isotropic opacity needs further confirmation by independent
diagnostics. Future modeling with a simultaneous analysis of processes from
the X-ray to the radio domain will certainly uncover the most appropriate
description and the ``real'' mass-loss rates and clarify the ``near-star
high-ion'' problem. Such modeling should include not only the X-ray
line-features, but also the SED in the complete X-ray band including the
wealth of the {\sc rosat} data. 

\section{Summary and further implications}
\label{sec:summary}

We reviewed the current status of both theoretical and observational aspects
of mass loss from hot massive stars, starting with the standard model and
the wind-momentum luminosity relation (WLR). Mass-loss predictions from line
statistics were discussed with particular emphasis on the temperature and
metallicity dependence of the wind momenta, emphasizing that the
spectral-type dependence of the WLR, especially with respect to the
bi-stability jump, is still a most active aspect of present-day hot-star wind
research.

We continued with the impact of rotation. Wind compression and the
predictions of predominantly prolate winds due to gravity darkening were
reviewed, whilst we concluded that the rotationally-induced bi-stability
concept for explaining the equatorially outflowing winds from B[e]
supergiants requires future 2D modeling. With respect to magnetic fields, we
made a clear distinction between internal and surface magnetic fields and
note that firm observational constraints on the strength and geometry of the
fields have become within reach of current instrumentation (e.g., using {\sc
espadons} or {\sc narval}). We reviewed the wind structure as a function of
the magnetic confinement parameter, and pointed out that weak magnetic
fields below present detection limits could have a large impact on weak winds.
It remains a possibility that the co-rotating interaction regions that are
the likely root of ultraviolet line profile variability of O-stars might
be related to an anchored field on the surface. 

A key topic concerned stationary models of radiation-driven winds. We 
described two basic methods in use to predicting the mass-loss rates of
massive stars and their pros and cons: hydrodynamical methods based on the
improved CAK approach and those based on Monte Carlo radiative transfer
methods. Predictions were provided for O and B supergiants, followed by
those for stars in closer proximity to the Eddington limit, including
Luminous Blue Variables (LBVs) and Wolf-Rayet (WR) stars. Furthermore, we
provided some basic insights into the possible existence of stellar winds at
very low and zero metallicity content (Population III) involving issues such
as the potential self-enrichment of metals through the evolution of the
first generations of massive stars themselves, and the potential importance
of continuum driven winds from super-Eddington stars, which do not require
the presence of metals in their atmospheres. Particularly, we considered the
predicted metallicity dependence of WR winds since this has relevant
implications for the angular momentum evolution of massive stars at low
metal content and the progenitor stars of long GRBs. 

As far as observational wind diagnostics are concerned, we discussed the
traditional UV, \Ha, and continuum bound-free/free-free emission, in
addition to modern non-LTE atmosphere analyses and emphasized the future
potential for NIR spectroscopy. One of the major steps in the last decade
has been the inclusion of metal line-blanketing, which has led to a new
temperature scale of OB-stars, with generally lower temperatures and
luminosities. Due to these lower luminosities the O-star mass-loss
predictions {\it per object} have been down-revised as a consequence of 
the changes in underlying stellar parameters -- without actually changing the predictions
themselves. Empirical mass-loss rates have been determined in parallel with
stellar parameters, by profile fitting methods (using the conventional ``by
eye'' method or optimization via genetic algorithms), for a variety of
spectral types, luminosity classes and environments, particularly within the
VLT-{\sc flames} survey of massive stars. In many cases, the derived mass-loss
rates of O-supergiants were a factor of $\sim$2 above predicted values. This
could either mean that the theoretical values are not large enough, or that
the empirical values are too high (or that both need revision). For mid and
late type B-supergiants, on the other hand, empirical \mdot-values were
found to be lower than the theoretical ones. Comparing the mass-loss rates
for the Galaxy, the LMC and the SMC, and accounting for an average clumping
correction, the {\it empirical} dependence of O-star mass-loss rates on
metallicity has been derived, which agrees quite well with the theoretical
predictions.

In addition to the minor problems outlined in the last paragraph, our
general understanding of radiation driven winds is challenged by two urgent
problems, the weak wind and the clumping problem. Regarding the former, we
have discussed that the presence of X-rays can lead to a degeneracy
of mass-loss rates derived from the UV, and suggested the NIR \Bra-line as a
promising tool for further progress on quantifying mass-loss rates for thin
winds. We have also outlined that just these X-rays might be the origin of
the weak wind problem, by modifying the ionization equilibrium and thus the
efficiency of radiative driving in the lower wind. Future work has to
provide tight constraints on the domain of occurrence of weak winds, and
particularly on the question which parameters discriminate objects with
``normal'' from those with weak winds. To be even more provocative: Can we
be sure that there are any late O-dwarfs with ``normal'' winds at all, or did we
over-interpret present mass-loss diagnostics, and the weak wind case is the
normal one?

In the last part of this review we discussed important issues related to
time-dependence and structure, featuring the clumping problem. We reported
that different diagnostics seem to suggest also different clumping
properties (with clumping factors differing by factors up to 20-100), and
summarized recent attempts to unify these results by invoking optical depth
effects (micro- vs. macro-clumping) and clumping in velocity space. One
particular aspect concerns the radial stratification of the clumping. Most 
analyses seem to have at least one result in common, namely the
indication of considerable clumping close to the wind-base. This is
in strong contrast to the conventional interpretation of wind-clumping as
caused by the line-driven instability (either self-excited or triggered),
since all corresponding simulations predict the onset of wind-clumping only
at rather large velocities (some hundreds of \kms). Thus, future effort
should be dedicated to improve our understanding of the physical cause of
the clumping, but also to continue with empirical studies, by combining
evidences from different diagnostics that are formed in different positions
in the wind. We have provided a promising ``toolkit'' for this objective. Of
course, the final goal is not only to ``measure'' clumping/volume filling
factors and porosity lengths, but also to accumulate information about the
shape of the clumps. Most spectral diagnostics (UV, \Ha, \Bra, IR-/radio
continuum) do not provide such geometric constraints, but we have pointed
out that linear polarimetry could help, as well as the diagnostics of X-ray
line emission.

Given the overwhelming observational evidence for wind clumping, it is
evident that those empirical rates which are based on $\rho^2$ diagnostics
and homogeneous wind models yield maximum values and that the {\it real}
mass-loss rates must be lower. Still, the most relevant question is {\it by
how much} these empirical rates must be down-revised. {\it If} the clumping
factor is only a factor of $\sim$4, mass-loss rates would need to be reduced
by a corresponding factor of $\sim$2. Then, at least for O- and early
B-stars, empirical rates would be in full agreement with {\it present}
theoretical rates (still assumed to be uncontaminated by clumping effects),
which are used in current models of massive star evolution. 

{\it If}, on the other hand, the empirical rates needed to be down-revised by
very large factors, of 10 or more, they would fall significantly below
theoretical rates, with some devastating effects on current massive star
evolution models. This would tie in with the downwards revision of WR
mass-loss rates about a decade ago, with implications involving that black
holes have larger masses than previously appreciated, and the possibility
that a significant fraction of the total mass loss would need to be lost in
super outbursts of LBVs. We note that for a variety of reasons including
stellar rotation rates and the origin of WR stars, present day stellar
evolution models could not survive with theoretical rates down-revised by
more than a factor of two, or, equivalently, empirical rates reduced by more
than a factor of four. 

\smallskip
\noindent
Only the future will tell!

\begin{acknowledgements}
We are indebted to Jean Surdej, Stan Owocki, David Cohen and Ken Gayley for
their insightful comments and suggestions. Many thanks to Jon Sundqvist for
carefully reading the manuscript and additional suggestions for
improvements. F.N. acknowledges a grant by the Spanish MEC through project
AYA2007-67546.
\end{acknowledgements}

\bibliographystyle{spbasic}
\bibliography{reviewaa_mdot_puls_vink_najarro}   

\end{document}